# Systematic transcriptome wide analysis of lncRNA-miRNA interactions.

Saakshi Jalali[1#], Deeksha Bhartiya[1#], Vinod Scaria[1]*

[1]GN Ramachandran Knowledge Center for Genome Informatics, CSIR Institute of Genomics and Integrative Biology (CSIR-IGIB), Mall Road, Delhi 110007, India

[#]These authors contributed equally to this work

*Corresponding author

Email addresses:

SJ: saakshi.jalali@igib.in

DB: deeksha.bhartiya@igib.in

VS: vinods@igib.in, vinods@igib.res.in

## Abstract

### Background
Long noncoding RNAs (lncRNAs) are a recently discovered class of non-protein coding RNAs which have now increasingly been shown to be involved in a wide variety of biological processes as regulatory molecules. The functional role of many of the members of this class has been an enigma, except a few of them like Malat and HOTAIR. Little is known regarding the regulatory interactions between noncoding RNA classes. Recent reports have suggested that lncRNAs could potentially interact with other noncoding RNAs including miroRNAs (miRNAs) and modulate their regulatory role through interactions. We hypothesized that long noncoding RNAs could participate as a layer of regulatory interactions with miRNAs. The availability of genome-scale datasets for argonaute targets across human transcriptome has prompted us to reconstruct a genome-scale network of interactions between miRNAs and lncRNAs.

### Results
We used well characterized experimental Photoactivatable-Ribonucleoside-Enhanced Crosslinking and Immunoprecipitation (PAR-CLIP) datasets and the recent genome-wide annotations for lncRNAs in public domain to construct a comprehensive transcriptome-wide map of miRNA regulatory elements. Comparative analysis revealed many of the miRNAs could target long noncoding RNAs, apart from the coding transcripts thus participating in a novel layer of regulatory interactions between noncoding RNA classes. We also find the miRNA regulatory elements have a positional preference, clustering towards the 3' and 5' ends of the long noncoding transcripts. We also further reconstruct a genome-wide map of miRNA interactions with lncRNAs as well as messenger RNAs.

### Conclusions
This analysis suggests widespread regulatory interactions between noncoding RNAs classes and suggests a novel functional role for lncRNAs. We also present the first transcriptome scale study on lncRNA-miRNA interactions and the first report of a genome-scale reconstruction of a noncoding RNA regulatory interactome involving lncRNAs.

## Introduction

Recent advances in sequencing technologies have made genome-scale mapping of transcriptional potential possible at single-nucleotide resolutions [1,2]. A number of reports in the immediate past have suggested the pervasive transcriptional potential of eukaryotic genomes [3], also suggesting a large number of novel genomic loci which were previously not known to encode any functional transcripts [4]. Though consistent evidence now suggests the existence of a large number of previously uncharacterized transcripts, a vast majority of the transcripts surprisingly do not code for a functional protein, and are grouped into a general class of noncoding RNAs (ncRNAs) [5-7]. The amenability of transcriptome mapping on a genome scale has enormously added to the number of ncRNAs, with members in the class presently outnumbering protein coding genes by a few folds. Recent efforts globally, including collaborative efforts like the ENCODE [8] have been successful in characterizing the transcript structures and their expression profiles in a wide number of cell types. This has also been supplemented by the availability of computational tools to map and identify transcript isoforms and quantify expression profiles in different cell types and conditions [9].

The ncRNA class now encompasses a wide diversity of subclasses organized based on their size, structure, function and conservation. miRNAs have been one of the recently discovered and well characterized classes of ncRNAs and are small regulatory RNA molecules processed from larger precursors through a highly co-ordinated pathway [10]. Presently they are known to mediate post-transcriptional control of gene expression by binding to the 3' untranslated regions of coding genes [11-13]. This is modulated through a ribonucleoprotein complex called the RNA induced silencing complex (RISC) [14]. Argonaute (Ago) proteins are the catalytic component of RISC complex and the small ncRNA bound to Ago guides it to sequence complementary target [15]. The regulatory role of miRNAs encompass a wide variety of biological processes including development, organogenesis [16,17] and pathophysiology of a number of diseases including of the neuronal [18], cardiovascular [19] and infectious diseases [20-22], just to name a few. Another major class of recently discovered ncRNAs is the long noncoding RNAs (lncRNAs) [23,24]. Unlike the miRNAs, lncRNAs are longer, and by definition > 200 nucleotides in length and usually show less sequence conservation. lncRNAs presently also encompass the previously known classes of processed pseudogenes, antisense transcripts and the recently discovered regulatory large intergenic noncoding RNAs (lincRNAs) [25]. Though a small number of lncRNAs have been functionally characterized like HOTAIR [26,27], Xist [28] and Malat [29], a large number of members in the class remain functionally uncharacterized. Presently well characterized members of the class have been shown to be involved in regulatory roles in diverse processes like imprinting, X-inactivation and development.

Though much of the focus in ncRNA research is directed towards understanding the regulation of protein-coding genes mediated by them, it has been suggested previously that ncRNAs could form a well orchestrated regulatory interaction network [30]. There have been reports previously which have suggested examples of such regulation, for example miRNA-miRNA interaction [31]. Recently there have also been reports suggesting the possibility of a widespread interaction network involving competitive endogenous RNAs (ceRNAs) where ncRNAs could modulate regulatory RNA by binding and titrating them off their binding sites on protein coding messengers [32]. An example for this type of regulation is exemplified by HULC, an lncRNA expressed in hepatocellular carcinoma which binds miR-372 and forms a regulatory interaction [33]. Similarly reports suggest linc-MD1 could interact with miR-133 and miR-135 and promote muscle differentiation [34]. Similar examples have been described in plant systems and encompass target mimicry [35,36]. The entire spectrum of ncRNA regulatory layer, especially the possibility of regulatory miRNAs-lncRNAs interactions remains unexplored.

We employed datasets in public domain for lncRNA annotations recently made available from the ENCODE project along with genome-wide cross linking immunoprecipitation (CLIP-Seq) datasets for four argonaute proteins [37,38]. We hypothesized that miRNAs and lncRNAs could form an interacting regulatory layer exemplifying the ceRNA hypothesis where multiple targets with multiple binding affinities compete for the regulatory RNAs and the members could modulate the regulation of each other by modulation of their relative concentrations (Supplementary File 1). We show that a subset of the annotated lncRNAs could indeed harbor miRNA recognition elements and be a part of the miRNA interaction network, thus playing a critical regulatory layer of interaction. We also integrate miRNA-target information to suggest a network of interactions between lncRNAs, miRNAs and protein-coding transcripts. To our knowledge this is the first transcriptome scale study on lncRNA-miRNA interactions and the first report of a genome-scale reconstruction of a noncoding RNA regulatory interactome involving lncRNAs.

## Results and Discussion

### Identification of high-confidence microRNA recognition elements mapping to long noncoding RNAs

The PAR-CLIP reads for HEK293 cells were downloaded from the NCBI Short Read Archive (ID SRA IDs: SRX020783, SRX020784, SRX020785, SRX020786) which amounted to a total of 1,93,01,715 CLIP-seq reads for Ago 1-4. The reads were further mapped to hg19 version of the human genome using a quality-aware reference mapping algorithm: Mapping and Assembly with qualities (MAQ). Mapping was performed with stringent criteria and allowed no mismatches. The Argonautes significantly overlap in their biological function. Ago1 has been shown to be involved in RNA-mediated post-transcriptional gene silencing, Ago2 shows slicer activity, Ago3 and Ago4 are required for RNA-mediated gene silencing (RNAi). Since there have been no significant differences reported between the mappings of reads for each of the Ago datasets, and since biologically their functions overlap, we merged the datasets as previously used for the analysis [39]. A total of 619208 hits were obtained for the Ago datasets. The number of reads mapping in each of the individual datasets is available as Table 2. To weed out potential noisy loci, we applied stringent criteria that all loci should be supported by at least 5 overlapping reads. The mappings were parsed to identify high-quality miRNA recognition elements using custom scripts. We defined the recognition element as a stretch of contiguous positions in the genome which matched the 5x read mapping criteria. A total of 15983 such consensus high confidence miRNA recognition element clusters were identified. These mapped to 113 lncRNA exons and 4480 mRNA exons from the GenCode annotation datasets [40]. We further also analyzed the relative contributions of each of the Ago datasets for each of the high-confidence miRNA recognition elements.

### Identification of the cognate microRNAs for the microRNA recognition elements

Argonaute proteins are a component of the RNA induced silencing complex and are intricate components modulating the function of the complex. CLIP-seq allows one to identify the miRNA recognition elements in the transcriptome. To understand the biological interactions between miRNAs and their cognate target transcripts, it is imperative to identify the miRNAs corresponding to each of the miRNA recognition elements. We used miRanda [41] a popular computational algorithm to identify miRNAs binding to cognate miRNA recognition elements. Analysis revealed 51 miRNA-lncRNA interactions with 29 miRNAs targeting 25 lncRNA transcripts (Figure 2) (Table 1). We also predicted miRNA targets for all gencode annotated transcripts predicted to be bound with Agos. Out of the total 4480 lncRNA clusters, 416 protein coding transcripts were found to be targeted by 1253 miRNA (Supplementary File 2). The paucity in the number of known human miRNAs from the present analysis could be potentially be an effect of the paucity in the biological understanding of the repertoire of smallRNA species in humans and their biological mechanisms. Apart from miRNAs, other smallRNA species also has potential to be associated with Argonaute proteins [42]. We hope future in-depth understanding of smallRNA species, their biogenesis and mechanism of action could potentially throw light into the biological relevance of the remaining regulatory motifs.

**Enrichment of microRNA regulatory elements in long noncoding RNA ends**

Analysis of the positional preference of miRNA regulatory elements in lncRNAs revealed a positional preference of the regulatory element to occur in both the 5' and 3' ends of the long noncoding RNAs (Figure 3). Such preferential positioning of the miRNA regulatory elements have been described previously for protein-coding transcripts [43], with clustering in the 3' un-translated regions of protein-coding genes. The biological significance of positional clustering in the ends of lncRNAs is uncertain at present though suggesting a possible pattern in organization of regulatory elements across transcripts.

**Long noncoding RNAs as a part of the microRNA interaction network**

The present analysis suggests lncRNAs also harbor potential miRNA regulatory elements and could participate in the miRNA regulatory network. We reconstituted a comprehensive genome-wide network of RNA mediated interactions putting together the interactions from genome-wide analyses and validated miRNA-mRNA interactions. The network brings to light hitherto unknown complexity of RNA regulatory interactions and how long noncoding RNAs could play regulatory roles in the miRNA mediated interactions with mRNAs (Figure 4). The reconstituted interaction network has 314 miRNAs targeting 35 lncRNAs and 946 mRNAs. The interaction network consists of several small and large clusters. To exemplify how lncRNAs could participate in the miRNA-mRNA interaction network, (Figure 4B2) shows, hsa-mir-196a, which is experimentally known to target ENST0000040584, ENST00000242159, ENST00000313173 transcripts, encoded by the HOX cluster genes HOXC8, HOXA7 and HOXD8 [44], was also found to target lncRNAs ENST00000519935, ENST00000523790, and ENST00000489695. The mir-196 miRNAs are known to directly cleave HOX mRNAs and modulates development of axial patterning [44,45]. The human miRNA hsa-mir-196a has been previously shown to be associated with the pathogenesis of cancers including colorectal cancer cells and has been shown to induce a pro-oncogenic behavior in human cancer cells [46]. Thus lncRNAs could potentially modulate the pathogenesis of the disease by modulating the key partner, mir-196a. We hope with the increasing understanding of the key roles in different diseases processes, a network approach would allow to prioritize the functional studies of lncRNAs in disease processes.

## Materials and Methods

**PAR-CLIP Datasets:** We used transcriptome wide interaction data for miRNA-AGO available from PAR-CLIP studies [47]. The dataset provides the RNA-binding proteins (miRNA-AGO) target sites identified from the human embryonic kidney (HEK) 293 cell lines stable expressing Ago1-4. The read sequences (SRA IDs: SRX020783, SRX020784, SRX020785, SRX020786) of this study were fetched from Sequence Read Archive (SRA) from NCBI and were assembled using MAQ [48] reference assembly software to the hg19 version of the human genome from UCSC Genome Browser [49]. To identify only high-confidence miRNA recognition elements, we used an arbitrary cutoff of 5x and all loci with equal to or more than 5X coverage were filtered for the further analysis.

**Long noncoding RNA Datasets:** The manually annotated lncRNAs from Gencode Version 9 (May 2011 freeze, GRCh37) (http://www.gencodegenes.org/) [40] were used for the analysis. The dataset contains both Ensembl and Havanna annotations and the transcripts fall into six biotypes, viz, processed transcripts, antisense, lincRNA, ncRNA_host, non_coding and retained_intron. The dataset had a total of 18878 long non-coding transcripts encompassing by 11004 genes. The former dataset of high-confidence miRNA recognition elements were mapped on the lncRNA exon positions using custom scripts.

**microRNA sequence datasets:** The dataset of mature miRNA sequences was derived from miRBase17 [50,51] for all annotated Human miRNAs. This dataset comprised of 1731 sequences. The dataset of validated miRNA targets were derived from miRecords [52].

**microRNA target predictions:** To identify the subset of miRNAs for the corresponding high-confidence AGO sites, we used a computational approach using a popular miRNA target prediction algorithm – miRanda [41]. miRanda uses dynamic programming and incorporates the free energy of the RNA duplex and miRNA binding rules to predict potential miRNA target sites. We used an empirical alignment score of 160 and minimum free energy of -20 kcal/mol. Similarly miRNA targets were predicted for all those messenger RNAs which harbored high confidence miRNA recognition elements.

**Positional preference of microRNA regulatory elements in long noncoding RNAs:** The positional preference of miRNAs across the lncRNA length was computed using in house scripts. We divided the entire length of the lncRNA into non-overlapping bins of 5 percent each. This was done to accommodate the diversity in lengths of the lncRNA. The frequencies of mapping miRNA regulatory elements were plotted for each of the bins

**microRNA-interaction network:** Both the interactions, viz, miRNA-lncRNA and miRNA-mRNA were compared and an unwieghted network displaying lncRNA-miRNA and mRNA-miRNA interactions were created using Cytoscape 2.8.2 [53]. The targets were then searched on miRecords to check whether any of the high-confidence miRNA recognition elements overlap with validated target sites. The miRNA-mRNA interactions were also taken from literature. The entire workflow of the computational analysis is summarized in Figure 1.

## Conclusion

The present analysis illustrates the potential of integration of genome-wide experimental datasets to provide insights into novel biological regulatory interactions in the cell. Interactions for miRNAs with lncRNAs have been suggested to play an important role in modulating regulation mediated through miRNAs. This analysis also provides a genome-scale report and reconstruction of miRNA interaction with mRNAs as well as lncRNAs. We also show how such a network approach could potentially suggest the lncRNAs for functional validation.

The study is not without caveats. The first being limitation in the diversity of CLIP-seq datasets available in public domain. The datasets used in the analysis are limited to HEK293 cells. The second major caveat is that majority of miRNA recognition elements could not be tracked back to their cognate miRNAs.

With the advent of faster and cheaper sequencing technologies [54,55] we hope newer genome-wide CLIP-seq datasets would emerge. This would potentially provide a better picture of the depth of interactions and a perspective on the spatio-temporal organization of these networks. We hope the advancements in technology and insights into the biological function and regulatory mechanism of small RNA classes and discovery of novel members of this class including miRNAs would provide a better insight into the regulatory ramifications and suggest the cognate RNAs for the regulatory elements discovered through genome-wide screens. We also hope the availability of technology to identify targets of RNA binding proteins would open up newer avenues of understanding the functions and targets of many of the RNA binding proteins. This would also provide for the much needed base to understand biological processes and regulatory interactions of ncRNAs with coding transcripts as well as between them and how these proteins modulate these through binding to their respective partners.

## Acknowledgments

Authors thank Dr. Sridhar Sivasubbu and Dr. Chetana Sachidanandan for scientific inputs, which helped in formulating the study and preparation of the manuscript. Authors acknowledge scientific discussions and inputs from Mr. Gopal Gunanathan Jayaraj which helped in formulating the study. DB acknowledges a Senior Research Fellowship from CSIR, India. The study was funded by CSIR, India through Grant "Comparative genomics and biology of noncoding RNA (NWP0036)".

## Figure Legends

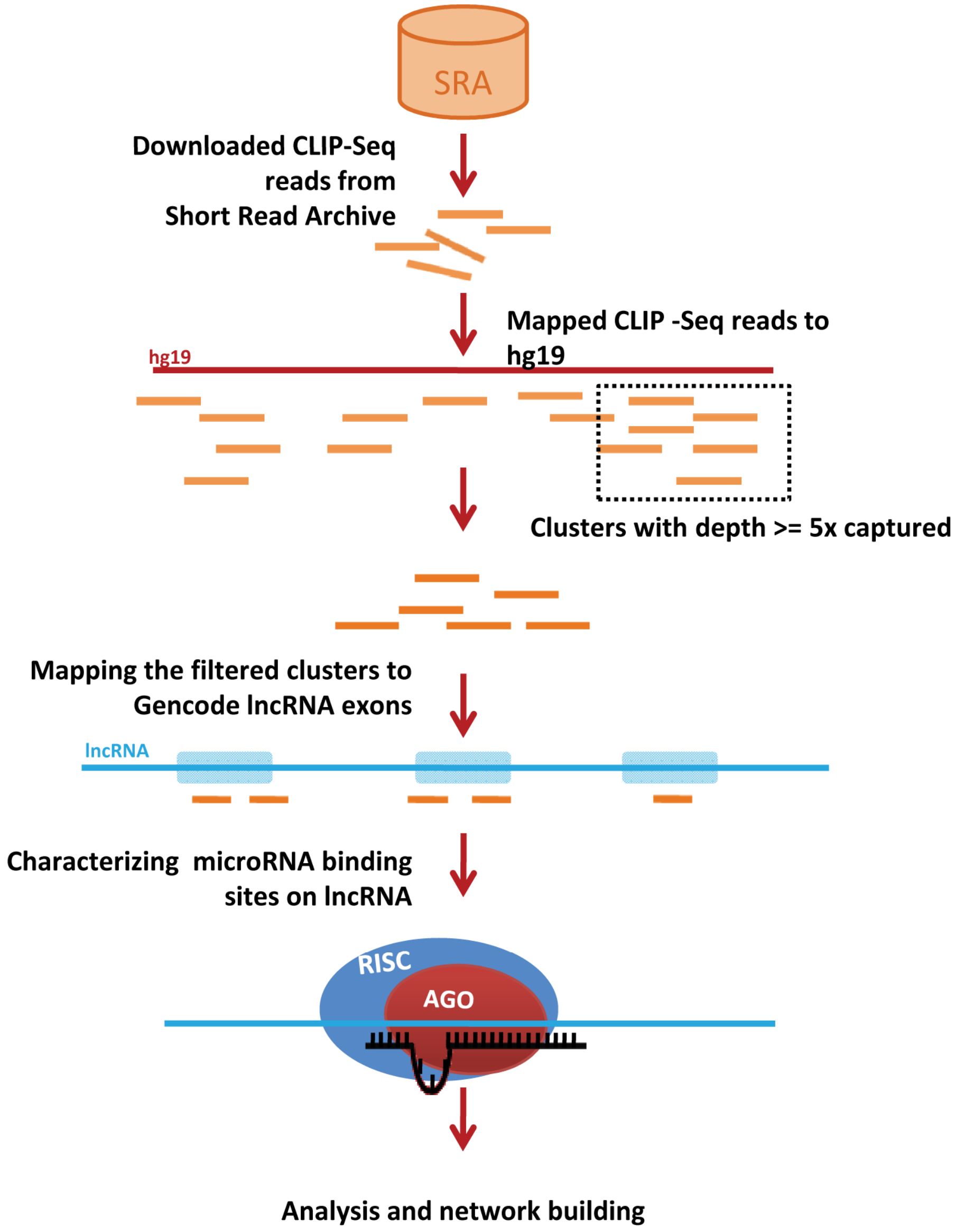

**Figure1:** Workflow for the Analysis of potential microRNA binding sites on lncRNAs and re-construction of the regulatory interactome.

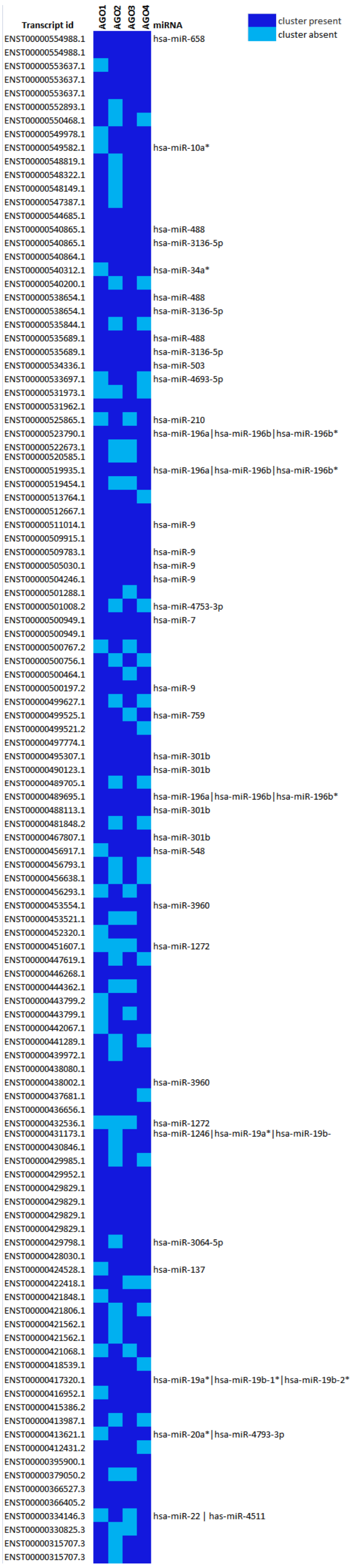

**Figure2:** Heatmap depicting the lncRNA transcripts targeted by miRNAs and the distribution of Ago types in the targeting complex. The presence and absence of specific Ago type in the RISC complex is represented by the difference in the colors in the heatmap. Also the targeting miRNA has also been mentioned.

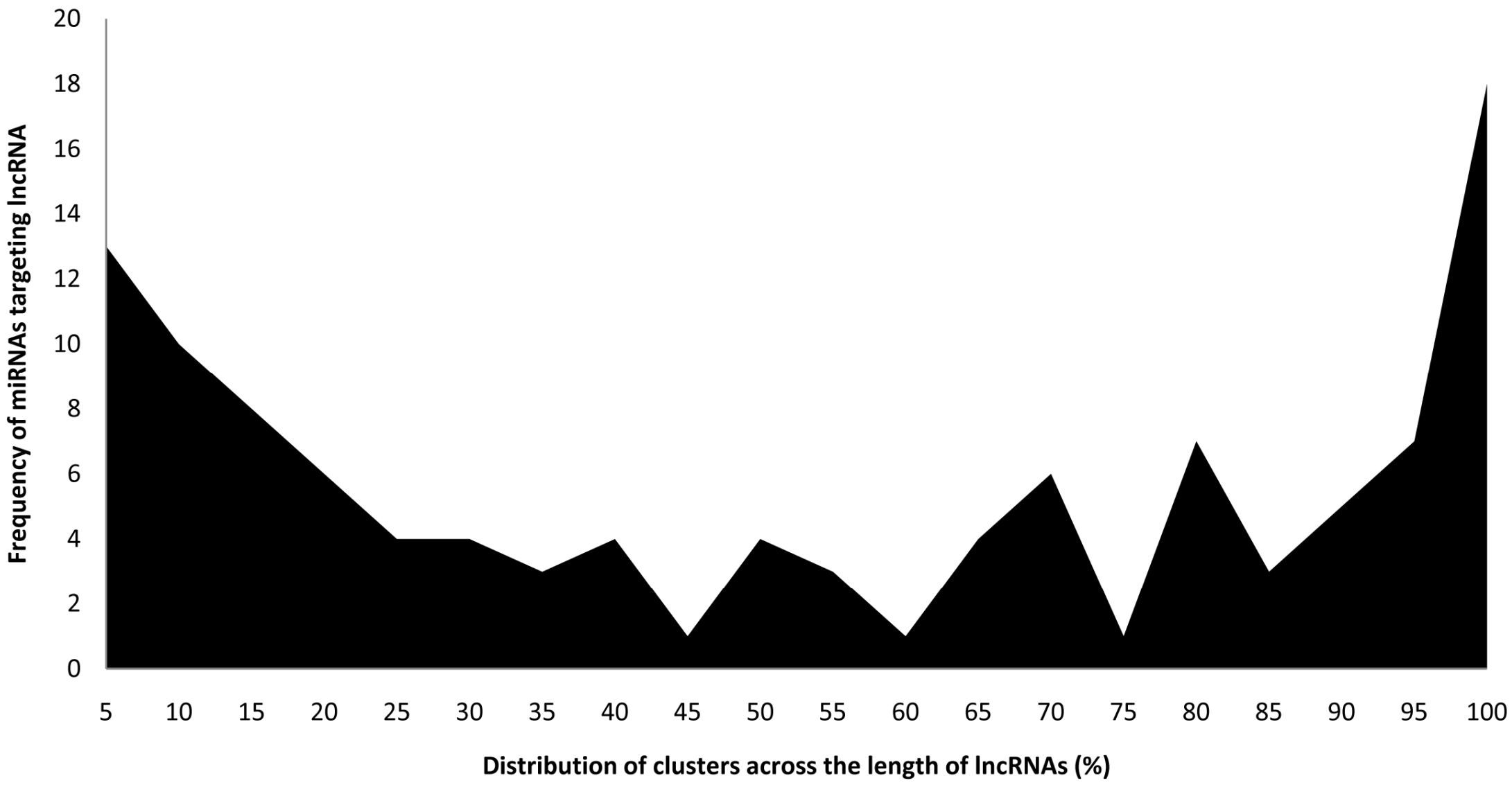

**Figure3:**The entire length of lncRNA was divided into frames of 5 percent each and was plotted against the mapping frequencies of miRNA showing the positional preference of microRNA regulatory elements in long noncoding RNAs.

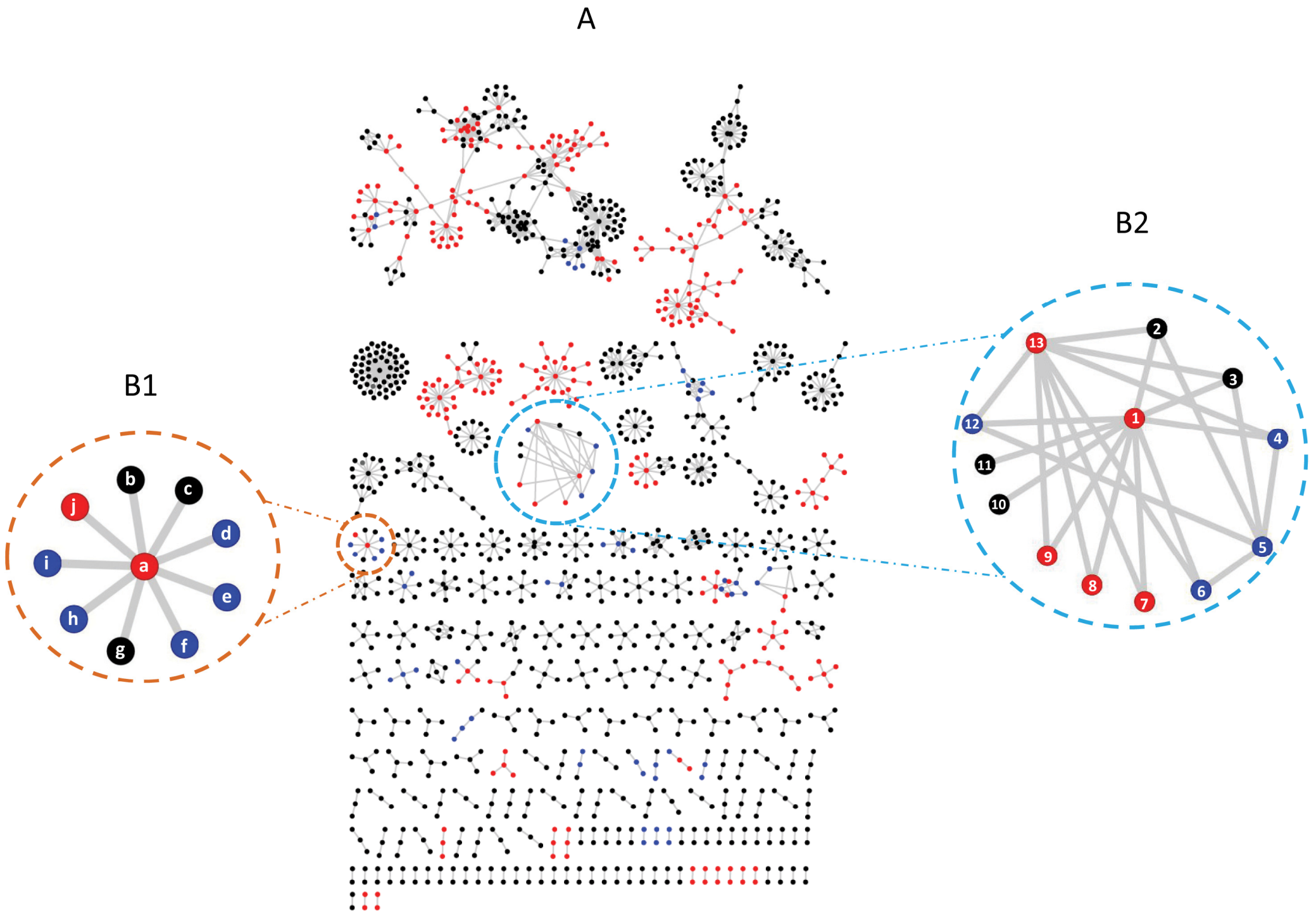

**Figure4: A:** Complete interaction network of lncRNA, miRNA and mRNA wherein the experimental miRNA-mRNA interactions are represented in red nodes, the predicted miRNA-lncRNA interactions are represented in blue nodes and total mRNA-miRNA interactions represented as black nodes.

**B1:** An interesting example from the network highlighted in orange showing interactions between An tion of network highlighting **miRNA:** (a) hsa-mir-9; **lncRNA:** (d) ENST00000500197.2, (e) ENST00000509783.1, (f) ENST00000511014.1, (h) ENST00000505030.1, (i) ENST00000504246.1; **mRNA:** (b) ENST00000384838.1, (c) ENST00000262095.2, (g) ENST00000491143.1, (j) ENST00000226574

**B2:** Another interesting example from the network highlighted in blue showing interactions between **miRNA**: (1) hsa-miR-196a, (5) hsa-miR-196b*, (13)hsa-miR-196b; **lncRNA**: (4) ENST00000523790.1; (6) ENST00000489695.1, (12) ENST00000519935.1; **mRNA**: (2) ENST00000354032.4, (3) ENST00000384852.1, (7) ENST00000313173, (8) ENST0000024215, (9) ENST00000040584, (10) ENST00000304786.7, (11) ENST00000366839.4

## Supplementary Files Legends

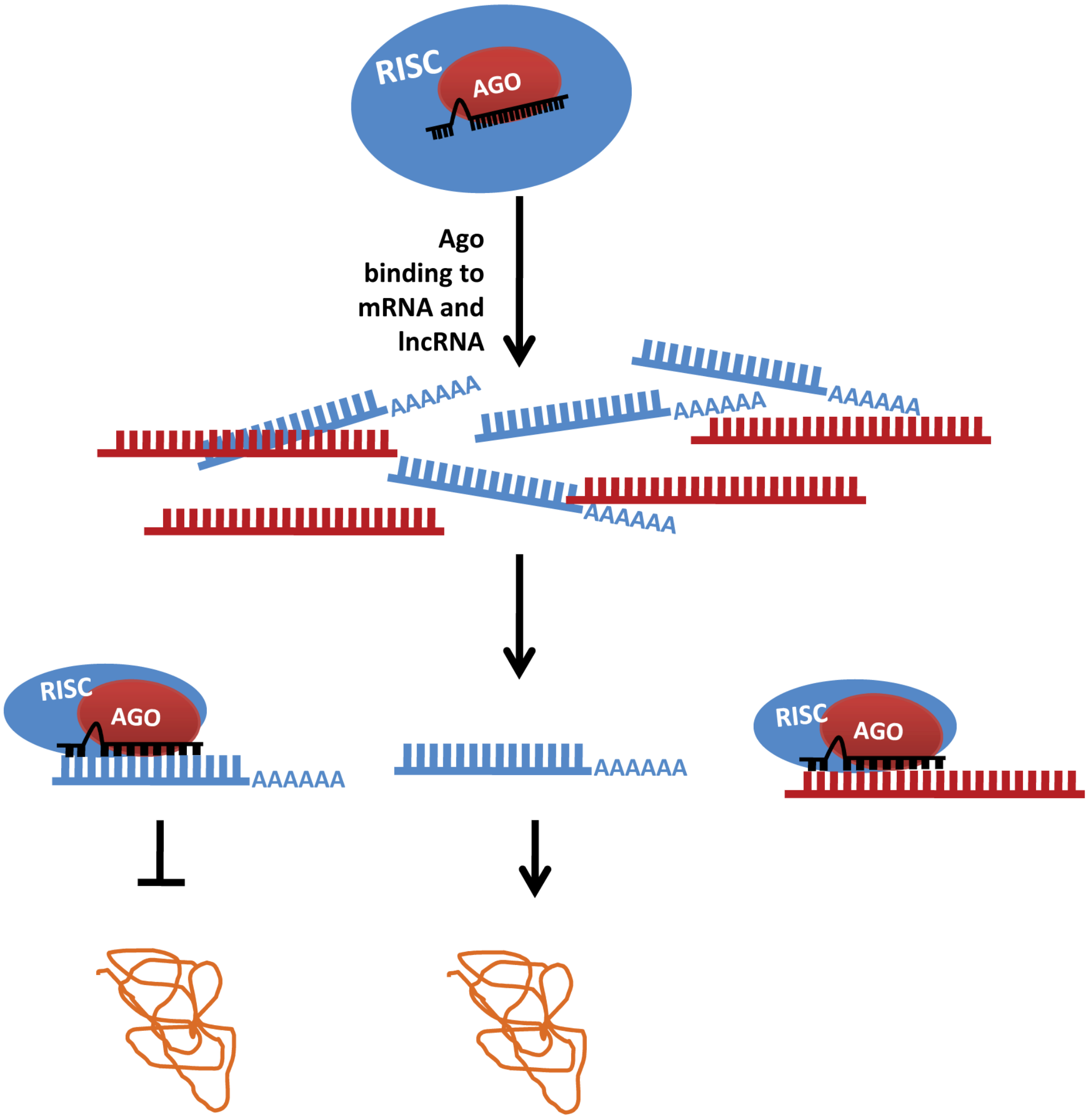

**Supplementary File 1:** Schematic of the proposed hypothesis.

**Supplementary File 2:** List of transcripts with the Ago binding sites.

| Transcript id | Chr No | AGO_sta | AGO_stp | Exon_sta | Exon_stp | strand | Annotation |
|---|---|---|---|---|---|---|---|
| ENST00000456638.1 | chr10 | 75556840 | 75556871 | 75556276 | 75557347 | - | antisense |
| ENST00000421806.1 | chr10 | 98754645 | 98754676 | 98752207 | 98755716 | + | antisense |
| ENST00000525865.1 | chr11 | 568102 | 568133 | 568089 | 568198 | - | processed_transcript |
| ENST00000531962.1 | chr11 | 32058241 | 32058278 | 32058202 | 32058542 | - | antisense |
| ENST00000395900.1 | chr11 | 32461619 | 32461651 | 32459391 | 32462950 | + | antisense |
| ENST00000538654.1 | chr11 | 62620803 | 62620827 | 62620283 | 62621062 | - | ncrna_host |
| ENST00000540865.1 | chr11 | 62620803 | 62620827 | 62620545 | 62621062 | - | ncrna_host |
| ENST00000535689.1 | chr11 | 62620803 | 62620827 | 62620826 | 62621062 | - | ncrna_host |
| ENST00000538654.1 | chr11 | 62620973 | 62621004 | 62620283 | 62621062 | - | ncrna_host |
| ENST00000540865.1 | chr11 | 62620973 | 62621004 | 62620545 | 62621062 | - | ncrna_host |
| ENST00000535689.1 | chr11 | 62620973 | 62621004 | 62620826 | 62621062 | - | ncrna_host |
| ENST00000534336.1 | chr11 | 65271852 | 65271883 | 65265233 | 65273940 | + | non_coding |
| ENST00000533697.1 | chr11 | 77727296 | 77727327 | 77727167 | 77727541 | + | antisense |
| ENST00000540312.1 | chr11 | 111384176 | 111384208 | 111383743 | 111384610 | + | lincRNA |
| ENST00000366405.2 | chr11 | 116646345 | 116646378 | 116646219 | 116646592 | - | lincRNA |
| ENST00000548149.1 | chr12 | 49521579 | 49521609 | 49521565 | 49521995 | + | antisense |
| ENST00000552893.1 | chr12 | 49521579 | 49521609 | 49521569 | 49521990 | + | antisense |
| ENST00000547387.1 | chr12 | 49522407 | 49522410 | 49521794 | 49522647 | + | antisense |
| ENST00000550468.1 | chr12 | 49667033 | 49667054 | 49666459 | 49667089 | - | antisense |
| ENST00000549978.1 | chr12 | 54427751 | 54427783 | 54427734 | 54427829 | + | non_coding |
| ENST00000501008.2 | chr12 | 92539403 | 92539434 | 92539349 | 92540538 | + | antisense |
| ENST00000535844.1 | chr12 | 122720362 | 122720393 | 122720333 | 122720470 | - | processed_transcript |
| ENST00000520585.1 | chr13 | 45964972 | 45965003 | 45964893 | 45965014 | + | processed_transcript |
| ENST00000522673.1 | chr13 | 45964972 | 45965003 | 45964849 | 45965037 | + | processed_transcript |
| ENST00000519454.1 | chr13 | 45964972 | 45965003 | 45964893 | 45965037 | + | processed_transcript |
| ENST00000330825.3 | chr13 | 45964972 | 45965003 | 45964849 | 45965037 | + | retained_intron |
| ENST00000379050.2 | chr13 | 45964972 | 45965003 | 45964880 | 45964993 | + | processed_transcript |
| ENST00000413621.1 | chr13 | 92002872 | 92002903 | 92002860 | 92002941 | + | non_coding |
| ENST00000417320.1 | chr13 | 92003193 | 92003224 | 92003146 | 92003225 | + | non_coding |
| ENST00000431173.1 | chr13 | 92003490 | 92003530 | 92003451 | 92003530 | + | non_coding |
| ENST00000446268.1 | chr13 | 92003615 | 92003646 | 92003568 | 92003645 | + | non_coding |
| ENST00000554988.1 | chr14 | 20811227 | 20811268 | 20811207 | 20811844 | - | antisense |
| ENST00000554988.1 | chr14 | 20811539 | 20811570 | 20811207 | 20811844 | - | antisense |
| ENST00000548819.1 | chr14 | 23398954 | 23398985 | 23398818 | 23398961 | + | antisense |
| ENST00000548322.1 | chr14 | 23398954 | 23398985 | 23398854 | 23398961 | + | antisense |
| ENST00000553637.1 | chr14 | 50053370 | 50053402 | 50053297 | 50053596 | + | antisense |
| ENST00000553637.1 | chr14 | 50053525 | 50053557 | 50053297 | 50053596 | + | antisense |
| ENST00000553637.1 | chr14 | 50053561 | 50053601 | 50053297 | 50053596 | + | antisense |
| ENST00000531973.1 | chr14 | 73958422 | 73958453 | 73957669 | 73960096 | + | processed_transcript |
| ENST00000500949.1 | chr15 | 41592533 | 41592564 | 41590239 | 41594168 | + | lincRNA |
| ENST00000500949.1 | chr15 | 41592772 | 41592803 | 41590239 | 41594168 | + | lincRNA |
| ENST00000544685.1 | chr15 | 83424787 | 83424820 | 83421645 | 83425831 | + | lincRNA |
| ENST00000334146.3 | chr17 | 1617198 | 1617229 | 1616997 | 1617308 | - | processed_transcript |
| ENST00000315707.3 | chr17 | 8125596 | 8125627 | 8125611 | 8126592 | - | antisense |
| ENST00000315707.3 | chr17 | 8126190 | 8126222 | 8125611 | 8126592 | - | antisense |
| ENST00000513764.1 | chr17 | 16285948 | 16285979 | 16285881 | 16286395 | + | processed_transcript |
| ENST00000497774.1 | chr17 | 16343391 | 16343420 | 16342641 | 16343567 | + | non_coding |
| ENST00000540200.1 | chr17 | 26673942 | 26673973 | 26673659 | 26675257 | - | antisense |
| ENST00000500767.2 | chr17 | 65877969 | 65877996 | 65877866 | 65878871 | - | lincRNA |
| ENST00000512667.1 | chr19 | 7855066 | 7855112 | 7854646 | 7855898 | + | lincRNA |
| ENST00000500464.1 | chr19 | 37266564 | 37266595 | 37266288 | 37267978 | + | lincRNA |
| ENST00000467807.1 | chr19 | 56889439 | 56889470 | 56888236 | 56890897 | + | antisense |
| ENST00000490123.1 | chr19 | 56889439 | 56889470 | 56888236 | 56890922 | + | antisense |
| ENST00000488113.1 | chr19 | 56889439 | 56889470 | 56888236 | 56889719 | + | antisense |
| ENST00000495307.1 | chr19 | 56889439 | 56889470 | 56888236 | 56891197 | + | antisense |
| ENST00000444362.1 | chr1 | 1337306 | 1337336 | 1335985 | 1337426 | + | processed_transcript |
| ENST00000453521.1 | chr1 | 1337306 | 1337336 | 1336623 | 1337426 | + | processed_transcript |
| ENST00000415386.2 | chr1 | 16861791 | 16861846 | 16860921 | 16862144 | - | processed_transcript |
| ENST00000438002.1 | chr1 | 17199077 | 17199132 | 17197440 | 17199602 | + | antisense |
| ENST00000453554.1 | chr1 | 17199077 | 17199132 | 17198779 | 17200002 | + | antisense |
| ENST00000413987.1 | chr1 | 28836079 | 28836109 | 28835342 | 28837404 | + | processed_transcript |
| ENST00000437681.1 | chr1 | 28836079 | 28836109 | 28835342 | 28836145 | + | processed_transcript |
| ENST00000499627.1 | chr1 | 28836079 | 28836109 | 28835913 | 28837109 | - | lincRNA |
| ENST00000424528.1 | chr1 | 98511638 | 98511669 | 98511564 | 98511952 | - | processed_transcript |
| ENST00000428030.1 | chr1 | 167164679 | 167164711 | 167164629 | 167164717 | - | antisense |
| ENST00000436656.1 | chr1 | 173834490 | 173834518 | 173834367 | 173834685 | - | non_coding |
| ENST00000442067.1 | chr1 | 173834490 | 173834518 | 173833769 | 173834685 | - | non_coding |
| ENST00000416952.1 | chr1 | 173834490 | 173834518 | 173834306 | 173834685 | - | non_coding |
| ENST00000443799.1 | chr1 | 173834490 | 173834518 | 173834491 | 173834685 | - | non_coding |
| ENST00000456293.1 | chr1 | 173835110 | 173835138 | 173834995 | 173835344 | - | non_coding |
| ENST00000443799.1 | chr1 | 173835110 | 173835138 | 173834995 | 173835344 | - | non_coding |
| ENST00000421068.1 | chr1 | 173835110 | 173835138 | 173834956 | 173835934 | - | non_coding |
| ENST00000451607.1 | chr1 | 173836817 | 173836844 | 173836293 | 173836867 | - | non_coding |
| ENST00000432536.1 | chr1 | 173836817 | 173836844 | 173836129 | 173836827 | - | non_coding |
| ENST00000438080.1 | chr1 | 179071132 | 179071166 | 179070798 | 179071334 | + | antisense |
| ENST00000500756.1 | chr1 | 205685302 | 205685333 | 205685228 | 205686154 | + | lincRNA |
| ENST00000366527.3 | chr1 | 245006518 | 245006549 | 245003940 | 245008646 | - | retained_intron |
| ENST00000489705.1 | chr1 | 245017704 | 245017730 | 245017620 | 245017805 | - | processed_transcript |
| ENST00000421562.1 | chr20 | 43708333 | 43708364 | 43707743 | 43708618 | + | processed_transcript |
| ENST00000421562.1 | chr20 | 43708531 | 43708557 | 43707743 | 43708618 | + | processed_transcript |
| ENST00000456917.1 | chr21 | 26946295 | 26946326 | 26946205 | 26947480 | + | non_coding |
| ENST00000441289.1 | chr22 | 31833266 | 31833297 | 31832963 | 31833328 | - | processed_transcript |
| ENST00000447619.1 | chr2 | 27294232 | 27294263 | 27294194 | 27294331 | - | antisense |
| ENST00000456793.1 | chr2 | 27294232 | 27294263 | 27294194 | 27294331 | - | antisense |
| ENST00000429985.1 | chr2 | 27294232 | 27294263 | 27294194 | 27294517 | - | retained_intron |
| ENST00000418539.1 | chr2 | 47562618 | 47562649 | 47558199 | 47571656 | + | lincRNA |
| ENST00000429952.1 | chr2 | 63271056 | 63271106 | 63271057 | 63272578 | - | antisense |
| ENST00000422418.1 | chr2 | 78020206 | 78020234 | 78019926 | 78020213 | + | lincRNA |
| ENST00000549582.1 | chr2 | 177015057 | 177015089 | 177015031 | 177015140 | + | non_coding |
| ENST00000429798.1 | chr3 | 45730477 | 45730522 | 45730512 | 45730626 | - | processed_transcript |
| ENST00000509915.1 | chr5 | 10250010 | 10250046 | 10249511 | 10250027 | + | antisense |
| ENST00000505030.1 | chr5 | 87962710 | 87962742 | 87960263 | 87963409 | - | processed_transcript |
| ENST00000500197.2 | chr5 | 87962710 | 87962742 | 87960263 | 87963409 | - | processed_transcript |
| ENST00000504246.1 | chr5 | 87962710 | 87962742 | 87962138 | 87963409 | - | processed_transcript |
| ENST00000511014.1 | chr5 | 87962710 | 87962742 | 87962139 | 87963409 | - | processed_transcript |
| ENST00000509783.1 | chr5 | 87962710 | 87962742 | 87962697 | 87963409 | - | processed_transcript |
| ENST00000412431.2 | chr5 | 148874686 | 148874717 | 148872949 | 148874782 | - | processed_transcript |
| ENST00000499521.2 | chr5 | 148874686 | 148874717 | 148873877 | 148878856 | - | processed_transcript |
| ENST00000501288.1 | chr6 | 26572948 | 26572977 | 26572256 | 26574925 | + | lincRNA |
| ENST00000499525.1 | chr6 | 28557156 | 28557187 | 28557006 | 28559524 | + | processed_transcript |
| ENST00000481848.2 | chr6 | 127764377 | 127764408 | 127759551 | 127765427 | - | processed_transcript |
| ENST00000489695.1 | chr7 | 27209112 | 27209168 | 27209086 | 27209356 | - | ncrna_host |
| ENST00000519935.1 | chr7 | 27209112 | 27209168 | 27208926 | 27209383 | + | antisense |
| ENST00000523790.1 | chr7 | 27209112 | 27209168 | 27208713 | 27209383 | + | antisense |
| ENST00000452320.1 | chr7 | 79088071 | 79088106 | 79088021 | 79088315 | + | retained_intron |
| ENST00000439972.1 | chr9 | 35450785 | 35450817 | 35449466 | 35451112 | + | processed_transcript |
| ENST00000430846.1 | chr9 | 35450785 | 35450817 | 35449619 | 35451112 | + | processed_transcript |
| ENST00000421848.1 | chr9 | 45023231 | 45023263 | 45022793 | 45023236 | + | lincRNA |
| ENST00000429829.1 | chrX | 73042781 | 73042812 | 73040491 | 73047819 | - | processed_transcript |
| ENST00000429829.1 | chrX | 73046431 | 73046436 | 73040491 | 73047819 | - | processed_transcript |
| ENST00000540864.1 | chrX | 73046431 | 73046436 | 73046289 | 73047729 | - | processed_transcript |
| ENST00000429829.1 | chrX | 73062555 | 73062586 | 73061217 | 73072588 | - | processed_transcript |
| ENST00000429829.1 | chrX | 73069870 | 73069901 | 73061217 | 73072588 | - | processed_transcript |
| ENST00000364991.1 | chr10 | 328034 | 328051 | 327955 | 328065 | - | rRNA |
| ENST00000497571.1 | chr10 | 3820006 | 3820037 | 3818188 | 3821782 | - | protein_coding |
| ENST00000542957.1 | chr10 | 3820006 | 3820037 | 3818189 | 3821782 | - | protein_coding |
| ENST00000355029.4 | chr10 | 5499087 | 5499119 | 5498551 | 5500426 | + | protein_coding |
| ENST00000542715.1 | chr10 | 5499087 | 5499119 | 5498551 | 5499258 | + | protein_coding |
| ENST00000380359.3 | chr10 | 5499087 | 5499119 | 5498551 | 5500426 | + | protein_coding |
| ENST00000380337.4 | chr10 | 5499087 | 5499119 | 5498551 | 5500401 | + | protein_coding |
| ENST00000328090.5 | chr10 | 5805214 | 5805245 | 5804992 | 5805703 | + | protein_coding |
| ENST00000442808.2 | chr10 | 5805214 | 5805245 | 5804992 | 5805695 | + | protein_coding |
| ENST00000459693.1 | chr10 | 5805214 | 5805245 | 5804495 | 5805703 | + | retained_intron |
| ENST00000328090.5 | chr10 | 5805517 | 5805548 | 5804992 | 5805703 | + | protein_coding |
| ENST00000442808.2 | chr10 | 5805517 | 5805548 | 5804992 | 5805695 | + | protein_coding |
| ENST00000459693.1 | chr10 | 5805517 | 5805548 | 5804495 | 5805703 | + | retained_intron |
| ENST00000378470.1 | chr10 | 14562330 | 14562361 | 14560556 | 14563305 | - | protein_coding |
| ENST00000181796.2 | chr10 | 14562330 | 14562361 | 14560559 | 14563305 | - | protein_coding |
| ENST00000468747.1 | chr10 | 14562330 | 14562361 | 14561158 | 14563305 | - | protein_coding |
| ENST00000378467.4 | chr10 | 14562330 | 14562361 | 14561512 | 14563305 | - | protein_coding |
| ENST00000487335.1 | chr10 | 14562330 | 14562361 | 14561901 | 14563305 | - | nonsense_mediated_decay |
| ENST00000378331.5 | chr10 | 14946307 | 14946341 | 14944405 | 14946314 | + | nonsense_mediated_decay |
| ENST00000313519.5 | chr10 | 14946307 | 14946341 | 14944405 | 14946314 | + | protein_coding |
| ENST00000377275.3 | chr10 | 18957526 | 18957557 | 18957459 | 18957606 | + | protein_coding |
| ENST00000377275.3 | chr10 | 18964484 | 18964515 | 18964097 | 18970568 | + | protein_coding |
| ENST00000377275.3 | chr10 | 18966932 | 18966963 | 18964097 | 18970568 | + | protein_coding |
| ENST00000377275.3 | chr10 | 18967626 | 18967653 | 18964097 | 18970568 | + | protein_coding |
| ENST00000377275.3 | chr10 | 18969368 | 18969399 | 18964097 | 18970568 | + | protein_coding |
| ENST00000376573.4 | chr10 | 22825924 | 22825955 | 22823778 | 22826210 | - | protein_coding |
| ENST00000323883.7 | chr10 | 22825924 | 22825955 | 22825667 | 22826210 | - | protein_coding |
| ENST00000474335.1 | chr10 | 22825924 | 22825955 | 22825822 | 22826210 | - | processed_transcript |
| ENST00000545335.1 | chr10 | 22825924 | 22825955 | 22825944 | 22826210 | - | protein_coding |
| ENST00000385195.1 | chr10 | 24564674 | 24564705 | 24564614 | 24564710 | + | miRNA |
| ENST00000376016.3 | chr10 | 27400812 | 27400843 | 27399383 | 27401049 | - | protein_coding |
| ENST00000326799.3 | chr10 | 27400812 | 27400843 | 27399383 | 27401049 | - | protein_coding |
| ENST00000375969.5 | chr10 | 27400812 | 27400843 | 27400663 | 27400816 | - | protein_coding |
| ENST00000375972.3 | chr10 | 27400812 | 27400843 | 27400663 | 27401049 | - | protein_coding |
| ENST00000375664.3 | chr10 | 28897139 | 28897170 | 28897115 | 28897360 | + | protein_coding |
| ENST00000375646.1 | chr10 | 28897139 | 28897170 | 28897115 | 28897360 | + | protein_coding |
| ENST00000347934.4 | chr10 | 28897139 | 28897170 | 28897115 | 28897360 | + | protein_coding |
| ENST00000354911.4 | chr10 | 28897139 | 28897170 | 28897115 | 28897360 | + | protein_coding |
| ENST00000428935.1 | chr10 | 28897139 | 28897170 | 28897115 | 28897360 | + | protein_coding |
| ENST00000424454.1 | chr10 | 28897139 | 28897170 | 28897115 | 28897360 | + | protein_coding |
| ENST00000538000.1 | chr10 | 28897139 | 28897170 | 28897115 | 28897219 | + | protein_coding |
| ENST00000439676.1 | chr10 | 28897139 | 28897170 | 28897115 | 28897360 | + | nonsense_mediated_decay |
| ENST00000345541.6 | chr10 | 28897139 | 28897170 | 28897115 | 28897360 | + | retained_intron |
| ENST00000476046.1 | chr10 | 28897139 | 28897170 | 28897115 | 28897522 | + | retained_intron |
| ENST00000540788.1 | chr10 | 29986870 | 29986901 | 29986832 | 29987157 | + | processed_pseudogene |
| ENST00000497342.1 | chr10 | 29986870 | 29986901 | 29986864 | 29987103 | + | processed_pseudogene |
| ENST00000302418.4 | chr10 | 32300190 | 32300221 | 32297938 | 32300444 | - | protein_coding |
| ENST00000302418.4 | chr10 | 32300256 | 32300275 | 32297938 | 32300444 | - | protein_coding |
| ENST00000265371.3 | chr10 | 33623352 | 33623383 | 33623238 | 33623614 | - | protein_coding |
| ENST00000374875.1 | chr10 | 33623352 | 33623383 | 33623238 | 33623564 | - | protein_coding |
| ENST00000374867.2 | chr10 | 33623352 | 33623383 | 33623238 | 33623833 | - | protein_coding |
| ENST00000374828.2 | chr10 | 33623352 | 33623383 | 33623238 | 33623570 | - | protein_coding |
| ENST00000374822.4 | chr10 | 33623352 | 33623383 | 33623238 | 33623570 | - | protein_coding |
| ENST00000374821.5 | chr10 | 33623352 | 33623383 | 33623238 | 33623450 | - | protein_coding |
| ENST00000374816.3 | chr10 | 33623352 | 33623383 | 33623238 | 33623450 | - | protein_coding |
| ENST00000469037.1 | chr10 | 38353936 | 38353967 | 38353016 | 38354023 | + | nonsense_mediated_decay |
| ENST00000277672.9 | chr10 | 38353936 | 38353967 | 38353016 | 38354016 | + | protein_coding |
| ENST00000374466.3 | chr10 | 43679031 | 43679062 | 43678698 | 43680756 | + | protein_coding |
| ENST00000544000.1 | chr10 | 43881550 | 43881581 | 43881065 | 43883384 | - | protein_coding |
| ENST00000443950.2 | chr10 | 43881550 | 43881581 | 43881065 | 43883384 | - | protein_coding |
| ENST00000356053.3 | chr10 | 43881550 | 43881581 | 43881065 | 43883384 | - | protein_coding |
| ENST00000357065.4 | chr10 | 43881550 | 43881581 | 43881065 | 43883384 | - | protein_coding |
| ENST00000337970.3 | chr10 | 43881550 | 43881581 | 43881470 | 43883384 | - | protein_coding |
| ENST00000544000.1 | chr10 | 43882391 | 43882422 | 43881065 | 43883384 | - | protein_coding |
| ENST00000443950.2 | chr10 | 43882391 | 43882422 | 43881065 | 43883384 | - | protein_coding |
| ENST00000356053.3 | chr10 | 43882391 | 43882422 | 43881065 | 43883384 | - | protein_coding |
| ENST00000357065.4 | chr10 | 43882391 | 43882422 | 43881065 | 43883384 | - | protein_coding |
| ENST00000337970.3 | chr10 | 43882391 | 43882422 | 43881470 | 43883384 | - | protein_coding |
| ENST00000540544.1 | chr10 | 43882391 | 43882422 | 43881756 | 43883139 | - | protein_coding |
| ENST00000478115.1 | chr10 | 50911096 | 50911127 | 50911035 | 50911419 | - | rRNA_pseudogene |
| ENST00000373944.3 | chr10 | 58117252 | 58117283 | 58116989 | 58117946 | - | protein_coding |
| ENST00000494312.1 | chr10 | 58117252 | 58117283 | 58117199 | 58117946 | - | processed_transcript |
| ENST00000395405.1 | chr10 | 58117252 | 58117283 | 58117199 | 58118220 | - | protein_coding |
| ENST00000318387.2 | chr10 | 58117252 | 58117283 | 58117206 | 58118220 | - | protein_coding |
| ENST00000373886.3 | chr10 | 60588951 | 60588981 | 60588521 | 60591195 | + | protein_coding |
| ENST00000263102.6 | chr10 | 61551804 | 61551835 | 61548521 | 61552869 | - | protein_coding |
| ENST00000491922.1 | chr10 | 61551804 | 61551835 | 61551510 | 61552869 | - | retained_intron |
| ENST00000519078.1 | chr10 | 62544491 | 62544522 | 62544463 | 62544619 | + | protein_coding |
| ENST00000475504.2 | chr10 | 62544491 | 62544522 | 62544463 | 62544619 | + | retained_intron |
| ENST00000395284.3 | chr10 | 62544491 | 62544522 | 62544463 | 62544619 | + | protein_coding |
| ENST00000316629.4 | chr10 | 62544491 | 62544522 | 62544463 | 62544619 | + | protein_coding |
| ENST00000448257.2 | chr10 | 62544491 | 62544522 | 62544463 | 62544619 | + | protein_coding |
| ENST00000373809.2 | chr10 | 62544491 | 62544522 | 62544463 | 62544619 | + | protein_coding |
| ENST00000519078.1 | chr10 | 62547820 | 62547851 | 62547818 | 62547992 | + | protein_coding |
| ENST00000395284.3 | chr10 | 62547820 | 62547851 | 62547818 | 62547988 | + | protein_coding |
| ENST00000448257.2 | chr10 | 62547820 | 62547851 | 62547818 | 62547992 | + | protein_coding |
| ENST00000487784.1 | chr10 | 62547820 | 62547851 | 62547429 | 62547992 | + | retained_intron |
| ENST00000395284.3 | chr10 | 62553780 | 62553812 | 62553635 | 62554610 | + | protein_coding |
| ENST00000316629.4 | chr10 | 62553780 | 62553812 | 62553635 | 62554601 | + | protein_coding |
| ENST00000448257.2 | chr10 | 62553780 | 62553812 | 62553635 | 62554586 | + | protein_coding |
| ENST00000373809.2 | chr10 | 62553780 | 62553812 | 62553635 | 62554610 | + | protein_coding |
| ENST00000395284.3 | chr10 | 62553853 | 62553878 | 62553635 | 62554610 | + | protein_coding |
| ENST00000316629.4 | chr10 | 62553853 | 62553878 | 62553635 | 62554601 | + | protein_coding |
| ENST00000448257.2 | chr10 | 62553853 | 62553878 | 62553635 | 62554586 | + | protein_coding |
| ENST00000373809.2 | chr10 | 62553853 | 62553878 | 62553635 | 62554610 | + | protein_coding |
| ENST00000395284.3 | chr10 | 62553884 | 62553884 | 62553635 | 62554610 | + | protein_coding |
| ENST00000316629.4 | chr10 | 62553884 | 62553884 | 62553635 | 62554601 | + | protein_coding |
| ENST00000448257.2 | chr10 | 62553884 | 62553884 | 62553635 | 62554586 | + | protein_coding |
| ENST00000373809.2 | chr10 | 62553884 | 62553884 | 62553635 | 62554610 | + | protein_coding |
| ENST00000469370.1 | chr10 | 68805299 | 68805339 | 68805211 | 68805467 | - | rRNA_pseudogene |
| ENST00000469172.1 | chr10 | 70098915 | 70098946 | 70098897 | 70098983 | + | processed_transcript |
| ENST00000461310.1 | chr10 | 70098915 | 70098946 | 70098897 | 70098983 | + | processed_transcript |
| ENST00000265866.7 | chr10 | 70098915 | 70098946 | 70098897 | 70098983 | + | protein_coding |
| ENST00000480987.1 | chr10 | 70098915 | 70098946 | 70098897 | 70098983 | + | processed_transcript |
| ENST00000441000.2 | chr10 | 70098915 | 70098946 | 70098897 | 70098983 | + | protein_coding |
| ENST00000486854.1 | chr10 | 70098915 | 70098946 | 70098897 | 70098958 | + | processed_transcript |
| ENST00000354695.5 | chr10 | 70098915 | 70098946 | 70098897 | 70098983 | + | protein_coding |
| ENST00000481819.1 | chr10 | 70098915 | 70098946 | 70098897 | 70098983 | + | processed_transcript |
| ENST00000491200.1 | chr10 | 70098915 | 70098946 | 70098897 | 70098983 | + | processed_transcript |
| ENST00000490442.1 | chr10 | 70098915 | 70098946 | 70098897 | 70098983 | + | processed_transcript |
| ENST00000478698.1 | chr10 | 70098915 | 70098946 | 70098563 | 70098983 | + | processed_transcript |
| ENST00000373232.2 | chr10 | 71962764 | 71962781 | 71962587 | 71962942 | - | protein_coding |
| ENST00000372837.3 | chr10 | 75541900 | 75541931 | 75541805 | 75541957 | + | protein_coding |
| ENST00000372833.5 | chr10 | 75541900 | 75541931 | 75541821 | 75541957 | + | protein_coding |
| ENST00000355264.4 | chr10 | 75881726 | 75881757 | 75881524 | 75883668 | - | protein_coding |
| ENST00000355264.4 | chr10 | 75883483 | 75883501 | 75881524 | 75883668 | - | protein_coding |
| ENST00000372745.1 | chr10 | 75883483 | 75883501 | 75882722 | 75883668 | - | protein_coding |
| ENST00000355264.4 | chr10 | 75897964 | 75897995 | 75897865 | 75898140 | - | protein_coding |
| ENST00000372745.1 | chr10 | 75897964 | 75897995 | 75897865 | 75898140 | - | protein_coding |
| ENST00000372134.3 | chr10 | 85912298 | 85912329 | 85912020 | 85913001 | + | protein_coding |
| ENST00000436406.2 | chr10 | 85912298 | 85912329 | 85912020 | 85913311 | + | protein_coding |
| ENST00000339736.6 | chr10 | 85912298 | 85912329 | 85912020 | 85912927 | + | protein_coding |
| ENST00000362234.2 | chr10 | 88024495 | 88024526 | 88024451 | 88024545 | - | miRNA |
| ENST00000371703.3 | chr10 | 92635813 | 92635844 | 92635781 | 92635855 | + | protein_coding |
| ENST00000413330.1 | chr10 | 92635813 | 92635844 | 92635781 | 92635855 | + | protein_coding |
| ENST00000487998.1 | chr10 | 92635813 | 92635844 | 92635781 | 92635855 | + | processed_transcript |
| ENST00000371705.4 | chr10 | 92635813 | 92635844 | 92635781 | 92635855 | + | protein_coding |
| ENST00000277882.3 | chr10 | 92635813 | 92635844 | 92635781 | 92635855 | + | protein_coding |
| ENST00000414836.1 | chr10 | 92635813 | 92635844 | 92635781 | 92635855 | + | protein_coding |
| ENST00000466462.1 | chr10 | 92635813 | 92635844 | 92635781 | 92635855 | + | processed_transcript |
| ENST00000480635.1 | chr10 | 92635813 | 92635844 | 92635822 | 92635855 | + | processed_transcript |
| ENST00000318333.4 | chr10 | 93567082 | 93567114 | 93566842 | 93567253 | - | processed_pseudogene |
| ENST00000445874.1 | chr10 | 93567082 | 93567114 | 93566936 | 93567196 | - | processed_pseudogene |
| ENST00000391094.1 | chr10 | 95270406 | 95270437 | 95270305 | 95270437 | - | rRNA |
| ENST00000371408.3 | chr10 | 95660689 | 95660719 | 95660509 | 95663581 | + | protein_coding |
| ENST00000427197.1 | chr10 | 95660689 | 95660719 | 95660509 | 95662490 | + | protein_coding |
| ENST00000371142.4 | chr10 | 98281190 | 98281221 | 98277866 | 98282087 | - | protein_coding |
| ENST00000540664.1 | chr10 | 98714964 | 98714986 | 98714710 | 98715591 | + | protein_coding |
| ENST00000371103.3 | chr10 | 98714964 | 98714986 | 98714710 | 98718690 | + | protein_coding |
| ENST00000371097.3 | chr10 | 98714964 | 98714986 | 98714710 | 98724198 | + | protein_coding |
| ENST00000356016.2 | chr10 | 98714964 | 98714986 | 98714710 | 98718648 | + | protein_coding |
| ENST00000371103.3 | chr10 | 98717230 | 98717261 | 98714710 | 98718690 | + | protein_coding |
| ENST00000371097.3 | chr10 | 98717230 | 98717261 | 98714710 | 98724198 | + | protein_coding |
| ENST00000356016.2 | chr10 | 98717230 | 98717261 | 98714710 | 98718648 | + | protein_coding |
| ENST00000453547.2 | chr10 | 99023182 | 99023213 | 99023177 | 99023386 | - | protein_coding |
| ENST00000316676.8 | chr10 | 99023182 | 99023213 | 99023177 | 99023386 | - | protein_coding |
| ENST00000355366.5 | chr10 | 99023182 | 99023213 | 99023177 | 99023386 | - | protein_coding |
| ENST00000358531.4 | chr10 | 99023182 | 99023213 | 99023177 | 99023386 | - | protein_coding |
| ENST00000492211.1 | chr10 | 99023182 | 99023213 | 99023177 | 99023386 | - | processed_transcript |
| ENST00000371027.1 | chr10 | 99023182 | 99023213 | 99023177 | 99023386 | - | protein_coding |
| ENST00000479633.1 | chr10 | 99023182 | 99023213 | 99023177 | 99023386 | - | processed_transcript |
| ENST00000393817.2 | chr10 | 99023182 | 99023213 | 99023177 | 99023386 | - | protein_coding |
| ENST00000358308.3 | chr10 | 99023182 | 99023213 | 99023177 | 99023386 | - | protein_coding |
| ENST00000466484.1 | chr10 | 99023182 | 99023213 | 99023177 | 99023386 | - | processed_transcript |
| ENST00000361832.1 | chr10 | 102034990 | 102035021 | 102033713 | 102035250 | - | protein_coding |
| ENST00000358848.1 | chr10 | 102034990 | 102035021 | 102033713 | 102035250 | - | protein_coding |
| ENST00000441611.1 | chr10 | 102034990 | 102035021 | 102033713 | 102035250 | - | protein_coding |
| ENST00000370372.1 | chr10 | 102034990 | 102035021 | 102034452 | 102035250 | - | protein_coding |
| ENST00000370355.2 | chr10 | 102124439 | 102124474 | 102120491 | 102124591 | + | protein_coding |
| ENST00000410482.1 | chr10 | 103124598 | 103124641 | 103124602 | 103124792 | - | snRNA |
| ENST00000410482.1 | chr10 | 103124741 | 103124772 | 103124602 | 103124792 | - | snRNA |
| ENST00000370110.5 | chr10 | 103541118 | 103541131 | 103541082 | 103541389 | - | protein_coding |
| ENST00000474993.1 | chr10 | 103541118 | 103541131 | 103541089 | 103541389 | - | processed_transcript |
| ENST00000405356.1 | chr10 | 103923582 | 103923613 | 103921868 | 103923627 | + | protein_coding |
| ENST00000370007.4 | chr10 | 103923582 | 103923613 | 103921868 | 103923623 | + | protein_coding |
| ENST00000477977.1 | chr10 | 103923582 | 103923613 | 103923528 | 103923623 | + | processed_transcript |
| ENST00000448841.1 | chr10 | 104575017 | 104575048 | 104572352 | 104575997 | + | protein_coding |
| ENST00000369889.4 | chr10 | 104575017 | 104575048 | 104572352 | 104576021 | + | protein_coding |
| ENST00000343289.5 | chr10 | 104846898 | 104846929 | 104845940 | 104849665 | - | protein_coding |
| ENST00000369825.1 | chr10 | 105154048 | 105154079 | 105153956 | 105154151 | - | protein_coding |
| ENST00000337003.4 | chr10 | 105154048 | 105154079 | 105153956 | 105154151 | - | protein_coding |
| ENST00000408840.1 | chr10 | 105154048 | 105154079 | 105154010 | 105154158 | - | miRNA |
| ENST00000369825.1 | chr10 | 105154087 | 105154118 | 105153956 | 105154151 | - | protein_coding |
| ENST00000337003.4 | chr10 | 105154087 | 105154118 | 105153956 | 105154151 | - | protein_coding |
| ENST00000408840.1 | chr10 | 105154087 | 105154118 | 105154010 | 105154158 | - | miRNA |
| ENST00000280154.7 | chr10 | 112657968 | 112657991 | 112657786 | 112659764 | + | protein_coding |
| ENST00000393104.2 | chr10 | 112657968 | 112657991 | 112657786 | 112659763 | + | protein_coding |
| ENST00000498367.1 | chr10 | 112657968 | 112657991 | 112657786 | 112658476 | + | processed_transcript |
| ENST00000369248.4 | chr10 | 116620549 | 116620580 | 116620508 | 116620648 | + | protein_coding |
| ENST00000369250.3 | chr10 | 116620549 | 116620580 | 116620508 | 116620648 | + | protein_coding |
| ENST00000411414.1 | chr10 | 116620549 | 116620580 | 116620550 | 116620648 | + | protein_coding |
| ENST00000439649.2 | chr10 | 117823989 | 117824020 | 117816444 | 117824055 | - | protein_coding |
| ENST00000369236.1 | chr10 | 117823989 | 117824020 | 117822953 | 117824055 | - | protein_coding |
| ENST00000355422.5 | chr10 | 117823989 | 117824020 | 117823275 | 117824055 | - | protein_coding |
| ENST00000544592.1 | chr10 | 117823989 | 117824020 | 117823727 | 117824055 | - | protein_coding |
| ENST00000369234.3 | chr10 | 117823989 | 117824020 | 117823912 | 117824055 | - | protein_coding |
| ENST00000334464.5 | chr10 | 119040863 | 119040902 | 119040000 | 119044982 | - | protein_coding |
| ENST00000355624.3 | chr10 | 119805360 | 119805385 | 119805322 | 119806114 | - | protein_coding |
| ENST00000369199.3 | chr10 | 119805360 | 119805385 | 119805322 | 119805674 | - | protein_coding |
| ENST00000369144.3 | chr10 | 120829011 | 120829030 | 120828958 | 120829166 | - | protein_coding |
| ENST00000541549.1 | chr10 | 120829011 | 120829030 | 120828958 | 120829166 | - | protein_coding |
| ENST00000479786.1 | chr10 | 124190993 | 124191024 | 124189140 | 124191867 | + | processed_transcript |
| ENST00000368990.3 | chr10 | 124190993 | 124191024 | 124189140 | 124191867 | + | protein_coding |
| ENST00000368989.1 | chr10 | 124190993 | 124191024 | 124189140 | 124191867 | + | protein_coding |
| ENST00000368988.1 | chr10 | 124190993 | 124191024 | 124189140 | 124191866 | + | protein_coding |
| ENST00000433307.1 | chr10 | 124190993 | 124191024 | 124189140 | 124191866 | + | protein_coding |
| ENST00000479786.1 | chr10 | 124191703 | 124191734 | 124189140 | 124191867 | + | processed_transcript |
| ENST00000368990.3 | chr10 | 124191703 | 124191734 | 124189140 | 124191867 | + | protein_coding |
| ENST00000368989.1 | chr10 | 124191703 | 124191734 | 124189140 | 124191867 | + | protein_coding |
| ENST00000368988.1 | chr10 | 124191703 | 124191734 | 124189140 | 124191866 | + | protein_coding |
| ENST00000433307.1 | chr10 | 124191703 | 124191734 | 124189140 | 124191866 | + | protein_coding |
| ENST00000368865.4 | chr10 | 124921891 | 124921922 | 124921752 | 124921929 | + | protein_coding |
| ENST00000538238.1 | chr10 | 124921891 | 124921922 | 124921752 | 124921929 | + | protein_coding |
| ENST00000368858.5 | chr10 | 124921891 | 124921922 | 124921752 | 124921929 | + | protein_coding |
| ENST00000407911.2 | chr10 | 124921891 | 124921922 | 124921752 | 124921929 | + | protein_coding |
| ENST00000481952.1 | chr10 | 124921891 | 124921922 | 124921770 | 124921929 | + | processed_transcript |
| ENST00000365617.1 | chr10 | 128609376 | 128609407 | 128609376 | 128609472 | + | misc_RNA |
| ENST00000368654.3 | chr10 | 129901620 | 129901651 | 129900843 | 129907687 | - | protein_coding |
| ENST00000368653.3 | chr10 | 129901620 | 129901651 | 129900843 | 129907687 | - | protein_coding |
| ENST00000537609.1 | chr10 | 129901620 | 129901651 | 129900843 | 129907687 | - | protein_coding |
| ENST00000368652.3 | chr10 | 129901620 | 129901651 | 129900843 | 129907687 | - | protein_coding |
| ENST00000368654.3 | chr10 | 129904518 | 129904550 | 129900843 | 129907687 | - | protein_coding |
| ENST00000368653.3 | chr10 | 129904518 | 129904550 | 129900843 | 129907687 | - | protein_coding |
| ENST00000537609.1 | chr10 | 129904518 | 129904550 | 129900843 | 129907687 | - | protein_coding |
| ENST00000368652.3 | chr10 | 129904518 | 129904550 | 129900843 | 129907687 | - | protein_coding |
| ENST00000368654.3 | chr10 | 129923939 | 129923970 | 129923840 | 129924020 | - | protein_coding |
| ENST00000368653.3 | chr10 | 129923939 | 129923970 | 129923840 | 129924020 | - | protein_coding |
| ENST00000537609.1 | chr10 | 129923939 | 129923970 | 129923840 | 129924020 | - | protein_coding |
| ENST00000368652.3 | chr10 | 129923939 | 129923970 | 129923840 | 129924020 | - | protein_coding |
| ENST00000362168.1 | chr11 | 568102 | 568133 | 568089 | 568198 | - | miRNA |
| ENST00000263645.4 | chr11 | 2418180 | 2418212 | 2418034 | 2418627 | + | protein_coding |
| ENST00000492627.1 | chr11 | 2418180 | 2418212 | 2418034 | 2418246 | + | protein_coding |
| ENST00000526072.1 | chr11 | 2418180 | 2418212 | 2418034 | 2418534 | + | protein_coding |
| ENST00000531840.1 | chr11 | 2418180 | 2418212 | 2416202 | 2418626 | + | retained_intron |
| ENST00000481687.1 | chr11 | 2418180 | 2418212 | 2418034 | 2418209 | + | protein_coding |
| ENST00000468153.1 | chr11 | 2418180 | 2418212 | 2418034 | 2418482 | + | retained_intron |
| ENST00000481386.1 | chr11 | 2418180 | 2418212 | 2418034 | 2418529 | + | retained_intron |
| ENST00000532170.1 | chr11 | 4156342 | 4156373 | 4156312 | 4156500 | + | nonsense_mediated_decay |
| ENST00000300738.5 | chr11 | 4156342 | 4156373 | 4156312 | 4156500 | + | protein_coding |
| ENST00000533349.1 | chr11 | 4156342 | 4156373 | 4156312 | 4156500 | + | nonsense_mediated_decay |
| ENST00000423050.2 | chr11 | 4156342 | 4156373 | 4156312 | 4156500 | + | protein_coding |
| ENST00000536894.1 | chr11 | 4156342 | 4156373 | 4156312 | 4156500 | + | protein_coding |
| ENST00000534285.1 | chr11 | 4156342 | 4156373 | 4156312 | 4156500 | + | protein_coding |
| ENST00000533495.1 | chr11 | 4156342 | 4156373 | 4156312 | 4156500 | + | nonsense_mediated_decay |
| ENST00000543838.1 | chr11 | 4156342 | 4156373 | 4156312 | 4156500 | + | protein_coding |
| ENST00000537197.1 | chr11 | 4156342 | 4156373 | 4156312 | 4156500 | + | protein_coding |
| ENST00000314138.6 | chr11 | 8710096 | 8710127 | 8707225 | 8711419 | + | protein_coding |
| ENST00000278175.5 | chr11 | 10328562 | 10328593 | 10327879 | 10328944 | + | protein_coding |
| ENST00000534464.1 | chr11 | 10328562 | 10328593 | 10327879 | 10328942 | + | protein_coding |
| ENST00000530439.1 | chr11 | 10328562 | 10328593 | 10327227 | 10328909 | + | protein_coding |
| ENST00000528655.1 | chr11 | 10328562 | 10328593 | 10327879 | 10328902 | + | protein_coding |
| ENST00000432999.2 | chr11 | 10875089 | 10875120 | 10874251 | 10876665 | - | protein_coding |
| ENST00000413761.2 | chr11 | 10875089 | 10875120 | 10874411 | 10876633 | - | protein_coding |
| ENST00000497848.1 | chr11 | 13347804 | 13347835 | 13347760 | 13347835 | - | scRNA_pseudogene |
| ENST00000250018.2 | chr11 | 18040214 | 18040255 | 18039111 | 18042712 | - | protein_coding |
| ENST00000527884.1 | chr11 | 19258897 | 19258928 | 19258861 | 19259017 | - | protein_coding |
| ENST00000531809.1 | chr11 | 19258897 | 19258928 | 19258861 | 19259017 | - | protein_coding |
| ENST00000396159.1 | chr11 | 19258897 | 19258928 | 19258861 | 19259017 | - | protein_coding |
| ENST00000250024.4 | chr11 | 19258897 | 19258928 | 19258861 | 19259017 | - | protein_coding |
| ENST00000532666.1 | chr11 | 19258897 | 19258928 | 19258894 | 19259017 | - | protein_coding |
| ENST00000364189.1 | chr11 | 21022421 | 21022459 | 21022427 | 21022543 | - | rRNA |
| ENST00000384522.1 | chr11 | 24477480 | 24477510 | 24477457 | 24477556 | + | misc_RNA |
| ENST00000278193.2 | chr11 | 27518614 | 27518645 | 27516123 | 27520351 | - | protein_coding |
| ENST00000524596.1 | chr11 | 27518614 | 27518645 | 27516445 | 27520351 | - | protein_coding |
| ENST00000530348.1 | chr11 | 32119926 | 32119957 | 32119896 | 32120074 | + | protein_coding |
| ENST00000532942.1 | chr11 | 32119926 | 32119957 | 32119896 | 32120074 | + | protein_coding |
| ENST00000054950.3 | chr11 | 32119926 | 32119957 | 32119896 | 32120074 | + | protein_coding |
| ENST00000527835.1 | chr11 | 32119926 | 32119957 | 32119946 | 32120074 | + | nonsense_mediated_decay |
| ENST00000532721.1 | chr11 | 32119926 | 32119957 | 32119896 | 32120004 | + | protein_coding |
| ENST00000533898.1 | chr11 | 32119926 | 32119957 | 32119896 | 32120074 | + | retained_intron |
| ENST00000303296.4 | chr11 | 33375250 | 33375281 | 33374638 | 33378569 | + | protein_coding |
| ENST00000379016.3 | chr11 | 33375250 | 33375281 | 33374638 | 33378569 | + | protein_coding |
| ENST00000456517.1 | chr11 | 33375250 | 33375281 | 33374638 | 33375645 | + | protein_coding |
| ENST00000525763.1 | chr11 | 33731624 | 33731655 | 33731452 | 33731889 | - | nonsense_mediated_decay |
| ENST00000533181.1 | chr11 | 33731624 | 33731655 | 33731452 | 33731829 | - | nonsense_mediated_decay |
| ENST00000395850.3 | chr11 | 33731624 | 33731655 | 33730246 | 33731889 | - | protein_coding |
| ENST00000533403.1 | chr11 | 33731624 | 33731655 | 33730379 | 33731889 | - | protein_coding |
| ENST00000351554.3 | chr11 | 33731624 | 33731655 | 33730411 | 33731889 | - | protein_coding |
| ENST00000528987.1 | chr11 | 33731624 | 33731655 | 33730416 | 33731889 | - | retained_intron |
| ENST00000415002.2 | chr11 | 33731624 | 33731655 | 33730446 | 33731889 | - | protein_coding |
| ENST00000445143.2 | chr11 | 33731624 | 33731655 | 33731344 | 33731889 | - | protein_coding |
| ENST00000426650.2 | chr11 | 33731624 | 33731655 | 33731372 | 33731889 | - | protein_coding |
| ENST00000437761.2 | chr11 | 33731624 | 33731655 | 33731475 | 33731889 | - | protein_coding |
| ENST00000527577.1 | chr11 | 33731624 | 33731655 | 33731551 | 33731889 | - | protein_coding |
| ENST00000528700.1 | chr11 | 33731624 | 33731655 | 33731627 | 33731889 | - | protein_coding |
| ENST00000526785.1 | chr11 | 33762777 | 33762802 | 33762485 | 33763630 | - | protein_coding |
| ENST00000265651.3 | chr11 | 33762777 | 33762802 | 33762492 | 33763630 | - | protein_coding |
| ENST00000321458.6 | chr11 | 33762777 | 33762802 | 33762492 | 33763630 | - | protein_coding |
| ENST00000530013.1 | chr11 | 33762777 | 33762802 | 33762502 | 33763630 | - | retained_intron |
| ENST00000530401.1 | chr11 | 33762777 | 33762802 | 33762511 | 33768866 | - | protein_coding |
| ENST00000341394.4 | chr11 | 34121096 | 34121117 | 34120850 | 34122703 | + | protein_coding |
| ENST00000532820.1 | chr11 | 34121096 | 34121117 | 34120850 | 34122011 | + | protein_coding |
| ENST00000533657.1 | chr11 | 34121096 | 34121117 | 34120850 | 34122698 | + | processed_transcript |
| ENST00000378852.3 | chr11 | 43364793 | 43364824 | 43363983 | 43366079 | + | protein_coding |
| ENST00000378852.3 | chr11 | 43365971 | 43366002 | 43363983 | 43366079 | + | protein_coding |
| ENST00000532962.1 | chr11 | 43923116 | 43923147 | 43923066 | 43923275 | + | nonsense_mediated_decay |
| ENST00000302708.4 | chr11 | 43923116 | 43923147 | 43923066 | 43923275 | + | protein_coding |
| ENST00000530803.1 | chr11 | 43923116 | 43923147 | 43923066 | 43923275 | + | nonsense_mediated_decay |
| ENST00000529366.1 | chr11 | 43923116 | 43923147 | 43923066 | 43923275 | + | protein_coding |
| ENST00000530754.1 | chr11 | 43923116 | 43923147 | 43923066 | 43923264 | + | retained_intron |
| ENST00000532129.1 | chr11 | 43923116 | 43923147 | 43923066 | 43923275 | + | protein_coding |
| ENST00000476780.1 | chr11 | 45848196 | 45848268 | 45848153 | 45848297 | - | rRNA_pseudogene |
| ENST00000529192.1 | chr11 | 46638383 | 46638415 | 46638053 | 46638399 | - | protein_coding |
| ENST00000390833.1 | chr11 | 46783943 | 46783966 | 46783939 | 46784049 | - | snoRNA |
| ENST00000363220.1 | chr11 | 47748513 | 47748544 | 47748446 | 47748544 | - | misc_RNA |
| ENST00000300146.9 | chr11 | 59421488 | 59421519 | 59421456 | 59421545 | - | protein_coding |
| ENST00000428532.2 | chr11 | 59421488 | 59421519 | 59421456 | 59421545 | - | protein_coding |
| ENST00000529605.1 | chr11 | 59421488 | 59421519 | 59421456 | 59421545 | - | retained_intron |
| ENST00000227525.3 | chr11 | 60690511 | 60690542 | 60689246 | 60690915 | + | protein_coding |
| ENST00000536171.1 | chr11 | 60690511 | 60690542 | 60689246 | 60690848 | + | protein_coding |
| ENST00000540280.1 | chr11 | 60690511 | 60690542 | 60690216 | 60690915 | + | processed_transcript |
| ENST00000340437.4 | chr11 | 61171103 | 61171135 | 61170121 | 61172176 | - | protein_coding |
| ENST00000394888.4 | chr11 | 61171103 | 61171135 | 61170121 | 61172176 | - | protein_coding |
| ENST00000439958.3 | chr11 | 61171103 | 61171135 | 61170203 | 61172176 | - | protein_coding |
| ENST00000533676.1 | chr11 | 61649569 | 61649596 | 61647513 | 61649998 | - | retained_intron |
| ENST00000525239.1 | chr11 | 62609092 | 62609150 | 62609041 | 62609281 | - | protein_coding |
| ENST00000538098.1 | chr11 | 62609092 | 62609150 | 62609041 | 62609281 | - | protein_coding |
| ENST00000410396.1 | chr11 | 62609092 | 62609150 | 62609091 | 62609281 | - | snRNA |
| ENST00000525239.1 | chr11 | 62609195 | 62609253 | 62609041 | 62609281 | - | protein_coding |
| ENST00000538098.1 | chr11 | 62609195 | 62609253 | 62609041 | 62609281 | - | protein_coding |
| ENST00000410396.1 | chr11 | 62609195 | 62609253 | 62609091 | 62609281 | - | snRNA |
| ENST00000364799.1 | chr11 | 62620803 | 62620827 | 62620797 | 62620867 | - | snoRNA |
| ENST00000314133.3 | chr11 | 63743712 | 63743738 | 63743697 | 63744015 | + | protein_coding |
| ENST00000406310.1 | chr11 | 64084178 | 64084215 | 64083179 | 64084210 | + | protein_coding |
| ENST00000000442.6 | chr11 | 64084178 | 64084215 | 64083179 | 64084210 | + | protein_coding |
| ENST00000405666.1 | chr11 | 64084178 | 64084215 | 64083179 | 64084215 | + | protein_coding |
| ENST00000535750.1 | chr11 | 64084178 | 64084215 | 64083932 | 64084440 | - | protein_coding |
| ENST00000535126.1 | chr11 | 64084178 | 64084215 | 64084163 | 64084619 | - | protein_coding |
| ENST00000544844.1 | chr11 | 64084178 | 64084215 | 64084163 | 64084440 | - | protein_coding |
| ENST00000537918.1 | chr11 | 64084178 | 64084215 | 64084166 | 64084440 | - | retained_intron |
| ENST00000308774.2 | chr11 | 64084178 | 64084215 | 64084167 | 64084440 | - | protein_coding |
| ENST00000539854.1 | chr11 | 64084178 | 64084215 | 64084168 | 64084619 | - | protein_coding |
| ENST00000384915.1 | chr11 | 64658633 | 64658664 | 64658609 | 64658718 | - | miRNA |
| ENST00000338369.2 | chr11 | 65653804 | 65653835 | 65653793 | 65653893 | - | protein_coding |
| ENST00000357519.4 | chr11 | 65653804 | 65653835 | 65653793 | 65653893 | - | protein_coding |
| ENST00000533045.1 | chr11 | 65653804 | 65653835 | 65653793 | 65653893 | - | protein_coding |
| ENST00000532679.1 | chr11 | 65653804 | 65653835 | 65653793 | 65653893 | - | retained_intron |
| ENST00000426652.2 | chr11 | 65653804 | 65653835 | 65653793 | 65653893 | - | processed_transcript |
| ENST00000311161.7 | chr11 | 66131743 | 66131774 | 66131696 | 66131895 | - | protein_coding |
| ENST00000357440.2 | chr11 | 66131743 | 66131774 | 66131696 | 66131895 | - | protein_coding |
| ENST00000540386.1 | chr11 | 66131743 | 66131774 | 66131696 | 66131895 | - | nonsense_mediated_decay |
| ENST00000544554.1 | chr11 | 66131743 | 66131774 | 66131696 | 66131895 | - | protein_coding |
| ENST00000541567.1 | chr11 | 66131743 | 66131774 | 66131696 | 66131895 | - | nonsense_mediated_decay |
| ENST00000546034.1 | chr11 | 66131743 | 66131774 | 66131696 | 66131895 | - | protein_coding |
| ENST00000310137.4 | chr11 | 66392615 | 66392646 | 66391685 | 66393149 | + | protein_coding |
| ENST00000227507.2 | chr11 | 69468065 | 69468096 | 69465886 | 69469242 | + | protein_coding |
| ENST00000542515.1 | chr11 | 69468065 | 69468096 | 69467844 | 69468084 | - | processed_transcript |
| ENST00000538554.1 | chr11 | 69468065 | 69468096 | 69467844 | 69468087 | - | protein_coding |
| ENST00000376587.3 | chr11 | 69468065 | 69468096 | 69467844 | 69468087 | - | protein_coding |
| ENST00000407721.2 | chr11 | 71146033 | 71146064 | 71145460 | 71146885 | - | protein_coding |
| ENST00000355527.3 | chr11 | 71146033 | 71146064 | 71145460 | 71146885 | - | protein_coding |
| ENST00000533800.1 | chr11 | 71146033 | 71146064 | 71146011 | 71146236 | - | protein_coding |
| ENST00000263681.2 | chr11 | 74353632 | 74353663 | 74351609 | 74354102 | + | protein_coding |
| ENST00000516522.1 | chr11 | 77597631 | 77597693 | 77597679 | 77597752 | + | miRNA |
| ENST00000516522.1 | chr11 | 77597716 | 77597756 | 77597679 | 77597752 | + | miRNA |
| ENST00000353172.5 | chr11 | 77727296 | 77727327 | 77726761 | 77728316 | - | protein_coding |
| ENST00000299626.5 | chr11 | 77825372 | 77825403 | 77825312 | 77825438 | - | protein_coding |
| ENST00000376156.3 | chr11 | 77825372 | 77825403 | 77825312 | 77825438 | - | protein_coding |
| ENST00000526737.1 | chr11 | 77825372 | 77825403 | 77825312 | 77825438 | - | nonsense_mediated_decay |
| ENST00000532306.1 | chr11 | 77825372 | 77825403 | 77825312 | 77825438 | - | protein_coding |
| ENST00000532440.1 | chr11 | 77825372 | 77825403 | 77825312 | 77825438 | - | protein_coding |
| ENST00000524925.1 | chr11 | 77825372 | 77825403 | 77825312 | 77825422 | - | processed_transcript |
| ENST00000529139.1 | chr11 | 77825372 | 77825403 | 77825312 | 77825438 | - | protein_coding |
| ENST00000525755.1 | chr11 | 77825372 | 77825403 | 77825312 | 77825438 | - | protein_coding |
| ENST00000530454.1 | chr11 | 77825372 | 77825403 | 77825312 | 77825438 | - | protein_coding |
| ENST00000525870.1 | chr11 | 77825372 | 77825403 | 77825312 | 77825438 | - | protein_coding |
| ENST00000532050.1 | chr11 | 77825372 | 77825403 | 77825312 | 77825438 | - | nonsense_mediated_decay |
| ENST00000527099.1 | chr11 | 77825372 | 77825403 | 77825312 | 77825438 | - | protein_coding |
| ENST00000390708.1 | chr11 | 79113066 | 79113097 | 79113066 | 79113153 | - | miRNA |
| ENST00000463796.1 | chr11 | 85195027 | 85195074 | 85195012 | 85195304 | + | rRNA_pseudogene |
| ENST00000463796.1 | chr11 | 85195080 | 85195151 | 85195012 | 85195304 | + | rRNA_pseudogene |
| ENST00000463796.1 | chr11 | 85195241 | 85195277 | 85195012 | 85195304 | + | rRNA_pseudogene |
| ENST00000536441.1 | chr11 | 94905284 | 94905314 | 94898704 | 94906505 | - | protein_coding |
| ENST00000278520.5 | chr11 | 96092256 | 96092288 | 96092157 | 96092342 | - | protein_coding |
| ENST00000542662.1 | chr11 | 96092256 | 96092288 | 96092157 | 96092342 | - | protein_coding |
| ENST00000545264.1 | chr11 | 96092256 | 96092288 | 96092157 | 96092446 | - | retained_intron |
| ENST00000423339.2 | chr11 | 96092256 | 96092288 | 96092157 | 96092342 | - | protein_coding |
| ENST00000361236.3 | chr11 | 102272857 | 102272888 | 102272647 | 102272937 | - | protein_coding |
| ENST00000398136.2 | chr11 | 102272857 | 102272888 | 102272647 | 102272937 | - | protein_coding |
| ENST00000532161.1 | chr11 | 102272857 | 102272888 | 102272647 | 102272937 | - | protein_coding |
| ENST00000528969.1 | chr11 | 102272857 | 102272888 | 102272647 | 102272937 | - | protein_coding |
| ENST00000529492.1 | chr11 | 102272857 | 102272888 | 102272475 | 102272937 | - | retained_intron |
| ENST00000525577.1 | chr11 | 102272857 | 102272888 | 102272656 | 102272937 | - | processed_transcript |
| ENST00000531103.1 | chr11 | 102272857 | 102272888 | 102272765 | 102272937 | - | protein_coding |
| ENST00000526676.1 | chr11 | 102272857 | 102272888 | 102272830 | 102272937 | - | protein_coding |
| ENST00000363985.1 | chr11 | 107955710 | 107955742 | 107955640 | 107955741 | + | misc_RNA |
| ENST00000533535.1 | chr11 | 108041842 | 108041874 | 108041859 | 108044578 | - | processed_transcript |
| ENST00000322536.3 | chr11 | 108788635 | 108788666 | 108788600 | 108788745 | + | protein_coding |
| ENST00000456020.1 | chr11 | 108788635 | 108788666 | 108788600 | 108788745 | + | protein_coding |
| ENST00000526794.1 | chr11 | 108788635 | 108788666 | 108788600 | 108792970 | + | protein_coding |
| ENST00000530116.1 | chr11 | 108788635 | 108788666 | 108788600 | 108788651 | + | processed_transcript |
| ENST00000524979.1 | chr11 | 108788635 | 108788666 | 108788600 | 108788745 | + | processed_transcript |
| ENST00000278590.3 | chr11 | 110036333 | 110036362 | 110035066 | 110042566 | + | protein_coding |
| ENST00000528673.1 | chr11 | 110036333 | 110036362 | 110035066 | 110036659 | + | protein_coding |
| ENST00000453089.2 | chr11 | 110036333 | 110036362 | 110035066 | 110042564 | + | protein_coding |
| ENST00000384831.1 | chr11 | 111384176 | 111384208 | 111384164 | 111384240 | + | miRNA |
| ENST00000264018.4 | chr11 | 118630647 | 118630678 | 118630631 | 118630753 | - | protein_coding |
| ENST00000534980.1 | chr11 | 118630647 | 118630678 | 118630631 | 118630753 | - | protein_coding |
| ENST00000526070.1 | chr11 | 118630647 | 118630678 | 118630631 | 118630753 | - | protein_coding |
| ENST00000533632.1 | chr11 | 118894606 | 118894629 | 118894031 | 118894844 | + | protein_coding |
| ENST00000530167.1 | chr11 | 118964720 | 118964751 | 118964564 | 118966177 | - | protein_coding |
| ENST00000375167.1 | chr11 | 118964720 | 118964751 | 118964589 | 118964971 | - | protein_coding |
| ENST00000527539.1 | chr11 | 119692832 | 119692863 | 119692013 | 119692838 | - | processed_pseudogene |
| ENST00000375095.2 | chr11 | 120202473 | 120202504 | 120200686 | 120204388 | + | protein_coding |
| ENST00000531346.1 | chr11 | 120202473 | 120202504 | 120200686 | 120204388 | + | processed_transcript |
| ENST00000529187.1 | chr11 | 120202473 | 120202504 | 120201116 | 120204391 | + | protein_coding |
| ENST00000362105.1 | chr11 | 122017266 | 122017297 | 122017229 | 122017301 | - | miRNA |
| ENST00000385259.1 | chr11 | 122022973 | 122023004 | 122022937 | 122023016 | - | miRNA |
| ENST00000524552.1 | chr11 | 122930038 | 122930072 | 122929767 | 122930156 | - | protein_coding |
| ENST00000527983.1 | chr11 | 122930038 | 122930072 | 122929941 | 122930736 | - | retained_intron |
| ENST00000365382.1 | chr11 | 122930038 | 122930072 | 122930043 | 122930130 | - | snoRNA |
| ENST00000530985.1 | chr11 | 125454125 | 125454156 | 125453416 | 125454575 | + | processed_transcript |
| ENST00000278903.6 | chr11 | 125454125 | 125454156 | 125453416 | 125454574 | + | protein_coding |
| ENST00000343678.4 | chr11 | 125454125 | 125454156 | 125453416 | 125454574 | + | protein_coding |
| ENST00000527628.1 | chr11 | 125454125 | 125454156 | 125453416 | 125454146 | + | processed_transcript |
| ENST00000535422.1 | chr11 | 125454125 | 125454156 | 125453416 | 125454137 | + | processed_transcript |
| ENST00000332118.5 | chr11 | 126133774 | 126133799 | 126132833 | 126133939 | - | protein_coding |
| ENST00000532259.1 | chr11 | 126133774 | 126133799 | 126133369 | 126133939 | - | protein_coding |
| ENST00000532268.1 | chr11 | 126133774 | 126133799 | 126133736 | 126134184 | - | retained_intron |
| ENST00000347735.6 | chr12 | 1599544 | 1599575 | 1599259 | 1600733 | + | protein_coding |
| ENST00000397203.2 | chr12 | 1599544 | 1599575 | 1599259 | 1605090 | + | protein_coding |
| ENST00000454194.2 | chr12 | 1599544 | 1599575 | 1599259 | 1605090 | + | protein_coding |
| ENST00000537890.1 | chr12 | 1599544 | 1599575 | 1599259 | 1605090 | + | protein_coding |
| ENST00000299183.8 | chr12 | 1599544 | 1599575 | 1599259 | 1605090 | + | protein_coding |
| ENST00000543086.1 | chr12 | 1599544 | 1599575 | 1599259 | 1600260 | + | protein_coding |
| ENST00000546231.1 | chr12 | 1599544 | 1599575 | 1599259 | 1599787 | + | protein_coding |
| ENST00000355446.5 | chr12 | 1599544 | 1599575 | 1599259 | 1601653 | + | protein_coding |
| ENST00000360905.3 | chr12 | 1599544 | 1599575 | 1599259 | 1601653 | + | protein_coding |
| ENST00000440394.2 | chr12 | 1599544 | 1599575 | 1599259 | 1602715 | + | protein_coding |
| ENST00000543151.1 | chr12 | 1599544 | 1599575 | 1599259 | 1600245 | + | processed_transcript |
| ENST00000261254.3 | chr12 | 4410632 | 4410663 | 4409026 | 4414516 | + | protein_coding |
| ENST00000237837.1 | chr12 | 4479137 | 4479165 | 4477393 | 4479949 | - | protein_coding |
| ENST00000544484.1 | chr12 | 6679368 | 6679399 | 6679249 | 6679859 | - | protein_coding |
| ENST00000544040.1 | chr12 | 6679368 | 6679399 | 6679249 | 6679859 | - | protein_coding |
| ENST00000309577.6 | chr12 | 6679368 | 6679399 | 6679249 | 6679859 | - | protein_coding |
| ENST00000357008.2 | chr12 | 6679368 | 6679399 | 6679249 | 6679859 | - | protein_coding |
| ENST00000537464.1 | chr12 | 6679368 | 6679399 | 6679249 | 6679859 | - | protein_coding |
| ENST00000545179.1 | chr12 | 7926144 | 7926174 | 7926148 | 7926175 | + | protein_coding |
| ENST00000477087.1 | chr12 | 12874037 | 12874068 | 12873970 | 12874138 | + | processed_transcript |
| ENST00000228872.4 | chr12 | 12874037 | 12874068 | 12873970 | 12875305 | + | protein_coding |
| ENST00000535674.1 | chr12 | 12874037 | 12874068 | 12873970 | 12874680 | + | protein_coding |
| ENST00000396340.1 | chr12 | 12874037 | 12874068 | 12873999 | 12874402 | + | protein_coding |
| ENST00000442489.1 | chr12 | 12874037 | 12874068 | 12873970 | 12874223 | + | protein_coding |
| ENST00000228872.4 | chr12 | 12874141 | 12874177 | 12873970 | 12875305 | + | protein_coding |
| ENST00000535674.1 | chr12 | 12874141 | 12874177 | 12873970 | 12874680 | + | protein_coding |
| ENST00000396340.1 | chr12 | 12874141 | 12874177 | 12873999 | 12874402 | + | protein_coding |
| ENST00000442489.1 | chr12 | 12874141 | 12874177 | 12873970 | 12874223 | + | protein_coding |
| ENST00000228872.4 | chr12 | 12874223 | 12874254 | 12873970 | 12875305 | + | protein_coding |
| ENST00000535674.1 | chr12 | 12874223 | 12874254 | 12873970 | 12874680 | + | protein_coding |
| ENST00000396340.1 | chr12 | 12874223 | 12874254 | 12873999 | 12874402 | + | protein_coding |
| ENST00000385248.1 | chr12 | 12917643 | 12917674 | 12917583 | 12917677 | + | miRNA |
| ENST00000014914.5 | chr12 | 13068816 | 13068847 | 13065381 | 13070871 | + | protein_coding |
| ENST00000385082.1 | chr12 | 13068816 | 13068847 | 13068763 | 13068852 | + | miRNA |
| ENST00000516536.1 | chr12 | 13199086 | 13199157 | 13199086 | 13199172 | + | miRNA |
| ENST00000197268.8 | chr12 | 13233665 | 13233696 | 13233559 | 13236383 | + | protein_coding |
| ENST00000416494.2 | chr12 | 13233665 | 13233696 | 13233559 | 13233848 | + | nonsense_mediated_decay |
| ENST00000541950.1 | chr12 | 13233665 | 13233696 | 13233559 | 13234246 | + | retained_intron |
| ENST00000459949.1 | chr12 | 20704361 | 20704498 | 20704358 | 20704522 | + | rRNA_pseudogene |
| ENST00000546136.1 | chr12 | 23685088 | 23685119 | 23682440 | 23687456 | - | protein_coding |
| ENST00000538081.1 | chr12 | 26173339 | 26173370 | 26173078 | 26173401 | + | processed_transcript |
| ENST00000538365.1 | chr12 | 26173339 | 26173370 | 26173078 | 26173433 | + | processed_transcript |
| ENST00000365317.1 | chr12 | 28658353 | 28658358 | 28658327 | 28658461 | + | rRNA |
| ENST00000360150.4 | chr12 | 29494020 | 29494041 | 29490285 | 29494151 | - | protein_coding |
| ENST00000549182.1 | chr12 | 29494020 | 29494041 | 29493790 | 29494151 | - | nonsense_mediated_decay |
| ENST00000548909.1 | chr12 | 29494020 | 29494041 | 29493951 | 29494151 | - | protein_coding |
| ENST00000551467.1 | chr12 | 29494020 | 29494041 | 29493953 | 29494151 | - | protein_coding |
| ENST00000548098.1 | chr12 | 29494020 | 29494041 | 29493966 | 29494151 | - | retained_intron |
| ENST00000201023.5 | chr12 | 29494020 | 29494041 | 29493966 | 29494151 | - | protein_coding |
| ENST00000256079.4 | chr12 | 30792536 | 30792567 | 30792449 | 30792669 | - | protein_coding |
| ENST00000545286.1 | chr12 | 30792536 | 30792567 | 30792449 | 30792669 | - | protein_coding |
| ENST00000544829.1 | chr12 | 30792536 | 30792567 | 30792449 | 30792669 | - | protein_coding |
| ENST00000448582.2 | chr12 | 31435755 | 31435786 | 31433518 | 31435805 | - | retained_intron |
| ENST00000539409.1 | chr12 | 31435755 | 31435786 | 31433518 | 31435805 | - | protein_coding |
| ENST00000337682.4 | chr12 | 31435755 | 31435786 | 31433518 | 31435805 | - | protein_coding |
| ENST00000454658.2 | chr12 | 31435755 | 31435786 | 31434996 | 31435805 | - | protein_coding |
| ENST00000398170.2 | chr12 | 31435755 | 31435786 | 31434996 | 31435805 | - | protein_coding |
| ENST00000395766.1 | chr12 | 31435755 | 31435786 | 31435232 | 31435805 | - | protein_coding |
| ENST00000544921.1 | chr12 | 31435755 | 31435786 | 31435573 | 31435805 | - | nonsense_mediated_decay |
| ENST00000542983.1 | chr12 | 31435755 | 31435786 | 31435622 | 31435805 | - | protein_coding |
| ENST00000340398.3 | chr12 | 31944339 | 31944378 | 31944123 | 31945175 | - | protein_coding |
| ENST00000531134.1 | chr12 | 32793529 | 32793560 | 32793210 | 32793673 | + | protein_coding |
| ENST00000493087.1 | chr12 | 32793529 | 32793560 | 32793210 | 32793584 | + | nonsense_mediated_decay |
| ENST00000551984.1 | chr12 | 32793529 | 32793560 | 32793210 | 32793682 | + | nonsense_mediated_decay |
| ENST00000427716.2 | chr12 | 32793529 | 32793560 | 32793210 | 32798984 | + | protein_coding |
| ENST00000546442.1 | chr12 | 32793529 | 32793560 | 32793210 | 32794453 | + | protein_coding |
| ENST00000525053.1 | chr12 | 32793529 | 32793560 | 32793210 | 32793671 | + | protein_coding |
| ENST00000362784.1 | chr12 | 34358728 | 34358766 | 34358634 | 34358752 | + | rRNA |
| ENST00000362973.1 | chr12 | 38555254 | 38555301 | 38555268 | 38555384 | - | rRNA |
| ENST00000384413.1 | chr12 | 42848593 | 42848615 | 42848522 | 42848623 | + | misc_RNA |
| ENST00000552521.1 | chr12 | 44187787 | 44187818 | 44187526 | 44189558 | - | protein_coding |
| ENST00000395510.2 | chr12 | 44187787 | 44187818 | 44187526 | 44189558 | - | protein_coding |
| ENST00000547459.1 | chr12 | 44187787 | 44187818 | 44187528 | 44189558 | - | nonsense_mediated_decay |
| ENST00000325127.4 | chr12 | 44187787 | 44187818 | 44187528 | 44189558 | - | protein_coding |
| ENST00000552521.1 | chr12 | 44188230 | 44188261 | 44187526 | 44189558 | - | protein_coding |
| ENST00000395510.2 | chr12 | 44188230 | 44188261 | 44187526 | 44189558 | - | protein_coding |
| ENST00000547459.1 | chr12 | 44188230 | 44188261 | 44187528 | 44189558 | - | nonsense_mediated_decay |
| ENST00000325127.4 | chr12 | 44188230 | 44188261 | 44187528 | 44189558 | - | protein_coding |
| ENST00000364808.1 | chr12 | 49321720 | 49321750 | 49321722 | 49321816 | - | misc_RNA |
| ENST00000548362.1 | chr12 | 49396059 | 49396090 | 49396057 | 49396653 | - | protein_coding |
| ENST00000395170.3 | chr12 | 49396059 | 49396090 | 49396058 | 49396789 | - | protein_coding |
| ENST00000547306.1 | chr12 | 49396059 | 49396090 | 49396061 | 49396789 | - | protein_coding |
| ENST00000316299.5 | chr12 | 49396059 | 49396090 | 49396061 | 49396789 | - | protein_coding |
| ENST00000548065.1 | chr12 | 49396059 | 49396090 | 49396064 | 49396789 | - | protein_coding |
| ENST00000395170.3 | chr12 | 49397607 | 49397638 | 49397540 | 49397705 | - | protein_coding |
| ENST00000547306.1 | chr12 | 49397607 | 49397638 | 49397540 | 49397705 | - | protein_coding |
| ENST00000316299.5 | chr12 | 49397607 | 49397638 | 49397540 | 49397705 | - | protein_coding |
| ENST00000548065.1 | chr12 | 49397607 | 49397638 | 49397540 | 49397705 | - | protein_coding |
| ENST00000546531.1 | chr12 | 49397607 | 49397638 | 49397540 | 49397705 | - | nonsense_mediated_decay |
| ENST00000551259.1 | chr12 | 49397607 | 49397638 | 49397540 | 49397705 | - | nonsense_mediated_decay |
| ENST00000552212.1 | chr12 | 49397607 | 49397638 | 49397540 | 49397705 | - | protein_coding |
| ENST00000551770.1 | chr12 | 49397607 | 49397638 | 49397540 | 49397705 | - | protein_coding |
| ENST00000551696.1 | chr12 | 49397607 | 49397638 | 49397540 | 49397705 | - | protein_coding |
| ENST00000550125.1 | chr12 | 49397607 | 49397638 | 49397540 | 49397705 | - | retained_intron |
| ENST00000552793.1 | chr12 | 49397607 | 49397638 | 49397545 | 49397705 | - | retained_intron |
| ENST00000548950.1 | chr12 | 49397607 | 49397638 | 49397555 | 49397705 | - | protein_coding |
| ENST00000549726.1 | chr12 | 49397607 | 49397638 | 49397593 | 49397705 | - | retained_intron |
| ENST00000336023.5 | chr12 | 49521579 | 49521609 | 49521565 | 49522721 | - | protein_coding |
| ENST00000332858.6 | chr12 | 49521579 | 49521609 | 49521569 | 49522721 | - | retained_intron |
| ENST00000547765.1 | chr12 | 49521579 | 49521609 | 49521601 | 49522721 | - | nonsense_mediated_decay |
| ENST00000336023.5 | chr12 | 49522407 | 49522410 | 49521565 | 49522721 | - | protein_coding |
| ENST00000332858.6 | chr12 | 49522407 | 49522410 | 49521569 | 49522721 | - | retained_intron |
| ENST00000547765.1 | chr12 | 49522407 | 49522410 | 49521601 | 49522721 | - | nonsense_mediated_decay |
| ENST00000551373.1 | chr12 | 49522407 | 49522410 | 49521625 | 49522721 | - | protein_coding |
| ENST00000550367.1 | chr12 | 49522407 | 49522410 | 49522237 | 49522568 | - | protein_coding |
| ENST00000301071.7 | chr12 | 49579326 | 49579354 | 49578579 | 49579773 | - | protein_coding |
| ENST00000552250.1 | chr12 | 49579326 | 49579354 | 49578598 | 49579773 | - | protein_coding |
| ENST00000548405.1 | chr12 | 49579326 | 49579354 | 49578601 | 49579773 | - | protein_coding |
| ENST00000295766.5 | chr12 | 49579326 | 49579354 | 49578742 | 49579773 | - | protein_coding |
| ENST00000550767.1 | chr12 | 49579326 | 49579354 | 49578793 | 49579773 | - | protein_coding |
| ENST00000541364.1 | chr12 | 49666323 | 49666350 | 49666036 | 49667114 | + | protein_coding |
| ENST00000552448.1 | chr12 | 49666323 | 49666350 | 49666036 | 49667111 | + | nonsense_mediated_decay |
| ENST00000301072.6 | chr12 | 49666323 | 49666350 | 49666036 | 49667109 | + | protein_coding |
| ENST00000548470.1 | chr12 | 49666323 | 49666350 | 49666036 | 49666949 | + | retained_intron |
| ENST00000541364.1 | chr12 | 49667033 | 49667054 | 49666036 | 49667114 | + | protein_coding |
| ENST00000552448.1 | chr12 | 49667033 | 49667054 | 49666036 | 49667111 | + | nonsense_mediated_decay |
| ENST00000301072.6 | chr12 | 49667033 | 49667054 | 49666036 | 49667109 | + | protein_coding |
| ENST00000321665.5 | chr12 | 49667033 | 49667054 | 49666939 | 49667116 | + | protein_coding |
| ENST00000267115.5 | chr12 | 50158567 | 50158598 | 50156656 | 50158713 | + | protein_coding |
| ENST00000423828.1 | chr12 | 50158567 | 50158598 | 50156656 | 50158717 | + | protein_coding |
| ENST00000427314.2 | chr12 | 50383057 | 50383088 | 50382945 | 50384126 | - | protein_coding |
| ENST00000312377.5 | chr12 | 50383057 | 50383088 | 50382945 | 50384126 | - | protein_coding |
| ENST00000434422.1 | chr12 | 50383057 | 50383088 | 50382945 | 50384126 | - | protein_coding |
| ENST00000454520.2 | chr12 | 50383057 | 50383088 | 50382947 | 50384126 | - | protein_coding |
| ENST00000427314.2 | chr12 | 50383177 | 50383208 | 50382945 | 50384126 | - | protein_coding |
| ENST00000312377.5 | chr12 | 50383177 | 50383208 | 50382945 | 50384126 | - | protein_coding |
| ENST00000434422.1 | chr12 | 50383177 | 50383208 | 50382945 | 50384126 | - | protein_coding |
| ENST00000454520.2 | chr12 | 50383177 | 50383208 | 50382947 | 50384126 | - | protein_coding |
| ENST00000293618.8 | chr12 | 50847367 | 50847399 | 50847243 | 50847455 | + | protein_coding |
| ENST00000429001.3 | chr12 | 50847367 | 50847399 | 50847243 | 50847455 | + | protein_coding |
| ENST00000398473.2 | chr12 | 50847367 | 50847399 | 50847243 | 50847455 | + | protein_coding |
| ENST00000522085.1 | chr12 | 50847367 | 50847399 | 50847243 | 50847455 | + | protein_coding |
| ENST00000398464.3 | chr12 | 50847367 | 50847399 | 50847243 | 50847455 | + | protein_coding |
| ENST00000518444.1 | chr12 | 50847367 | 50847399 | 50847243 | 50847455 | + | protein_coding |
| ENST00000518561.1 | chr12 | 50847367 | 50847399 | 50847243 | 50847455 | + | protein_coding |
| ENST00000520064.1 | chr12 | 50847367 | 50847399 | 50847243 | 50847455 | + | protein_coding |
| ENST00000521120.1 | chr12 | 50847367 | 50847399 | 50847243 | 50847455 | + | retained_intron |
| ENST00000293618.8 | chr12 | 50871299 | 50871330 | 50869309 | 50873779 | + | protein_coding |
| ENST00000429001.3 | chr12 | 50871299 | 50871330 | 50869309 | 50871300 | + | protein_coding |
| ENST00000398473.2 | chr12 | 50871299 | 50871330 | 50869309 | 50873787 | + | protein_coding |
| ENST00000412716.2 | chr12 | 51637465 | 51637496 | 51636114 | 51637569 | + | protein_coding |
| ENST00000449723.2 | chr12 | 51637465 | 51637496 | 51636114 | 51637568 | + | protein_coding |
| ENST00000436900.2 | chr12 | 51637465 | 51637496 | 51636114 | 51637569 | + | nonsense_mediated_decay |
| ENST00000257963.4 | chr12 | 52390735 | 52390759 | 52387769 | 52390862 | + | protein_coding |
| ENST00000360284.3 | chr12 | 52452629 | 52452660 | 52452472 | 52452833 | + | protein_coding |
| ENST00000545748.1 | chr12 | 52452629 | 52452660 | 52452472 | 52453285 | + | protein_coding |
| ENST00000550082.1 | chr12 | 52452629 | 52452660 | 52452472 | 52452833 | + | protein_coding |
| ENST00000243050.1 | chr12 | 52452629 | 52452660 | 52452472 | 52453285 | + | protein_coding |
| ENST00000394825.1 | chr12 | 52452629 | 52452660 | 52452472 | 52453285 | + | protein_coding |
| ENST00000394824.2 | chr12 | 52452629 | 52452660 | 52452472 | 52453291 | + | protein_coding |
| ENST00000550557.1 | chr12 | 52452629 | 52452660 | 52452472 | 52453287 | + | retained_intron |
| ENST00000388837.2 | chr12 | 53345544 | 53345575 | 53345515 | 53345640 | + | protein_coding |
| ENST00000550600.1 | chr12 | 53345544 | 53345575 | 53345515 | 53345640 | + | protein_coding |
| ENST00000388835.3 | chr12 | 53345544 | 53345575 | 53345515 | 53345640 | + | protein_coding |
| ENST00000549078.1 | chr12 | 53345544 | 53345575 | 53345515 | 53345640 | + | retained_intron |
| ENST00000546656.1 | chr12 | 53345544 | 53345575 | 53345280 | 53345640 | + | retained_intron |
| ENST00000548496.1 | chr12 | 53345544 | 53345575 | 53345526 | 53346126 | + | retained_intron |
| ENST00000551660.1 | chr12 | 53646959 | 53646993 | 53646602 | 53647536 | + | protein_coding |
| ENST00000534842.1 | chr12 | 53646959 | 53646993 | 53646602 | 53648189 | + | protein_coding |
| ENST00000328704.4 | chr12 | 53646959 | 53646993 | 53646602 | 53647536 | + | protein_coding |
| ENST00000329548.4 | chr12 | 53646959 | 53646993 | 53646602 | 53648188 | + | protein_coding |
| ENST00000552097.1 | chr12 | 53646959 | 53646993 | 53646602 | 53646993 | + | processed_transcript |
| ENST00000359282.5 | chr12 | 53862584 | 53862615 | 53862561 | 53862616 | + | protein_coding |
| ENST00000447282.1 | chr12 | 53862584 | 53862615 | 53862561 | 53862616 | + | protein_coding |
| ENST00000437231.1 | chr12 | 53862584 | 53862615 | 53862561 | 53862616 | + | protein_coding |
| ENST00000439930.2 | chr12 | 53862584 | 53862615 | 53862564 | 53862616 | + | protein_coding |
| ENST00000549863.1 | chr12 | 53862584 | 53862615 | 53862561 | 53862616 | + | protein_coding |
| ENST00000359462.5 | chr12 | 53862584 | 53862615 | 53862561 | 53862616 | + | protein_coding |
| ENST00000550927.1 | chr12 | 53862584 | 53862615 | 53862561 | 53862616 | + | protein_coding |
| ENST00000546463.1 | chr12 | 53862584 | 53862615 | 53862561 | 53862616 | + | protein_coding |
| ENST00000552296.1 | chr12 | 53862584 | 53862615 | 53862564 | 53862616 | + | protein_coding |
| ENST00000552083.1 | chr12 | 53862584 | 53862615 | 53862561 | 53862616 | + | protein_coding |
| ENST00000552819.1 | chr12 | 53862584 | 53862615 | 53862561 | 53862616 | + | protein_coding |
| ENST00000455667.2 | chr12 | 53862584 | 53862615 | 53862561 | 53862616 | + | protein_coding |
| ENST00000548933.1 | chr12 | 53862584 | 53862615 | 53862561 | 53862616 | + | protein_coding |
| ENST00000379777.5 | chr12 | 53862584 | 53862615 | 53862564 | 53862616 | + | protein_coding |
| ENST00000550585.1 | chr12 | 53862584 | 53862615 | 53862561 | 53862616 | + | retained_intron |
| ENST00000553064.1 | chr12 | 53862584 | 53862615 | 53862561 | 53862616 | + | protein_coding |
| ENST00000547859.1 | chr12 | 53862584 | 53862615 | 53862561 | 53862616 | + | processed_transcript |
| ENST00000547048.1 | chr12 | 53862584 | 53862615 | 53861507 | 53862616 | + | retained_intron |
| ENST00000550733.1 | chr12 | 53862584 | 53862615 | 53862252 | 53862616 | + | processed_transcript |
| ENST00000384839.1 | chr12 | 54427751 | 54427783 | 54427734 | 54427829 | + | miRNA |
| ENST00000209875.4 | chr12 | 54628758 | 54628789 | 54624724 | 54635689 | - | protein_coding |
| ENST00000553116.1 | chr12 | 56386938 | 56386969 | 56385881 | 56388490 | + | protein_coding |
| ENST00000360299.5 | chr12 | 56386938 | 56386969 | 56385881 | 56388489 | + | protein_coding |
| ENST00000549505.1 | chr12 | 56386938 | 56386969 | 56385881 | 56388490 | + | nonsense_mediated_decay |
| ENST00000541217.1 | chr12 | 56386938 | 56386969 | 56386224 | 56388469 | + | pseudogene |
| ENST00000547408.1 | chr12 | 56556715 | 56556746 | 56556639 | 56556792 | + | protein_coding |
| ENST00000546845.1 | chr12 | 56556715 | 56556746 | 56556639 | 56556754 | + | nonsense_mediated_decay |
| ENST00000551954.1 | chr12 | 56556715 | 56556746 | 56556639 | 56557280 | + | processed_transcript |
| ENST00000345093.4 | chr12 | 56598459 | 56598490 | 56598285 | 56600582 | - | protein_coding |
| ENST00000345093.4 | chr12 | 56600014 | 56600045 | 56598285 | 56600582 | - | protein_coding |
| ENST00000547967.1 | chr12 | 56600014 | 56600045 | 56599234 | 56600582 | - | nonsense_mediated_decay |
| ENST00000394013.2 | chr12 | 56600014 | 56600045 | 56599234 | 56600582 | - | protein_coding |
| ENST00000549143.1 | chr12 | 56676294 | 56676325 | 56676204 | 56676392 | - | nonsense_mediated_decay |
| ENST00000548567.1 | chr12 | 56676294 | 56676325 | 56676204 | 56676392 | - | protein_coding |
| ENST00000351328.3 | chr12 | 56676294 | 56676325 | 56676204 | 56676392 | - | protein_coding |
| ENST00000542324.2 | chr12 | 56676294 | 56676325 | 56676204 | 56676392 | - | protein_coding |
| ENST00000546621.1 | chr12 | 56676294 | 56676325 | 56676204 | 56676392 | - | retained_intron |
| ENST00000546891.1 | chr12 | 56676294 | 56676325 | 56676204 | 56676392 | - | nonsense_mediated_decay |
| ENST00000548849.1 | chr12 | 56676294 | 56676325 | 56676204 | 56676392 | - | nonsense_mediated_decay |
| ENST00000549221.1 | chr12 | 56676294 | 56676325 | 56676204 | 56676392 | - | protein_coding |
| ENST00000546930.1 | chr12 | 56676294 | 56676325 | 56676204 | 56676392 | - | protein_coding |
| ENST00000551936.1 | chr12 | 56676294 | 56676325 | 56676204 | 56676392 | - | protein_coding |
| ENST00000550734.1 | chr12 | 56676294 | 56676325 | 56676204 | 56676392 | - | protein_coding |
| ENST00000551253.1 | chr12 | 56676294 | 56676325 | 56676204 | 56676392 | - | protein_coding |
| ENST00000547449.1 | chr12 | 56676294 | 56676325 | 56676121 | 56676392 | - | retained_intron |
| ENST00000547912.1 | chr12 | 56676294 | 56676325 | 56676169 | 56676775 | - | retained_intron |
| ENST00000552688.1 | chr12 | 56676294 | 56676325 | 56676211 | 56676392 | - | protein_coding |
| ENST00000546554.1 | chr12 | 56676294 | 56676325 | 56676223 | 56676392 | - | protein_coding |
| ENST00000551473.1 | chr12 | 56676294 | 56676325 | 56676232 | 56676392 | - | protein_coding |
| ENST00000547298.1 | chr12 | 56676294 | 56676325 | 56676241 | 56676392 | - | protein_coding |
| ENST00000551137.1 | chr12 | 56676294 | 56676325 | 56676250 | 56676392 | - | protein_coding |
| ENST00000550655.1 | chr12 | 56676294 | 56676325 | 56676256 | 56676392 | - | protein_coding |
| ENST00000551968.1 | chr12 | 56676294 | 56676325 | 56676281 | 56676392 | - | protein_coding |
| ENST00000555375.1 | chr12 | 57496750 | 57496796 | 57496612 | 57496773 | - | protein_coding |
| ENST00000385293.1 | chr12 | 57912996 | 57913027 | 57912946 | 57913042 | - | miRNA |
| ENST00000547992.1 | chr12 | 58142360 | 58142391 | 58140999 | 58143994 | + | protein_coding |
| ENST00000257904.5 | chr12 | 58142360 | 58142391 | 58142034 | 58142400 | - | protein_coding |
| ENST00000312990.6 | chr12 | 58142360 | 58142391 | 58142045 | 58142400 | - | protein_coding |
| ENST00000552713.1 | chr12 | 58142360 | 58142391 | 58142194 | 58142400 | - | retained_intron |
| ENST00000549606.1 | chr12 | 58142360 | 58142391 | 58142209 | 58142400 | - | protein_coding |
| ENST00000553237.1 | chr12 | 58142360 | 58142391 | 58142303 | 58142400 | - | nonsense_mediated_decay |
| ENST00000540325.1 | chr12 | 58142360 | 58142391 | 58142303 | 58142400 | - | protein_coding |
| ENST00000546489.1 | chr12 | 58142360 | 58142391 | 58142374 | 58142400 | - | protein_coding |
| ENST00000385054.1 | chr12 | 58218431 | 58218462 | 58218392 | 58218475 | - | miRNA |
| ENST00000379141.4 | chr12 | 59266167 | 59266197 | 59265931 | 59266598 | - | protein_coding |
| ENST00000320743.3 | chr12 | 59266167 | 59266197 | 59265931 | 59266598 | - | protein_coding |
| ENST00000353364.3 | chr12 | 62799352 | 62799383 | 62797973 | 62799899 | + | protein_coding |
| ENST00000280377.5 | chr12 | 62799352 | 62799383 | 62797973 | 62799899 | + | protein_coding |
| ENST00000393654.3 | chr12 | 62799352 | 62799383 | 62797973 | 62799898 | + | protein_coding |
| ENST00000362309.1 | chr12 | 62997471 | 62997502 | 62997466 | 62997550 | + | miRNA |
| ENST00000403681.2 | chr12 | 66358305 | 66358336 | 66357025 | 66360075 | + | protein_coding |
| ENST00000403681.2 | chr12 | 66359575 | 66359606 | 66357025 | 66360075 | + | protein_coding |
| ENST00000344096.3 | chr12 | 68052363 | 68052394 | 68050886 | 68059186 | + | protein_coding |
| ENST00000393555.3 | chr12 | 68052363 | 68052394 | 68050886 | 68052689 | + | protein_coding |
| ENST00000462284.1 | chr12 | 69235803 | 69235834 | 69233054 | 69239214 | + | protein_coding |
| ENST00000544648.1 | chr12 | 69235803 | 69235834 | 69233054 | 69239211 | + | protein_coding |
| ENST00000258149.5 | chr12 | 69235803 | 69235834 | 69233054 | 69239211 | + | protein_coding |
| ENST00000356290.4 | chr12 | 69235803 | 69235834 | 69233054 | 69239211 | + | protein_coding |
| ENST00000540827.1 | chr12 | 69235803 | 69235834 | 69233054 | 69239211 | + | protein_coding |
| ENST00000428863.2 | chr12 | 69235803 | 69235834 | 69233054 | 69239211 | + | protein_coding |
| ENST00000537570.1 | chr12 | 69235803 | 69235834 | 69235726 | 69236164 | + | processed_pseudogene |
| ENST00000299293.2 | chr12 | 69968334 | 69968348 | 69967785 | 69973559 | + | protein_coding |
| ENST00000549921.1 | chr12 | 69968334 | 69968348 | 69967785 | 69968744 | + | protein_coding |
| ENST00000550389.1 | chr12 | 69968334 | 69968348 | 69967785 | 69973562 | + | protein_coding |
| ENST00000397997.2 | chr12 | 69968334 | 69968348 | 69967785 | 69973559 | + | protein_coding |
| ENST00000299293.2 | chr12 | 69968798 | 69968828 | 69967785 | 69973559 | + | protein_coding |
| ENST00000550389.1 | chr12 | 69968798 | 69968828 | 69967785 | 69973562 | + | protein_coding |
| ENST00000397997.2 | chr12 | 69968798 | 69968828 | 69967785 | 69973559 | + | protein_coding |
| ENST00000299293.2 | chr12 | 69969959 | 69969990 | 69967785 | 69973559 | + | protein_coding |
| ENST00000550389.1 | chr12 | 69969959 | 69969990 | 69967785 | 69973562 | + | protein_coding |
| ENST00000397997.2 | chr12 | 69969959 | 69969990 | 69967785 | 69973559 | + | protein_coding |
| ENST00000299293.2 | chr12 | 69971848 | 69971879 | 69967785 | 69973559 | + | protein_coding |
| ENST00000550389.1 | chr12 | 69971848 | 69971879 | 69967785 | 69973562 | + | protein_coding |
| ENST00000397997.2 | chr12 | 69971848 | 69971879 | 69967785 | 69973559 | + | protein_coding |
| ENST00000266673.5 | chr12 | 72096969 | 72097005 | 72094612 | 72097836 | + | protein_coding |
| ENST00000547816.1 | chr12 | 72096969 | 72097005 | 72096766 | 72096987 | + | nonsense_mediated_decay |
| ENST00000519948.2 | chr12 | 74933541 | 74933572 | 74931551 | 74935223 | + | protein_coding |
| ENST00000393262.3 | chr12 | 76739407 | 76739438 | 76738254 | 76741567 | - | protein_coding |
| ENST00000393249.2 | chr12 | 76747313 | 76747344 | 76745577 | 76749801 | - | protein_coding |
| ENST00000261183.3 | chr12 | 76747313 | 76747344 | 76745580 | 76749801 | - | protein_coding |
| ENST00000446075.2 | chr12 | 76747313 | 76747344 | 76745589 | 76749801 | - | protein_coding |
| ENST00000548926.1 | chr12 | 77151217 | 77151247 | 77150971 | 77151694 | + | processed_pseudogene |
| ENST00000322886.7 | chr12 | 77415339 | 77415370 | 77415027 | 77417965 | - | protein_coding |
| ENST00000339887.6 | chr12 | 77415339 | 77415370 | 77415027 | 77417965 | - | protein_coding |
| ENST00000416496.2 | chr12 | 77415339 | 77415370 | 77415028 | 77417965 | - | protein_coding |
| ENST00000328827.4 | chr12 | 79986288 | 79986319 | 79978654 | 79986473 | - | protein_coding |
| ENST00000547699.1 | chr12 | 79986288 | 79986319 | 79986295 | 79986473 | - | retained_intron |
| ENST00000492623.1 | chr12 | 80497043 | 80497074 | 80496679 | 80497113 | - | processed_pseudogene |
| ENST00000266712.6 | chr12 | 88590079 | 88590110 | 88588615 | 88593664 | + | protein_coding |
| ENST00000256015.3 | chr12 | 92539403 | 92539434 | 92539164 | 92539673 | - | protein_coding |
| ENST00000550187.1 | chr12 | 95365120 | 95365151 | 95365108 | 95365396 | - | processed_transcript |
| ENST00000547157.1 | chr12 | 95365120 | 95365151 | 95365108 | 95365261 | - | protein_coding |
| ENST00000538372.2 | chr12 | 95365120 | 95365151 | 95365109 | 95365396 | - | retained_intron |
| ENST00000551991.1 | chr12 | 95365120 | 95365151 | 95365109 | 95365396 | - | nonsense_mediated_decay |
| ENST00000327772.2 | chr12 | 95365120 | 95365151 | 95365109 | 95365396 | - | protein_coding |
| ENST00000547986.1 | chr12 | 95365120 | 95365151 | 95365112 | 95365396 | - | protein_coding |
| ENST00000550187.1 | chr12 | 95365336 | 95365367 | 95365108 | 95365396 | - | processed_transcript |
| ENST00000538372.2 | chr12 | 95365336 | 95365367 | 95365109 | 95365396 | - | retained_intron |
| ENST00000551991.1 | chr12 | 95365336 | 95365367 | 95365109 | 95365396 | - | nonsense_mediated_decay |
| ENST00000327772.2 | chr12 | 95365336 | 95365367 | 95365109 | 95365396 | - | protein_coding |
| ENST00000547986.1 | chr12 | 95365336 | 95365367 | 95365112 | 95365396 | - | protein_coding |
| ENST00000546788.1 | chr12 | 95365336 | 95365367 | 95365254 | 95365396 | - | nonsense_mediated_decay |
| ENST00000356859.4 | chr12 | 95693467 | 95693498 | 95692870 | 95694948 | + | processed_transcript |
| ENST00000362302.1 | chr12 | 95702221 | 95702252 | 95702196 | 95702289 | + | miRNA |
| ENST00000535095.1 | chr12 | 95888819 | 95888842 | 95888733 | 95888914 | + | nonsense_mediated_decay |
| ENST00000323666.5 | chr12 | 95888819 | 95888842 | 95888733 | 95888914 | + | protein_coding |
| ENST00000546753.1 | chr12 | 95888819 | 95888842 | 95888733 | 95888914 | + | protein_coding |
| ENST00000261220.9 | chr12 | 95888819 | 95888842 | 95888733 | 95888914 | + | protein_coding |
| ENST00000550777.1 | chr12 | 95888819 | 95888842 | 95888733 | 95888914 | + | protein_coding |
| ENST00000551840.1 | chr12 | 95888819 | 95888842 | 95888733 | 95888914 | + | protein_coding |
| ENST00000549808.1 | chr12 | 95888819 | 95888842 | 95888733 | 95888925 | + | protein_coding |
| ENST00000391146.1 | chr12 | 96011536 | 96011564 | 96011465 | 96011562 | + | misc_RNA |
| ENST00000384854.1 | chr12 | 97957612 | 97957643 | 97957590 | 97957689 | + | miRNA |
| ENST00000266732.4 | chr12 | 98928140 | 98928171 | 98926601 | 98929412 | + | protein_coding |
| ENST00000556029.1 | chr12 | 98941778 | 98941804 | 98941351 | 98944157 | + | protein_coding |
| ENST00000343315.5 | chr12 | 98941778 | 98941804 | 98941351 | 98942606 | + | protein_coding |
| ENST00000393053.2 | chr12 | 98941778 | 98941804 | 98941351 | 98942735 | + | protein_coding |
| ENST00000556029.1 | chr12 | 98941886 | 98941917 | 98941351 | 98944157 | + | protein_coding |
| ENST00000343315.5 | chr12 | 98941886 | 98941917 | 98941351 | 98942606 | + | protein_coding |
| ENST00000393053.2 | chr12 | 98941886 | 98941917 | 98941351 | 98942735 | + | protein_coding |
| ENST00000556029.1 | chr12 | 98943725 | 98943756 | 98941351 | 98944157 | + | protein_coding |
| ENST00000548480.1 | chr12 | 98993629 | 98993646 | 98993636 | 98993905 | + | retained_intron |
| ENST00000391141.1 | chr12 | 98993629 | 98993646 | 98993413 | 98993661 | + | snoRNA |
| ENST00000362545.1 | chr12 | 104519014 | 104519064 | 104519014 | 104519134 | + | rRNA |
| ENST00000242576.2 | chr12 | 109547844 | 109547873 | 109547634 | 109548797 | + | protein_coding |
| ENST00000336865.2 | chr12 | 109547844 | 109547873 | 109547634 | 109548780 | + | protein_coding |
| ENST00000539287.1 | chr12 | 109547844 | 109547873 | 109547637 | 109548797 | + | nonsense_mediated_decay |
| ENST00000446767.2 | chr12 | 109547844 | 109547873 | 109547634 | 109548797 | + | nonsense_mediated_decay |
| ENST00000537518.1 | chr12 | 109547844 | 109547873 | 109547634 | 109548132 | + | protein_coding |
| ENST00000308664.6 | chr12 | 110783118 | 110783149 | 110783054 | 110783187 | + | protein_coding |
| ENST00000395494.2 | chr12 | 110783118 | 110783149 | 110783054 | 110783187 | + | protein_coding |
| ENST00000377685.4 | chr12 | 110783118 | 110783149 | 110783054 | 110783187 | + | nonsense_mediated_decay |
| ENST00000539276.2 | chr12 | 110783118 | 110783149 | 110783054 | 110783187 | + | protein_coding |
| ENST00000548169.2 | chr12 | 110783118 | 110783149 | 110783054 | 110783187 | + | protein_coding |
| ENST00000354300.3 | chr12 | 110972895 | 110972920 | 110969120 | 110974900 | - | protein_coding |
| ENST00000261735.3 | chr12 | 112451297 | 112451326 | 112451120 | 112451413 | + | protein_coding |
| ENST00000455836.1 | chr12 | 112451297 | 112451326 | 112451152 | 112451413 | + | protein_coding |
| ENST00000553161.1 | chr12 | 112451297 | 112451326 | 112451233 | 112451413 | + | retained_intron |
| ENST00000552052.1 | chr12 | 112451297 | 112451326 | 112451284 | 112451413 | + | protein_coding |
| ENST00000553205.1 | chr12 | 112843753 | 112843783 | 112842994 | 112843841 | - | retained_intron |
| ENST00000202773.9 | chr12 | 112843753 | 112843783 | 112843657 | 112843841 | - | protein_coding |
| ENST00000424576.2 | chr12 | 112843753 | 112843783 | 112843657 | 112843841 | - | protein_coding |
| ENST00000549923.1 | chr12 | 112843753 | 112843783 | 112843657 | 112843841 | - | protein_coding |
| ENST00000351677.2 | chr12 | 112944769 | 112944806 | 112943629 | 112947717 | + | protein_coding |
| ENST00000314045.7 | chr12 | 113595109 | 113595140 | 113594979 | 113596917 | - | protein_coding |
| ENST00000306014.5 | chr12 | 113595109 | 113595140 | 113594979 | 113596914 | - | protein_coding |
| ENST00000365668.1 | chr12 | 120729573 | 120729604 | 120729566 | 120729706 | - | snRNA |
| ENST00000363925.1 | chr12 | 120730949 | 120730970 | 120730900 | 120731040 | - | snRNA |
| ENST00000229379.2 | chr12 | 120878277 | 120878308 | 120878257 | 120878545 | + | protein_coding |
| ENST00000549525.1 | chr12 | 120878277 | 120878308 | 120877774 | 120878540 | + | retained_intron |
| ENST00000267199.3 | chr12 | 122720362 | 122720393 | 122720333 | 122720470 | - | protein_coding |
| ENST00000543633.1 | chr12 | 122720362 | 122720393 | 122720333 | 122720470 | - | nonsense_mediated_decay |
| ENST00000541169.1 | chr12 | 122720362 | 122720393 | 122720333 | 122720470 | - | processed_transcript |
| ENST00000361654.3 | chr12 | 122756772 | 122756802 | 122755979 | 122757647 | - | nonsense_mediated_decay |
| ENST00000545889.1 | chr12 | 122756772 | 122756802 | 122755981 | 122757647 | - | protein_coding |
| ENST00000302528.7 | chr12 | 122756772 | 122756802 | 122755981 | 122757647 | - | protein_coding |
| ENST00000358808.2 | chr12 | 122756772 | 122756802 | 122755981 | 122757647 | - | protein_coding |
| ENST00000542885.1 | chr12 | 122756772 | 122756802 | 122755981 | 122757647 | - | protein_coding |
| ENST00000392458.3 | chr12 | 122756772 | 122756802 | 122756019 | 122757647 | - | protein_coding |
| ENST00000537178.1 | chr12 | 122756772 | 122756802 | 122756590 | 122757647 | - | protein_coding |
| ENST00000514271.2 | chr12 | 122826308 | 122826347 | 122825300 | 122828591 | - | retained_intron |
| ENST00000331738.6 | chr12 | 122989862 | 122989893 | 122989240 | 122990253 | - | protein_coding |
| ENST00000354654.2 | chr12 | 122989862 | 122989893 | 122989273 | 122990253 | - | protein_coding |
| ENST00000433877.2 | chr12 | 122989862 | 122989893 | 122989606 | 122990253 | - | nonsense_mediated_decay |
| ENST00000532695.1 | chr12 | 122989862 | 122989893 | 122989606 | 122990253 | - | nonsense_mediated_decay |
| ENST00000527173.2 | chr12 | 122989862 | 122989893 | 122989853 | 122990253 | - | retained_intron |
| ENST00000392442.2 | chr12 | 122989862 | 122989893 | 122989877 | 122990253 | - | processed_transcript |
| ENST00000331738.6 | chr12 | 122989937 | 122989968 | 122989240 | 122990253 | - | protein_coding |
| ENST00000354654.2 | chr12 | 122989937 | 122989968 | 122989273 | 122990253 | - | protein_coding |
| ENST00000433877.2 | chr12 | 122989937 | 122989968 | 122989606 | 122990253 | - | nonsense_mediated_decay |
| ENST00000532695.1 | chr12 | 122989937 | 122989968 | 122989606 | 122990253 | - | nonsense_mediated_decay |
| ENST00000527173.2 | chr12 | 122989937 | 122989968 | 122989853 | 122990253 | - | retained_intron |
| ENST00000392442.2 | chr12 | 122989937 | 122989968 | 122989877 | 122990253 | - | processed_transcript |
| ENST00000525332.2 | chr12 | 122989937 | 122989968 | 122989927 | 122990253 | - | processed_transcript |
| ENST00000420886.2 | chr12 | 123779185 | 123779214 | 123773656 | 123780597 | - | protein_coding |
| ENST00000267176.4 | chr12 | 123779185 | 123779214 | 123773656 | 123780597 | - | protein_coding |
| ENST00000228955.7 | chr12 | 124145106 | 124145137 | 124144713 | 124145376 | + | protein_coding |
| ENST00000543341.1 | chr12 | 124145106 | 124145137 | 124144713 | 124146479 | + | protein_coding |
| ENST00000542231.1 | chr12 | 124145106 | 124145137 | 124144713 | 124146187 | + | protein_coding |
| ENST00000539429.1 | chr12 | 124145106 | 124145137 | 124144713 | 124145175 | + | retained_intron |
| ENST00000536769.1 | chr12 | 125396210 | 125396243 | 125396150 | 125399894 | - | protein_coding |
| ENST00000535460.1 | chr12 | 125396210 | 125396243 | 125396192 | 125396726 | - | protein_coding |
| ENST00000538617.1 | chr12 | 125396210 | 125396243 | 125396192 | 125396726 | - | protein_coding |
| ENST00000536834.1 | chr12 | 125396210 | 125396243 | 125396194 | 125396434 | - | nonsense_mediated_decay |
| ENST00000541046.1 | chr12 | 125396210 | 125396243 | 125396194 | 125397813 | - | protein_coding |
| ENST00000339647.5 | chr12 | 125396210 | 125396243 | 125396194 | 125398320 | - | protein_coding |
| ENST00000546120.1 | chr12 | 125396210 | 125396243 | 125396194 | 125397780 | - | protein_coding |
| ENST00000544656.1 | chr12 | 125396210 | 125396243 | 125396206 | 125396445 | - | protein_coding |
| ENST00000488123.1 | chr12 | 127650495 | 127650535 | 127650453 | 127650987 | - | rRNA_pseudogene |
| ENST00000488123.1 | chr12 | 127650540 | 127650584 | 127650453 | 127650987 | - | rRNA_pseudogene |
| ENST00000488123.1 | chr12 | 127650893 | 127650975 | 127650453 | 127650987 | - | rRNA_pseudogene |
| ENST00000266771.5 | chr12 | 129278847 | 129278878 | 129277739 | 129278901 | - | protein_coding |
| ENST00000376744.4 | chr12 | 129278847 | 129278878 | 129277740 | 129278901 | - | nonsense_mediated_decay |
| ENST00000545031.1 | chr12 | 129278847 | 129278878 | 129278440 | 129278901 | - | protein_coding |
| ENST00000544112.1 | chr12 | 129278847 | 129278878 | 129278718 | 129278901 | - | protein_coding |
| ENST00000245255.3 | chr12 | 130857144 | 130857184 | 130856027 | 130857182 | + | protein_coding |
| ENST00000543796.1 | chr12 | 131360730 | 131360761 | 131360427 | 131362223 | + | protein_coding |
| ENST00000448750.3 | chr12 | 131360730 | 131360761 | 131360427 | 131362221 | + | protein_coding |
| ENST00000541630.1 | chr12 | 131360730 | 131360761 | 131360427 | 131361278 | + | protein_coding |
| ENST00000392369.2 | chr12 | 131360730 | 131360761 | 131360427 | 131360826 | + | protein_coding |
| ENST00000254675.3 | chr12 | 131360730 | 131360761 | 131360427 | 131360795 | + | protein_coding |
| ENST00000541679.1 | chr12 | 131360730 | 131360761 | 131360427 | 131360779 | + | retained_intron |
| ENST00000543796.1 | chr12 | 131361293 | 131361326 | 131360427 | 131362223 | + | protein_coding |
| ENST00000448750.3 | chr12 | 131361293 | 131361326 | 131360427 | 131362221 | + | protein_coding |
| ENST00000543796.1 | chr12 | 131361617 | 131361649 | 131360427 | 131362223 | + | protein_coding |
| ENST00000448750.3 | chr12 | 131361617 | 131361649 | 131360427 | 131362221 | + | protein_coding |
| ENST00000204726.3 | chr12 | 133345510 | 133345541 | 133345495 | 133349880 | - | protein_coding |
| ENST00000450791.2 | chr12 | 133345510 | 133345541 | 133345501 | 133349880 | - | protein_coding |
| ENST00000488441.2 | chr12 | 133402546 | 133402577 | 133402287 | 133402770 | + | processed_pseudogene |
| ENST00000425052.1 | chr12 | 133402546 | 133402577 | 133402287 | 133402772 | + | processed_pseudogene |
| ENST00000535942.1 | chr13 | 20426131 | 20426162 | 20425495 | 20426330 | - | retained_intron |
| ENST00000337963.4 | chr13 | 20426131 | 20426162 | 20425829 | 20426330 | - | protein_coding |
| ENST00000502168.2 | chr13 | 20426131 | 20426162 | 20425829 | 20426290 | - | protein_coding |
| ENST00000382907.4 | chr13 | 20426131 | 20426162 | 20425829 | 20426330 | - | protein_coding |
| ENST00000382905.4 | chr13 | 20426131 | 20426162 | 20425829 | 20426330 | - | protein_coding |
| ENST00000467542.1 | chr13 | 20426131 | 20426162 | 20425495 | 20426330 | - | retained_intron |
| ENST00000495534.1 | chr13 | 20426131 | 20426162 | 20425495 | 20426330 | - | retained_intron |
| ENST00000382533.4 | chr13 | 21721426 | 21721457 | 21721325 | 21723221 | + | protein_coding |
| ENST00000492245.1 | chr13 | 21721426 | 21721457 | 21721325 | 21721552 | + | processed_transcript |
| ENST00000450573.1 | chr13 | 21721426 | 21721457 | 21721325 | 21721458 | + | protein_coding |
| ENST00000467636.1 | chr13 | 21721426 | 21721457 | 21721306 | 21721742 | + | processed_transcript |
| ENST00000471009.1 | chr13 | 21721426 | 21721457 | 21721325 | 21721530 | + | processed_transcript |
| ENST00000463407.1 | chr13 | 25908473 | 25908504 | 25905495 | 25910478 | + | protein_coding |
| ENST00000380752.4 | chr13 | 30086759 | 30086778 | 30083547 | 30088720 | - | protein_coding |
| ENST00000380615.3 | chr13 | 30777597 | 30777628 | 30776767 | 30782875 | - | protein_coding |
| ENST00000470258.1 | chr13 | 39623011 | 39623044 | 39621811 | 39624246 | + | protein_coding |
| ENST00000379600.3 | chr13 | 39623011 | 39623044 | 39621811 | 39624246 | + | protein_coding |
| ENST00000379599.2 | chr13 | 39623011 | 39623044 | 39621811 | 39624246 | + | protein_coding |
| ENST00000537150.1 | chr13 | 39623011 | 39623044 | 39621811 | 39624240 | + | protein_coding |
| ENST00000379483.3 | chr13 | 41764113 | 41764139 | 41763969 | 41768702 | - | protein_coding |
| ENST00000545120.1 | chr13 | 45964972 | 45965003 | 45964486 | 45964983 | - | retrotransposed |
| ENST00000398357.2 | chr13 | 45964972 | 45965003 | 45964488 | 45965402 | - | processed_pseudogene |
| ENST00000523988.1 | chr13 | 45969533 | 45969555 | 45967451 | 45973213 | - | retained_intron |
| ENST00000519676.1 | chr13 | 45969533 | 45969555 | 45967451 | 45970150 | - | protein_coding |
| ENST00000536510.1 | chr13 | 45969533 | 45969555 | 45969432 | 45970150 | - | protein_coding |
| ENST00000242848.4 | chr13 | 46530075 | 46530106 | 46528600 | 46531432 | - | protein_coding |
| ENST00000378921.2 | chr13 | 46530075 | 46530106 | 46528602 | 46531432 | - | protein_coding |
| ENST00000242848.4 | chr13 | 46559631 | 46559662 | 46559432 | 46559896 | - | protein_coding |
| ENST00000282007.3 | chr13 | 46559631 | 46559662 | 46559432 | 46559896 | - | protein_coding |
| ENST00000431251.2 | chr13 | 46559631 | 46559662 | 46559432 | 46559896 | - | protein_coding |
| ENST00000378720.3 | chr13 | 47355719 | 47355750 | 47355631 | 47355750 | - | protein_coding |
| ENST00000378697.1 | chr13 | 47355719 | 47355750 | 47355631 | 47355750 | - | protein_coding |
| ENST00000412582.1 | chr13 | 47355719 | 47355750 | 47355631 | 47355750 | - | protein_coding |
| ENST00000495654.1 | chr13 | 47355719 | 47355750 | 47355631 | 47355750 | - | processed_transcript |
| ENST00000378565.4 | chr13 | 48835370 | 48835401 | 48835275 | 48836232 | + | protein_coding |
| ENST00000378549.5 | chr13 | 48835370 | 48835401 | 48835275 | 48835524 | + | protein_coding |
| ENST00000463839.1 | chr13 | 48835370 | 48835401 | 48835273 | 48835457 | + | processed_transcript |
| ENST00000495963.1 | chr13 | 50235073 | 50235100 | 50234859 | 50235344 | - | processed_transcript |
| ENST00000473576.1 | chr13 | 50235073 | 50235100 | 50234859 | 50235344 | - | processed_transcript |
| ENST00000378272.5 | chr13 | 50235073 | 50235100 | 50234859 | 50235344 | - | protein_coding |
| ENST00000378284.1 | chr13 | 50235073 | 50235100 | 50234859 | 50235344 | - | protein_coding |
| ENST00000242827.6 | chr13 | 50235073 | 50235100 | 50234862 | 50235344 | - | protein_coding |
| ENST00000378274.5 | chr13 | 50235073 | 50235100 | 50234867 | 50235344 | - | protein_coding |
| ENST00000378270.5 | chr13 | 50235073 | 50235100 | 50234970 | 50235344 | - | protein_coding |
| ENST00000261667.3 | chr13 | 50274104 | 50274135 | 50273447 | 50276034 | - | protein_coding |
| ENST00000385271.1 | chr13 | 50623153 | 50623184 | 50623109 | 50623197 | - | miRNA |
| ENST00000384985.1 | chr13 | 50623293 | 50623324 | 50623255 | 50623337 | - | miRNA |
| ENST00000400324.3 | chr13 | 60240638 | 60240669 | 60239717 | 60240980 | - | protein_coding |
| ENST00000400330.1 | chr13 | 60240638 | 60240669 | 60239725 | 60240980 | - | protein_coding |
| ENST00000400327.1 | chr13 | 60240638 | 60240669 | 60239725 | 60240980 | - | protein_coding |
| ENST00000413168.1 | chr13 | 60240638 | 60240669 | 60239725 | 60240980 | - | protein_coding |
| ENST00000400329.3 | chr13 | 60240638 | 60240669 | 60239725 | 60240980 | - | protein_coding |
| ENST00000377818.3 | chr13 | 73284238 | 73284269 | 73282495 | 73284483 | - | protein_coding |
| ENST00000377818.3 | chr13 | 73284323 | 73284333 | 73282495 | 73284483 | - | protein_coding |
| ENST00000427733.1 | chr13 | 90883496 | 90883527 | 90882638 | 90883936 | + | processed_pseudogene |
| ENST00000385123.1 | chr13 | 90883496 | 90883527 | 90883436 | 90883531 | + | miRNA |
| ENST00000385012.1 | chr13 | 92002872 | 92002903 | 92002859 | 92002942 | + | miRNA |
| ENST00000384878.1 | chr13 | 92003193 | 92003224 | 92003145 | 92003226 | + | miRNA |
| ENST00000362279.1 | chr13 | 92003326 | 92003387 | 92003319 | 92003389 | + | miRNA |
| ENST00000384829.1 | chr13 | 92003490 | 92003530 | 92003446 | 92003532 | + | miRNA |
| ENST00000385233.1 | chr13 | 92003615 | 92003646 | 92003568 | 92003645 | + | miRNA |
| ENST00000384986.1 | chr13 | 100008400 | 100008431 | 100008385 | 100008482 | + | miRNA |
| ENST00000390895.1 | chr13 | 100188619 | 100188650 | 100188573 | 100188650 | - | misc_RNA |
| ENST00000397451.1 | chr13 | 100616938 | 100616969 | 100615275 | 100618073 | - | protein_coding |
| ENST00000376335.3 | chr13 | 100638228 | 100638259 | 100637577 | 100639018 | + | protein_coding |
| ENST00000397444.1 | chr13 | 100638228 | 100638259 | 100637577 | 100639018 | + | protein_coding |
| ENST00000425702.1 | chr13 | 100638228 | 100638259 | 100638257 | 100638325 | + | protein_coding |
| ENST00000375898.3 | chr13 | 108882278 | 108882305 | 108881547 | 108886603 | + | protein_coding |
| ENST00000540629.1 | chr13 | 111364571 | 111364602 | 111364579 | 111364833 | - | processed_transcript |
| ENST00000516869.1 | chr14 | 20811227 | 20811268 | 20811234 | 20811566 | - | misc_RNA |
| ENST00000516869.1 | chr14 | 20811539 | 20811570 | 20811234 | 20811566 | - | misc_RNA |
| ENST00000363355.1 | chr14 | 20883146 | 20883177 | 20883146 | 20883257 | + | rRNA |
| ENST00000361505.5 | chr14 | 20944714 | 20944745 | 20944543 | 20945253 | + | protein_coding |
| ENST00000554056.1 | chr14 | 20944714 | 20944745 | 20944543 | 20945217 | + | retained_intron |
| ENST00000556754.1 | chr14 | 20944714 | 20944745 | 20944543 | 20945246 | + | retained_intron |
| ENST00000362591.1 | chr14 | 21147632 | 21147663 | 21147570 | 21147671 | + | misc_RNA |
| ENST00000336053.6 | chr14 | 21702341 | 21702372 | 21702112 | 21702388 | - | protein_coding |
| ENST00000320084.7 | chr14 | 21702341 | 21702372 | 21702112 | 21702410 | - | protein_coding |
| ENST00000449098.1 | chr14 | 21702341 | 21702372 | 21702112 | 21702388 | - | protein_coding |
| ENST00000554969.1 | chr14 | 21702341 | 21702372 | 21702112 | 21702388 | - | protein_coding |
| ENST00000556142.1 | chr14 | 21702341 | 21702372 | 21702112 | 21702388 | - | protein_coding |
| ENST00000554455.1 | chr14 | 21702341 | 21702372 | 21702112 | 21702388 | - | protein_coding |
| ENST00000430246.2 | chr14 | 21702341 | 21702372 | 21702112 | 21702388 | - | protein_coding |
| ENST00000400042.4 | chr14 | 21702341 | 21702372 | 21702237 | 21702410 | - | protein_coding |
| ENST00000553753.1 | chr14 | 21702341 | 21702372 | 21702112 | 21702388 | - | protein_coding |
| ENST00000555914.1 | chr14 | 21702341 | 21702372 | 21702112 | 21702388 | - | protein_coding |
| ENST00000555309.1 | chr14 | 21702341 | 21702372 | 21702112 | 21702388 | - | protein_coding |
| ENST00000557442.1 | chr14 | 21702341 | 21702372 | 21702112 | 21702388 | - | nonsense_mediated_decay |
| ENST00000452166.2 | chr14 | 21702341 | 21702372 | 21702112 | 21702410 | - | protein_coding |
| ENST00000556628.1 | chr14 | 21702341 | 21702372 | 21702237 | 21702388 | - | protein_coding |
| ENST00000555883.1 | chr14 | 21702341 | 21702372 | 21702112 | 21702388 | - | protein_coding |
| ENST00000556513.1 | chr14 | 21702341 | 21702372 | 21702112 | 21702388 | - | protein_coding |
| ENST00000557201.1 | chr14 | 21702341 | 21702372 | 21702112 | 21702388 | - | protein_coding |
| ENST00000553300.1 | chr14 | 21702341 | 21702372 | 21702112 | 21702388 | - | protein_coding |
| ENST00000216296.6 | chr14 | 21702341 | 21702372 | 21702319 | 21702410 | - | protein_coding |
| ENST00000556897.1 | chr14 | 21702341 | 21702372 | 21702112 | 21702388 | - | protein_coding |
| ENST00000420743.2 | chr14 | 21702341 | 21702372 | 21702112 | 21702388 | - | protein_coding |
| ENST00000445284.2 | chr14 | 21702341 | 21702372 | 21702112 | 21702388 | - | protein_coding |
| ENST00000554383.1 | chr14 | 21702341 | 21702372 | 21702112 | 21702388 | - | protein_coding |
| ENST00000555215.1 | chr14 | 21702341 | 21702372 | 21702112 | 21702388 | - | protein_coding |
| ENST00000555137.1 | chr14 | 21702341 | 21702372 | 21702112 | 21702388 | - | protein_coding |
| ENST00000554891.1 | chr14 | 21702341 | 21702372 | 21702112 | 21702388 | - | protein_coding |
| ENST00000556226.1 | chr14 | 21702341 | 21702372 | 21702112 | 21702388 | - | protein_coding |
| ENST00000555176.1 | chr14 | 21702341 | 21702372 | 21702112 | 21702388 | - | protein_coding |
| ENST00000553614.1 | chr14 | 21702341 | 21702372 | 21702112 | 21702388 | - | retained_intron |
| ENST00000557336.1 | chr14 | 21702341 | 21702372 | 21702154 | 21702388 | - | protein_coding |
| ENST00000557768.1 | chr14 | 21702341 | 21702372 | 21702190 | 21702388 | - | protein_coding |
| ENST00000361611.6 | chr14 | 23495541 | 23495552 | 23495060 | 23495584 | - | protein_coding |
| ENST00000493471.2 | chr14 | 23495541 | 23495552 | 23495067 | 23495584 | - | protein_coding |
| ENST00000460922.2 | chr14 | 23495541 | 23495552 | 23495067 | 23495584 | - | protein_coding |
| ENST00000425762.2 | chr14 | 23495541 | 23495552 | 23495111 | 23495584 | - | protein_coding |
| ENST00000334454.10 | chr14 | 23495541 | 23495552 | 23495506 | 23495584 | - | nonsense_mediated_decay |
| ENST00000319074.4 | chr14 | 23568754 | 23568781 | 23566867 | 23569665 | + | protein_coding |
| ENST00000485514.1 | chr14 | 23987058 | 23987059 | 23986946 | 23987095 | - | rRNA_pseudogene |
| ENST00000385217.1 | chr14 | 31483864 | 31483895 | 31483852 | 31483948 | - | miRNA |
| ENST00000355683.5 | chr14 | 31495174 | 31495205 | 31495110 | 31495607 | - | protein_coding |
| ENST00000357479.5 | chr14 | 31495174 | 31495205 | 31495110 | 31495588 | - | protein_coding |
| ENST00000555358.1 | chr14 | 31495174 | 31495205 | 31495110 | 31495525 | - | nonsense_mediated_decay |
| ENST00000383869.1 | chr14 | 35016052 | 35016053 | 35015920 | 35016083 | + | snRNA |
| ENST00000383861.1 | chr14 | 35025432 | 35025459 | 35025432 | 35025595 | - | snRNA |
| ENST00000383861.1 | chr14 | 35025462 | 35025466 | 35025432 | 35025595 | - | snRNA |
| ENST00000554037.1 | chr14 | 39560300 | 39560332 | 39560105 | 39560302 | + | processed_pseudogene |
| ENST00000216832.4 | chr14 | 39651323 | 39651354 | 39649707 | 39652422 | + | protein_coding |
| ENST00000429476.2 | chr14 | 45468577 | 45468608 | 45468566 | 45468700 | + | protein_coding |
| ENST00000361577.3 | chr14 | 45468577 | 45468608 | 45468566 | 45468700 | + | protein_coding |
| ENST00000557423.1 | chr14 | 45468577 | 45468608 | 45468566 | 45468700 | + | nonsense_mediated_decay |
| ENST00000555607.1 | chr14 | 45468577 | 45468608 | 45468566 | 45468700 | + | retained_intron |
| ENST00000361462.2 | chr14 | 45468577 | 45468608 | 45468566 | 45468700 | + | protein_coding |
| ENST00000382233.2 | chr14 | 45468577 | 45468608 | 45468566 | 45468700 | + | protein_coding |
| ENST00000555874.1 | chr14 | 45468577 | 45468608 | 45468566 | 45468700 | + | protein_coding |
| ENST00000557250.1 | chr14 | 45468577 | 45468608 | 45468566 | 45468700 | + | protein_coding |
| ENST00000245458.6 | chr14 | 50050350 | 50050355 | 50050290 | 50050393 | - | protein_coding |
| ENST00000557367.1 | chr14 | 50050350 | 50050355 | 50050297 | 50050393 | - | retained_intron |
| ENST00000557111.1 | chr14 | 50050350 | 50050355 | 50050297 | 50050393 | - | processed_transcript |
| ENST00000245441.5 | chr14 | 51187948 | 51187979 | 51186481 | 51190390 | - | protein_coding |
| ENST00000311149.8 | chr14 | 51187948 | 51187979 | 51186481 | 51190390 | - | protein_coding |
| ENST00000389868.3 | chr14 | 51187948 | 51187979 | 51186481 | 51190390 | - | protein_coding |
| ENST00000476352.1 | chr14 | 51187948 | 51187979 | 51186538 | 51190390 | - | nonsense_mediated_decay |
| ENST00000261700.3 | chr14 | 52471095 | 52471126 | 52471080 | 52471414 | + | protein_coding |
| ENST00000556760.1 | chr14 | 52471095 | 52471126 | 52471080 | 52471371 | + | protein_coding |
| ENST00000553362.1 | chr14 | 52471095 | 52471126 | 52471080 | 52471420 | + | protein_coding |
| ENST00000553479.1 | chr14 | 52471095 | 52471126 | 52471080 | 52471410 | + | retained_intron |
| ENST00000358056.3 | chr14 | 54955840 | 54955871 | 54955643 | 54955914 | - | protein_coding |
| ENST00000325658.3 | chr14 | 60754144 | 60754176 | 60752342 | 60755273 | + | protein_coding |
| ENST00000216513.4 | chr14 | 61186671 | 61186702 | 61186478 | 61187163 | - | protein_coding |
| ENST00000554079.1 | chr14 | 61186671 | 61186702 | 61186478 | 61187163 | - | protein_coding |
| ENST00000556952.1 | chr14 | 61186671 | 61186702 | 61186478 | 61187163 | - | protein_coding |
| ENST00000247225.6 | chr14 | 64151054 | 64151085 | 64150932 | 64153374 | - | protein_coding |
| ENST00000247225.6 | chr14 | 64151386 | 64151417 | 64150932 | 64153374 | - | protein_coding |
| ENST00000365312.1 | chr14 | 64315027 | 64315058 | 64314979 | 64315080 | - | misc_RNA |
| ENST00000381346.4 | chr14 | 68143732 | 68143761 | 68143518 | 68145140 | - | protein_coding |
| ENST00000347230.4 | chr14 | 68213484 | 68213515 | 68213237 | 68215356 | - | protein_coding |
| ENST00000411699.1 | chr14 | 68213484 | 68213515 | 68213237 | 68215356 | - | protein_coding |
| ENST00000031146.4 | chr14 | 69928016 | 69928036 | 69925080 | 69929098 | + | processed_transcript |
| ENST00000410708.1 | chr14 | 70827629 | 70827651 | 70827609 | 70827805 | + | snRNA |
| ENST00000554315.1 | chr14 | 73752899 | 73752931 | 73752901 | 73754022 | - | retained_intron |
| ENST00000304061.5 | chr14 | 73958422 | 73958453 | 73957983 | 73960105 | + | pseudogene |
| ENST00000489070.1 | chr14 | 74055540 | 74055563 | 74055529 | 74055601 | - | tRNA_pseudogene |
| ENST00000394071.2 | chr14 | 74183387 | 74183421 | 74181825 | 74186204 | - | protein_coding |
| ENST00000476562.1 | chr14 | 74183387 | 74183421 | 74181827 | 74186204 | - | retained_intron |
| ENST00000286523.5 | chr14 | 74183387 | 74183421 | 74181834 | 74186204 | - | protein_coding |
| ENST00000356357.4 | chr14 | 75128184 | 75128214 | 75127955 | 75130525 | - | protein_coding |
| ENST00000555330.1 | chr14 | 75128184 | 75128214 | 75127975 | 75130525 | - | nonsense_mediated_decay |
| ENST00000303575.4 | chr14 | 75643160 | 75643191 | 75643058 | 75643334 | - | protein_coding |
| ENST00000555873.1 | chr14 | 75643160 | 75643191 | 75643058 | 75643300 | - | nonsense_mediated_decay |
| ENST00000555085.1 | chr14 | 75643160 | 75643191 | 75643058 | 75643315 | - | retained_intron |
| ENST00000238647.3 | chr14 | 77491258 | 77491292 | 77490888 | 77495034 | - | protein_coding |
| ENST00000281127.7 | chr14 | 80330578 | 80330609 | 80328006 | 80330762 | + | protein_coding |
| ENST00000428277.2 | chr14 | 80330578 | 80330609 | 80328006 | 80330760 | + | protein_coding |
| ENST00000553612.1 | chr14 | 81643693 | 81643724 | 81641796 | 81646674 | - | protein_coding |
| ENST00000471006.1 | chr14 | 90341389 | 90341432 | 90341365 | 90341577 | - | rRNA_pseudogene |
| ENST00000354366.3 | chr14 | 90743653 | 90743684 | 90742580 | 90744805 | - | protein_coding |
| ENST00000554684.1 | chr14 | 91957095 | 91957126 | 91957091 | 91957146 | - | protein_coding |
| ENST00000337238.4 | chr14 | 91957095 | 91957126 | 91957091 | 91957146 | - | protein_coding |
| ENST00000428424.2 | chr14 | 91957095 | 91957126 | 91957091 | 91957146 | - | protein_coding |
| ENST00000554943.1 | chr14 | 91957095 | 91957126 | 91957091 | 91957146 | - | protein_coding |
| ENST00000555462.1 | chr14 | 91957095 | 91957126 | 91957091 | 91957146 | - | protein_coding |
| ENST00000554308.1 | chr14 | 91957095 | 91957126 | 91957091 | 91957146 | - | nonsense_mediated_decay |
| ENST00000554390.1 | chr14 | 91957095 | 91957126 | 91957091 | 91957146 | - | protein_coding |
| ENST00000555029.1 | chr14 | 91957095 | 91957126 | 91957091 | 91957146 | - | protein_coding |
| ENST00000557018.1 | chr14 | 91957095 | 91957126 | 91957091 | 91957146 | - | protein_coding |
| ENST00000554511.1 | chr14 | 91957095 | 91957126 | 91957091 | 91957146 | - | protein_coding |
| ENST00000516679.1 | chr14 | 92270232 | 92270263 | 92270233 | 92270316 | - | miRNA |
| ENST00000298894.4 | chr14 | 93649444 | 93649460 | 93648541 | 93650707 | - | protein_coding |
| ENST00000556883.1 | chr14 | 93649444 | 93649460 | 93648544 | 93650707 | - | protein_coding |
| ENST00000343455.3 | chr14 | 95555349 | 95555380 | 95552565 | 95557000 | - | protein_coding |
| ENST00000526495.1 | chr14 | 95555349 | 95555380 | 95552565 | 95557000 | - | protein_coding |
| ENST00000393063.1 | chr14 | 95555349 | 95555380 | 95552566 | 95557000 | - | protein_coding |
| ENST00000343455.3 | chr14 | 95569724 | 95569756 | 95569683 | 95570463 | - | protein_coding |
| ENST00000526495.1 | chr14 | 95569724 | 95569756 | 95569683 | 95570463 | - | protein_coding |
| ENST00000393063.1 | chr14 | 95569724 | 95569756 | 95569683 | 95570463 | - | protein_coding |
| ENST00000527414.1 | chr14 | 95569724 | 95569756 | 95569683 | 95570463 | - | protein_coding |
| ENST00000556045.1 | chr14 | 95569724 | 95569756 | 95569683 | 95570386 | - | protein_coding |
| ENST00000541352.1 | chr14 | 95569724 | 95569756 | 95569683 | 95570463 | - | protein_coding |
| ENST00000532939.1 | chr14 | 95569724 | 95569756 | 95569683 | 95569767 | - | protein_coding |
| ENST00000363636.1 | chr14 | 100049424 | 100049456 | 100049354 | 100049455 | + | misc_RNA |
| ENST00000390219.1 | chr14 | 101318749 | 101318780 | 101318727 | 101318824 | + | miRNA |
| ENST00000534062.1 | chr14 | 101349344 | 101349375 | 101346992 | 101351184 | - | protein_coding |
| ENST00000384876.1 | chr14 | 101349344 | 101349375 | 101349316 | 101349412 | + | miRNA |
| ENST00000385199.1 | chr14 | 101506571 | 101506602 | 101506556 | 101506636 | + | miRNA |
| ENST00000385199.1 | chr14 | 101506606 | 101506637 | 101506556 | 101506636 | + | miRNA |
| ENST00000362150.1 | chr14 | 101512305 | 101512337 | 101512257 | 101512331 | + | miRNA |
| ENST00000385021.1 | chr14 | 101512842 | 101512873 | 101512792 | 101512875 | + | miRNA |
| ENST00000401280.1 | chr14 | 101514286 | 101514317 | 101514238 | 101514316 | + | miRNA |
| ENST00000362159.2 | chr14 | 101515947 | 101515978 | 101515887 | 101515983 | + | miRNA |
| ENST00000385009.1 | chr14 | 101520653 | 101520684 | 101520643 | 101520718 | + | miRNA |
| ENST00000385292.1 | chr14 | 101521764 | 101521795 | 101521756 | 101521828 | + | miRNA |
| ENST00000385269.1 | chr14 | 101522570 | 101522601 | 101522527 | 101522606 | + | miRNA |
| ENST00000401360.1 | chr14 | 101530841 | 101530872 | 101530832 | 101530915 | + | miRNA |
| ENST00000216281.7 | chr14 | 102547968 | 102548031 | 102547106 | 102548158 | - | protein_coding |
| ENST00000334701.7 | chr14 | 102547968 | 102548031 | 102547386 | 102548158 | - | protein_coding |
| ENST00000216554.2 | chr14 | 103802065 | 103802096 | 103801990 | 103802269 | + | protein_coding |
| ENST00000392715.1 | chr14 | 103802065 | 103802096 | 103801990 | 103802269 | + | protein_coding |
| ENST00000216602.6 | chr14 | 104199915 | 104199946 | 104199320 | 104200005 | + | protein_coding |
| ENST00000311141.2 | chr14 | 104199915 | 104199946 | 104199320 | 104199999 | + | protein_coding |
| ENST00000555501.1 | chr14 | 104199915 | 104199946 | 104199320 | 104200001 | + | retained_intron |
| ENST00000554757.1 | chr14 | 104199915 | 104199946 | 104199320 | 104200000 | + | retained_intron |
| ENST00000336219.3 | chr14 | 105477059 | 105477097 | 105475910 | 105478272 | - | protein_coding |
| ENST00000362610.1 | chr14 | 107092677 | 107092729 | 107092677 | 107092791 | + | rRNA |
| ENST00000313077.7 | chr15 | 23005493 | 23005524 | 23002876 | 23006016 | + | protein_coding |
| ENST00000337451.3 | chr15 | 23005493 | 23005524 | 23004684 | 23006855 | - | protein_coding |
| ENST00000398014.2 | chr15 | 23005493 | 23005524 | 23004684 | 23006855 | - | protein_coding |
| ENST00000359727.3 | chr15 | 23005493 | 23005524 | 23004684 | 23006855 | - | protein_coding |
| ENST00000539711.1 | chr15 | 23005493 | 23005524 | 23004684 | 23006855 | - | protein_coding |
| ENST00000398013.2 | chr15 | 23005493 | 23005524 | 23005349 | 23006855 | - | protein_coding |
| ENST00000397609.2 | chr15 | 38776918 | 38776936 | 38776456 | 38779911 | + | protein_coding |
| ENST00000260356.4 | chr15 | 39889477 | 39889500 | 39887563 | 39891119 | + | protein_coding |
| ENST00000528967.1 | chr15 | 40939058 | 40939089 | 40939084 | 40939272 | + | processed_transcript |
| ENST00000384979.1 | chr15 | 42491792 | 42491849 | 42491768 | 42491864 | - | miRNA |
| ENST00000300289.5 | chr15 | 44063425 | 44063456 | 44063303 | 44065477 | + | protein_coding |
| ENST00000538826.1 | chr15 | 44063425 | 44063456 | 44063303 | 44063735 | + | protein_coding |
| ENST00000537673.1 | chr15 | 44063425 | 44063456 | 44063303 | 44063649 | + | protein_coding |
| ENST00000434494.1 | chr15 | 44063425 | 44063456 | 44063303 | 44063743 | + | nonsense_mediated_decay |
| ENST00000538521.1 | chr15 | 44063425 | 44063456 | 44063303 | 44063735 | + | protein_coding |
| ENST00000497349.1 | chr15 | 44063425 | 44063456 | 44063303 | 44063743 | + | retained_intron |
| ENST00000260327.4 | chr15 | 44818855 | 44818886 | 44816307 | 44821236 | + | protein_coding |
| ENST00000396780.1 | chr15 | 44818855 | 44818886 | 44816307 | 44819430 | + | protein_coding |
| ENST00000349264.5 | chr15 | 45009874 | 45009905 | 45009805 | 45010359 | + | protein_coding |
| ENST00000290894.7 | chr15 | 45493348 | 45493384 | 45492925 | 45493373 | - | protein_coding |
| ENST00000361989.2 | chr15 | 45493348 | 45493384 | 45492925 | 45493373 | - | protein_coding |
| ENST00000299259.5 | chr15 | 49419199 | 49419225 | 49417473 | 49420291 | - | protein_coding |
| ENST00000388901.4 | chr15 | 49419199 | 49419225 | 49417473 | 49420291 | - | protein_coding |
| ENST00000299259.5 | chr15 | 49419243 | 49419255 | 49417473 | 49420291 | - | protein_coding |
| ENST00000388901.4 | chr15 | 49419243 | 49419255 | 49417473 | 49420291 | - | protein_coding |
| ENST00000299259.5 | chr15 | 49419750 | 49419781 | 49417473 | 49420291 | - | protein_coding |
| ENST00000388901.4 | chr15 | 49419750 | 49419781 | 49417473 | 49420291 | - | protein_coding |
| ENST00000542928.1 | chr15 | 49419750 | 49419781 | 49419733 | 49420291 | - | protein_coding |
| ENST00000251076.4 | chr15 | 51740544 | 51740574 | 51739908 | 51741390 | - | protein_coding |
| ENST00000543779.1 | chr15 | 51740544 | 51740574 | 51739921 | 51741390 | - | protein_coding |
| ENST00000449909.2 | chr15 | 51740544 | 51740574 | 51739921 | 51741390 | - | protein_coding |
| ENST00000436119.2 | chr15 | 51740544 | 51740574 | 51739921 | 51741390 | - | protein_coding |
| ENST00000249822.3 | chr15 | 52842981 | 52843012 | 52839432 | 52844301 | - | protein_coding |
| ENST00000249822.3 | chr15 | 52843415 | 52843446 | 52839432 | 52844301 | - | protein_coding |
| ENST00000249822.3 | chr15 | 52843653 | 52843684 | 52839432 | 52844301 | - | protein_coding |
| ENST00000249822.3 | chr15 | 52844076 | 52844076 | 52839432 | 52844301 | - | protein_coding |
| ENST00000261844.6 | chr15 | 52902300 | 52902322 | 52900765 | 52902580 | - | protein_coding |
| ENST00000399202.3 | chr15 | 52902300 | 52902322 | 52900765 | 52902580 | - | protein_coding |
| ENST00000534964.1 | chr15 | 52902300 | 52902322 | 52900765 | 52902580 | - | protein_coding |
| ENST00000546305.1 | chr15 | 52902300 | 52902322 | 52900765 | 52902580 | - | protein_coding |
| ENST00000385229.1 | chr15 | 55665141 | 55665172 | 55665138 | 55665232 | - | miRNA |
| ENST00000348370.3 | chr15 | 59350609 | 59350640 | 59350555 | 59350749 | + | protein_coding |
| ENST00000434298.1 | chr15 | 59350609 | 59350640 | 59350555 | 59350749 | + | protein_coding |
| ENST00000267859.3 | chr15 | 59955293 | 59955324 | 59951345 | 59956319 | - | protein_coding |
| ENST00000415213.1 | chr15 | 59955293 | 59955324 | 59955064 | 59956319 | - | protein_coding |
| ENST00000478981.1 | chr15 | 59955293 | 59955324 | 59955092 | 59956319 | - | processed_transcript |
| ENST00000335670.6 | chr15 | 60788969 | 60789000 | 60780483 | 60789818 | - | protein_coding |
| ENST00000449337.1 | chr15 | 60788969 | 60789000 | 60780483 | 60789818 | - | protein_coding |
| ENST00000335670.6 | chr15 | 61521505 | 61521545 | 61521252 | 61521518 | - | protein_coding |
| ENST00000388402.1 | chr15 | 63116170 | 63116201 | 63116156 | 63116240 | + | miRNA |
| ENST00000178638.3 | chr15 | 63617959 | 63617990 | 63613577 | 63618556 | - | protein_coding |
| ENST00000344366.3 | chr15 | 63617959 | 63617990 | 63616887 | 63618556 | - | protein_coding |
| ENST00000422263.2 | chr15 | 63617959 | 63617990 | 63616906 | 63618556 | - | protein_coding |
| ENST00000300035.4 | chr15 | 64673172 | 64673203 | 64673157 | 64673237 | - | protein_coding |
| ENST00000380258.1 | chr15 | 64673172 | 64673203 | 64673157 | 64673237 | - | protein_coding |
| ENST00000204566.2 | chr15 | 65261618 | 65261649 | 65261592 | 65261699 | - | protein_coding |
| ENST00000416889.1 | chr15 | 65261618 | 65261649 | 65261592 | 65261699 | - | protein_coding |
| ENST00000433215.1 | chr15 | 65261618 | 65261649 | 65261592 | 65261699 | - | protein_coding |
| ENST00000362698.1 | chr15 | 65588432 | 65588466 | 65588389 | 65588504 | + | snRNA |
| ENST00000363286.1 | chr15 | 65597058 | 65597091 | 65597015 | 65597130 | + | snRNA |
| ENST00000363286.1 | chr15 | 65597102 | 65597108 | 65597015 | 65597130 | + | snRNA |
| ENST00000352385.2 | chr15 | 65675492 | 65675524 | 65673802 | 65676757 | - | protein_coding |
| ENST00000356152.1 | chr15 | 65675492 | 65675524 | 65673825 | 65676757 | - | protein_coding |
| ENST00000261875.4 | chr15 | 65870147 | 65870178 | 65868641 | 65870693 | + | protein_coding |
| ENST00000261890.2 | chr15 | 66169800 | 66169831 | 66169670 | 66169865 | + | protein_coding |
| ENST00000435304.2 | chr15 | 66169800 | 66169831 | 66169670 | 66169865 | + | protein_coding |
| ENST00000307102.3 | chr15 | 66783334 | 66783364 | 66782840 | 66783882 | + | protein_coding |
| ENST00000365659.1 | chr15 | 66794366 | 66794393 | 66794358 | 66794429 | - | snoRNA |
| ENST00000307961.5 | chr15 | 66794994 | 66795025 | 66794950 | 66795088 | - | protein_coding |
| ENST00000449253.1 | chr15 | 66794994 | 66795025 | 66795014 | 66795088 | - | protein_coding |
| ENST00000432669.1 | chr15 | 66794994 | 66795025 | 66794950 | 66795088 | - | protein_coding |
| ENST00000260363.4 | chr15 | 69740564 | 69740595 | 69740131 | 69740756 | + | protein_coding |
| ENST00000352331.3 | chr15 | 69740564 | 69740595 | 69740131 | 69740756 | + | protein_coding |
| ENST00000260379.5 | chr15 | 69745286 | 69745300 | 69745123 | 69745359 | + | protein_coding |
| ENST00000357790.4 | chr15 | 69745286 | 69745300 | 69745157 | 69745359 | + | protein_coding |
| ENST00000487304.1 | chr15 | 69745286 | 69745300 | 69745158 | 69745359 | + | retained_intron |
| ENST00000260379.5 | chr15 | 69747834 | 69747873 | 69747767 | 69748172 | + | protein_coding |
| ENST00000357790.4 | chr15 | 69747834 | 69747873 | 69747767 | 69747885 | + | protein_coding |
| ENST00000488122.1 | chr15 | 69747834 | 69747873 | 69747767 | 69747885 | + | retained_intron |
| ENST00000442299.2 | chr15 | 70368445 | 70368476 | 70368435 | 70368497 | - | protein_coding |
| ENST00000451782.1 | chr15 | 70368445 | 70368476 | 70368435 | 70368497 | - | protein_coding |
| ENST00000317509.7 | chr15 | 70368445 | 70368476 | 70368435 | 70368497 | - | protein_coding |
| ENST00000440567.2 | chr15 | 70368445 | 70368476 | 70368435 | 70368497 | - | protein_coding |
| ENST00000539550.1 | chr15 | 70368445 | 70368476 | 70368435 | 70368497 | - | protein_coding |
| ENST00000385230.1 | chr15 | 70371716 | 70371747 | 70371711 | 70371807 | - | miRNA |
| ENST00000319622.6 | chr15 | 72492930 | 72492961 | 72492815 | 72492996 | - | protein_coding |
| ENST00000335181.4 | chr15 | 72492930 | 72492961 | 72492815 | 72492996 | - | protein_coding |
| ENST00000434220.1 | chr15 | 72492930 | 72492961 | 72492815 | 72492996 | - | protein_coding |
| ENST00000327974.8 | chr15 | 72492930 | 72492961 | 72492815 | 72492996 | - | protein_coding |
| ENST00000389093.3 | chr15 | 72492930 | 72492961 | 72492815 | 72492996 | - | protein_coding |
| ENST00000449901.2 | chr15 | 72492930 | 72492961 | 72492815 | 72492996 | - | protein_coding |
| ENST00000319622.6 | chr15 | 72511301 | 72511332 | 72511285 | 72511451 | - | protein_coding |
| ENST00000335181.4 | chr15 | 72511301 | 72511332 | 72511285 | 72511451 | - | protein_coding |
| ENST00000434220.1 | chr15 | 72511301 | 72511332 | 72511285 | 72511451 | - | protein_coding |
| ENST00000327974.8 | chr15 | 72511301 | 72511332 | 72511320 | 72511451 | - | protein_coding |
| ENST00000389093.3 | chr15 | 72511301 | 72511332 | 72511285 | 72511451 | - | protein_coding |
| ENST00000449901.2 | chr15 | 72511301 | 72511332 | 72511285 | 72511451 | - | protein_coding |
| ENST00000379887.4 | chr15 | 72879618 | 72879649 | 72875549 | 72879692 | + | protein_coding |
| ENST00000384957.1 | chr15 | 72879618 | 72879649 | 72879558 | 72879654 | + | miRNA |
| ENST00000268099.8 | chr15 | 75136696 | 75136698 | 75136071 | 75137558 | - | protein_coding |
| ENST00000543345.1 | chr15 | 75136696 | 75136698 | 75136073 | 75137934 | - | protein_coding |
| ENST00000384904.1 | chr15 | 75645980 | 75646011 | 75645952 | 75646026 | - | miRNA |
| ENST00000338677.4 | chr15 | 76192572 | 76192604 | 76191768 | 76193380 | + | protein_coding |
| ENST00000267938.4 | chr15 | 76192572 | 76192604 | 76191768 | 76193419 | + | protein_coding |
| ENST00000426727.1 | chr15 | 76192572 | 76192604 | 76191768 | 76193380 | + | protein_coding |
| ENST00000426013.1 | chr15 | 79165348 | 79165379 | 79165172 | 79165399 | + | protein_coding |
| ENST00000331268.4 | chr15 | 79165348 | 79165379 | 79165211 | 79165399 | + | protein_coding |
| ENST00000261749.6 | chr15 | 80430401 | 80430433 | 80429822 | 80430735 | + | protein_coding |
| ENST00000385268.1 | chr15 | 81134323 | 81134354 | 81134319 | 81134414 | - | miRNA |
| ENST00000516881.1 | chr15 | 83424787 | 83424820 | 83424697 | 83424823 | + | snoRNA |
| ENST00000361243.1 | chr15 | 86289902 | 86289933 | 86287859 | 86292586 | + | protein_coding |
| ENST00000394518.1 | chr15 | 86289902 | 86289933 | 86287859 | 86292586 | + | protein_coding |
| ENST00000394516.1 | chr15 | 86289902 | 86289933 | 86287859 | 86292586 | + | protein_coding |
| ENST00000458540.1 | chr15 | 86289902 | 86289933 | 86287859 | 86292586 | + | protein_coding |
| ENST00000394510.2 | chr15 | 86289902 | 86289933 | 86287859 | 86292586 | + | protein_coding |
| ENST00000384970.1 | chr15 | 89155087 | 89155118 | 89155056 | 89155165 | + | miRNA |
| ENST00000336418.4 | chr15 | 90378434 | 90378459 | 90373831 | 90378875 | - | protein_coding |
| ENST00000423566.2 | chr15 | 90378434 | 90378459 | 90378214 | 90378875 | - | protein_coding |
| ENST00000398333.3 | chr15 | 90378434 | 90378459 | 90377540 | 90378875 | - | protein_coding |
| ENST00000410144.1 | chr15 | 96289190 | 96289219 | 96289033 | 96289223 | + | snRNA |
| ENST00000421109.1 | chr15 | 96881521 | 96881551 | 96880577 | 96883492 | + | protein_coding |
| ENST00000394166.3 | chr15 | 96881521 | 96881551 | 96880577 | 96883492 | + | protein_coding |
| ENST00000394171.2 | chr15 | 96881521 | 96881551 | 96880577 | 96883492 | + | protein_coding |
| ENST00000453270.1 | chr15 | 96881521 | 96881551 | 96880577 | 96883492 | + | protein_coding |
| ENST00000421109.1 | chr15 | 96881823 | 96881854 | 96880577 | 96883492 | + | protein_coding |
| ENST00000394166.3 | chr15 | 96881823 | 96881854 | 96880577 | 96883492 | + | protein_coding |
| ENST00000394171.2 | chr15 | 96881823 | 96881854 | 96880577 | 96883492 | + | protein_coding |
| ENST00000453270.1 | chr15 | 96881823 | 96881854 | 96880577 | 96883492 | + | protein_coding |
| ENST00000268035.6 | chr15 | 99503285 | 99503298 | 99500290 | 99507759 | + | protein_coding |
| ENST00000383903.1 | chr16 | 2205074 | 2205106 | 2205024 | 2205106 | - | snoRNA |
| ENST00000320225.4 | chr16 | 2303456 | 2303471 | 2303117 | 2304064 | - | protein_coding |
| ENST00000397086.1 | chr16 | 2303456 | 2303471 | 2303124 | 2304064 | - | protein_coding |
| ENST00000301740.7 | chr16 | 2809615 | 2809646 | 2809570 | 2809662 | + | protein_coding |
| ENST00000382301.2 | chr16 | 2809615 | 2809646 | 2809570 | 2809662 | + | protein_coding |
| ENST00000396975.1 | chr16 | 2809615 | 2809646 | 2809570 | 2809662 | + | protein_coding |
| ENST00000544933.1 | chr16 | 2809615 | 2809646 | 2809570 | 2809662 | + | protein_coding |
| ENST00000426305.2 | chr16 | 2809615 | 2809646 | 2809570 | 2809662 | + | protein_coding |
| ENST00000262300.7 | chr16 | 3022915 | 3022933 | 3022792 | 3023065 | - | protein_coding |
| ENST00000440027.1 | chr16 | 3022915 | 3022933 | 3022794 | 3023065 | - | protein_coding |
| ENST00000402679.2 | chr16 | 3022915 | 3022933 | 3022820 | 3023065 | - | protein_coding |
| ENST00000318782.8 | chr16 | 3022915 | 3022933 | 3021516 | 3023490 | + | protein_coding |
| ENST00000293978.7 | chr16 | 3022915 | 3022933 | 3021516 | 3023483 | + | protein_coding |
| ENST00000470337.1 | chr16 | 3221000 | 3221028 | 3220961 | 3221031 | - | tRNA_pseudogene |
| ENST00000396870.3 | chr16 | 3273682 | 3273713 | 3272325 | 3274613 | - | protein_coding |
| ENST00000396868.3 | chr16 | 3273682 | 3273713 | 3272325 | 3274613 | - | protein_coding |
| ENST00000396871.3 | chr16 | 3273682 | 3273713 | 3272325 | 3274610 | - | protein_coding |
| ENST00000414144.1 | chr16 | 3273682 | 3273713 | 3272325 | 3274610 | - | protein_coding |
| ENST00000431561.2 | chr16 | 3273682 | 3273713 | 3272325 | 3274613 | - | protein_coding |
| ENST00000293995.3 | chr16 | 3367456 | 3367487 | 3367190 | 3368574 | + | protein_coding |
| ENST00000498240.1 | chr16 | 3367456 | 3367487 | 3367190 | 3368574 | + | processed_transcript |
| ENST00000322048.5 | chr16 | 4848111 | 4848142 | 4848072 | 4848185 | - | protein_coding |
| ENST00000304414.6 | chr16 | 18803208 | 18803241 | 18802991 | 18804692 | - | protein_coding |
| ENST00000304414.6 | chr16 | 18804470 | 18804501 | 18802991 | 18804692 | - | protein_coding |
| ENST00000545430.1 | chr16 | 18804470 | 18804501 | 18803775 | 18804692 | - | protein_coding |
| ENST00000546206.1 | chr16 | 18804470 | 18804501 | 18804269 | 18804692 | - | protein_coding |
| ENST00000381039.2 | chr16 | 24583140 | 24583170 | 24582197 | 24584181 | + | protein_coding |
| ENST00000319715.4 | chr16 | 24583140 | 24583170 | 24582197 | 24584184 | + | protein_coding |
| ENST00000348022.2 | chr16 | 24583140 | 24583170 | 24582197 | 24584184 | + | protein_coding |
| ENST00000304516.6 | chr16 | 30016678 | 30016709 | 30016542 | 30017114 | + | protein_coding |
| ENST00000540562.1 | chr16 | 30016678 | 30016709 | 30016542 | 30017111 | + | protein_coding |
| ENST00000263025.4 | chr16 | 30129020 | 30129051 | 30128991 | 30129105 | - | protein_coding |
| ENST00000466521.1 | chr16 | 30129020 | 30129051 | 30128991 | 30129105 | - | nonsense_mediated_decay |
| ENST00000484663.1 | chr16 | 30129020 | 30129051 | 30128991 | 30129105 | - | protein_coding |
| ENST00000322266.5 | chr16 | 30129020 | 30129051 | 30128991 | 30129105 | - | protein_coding |
| ENST00000403394.1 | chr16 | 30129020 | 30129051 | 30128991 | 30129105 | - | protein_coding |
| ENST00000490298.1 | chr16 | 30129020 | 30129051 | 30128991 | 30129105 | - | nonsense_mediated_decay |
| ENST00000395200.1 | chr16 | 30129020 | 30129051 | 30128991 | 30129105 | - | protein_coding |
| ENST00000395202.1 | chr16 | 30129020 | 30129051 | 30128991 | 30129105 | - | protein_coding |
| ENST00000478356.1 | chr16 | 30129020 | 30129051 | 30128991 | 30129105 | - | protein_coding |
| ENST00000395199.3 | chr16 | 30129020 | 30129051 | 30128991 | 30129105 | - | protein_coding |
| ENST00000495629.1 | chr16 | 30129020 | 30129051 | 30128991 | 30129105 | - | protein_coding |
| ENST00000262519.7 | chr16 | 30977075 | 30977106 | 30976922 | 30977707 | + | protein_coding |
| ENST00000300870.8 | chr16 | 31928412 | 31928443 | 31925797 | 31928622 | + | protein_coding |
| ENST00000394846.2 | chr16 | 31928412 | 31928443 | 31927639 | 31928622 | + | protein_coding |
| ENST00000356559.1 | chr16 | 31985553 | 31985588 | 31985543 | 31985682 | + | protein_coding |
| ENST00000262384.2 | chr16 | 48576616 | 48576647 | 48572639 | 48577172 | - | protein_coding |
| ENST00000436909.2 | chr16 | 50265629 | 50265647 | 50263063 | 50269221 | + | protein_coding |
| ENST00000357464.3 | chr16 | 50265629 | 50265647 | 50263063 | 50269221 | + | protein_coding |
| ENST00000251020.3 | chr16 | 51175596 | 51175619 | 51172599 | 51176056 | - | protein_coding |
| ENST00000440970.1 | chr16 | 51175596 | 51175619 | 51172599 | 51176056 | - | protein_coding |
| ENST00000457559.2 | chr16 | 51175596 | 51175619 | 51172599 | 51175710 | - | protein_coding |
| ENST00000264010.4 | chr16 | 67671876 | 67671907 | 67671591 | 67673086 | + | protein_coding |
| ENST00000401394.1 | chr16 | 67671876 | 67671907 | 67671591 | 67673086 | + | protein_coding |
| ENST00000363688.1 | chr16 | 68776404 | 68776440 | 68776383 | 68776491 | + | rRNA |
| ENST00000512062.1 | chr16 | 69497635 | 69497666 | 69496333 | 69500169 | + | protein_coding |
| ENST00000307892.8 | chr16 | 69497635 | 69497666 | 69496333 | 69500167 | + | protein_coding |
| ENST00000515314.1 | chr16 | 69497635 | 69497666 | 69496333 | 69497681 | + | protein_coding |
| ENST00000302516.5 | chr16 | 70606022 | 70606054 | 70605576 | 70608820 | + | protein_coding |
| ENST00000310750.4 | chr16 | 70606022 | 70606054 | 70605964 | 70606044 | + | protein_coding |
| ENST00000338099.4 | chr16 | 71318613 | 71318644 | 71316203 | 71319842 | - | protein_coding |
| ENST00000434935.1 | chr16 | 71318613 | 71318644 | 71316206 | 71319842 | - | protein_coding |
| ENST00000356272.3 | chr16 | 71679003 | 71679034 | 71678852 | 71683947 | - | protein_coding |
| ENST00000393524.1 | chr16 | 71679003 | 71679034 | 71678852 | 71683947 | - | protein_coding |
| ENST00000411292.1 | chr16 | 71792309 | 71792340 | 71792305 | 71792390 | - | snoRNA |
| ENST00000248272.2 | chr16 | 81413595 | 81413626 | 81411020 | 81413940 | + | protein_coding |
| ENST00000301030.3 | chr16 | 89334072 | 89334103 | 89334035 | 89335071 | - | protein_coding |
| ENST00000378330.2 | chr16 | 89334072 | 89334103 | 89334038 | 89335071 | - | protein_coding |
| ENST00000311528.4 | chr16 | 89629594 | 89629625 | 89629292 | 89630950 | + | protein_coding |
| ENST00000452368.2 | chr16 | 89629594 | 89629625 | 89629292 | 89630360 | + | protein_coding |
| ENST00000393099.3 | chr16 | 89629594 | 89629625 | 89629292 | 89630950 | + | protein_coding |
| ENST00000472354.1 | chr16 | 89629594 | 89629625 | 89629292 | 89629846 | + | processed_transcript |
| ENST00000304992.5 | chr17 | 1576378 | 1576409 | 1576375 | 1576491 | - | protein_coding |
| ENST00000540177.1 | chr17 | 1576378 | 1576409 | 1576375 | 1576491 | - | protein_coding |
| ENST00000304992.5 | chr17 | 1580900 | 1580931 | 1580859 | 1580988 | - | protein_coding |
| ENST00000540177.1 | chr17 | 1580900 | 1580931 | 1580859 | 1580988 | - | protein_coding |
| ENST00000304992.5 | chr17 | 1584225 | 1584268 | 1584223 | 1584348 | - | protein_coding |
| ENST00000540177.1 | chr17 | 1584225 | 1584268 | 1584223 | 1584348 | - | protein_coding |
| ENST00000362190.1 | chr17 | 1617198 | 1617229 | 1617191 | 1617285 | - | miRNA |
| ENST00000516003.1 | chr17 | 4089643 | 4089674 | 4089621 | 4089719 | + | misc_RNA |
| ENST00000262482.5 | chr17 | 4846179 | 4846210 | 4846146 | 4846236 | + | protein_coding |
| ENST00000225655.5 | chr17 | 4849076 | 4849107 | 4848947 | 4849292 | - | protein_coding |
| ENST00000225698.3 | chr17 | 5337047 | 5337078 | 5336989 | 5337087 | - | protein_coding |
| ENST00000381165.3 | chr17 | 5394091 | 5394108 | 5392143 | 5394134 | + | protein_coding |
| ENST00000311403.3 | chr17 | 7364943 | 7364974 | 7362685 | 7367209 | - | protein_coding |
| ENST00000380599.3 | chr17 | 7364943 | 7364974 | 7362688 | 7367209 | - | protein_coding |
| ENST00000315684.7 | chr17 | 8129943 | 8129945 | 8128140 | 8131637 | - | protein_coding |
| ENST00000449476.1 | chr17 | 8129943 | 8129945 | 8128140 | 8131637 | - | protein_coding |
| ENST00000315684.7 | chr17 | 8129983 | 8130014 | 8128140 | 8131637 | - | protein_coding |
| ENST00000449476.1 | chr17 | 8129983 | 8130014 | 8128140 | 8131637 | - | protein_coding |
| ENST00000390234.1 | chr17 | 11985226 | 11985257 | 11985216 | 11985313 | + | miRNA |
| ENST00000302182.3 | chr17 | 16285948 | 16285979 | 16285216 | 16286054 | + | protein_coding |
| ENST00000535788.1 | chr17 | 16285948 | 16285979 | 16285673 | 16286053 | + | protein_coding |
| ENST00000395839.1 | chr17 | 16285948 | 16285979 | 16285216 | 16286054 | + | protein_coding |
| ENST00000395837.1 | chr17 | 16285948 | 16285979 | 16285216 | 16286054 | + | protein_coding |
| ENST00000384229.1 | chr17 | 16343391 | 16343420 | 16343350 | 16343420 | + | snoRNA |
| ENST00000268717.4 | chr17 | 17165290 | 17165321 | 17165280 | 17165420 | - | protein_coding |
| ENST00000539941.1 | chr17 | 17165290 | 17165321 | 17165280 | 17165420 | - | protein_coding |
| ENST00000486810.1 | chr17 | 17165290 | 17165321 | 17165280 | 17165420 | - | processed_transcript |
| ENST00000417352.1 | chr17 | 17165290 | 17165321 | 17165280 | 17165420 | - | protein_coding |
| ENST00000439936.1 | chr17 | 17165290 | 17165321 | 17165283 | 17165420 | - | protein_coding |
| ENST00000492672.1 | chr17 | 17165290 | 17165321 | 17165312 | 17165420 | - | processed_transcript |
| ENST00000385104.1 | chr17 | 17717199 | 17717230 | 17717150 | 17717245 | - | miRNA |
| ENST00000363359.1 | chr17 | 18965409 | 18965444 | 18965225 | 18965440 | + | snoRNA |
| ENST00000364880.1 | chr17 | 18967229 | 18967265 | 18967234 | 18967449 | - | snoRNA |
| ENST00000362793.1 | chr17 | 19015729 | 19015765 | 19015734 | 19015949 | - | snoRNA |
| ENST00000365494.1 | chr17 | 19091512 | 19091548 | 19091329 | 19091544 | + | snoRNA |
| ENST00000362428.1 | chr17 | 19093339 | 19093375 | 19093343 | 19093558 | - | snoRNA |
| ENST00000474898.1 | chr17 | 19355690 | 19355720 | 19355539 | 19356104 | + | rRNA_pseudogene |
| ENST00000455244.1 | chr17 | 20688182 | 20688213 | 20687958 | 20689040 | + | processed_pseudogene |
| ENST00000226230.5 | chr17 | 26654840 | 26654871 | 26653560 | 26655711 | + | protein_coding |
| ENST00000336687.5 | chr17 | 26654840 | 26654871 | 26653560 | 26655250 | + | protein_coding |
| ENST00000226225.2 | chr17 | 26673942 | 26673973 | 26671390 | 26674035 | + | protein_coding |
| ENST00000473896.1 | chr17 | 27049638 | 27049667 | 27049439 | 27049917 | + | processed_transcript |
| ENST00000459174.1 | chr17 | 27049638 | 27049667 | 27049600 | 27049671 | + | snoRNA |
| ENST00000254928.4 | chr17 | 27182143 | 27182174 | 27182020 | 27182335 | + | protein_coding |
| ENST00000412138.1 | chr17 | 27182143 | 27182174 | 27182054 | 27182335 | + | protein_coding |
| ENST00000461894.1 | chr17 | 27182143 | 27182174 | 27182054 | 27182335 | + | processed_transcript |
| ENST00000385059.1 | chr17 | 27188411 | 27188442 | 27188387 | 27188458 | - | miRNA |
| ENST00000225388.2 | chr17 | 27584551 | 27584582 | 27582855 | 27591609 | - | protein_coding |
| ENST00000225388.2 | chr17 | 27589658 | 27589688 | 27582855 | 27591609 | - | protein_coding |
| ENST00000225388.2 | chr17 | 27613668 | 27613699 | 27613010 | 27614734 | - | protein_coding |
| ENST00000225388.2 | chr17 | 27613855 | 27613886 | 27613010 | 27614734 | - | protein_coding |
| ENST00000261716.2 | chr17 | 27869997 | 27870028 | 27869579 | 27871502 | + | protein_coding |
| ENST00000536202.1 | chr17 | 27869997 | 27870028 | 27869579 | 27871358 | + | protein_coding |
| ENST00000307201.2 | chr17 | 27889824 | 27889855 | 27887691 | 27890104 | - | protein_coding |
| ENST00000269033.3 | chr17 | 27956908 | 27956939 | 27952956 | 27959949 | - | protein_coding |
| ENST00000467446.1 | chr17 | 28444148 | 28444180 | 28444172 | 28444350 | + | processed_transcript |
| ENST00000362201.2 | chr17 | 28444148 | 28444180 | 28444097 | 28444190 | + | miRNA |
| ENST00000394670.3 | chr17 | 30696875 | 30696906 | 30696618 | 30708905 | + | protein_coding |
| ENST00000394673.1 | chr17 | 30696875 | 30696906 | 30696618 | 30697362 | + | protein_coding |
| ENST00000321233.5 | chr17 | 30696875 | 30696906 | 30696618 | 30697467 | + | protein_coding |
| ENST00000341711.6 | chr17 | 30696875 | 30696906 | 30696618 | 30697467 | + | protein_coding |
| ENST00000342555.5 | chr17 | 30696875 | 30696906 | 30696618 | 30696956 | + | protein_coding |
| ENST00000394670.3 | chr17 | 30696979 | 30697010 | 30696618 | 30708905 | + | protein_coding |
| ENST00000394673.1 | chr17 | 30696979 | 30697010 | 30696618 | 30697362 | + | protein_coding |
| ENST00000321233.5 | chr17 | 30696979 | 30697010 | 30696618 | 30697467 | + | protein_coding |
| ENST00000341711.6 | chr17 | 30696979 | 30697010 | 30696618 | 30697467 | + | protein_coding |
| ENST00000394670.3 | chr17 | 30701221 | 30701249 | 30696618 | 30708905 | + | protein_coding |
| ENST00000261712.2 | chr17 | 30806303 | 30806334 | 30806269 | 30806394 | + | protein_coding |
| ENST00000457654.1 | chr17 | 30806303 | 30806334 | 30806269 | 30806394 | + | protein_coding |
| ENST00000493026.1 | chr17 | 30806303 | 30806334 | 30806269 | 30807215 | + | retained_intron |
| ENST00000318217.4 | chr17 | 31048156 | 31048187 | 31048041 | 31048207 | - | protein_coding |
| ENST00000394649.3 | chr17 | 31048156 | 31048187 | 31048041 | 31048207 | - | protein_coding |
| ENST00000442241.2 | chr17 | 33462333 | 33462347 | 33462268 | 33462470 | - | protein_coding |
| ENST00000360831.4 | chr17 | 33462333 | 33462347 | 33462268 | 33462470 | - | protein_coding |
| ENST00000436188.1 | chr17 | 33462333 | 33462347 | 33462268 | 33462470 | - | protein_coding |
| ENST00000537697.1 | chr17 | 33462333 | 33462347 | 33462268 | 33462470 | - | protein_coding |
| ENST00000384567.1 | chr17 | 33900676 | 33900707 | 33900676 | 33900772 | + | snoRNA |
| ENST00000262325.6 | chr17 | 34052435 | 34052466 | 34050644 | 34053434 | + | protein_coding |
| ENST00000312678.7 | chr17 | 34052435 | 34052466 | 34050644 | 34053434 | + | protein_coding |
| ENST00000225430.4 | chr17 | 37360897 | 37360928 | 37360778 | 37360980 | + | protein_coding |
| ENST00000302584.3 | chr17 | 37761009 | 37761048 | 37760021 | 37762857 | - | protein_coding |
| ENST00000264639.4 | chr17 | 38140683 | 38140714 | 38140547 | 38140737 | + | protein_coding |
| ENST00000415039.2 | chr17 | 38140683 | 38140714 | 38140547 | 38140737 | + | protein_coding |
| ENST00000541736.1 | chr17 | 38140683 | 38140714 | 38140712 | 38140737 | + | protein_coding |
| ENST00000339569.4 | chr17 | 38285566 | 38285585 | 38285498 | 38285880 | + | protein_coding |
| ENST00000398532.2 | chr17 | 38285566 | 38285585 | 38285498 | 38285880 | + | protein_coding |
| ENST00000363865.1 | chr17 | 38399542 | 38399569 | 38399475 | 38399571 | + | misc_RNA |
| ENST00000323571.4 | chr17 | 38439407 | 38439443 | 38434437 | 38440388 | + | protein_coding |
| ENST00000462917.1 | chr17 | 39847368 | 39847400 | 39847034 | 39847898 | + | retained_intron |
| ENST00000469257.1 | chr17 | 39847368 | 39847400 | 39847034 | 39848920 | + | protein_coding |
| ENST00000482111.1 | chr17 | 39847368 | 39847400 | 39847034 | 39847874 | + | retained_intron |
| ENST00000310837.4 | chr17 | 39847368 | 39847400 | 39847034 | 39847882 | + | processed_transcript |
| ENST00000479972.1 | chr17 | 39847368 | 39847400 | 39847034 | 39847440 | + | retained_intron |
| ENST00000365050.1 | chr17 | 39874406 | 39874437 | 39874406 | 39874512 | + | rRNA |
| ENST00000363640.1 | chr17 | 41084928 | 41084959 | 41084865 | 41084959 | - | misc_RNA |
| ENST00000253788.4 | chr17 | 41152091 | 41152122 | 41151950 | 41152119 | + | protein_coding |
| ENST00000411012.1 | chr17 | 41464717 | 41464741 | 41464594 | 41464785 | - | snRNA |
| ENST00000262419.5 | chr17 | 44108742 | 44108773 | 44107334 | 44109069 | - | protein_coding |
| ENST00000432791.1 | chr17 | 44108742 | 44108773 | 44107334 | 44109069 | - | protein_coding |
| ENST00000393476.3 | chr17 | 44108742 | 44108773 | 44107334 | 44109069 | - | protein_coding |
| ENST00000290158.3 | chr17 | 45760463 | 45760489 | 45759770 | 45760998 | + | protein_coding |
| ENST00000225573.3 | chr17 | 46024836 | 46024867 | 46023980 | 46026673 | + | protein_coding |
| ENST00000484302.2 | chr17 | 46675591 | 46675622 | 46675098 | 46675622 | - | protein_coding |
| ENST00000490419.1 | chr17 | 46675591 | 46675622 | 46675573 | 46675622 | - | processed_transcript |
| ENST00000239144.3 | chr17 | 46690125 | 46690157 | 46688446 | 46690871 | - | protein_coding |
| ENST00000384097.1 | chr17 | 48463515 | 48463545 | 48463506 | 48463607 | - | misc_RNA |
| ENST00000393196.3 | chr17 | 49239227 | 49239258 | 49239089 | 49239556 | + | protein_coding |
| ENST00000336097.3 | chr17 | 49239227 | 49239258 | 49239089 | 49239422 | + | protein_coding |
| ENST00000511355.1 | chr17 | 49239227 | 49239258 | 49238459 | 49239789 | + | protein_coding |
| ENST00000475573.1 | chr17 | 49239227 | 49239258 | 49239089 | 49239396 | + | nonsense_mediated_decay |
| ENST00000013034.3 | chr17 | 49239227 | 49239258 | 49239089 | 49239259 | + | protein_coding |
| ENST00000405860.1 | chr17 | 49281216 | 49281247 | 49281152 | 49281286 | - | protein_coding |
| ENST00000415868.1 | chr17 | 49281216 | 49281247 | 49281152 | 49281286 | - | protein_coding |
| ENST00000436971.1 | chr17 | 49281216 | 49281247 | 49281152 | 49281286 | - | nonsense_mediated_decay |
| ENST00000486816.1 | chr17 | 49281216 | 49281247 | 49281152 | 49281286 | - | retained_intron |
| ENST00000376381.2 | chr17 | 49281216 | 49281247 | 49281152 | 49281286 | - | protein_coding |
| ENST00000299377.4 | chr17 | 53797270 | 53797301 | 53796990 | 53798496 | - | protein_coding |
| ENST00000424486.1 | chr17 | 53797270 | 53797301 | 53796990 | 53798496 | - | protein_coding |
| ENST00000240316.3 | chr17 | 55015632 | 55015663 | 55015563 | 55016515 | - | protein_coding |
| ENST00000258962.3 | chr17 | 56084362 | 56084392 | 56084305 | 56084707 | - | protein_coding |
| ENST00000384835.1 | chr17 | 56408597 | 56408628 | 56408593 | 56408679 | - | miRNA |
| ENST00000516403.1 | chr17 | 57206768 | 57206799 | 57206637 | 57206819 | - | snRNA |
| ENST00000362134.1 | chr17 | 57918634 | 57918665 | 57918627 | 57918698 | + | miRNA |
| ENST00000397786.1 | chr17 | 60112873 | 60112904 | 60112824 | 60112969 | - | protein_coding |
| ENST00000262436.4 | chr17 | 60112873 | 60112904 | 60112824 | 60112969 | - | protein_coding |
| ENST00000310827.4 | chr17 | 61667088 | 61667119 | 61666362 | 61671639 | + | protein_coding |
| ENST00000415273.2 | chr17 | 61667088 | 61667119 | 61666362 | 61667147 | + | protein_coding |
| ENST00000540698.1 | chr17 | 62496112 | 62496144 | 62494374 | 62496444 | - | protein_coding |
| ENST00000450599.1 | chr17 | 62496112 | 62496144 | 62495741 | 62496444 | - | protein_coding |
| ENST00000225792.4 | chr17 | 62496112 | 62496144 | 62495909 | 62496444 | - | protein_coding |
| ENST00000400873.3 | chr17 | 62745953 | 62745976 | 62745781 | 62746126 | - | protein_coding |
| ENST00000439174.1 | chr17 | 63009170 | 63009200 | 63005407 | 63010947 | - | protein_coding |
| ENST00000330459.2 | chr17 | 66039230 | 66039261 | 66039216 | 66039479 | + | protein_coding |
| ENST00000537025.1 | chr17 | 66039230 | 66039261 | 66039216 | 66039479 | + | protein_coding |
| ENST00000330459.2 | chr17 | 66040532 | 66040563 | 66040437 | 66040619 | + | protein_coding |
| ENST00000537025.1 | chr17 | 66040532 | 66040563 | 66040437 | 66040619 | + | protein_coding |
| ENST00000330459.2 | chr17 | 66042815 | 66042837 | 66042620 | 66042969 | + | protein_coding |
| ENST00000537025.1 | chr17 | 66042815 | 66042837 | 66042620 | 66042958 | + | protein_coding |
| ENST00000384830.1 | chr17 | 66420643 | 66420674 | 66420592 | 66420689 | - | miRNA |
| ENST00000243457.2 | chr17 | 68175206 | 68175238 | 68170965 | 68176160 | + | protein_coding |
| ENST00000316804.5 | chr17 | 73314329 | 73314360 | 73314157 | 73316634 | - | protein_coding |
| ENST00000392562.1 | chr17 | 73314329 | 73314360 | 73314157 | 73316634 | - | protein_coding |
| ENST00000392564.1 | chr17 | 73314329 | 73314360 | 73314157 | 73316634 | - | protein_coding |
| ENST00000392563.1 | chr17 | 73314329 | 73314360 | 73314157 | 73316634 | - | protein_coding |
| ENST00000316615.5 | chr17 | 73314329 | 73314360 | 73314157 | 73316634 | - | protein_coding |
| ENST00000535682.1 | chr17 | 73314329 | 73314360 | 73314159 | 73315037 | - | pseudogene |
| ENST00000254810.3 | chr17 | 73774322 | 73774361 | 73772517 | 73774804 | - | protein_coding |
| ENST00000254810.3 | chr17 | 73774582 | 73774593 | 73772517 | 73774804 | - | protein_coding |
| ENST00000254806.3 | chr17 | 73842672 | 73842703 | 73841780 | 73842868 | - | protein_coding |
| ENST00000344296.4 | chr17 | 73842672 | 73842703 | 73842404 | 73842868 | - | protein_coding |
| ENST00000433525.2 | chr17 | 73842672 | 73842703 | 73842576 | 73842868 | - | protein_coding |
| ENST00000431190.1 | chr17 | 73842672 | 73842703 | 73842663 | 73842827 | - | protein_coding |
| ENST00000269383.3 | chr17 | 73885931 | 73885962 | 73885041 | 73887428 | - | protein_coding |
| ENST00000494879.1 | chr17 | 75158336 | 75158380 | 75158032 | 75158389 | - | rRNA_pseudogene |
| ENST00000455761.2 | chr17 | 76045714 | 76045745 | 76044926 | 76047538 | + | protein_coding |
| ENST00000395801.3 | chr17 | 76045714 | 76045745 | 76044926 | 76047529 | + | protein_coding |
| ENST00000335749.4 | chr17 | 76045714 | 76045745 | 76044926 | 76047529 | + | protein_coding |
| ENST00000301624.4 | chr17 | 76045714 | 76045745 | 76044926 | 76047538 | + | protein_coding |
| ENST00000541771.1 | chr17 | 76045714 | 76045745 | 76044994 | 76047538 | + | protein_coding |
| ENST00000544502.1 | chr17 | 76045714 | 76045745 | 76045144 | 76047529 | + | protein_coding |
| ENST00000455761.2 | chr17 | 76083141 | 76083172 | 76082934 | 76083174 | + | protein_coding |
| ENST00000395801.3 | chr17 | 76083141 | 76083172 | 76082934 | 76083174 | + | protein_coding |
| ENST00000335749.4 | chr17 | 76083141 | 76083172 | 76082934 | 76083174 | + | protein_coding |
| ENST00000301624.4 | chr17 | 76083141 | 76083172 | 76082934 | 76083174 | + | protein_coding |
| ENST00000541771.1 | chr17 | 76083141 | 76083172 | 76082934 | 76083174 | + | protein_coding |
| ENST00000544502.1 | chr17 | 76083141 | 76083172 | 76082934 | 76083174 | + | protein_coding |
| ENST00000335749.4 | chr17 | 76101603 | 76101634 | 76100566 | 76104916 | + | protein_coding |
| ENST00000301624.4 | chr17 | 76101603 | 76101634 | 76100566 | 76104916 | + | protein_coding |
| ENST00000544529.1 | chr17 | 76101603 | 76101634 | 76101532 | 76101832 | + | pseudogene |
| ENST00000374948.1 | chr17 | 76220266 | 76220302 | 76219546 | 76221715 | + | protein_coding |
| ENST00000301633.3 | chr17 | 76220266 | 76220302 | 76219546 | 76221715 | + | protein_coding |
| ENST00000350051.2 | chr17 | 76220266 | 76220302 | 76219546 | 76221715 | + | protein_coding |
| ENST00000310942.3 | chr17 | 77761354 | 77761385 | 77757531 | 77761449 | + | protein_coding |
| ENST00000269349.3 | chr17 | 78109185 | 78109215 | 78109013 | 78109305 | - | protein_coding |
| ENST00000331925.1 | chr17 | 79479763 | 79479786 | 79479760 | 79479827 | - | protein_coding |
| ENST00000447294.2 | chr17 | 79479763 | 79479786 | 79479760 | 79479827 | - | protein_coding |
| ENST00000331483.3 | chr17 | 79801601 | 79801632 | 79801037 | 79801968 | - | protein_coding |
| ENST00000415593.1 | chr17 | 79801601 | 79801632 | 79801042 | 79801968 | - | protein_coding |
| ENST00000473021.1 | chr17 | 79801601 | 79801632 | 79801368 | 79801968 | - | retained_intron |
| ENST00000436463.2 | chr17 | 79801601 | 79801632 | 79801410 | 79801968 | - | protein_coding |
| ENST00000269321.6 | chr17 | 79826523 | 79826554 | 79825598 | 79826951 | - | protein_coding |
| ENST00000541078.1 | chr17 | 79826523 | 79826554 | 79825598 | 79826951 | - | protein_coding |
| ENST00000400721.3 | chr17 | 79826523 | 79826554 | 79825887 | 79826845 | - | protein_coding |
| ENST00000331204.4 | chr17 | 79845875 | 79845892 | 79845713 | 79846021 | - | protein_coding |
| ENST00000505490.1 | chr17 | 79845875 | 79845892 | 79845717 | 79846021 | - | protein_coding |
| ENST00000329286.5 | chr18 | 8639143 | 8639189 | 8638147 | 8639380 | + | protein_coding |
| ENST00000383314.2 | chr18 | 13763149 | 13763180 | 13759941 | 13764557 | + | protein_coding |
| ENST00000544744.1 | chr18 | 13763149 | 13763180 | 13763092 | 13763494 | + | protein_coding |
| ENST00000262173.3 | chr18 | 13763149 | 13763180 | 13759941 | 13764554 | + | protein_coding |
| ENST00000304621.5 | chr18 | 21714221 | 21714254 | 21712449 | 21715573 | + | protein_coding |
| ENST00000317571.2 | chr18 | 21714221 | 21714254 | 21712449 | 21715573 | + | protein_coding |
| ENST00000363004.1 | chr18 | 21750469 | 21750498 | 21750469 | 21750587 | - | rRNA |
| ENST00000355632.4 | chr18 | 32838834 | 32838865 | 32837892 | 32839191 | + | protein_coding |
| ENST00000361795.3 | chr18 | 35145921 | 35145945 | 35145319 | 35146000 | - | protein_coding |
| ENST00000412753.1 | chr18 | 35145921 | 35145945 | 35145319 | 35146000 | - | protein_coding |
| ENST00000420428.1 | chr18 | 35145921 | 35145945 | 35145319 | 35146000 | - | protein_coding |
| ENST00000334919.4 | chr18 | 35145921 | 35145945 | 35145319 | 35146000 | - | protein_coding |
| ENST00000361683.4 | chr18 | 35145921 | 35145945 | 35145319 | 35146000 | - | protein_coding |
| ENST00000545051.1 | chr18 | 46447668 | 46447705 | 46446223 | 46448280 | - | protein_coding |
| ENST00000262158.2 | chr18 | 46447668 | 46447705 | 46446223 | 46448280 | - | protein_coding |
| ENST00000491143.1 | chr18 | 55145969 | 55146000 | 55143669 | 55158530 | + | protein_coding |
| ENST00000262095.2 | chr18 | 55145969 | 55146000 | 55143669 | 55158529 | + | protein_coding |
| ENST00000256854.4 | chr18 | 55268592 | 55268621 | 55267896 | 55269015 | - | protein_coding |
| ENST00000423481.2 | chr18 | 55268592 | 55268621 | 55268582 | 55269015 | - | protein_coding |
| ENST00000256854.4 | chr18 | 55268641 | 55268662 | 55267896 | 55269015 | - | protein_coding |
| ENST00000423481.2 | chr18 | 55268641 | 55268662 | 55268582 | 55269015 | - | protein_coding |
| ENST00000256854.4 | chr18 | 55274533 | 55274564 | 55274366 | 55274565 | - | protein_coding |
| ENST00000423481.2 | chr18 | 55274533 | 55274564 | 55274366 | 55274565 | - | protein_coding |
| ENST00000411676.2 | chr18 | 55274533 | 55274564 | 55274388 | 55274897 | - | protein_coding |
| ENST00000540592.1 | chr18 | 55274533 | 55274564 | 55274434 | 55274565 | - | protein_coding |
| ENST00000400345.2 | chr18 | 56065390 | 56065426 | 56063399 | 56068772 | + | protein_coding |
| ENST00000382850.3 | chr18 | 56065390 | 56065426 | 56063399 | 56068772 | + | protein_coding |
| ENST00000356462.5 | chr18 | 56065390 | 56065426 | 56063399 | 56068772 | + | protein_coding |
| ENST00000256830.8 | chr18 | 56065390 | 56065426 | 56063399 | 56068772 | + | protein_coding |
| ENST00000256832.7 | chr18 | 56065390 | 56065426 | 56063399 | 56068772 | + | protein_coding |
| ENST00000456986.1 | chr18 | 56065390 | 56065426 | 56063399 | 56068772 | + | protein_coding |
| ENST00000357895.4 | chr18 | 56065390 | 56065426 | 56063399 | 56068772 | + | protein_coding |
| ENST00000435432.1 | chr18 | 56065390 | 56065426 | 56063399 | 56068772 | + | protein_coding |
| ENST00000456173.1 | chr18 | 56065390 | 56065426 | 56063399 | 56068772 | + | protein_coding |
| ENST00000431212.1 | chr18 | 56065390 | 56065426 | 56063399 | 56068772 | + | protein_coding |
| ENST00000316660.6 | chr18 | 57570699 | 57570700 | 57569879 | 57571538 | + | protein_coding |
| ENST00000269518.9 | chr18 | 57570699 | 57570700 | 57569879 | 57570769 | + | protein_coding |
| ENST00000316660.6 | chr18 | 57571039 | 57571071 | 57569879 | 57571538 | + | protein_coding |
| ENST00000262198.3 | chr18 | 77893531 | 77893561 | 77893495 | 77898226 | + | protein_coding |
| ENST00000361574.4 | chr19 | 896815 | 896846 | 896503 | 897540 | - | protein_coding |
| ENST00000263620.2 | chr19 | 972024 | 972055 | 971878 | 972781 | + | protein_coding |
| ENST00000300952.2 | chr19 | 1257579 | 1257610 | 1256994 | 1259139 | + | protein_coding |
| ENST00000300952.2 | chr19 | 1257702 | 1257702 | 1256994 | 1259139 | + | protein_coding |
| ENST00000233609.3 | chr19 | 1440443 | 1440458 | 1440348 | 1440492 | + | protein_coding |
| ENST00000255608.3 | chr19 | 1986103 | 1986134 | 1985447 | 1986648 | - | protein_coding |
| ENST00000357066.2 | chr19 | 2078329 | 2078345 | 2078139 | 2078678 | - | protein_coding |
| ENST00000326631.2 | chr19 | 2235825 | 2235856 | 2235761 | 2235827 | - | protein_coding |
| ENST00000325327.3 | chr19 | 2428173 | 2428225 | 2428164 | 2430950 | - | protein_coding |
| ENST00000524597.1 | chr19 | 2428173 | 2428225 | 2428179 | 2429492 | - | processed_transcript |
| ENST00000325327.3 | chr19 | 2429211 | 2429242 | 2428164 | 2430950 | - | protein_coding |
| ENST00000524597.1 | chr19 | 2429211 | 2429242 | 2428179 | 2429492 | - | processed_transcript |
| ENST00000534597.1 | chr19 | 2429211 | 2429242 | 2428225 | 2429486 | - | processed_transcript |
| ENST00000325327.3 | chr19 | 2434909 | 2434942 | 2434786 | 2434911 | - | protein_coding |
| ENST00000490554.1 | chr19 | 2434909 | 2434942 | 2434786 | 2434957 | - | retained_intron |
| ENST00000527409.1 | chr19 | 2434909 | 2434942 | 2434786 | 2434911 | - | processed_transcript |
| ENST00000335312.2 | chr19 | 3630727 | 3630750 | 3630182 | 3633167 | - | protein_coding |
| ENST00000385000.1 | chr19 | 3961419 | 3961450 | 3961412 | 3961510 | - | miRNA |
| ENST00000309311.5 | chr19 | 3976225 | 3976256 | 3976055 | 3976745 | - | protein_coding |
| ENST00000543343.1 | chr19 | 3980638 | 3980669 | 3980512 | 3980707 | - | protein_coding |
| ENST00000309311.5 | chr19 | 3980638 | 3980669 | 3980512 | 3980707 | - | protein_coding |
| ENST00000357909.2 | chr19 | 4679889 | 4679920 | 4679847 | 4679958 | - | protein_coding |
| ENST00000381797.2 | chr19 | 4679889 | 4679920 | 4679847 | 4679958 | - | protein_coding |
| ENST00000262960.8 | chr19 | 4679889 | 4679920 | 4679847 | 4679958 | - | protein_coding |
| ENST00000384898.1 | chr19 | 4770712 | 4770744 | 4770682 | 4770791 | + | miRNA |
| ENST00000221515.1 | chr19 | 7735323 | 7735355 | 7735105 | 7735340 | + | protein_coding |
| ENST00000363030.1 | chr19 | 7951852 | 7951881 | 7951862 | 7951981 | - | rRNA |
| ENST00000407627.1 | chr19 | 8028265 | 8028296 | 8023458 | 8028691 | - | protein_coding |
| ENST00000351593.5 | chr19 | 8028265 | 8028296 | 8028043 | 8028691 | - | protein_coding |
| ENST00000247956.5 | chr19 | 9272986 | 9273017 | 9270790 | 9274090 | + | protein_coding |
| ENST00000360385.2 | chr19 | 9272986 | 9273017 | 9270790 | 9274090 | + | protein_coding |
| ENST00000317726.3 | chr19 | 10333845 | 10333881 | 10332109 | 10335623 | - | protein_coding |
| ENST00000222005.1 | chr19 | 10501933 | 10501964 | 10501809 | 10502382 | - | protein_coding |
| ENST00000419866.2 | chr19 | 10577796 | 10577826 | 10577563 | 10578325 | + | protein_coding |
| ENST00000380702.2 | chr19 | 10577796 | 10577826 | 10577563 | 10580305 | + | protein_coding |
| ENST00000352831.5 | chr19 | 10577796 | 10577826 | 10577563 | 10580305 | + | protein_coding |
| ENST00000293683.4 | chr19 | 10577796 | 10577826 | 10577563 | 10580305 | + | protein_coding |
| ENST00000440014.1 | chr19 | 10577796 | 10577826 | 10577563 | 10580305 | + | protein_coding |
| ENST00000344979.2 | chr19 | 10577796 | 10577826 | 10577563 | 10580305 | + | protein_coding |
| ENST00000407327.3 | chr19 | 10745674 | 10745705 | 10745650 | 10745749 | + | protein_coding |
| ENST00000335757.4 | chr19 | 10745674 | 10745705 | 10745650 | 10745749 | + | protein_coding |
| ENST00000380614.1 | chr19 | 10745674 | 10745705 | 10745650 | 10745749 | + | protein_coding |
| ENST00000449870.1 | chr19 | 10800006 | 10800027 | 10799855 | 10803093 | + | protein_coding |
| ENST00000318511.3 | chr19 | 10800006 | 10800027 | 10799855 | 10803093 | + | protein_coding |
| ENST00000385237.1 | chr19 | 10829095 | 10829126 | 10829080 | 10829179 | + | miRNA |
| ENST00000254321.4 | chr19 | 12059818 | 12059849 | 12059091 | 12061578 | + | protein_coding |
| ENST00000482090.1 | chr19 | 12059818 | 12059849 | 12059091 | 12061588 | + | retained_intron |
| ENST00000254321.4 | chr19 | 12061226 | 12061257 | 12059091 | 12061578 | + | protein_coding |
| ENST00000482090.1 | chr19 | 12061226 | 12061257 | 12059091 | 12061588 | + | retained_intron |
| ENST00000316448.4 | chr19 | 13055263 | 13055294 | 13054527 | 13055304 | + | protein_coding |
| ENST00000384853.1 | chr19 | 13985724 | 13985755 | 13985689 | 13985825 | + | miRNA |
| ENST00000384974.1 | chr19 | 14640415 | 14640446 | 14640355 | 14640452 | + | miRNA |
| ENST00000215567.4 | chr19 | 14640415 | 14640446 | 14640382 | 14640523 | + | protein_coding |
| ENST00000436007.2 | chr19 | 14640415 | 14640446 | 14640407 | 14640523 | + | protein_coding |
| ENST00000389282.4 | chr19 | 15533104 | 15533135 | 15532319 | 15534046 | - | protein_coding |
| ENST00000263381.5 | chr19 | 15533104 | 15533135 | 15532319 | 15534046 | - | protein_coding |
| ENST00000416927.2 | chr19 | 15533104 | 15533135 | 15532910 | 15534046 | - | protein_coding |
| ENST00000338128.6 | chr19 | 18547056 | 18547087 | 18546842 | 18547057 | - | protein_coding |
| ENST00000457269.2 | chr19 | 18547056 | 18547087 | 18546842 | 18547057 | - | protein_coding |
| ENST00000545187.1 | chr19 | 18547056 | 18547087 | 18546842 | 18547057 | - | protein_coding |
| ENST00000317018.6 | chr19 | 18547056 | 18547087 | 18546842 | 18547057 | - | protein_coding |
| ENST00000337018.5 | chr19 | 19136134 | 19136165 | 19135428 | 19137035 | - | protein_coding |
| ENST00000330854.10 | chr19 | 19136134 | 19136165 | 19135428 | 19137035 | - | protein_coding |
| ENST00000452918.1 | chr19 | 19136134 | 19136165 | 19135428 | 19137035 | - | protein_coding |
| ENST00000456085.2 | chr19 | 19136134 | 19136165 | 19135428 | 19136265 | - | protein_coding |
| ENST00000364165.1 | chr19 | 21295925 | 21295962 | 21295934 | 21296042 | - | rRNA |
| ENST00000516139.1 | chr19 | 24184082 | 24184148 | 24184075 | 24184165 | - | miRNA |
| ENST00000365096.1 | chr19 | 24187278 | 24187309 | 24187160 | 24187309 | - | rRNA |
| ENST00000360605.3 | chr19 | 30500211 | 30500242 | 30499912 | 30500260 | + | protein_coding |
| ENST00000392271.1 | chr19 | 30500211 | 30500242 | 30499912 | 30500260 | + | protein_coding |
| ENST00000542441.1 | chr19 | 30500211 | 30500242 | 30499912 | 30500260 | + | protein_coding |
| ENST00000312051.6 | chr19 | 30500211 | 30500242 | 30499912 | 30500260 | + | protein_coding |
| ENST00000355898.4 | chr19 | 32844281 | 32844312 | 32843735 | 32845863 | + | protein_coding |
| ENST00000311921.2 | chr19 | 32844281 | 32844312 | 32843735 | 32845863 | + | protein_coding |
| ENST00000544431.1 | chr19 | 32844281 | 32844312 | 32843735 | 32845863 | + | protein_coding |
| ENST00000392250.2 | chr19 | 32976288 | 32976319 | 32972985 | 32976795 | + | protein_coding |
| ENST00000342179.4 | chr19 | 32976288 | 32976319 | 32972985 | 32976795 | + | protein_coding |
| ENST00000544216.2 | chr19 | 34687539 | 34687570 | 34687539 | 34687668 | + | protein_coding |
| ENST00000433627.3 | chr19 | 34687539 | 34687570 | 34687539 | 34687668 | + | protein_coding |
| ENST00000540746.1 | chr19 | 34687539 | 34687570 | 34687539 | 34687668 | + | protein_coding |
| ENST00000544216.2 | chr19 | 34718659 | 34718690 | 34718270 | 34720420 | + | protein_coding |
| ENST00000433627.3 | chr19 | 34718659 | 34718690 | 34718270 | 34720420 | + | protein_coding |
| ENST00000540746.1 | chr19 | 34718659 | 34718690 | 34718270 | 34718689 | + | protein_coding |
| ENST00000507959.1 | chr19 | 35175077 | 35175125 | 35174897 | 35175310 | + | protein_coding |
| ENST00000505365.1 | chr19 | 35175077 | 35175125 | 35175098 | 35177297 | + | protein_coding |
| ENST00000434389.1 | chr19 | 35996636 | 35996667 | 35996620 | 35996667 | - | protein_coding |
| ENST00000402589.2 | chr19 | 35996636 | 35996667 | 35996620 | 35996667 | - | protein_coding |
| ENST00000462721.1 | chr19 | 35996636 | 35996667 | 35996620 | 35996667 | - | processed_transcript |
| ENST00000339686.3 | chr19 | 35996636 | 35996667 | 35996620 | 35996667 | - | protein_coding |
| ENST00000488762.1 | chr19 | 35996636 | 35996667 | 35996620 | 35996667 | - | processed_transcript |
| ENST00000490622.1 | chr19 | 35996636 | 35996667 | 35996620 | 35996667 | - | retained_intron |
| ENST00000392207.3 | chr19 | 35996636 | 35996667 | 35996620 | 35996667 | - | protein_coding |
| ENST00000436012.1 | chr19 | 35996636 | 35996667 | 35996620 | 35996667 | - | protein_coding |
| ENST00000467532.1 | chr19 | 35996636 | 35996667 | 35996620 | 35996667 | - | processed_transcript |
| ENST00000414866.2 | chr19 | 35996636 | 35996667 | 35996620 | 35996667 | - | protein_coding |
| ENST00000480502.1 | chr19 | 35996636 | 35996667 | 35996620 | 35996667 | - | processed_transcript |
| ENST00000467637.1 | chr19 | 35996636 | 35996667 | 35996620 | 35996667 | - | processed_transcript |
| ENST00000429837.1 | chr19 | 35996636 | 35996667 | 35996620 | 35996667 | - | protein_coding |
| ENST00000419602.1 | chr19 | 35996636 | 35996667 | 35996620 | 35996667 | - | protein_coding |
| ENST00000462126.1 | chr19 | 35996636 | 35996667 | 35996620 | 35996667 | - | processed_transcript |
| ENST00000443857.1 | chr19 | 35996636 | 35996667 | 35996620 | 35996667 | - | protein_coding |
| ENST00000443640.1 | chr19 | 35996636 | 35996667 | 35996620 | 35996667 | - | protein_coding |
| ENST00000488542.1 | chr19 | 35996636 | 35996667 | 35996620 | 35996667 | - | retained_intron |
| ENST00000460051.1 | chr19 | 35996636 | 35996667 | 35996620 | 35996667 | - | processed_transcript |
| ENST00000498211.1 | chr19 | 35996636 | 35996667 | 35996620 | 35996667 | - | processed_transcript |
| ENST00000483855.1 | chr19 | 35996636 | 35996667 | 35996620 | 35996667 | - | processed_transcript |
| ENST00000493979.1 | chr19 | 35996636 | 35996667 | 35996620 | 35996667 | - | processed_transcript |
| ENST00000464709.1 | chr19 | 35996636 | 35996667 | 35996620 | 35996667 | - | processed_transcript |
| ENST00000463292.1 | chr19 | 35996636 | 35996667 | 35996620 | 35996667 | - | processed_transcript |
| ENST00000465927.1 | chr19 | 35996636 | 35996667 | 35996620 | 35996667 | - | retained_intron |
| ENST00000464894.1 | chr19 | 35996636 | 35996667 | 35996620 | 35996667 | - | processed_transcript |
| ENST00000493517.1 | chr19 | 35996636 | 35996667 | 35996620 | 35996667 | - | processed_transcript |
| ENST00000492341.1 | chr19 | 35996636 | 35996667 | 35996620 | 35996667 | - | retained_intron |
| ENST00000482321.1 | chr19 | 35996636 | 35996667 | 35996620 | 35996667 | - | retained_intron |
| ENST00000476246.1 | chr19 | 35996636 | 35996667 | 35996620 | 35996667 | - | retained_intron |
| ENST00000471786.1 | chr19 | 35996636 | 35996667 | 35996620 | 35996667 | - | retained_intron |
| ENST00000486450.1 | chr19 | 35996636 | 35996667 | 35996620 | 35996667 | - | retained_intron |
| ENST00000498593.1 | chr19 | 35996636 | 35996667 | 35996620 | 35996667 | - | retained_intron |
| ENST00000471017.1 | chr19 | 35996636 | 35996667 | 35996620 | 35996667 | - | retained_intron |
| ENST00000498269.1 | chr19 | 35996636 | 35996667 | 35996620 | 35996667 | - | retained_intron |
| ENST00000470746.1 | chr19 | 35996636 | 35996667 | 35996620 | 35996667 | - | retained_intron |
| ENST00000474992.1 | chr19 | 35996636 | 35996667 | 35996620 | 35996667 | - | processed_transcript |
| ENST00000476051.1 | chr19 | 35996636 | 35996667 | 35996620 | 35996667 | - | retained_intron |
| ENST00000474928.1 | chr19 | 35996636 | 35996667 | 35996620 | 35996667 | - | retained_intron |
| ENST00000488892.1 | chr19 | 35996636 | 35996667 | 35996620 | 35996667 | - | processed_transcript |
| ENST00000462538.1 | chr19 | 35996636 | 35996667 | 35996620 | 35996667 | - | processed_transcript |
| ENST00000480507.1 | chr19 | 35996636 | 35996667 | 35996620 | 35996667 | - | processed_transcript |
| ENST00000461300.1 | chr19 | 35996636 | 35996667 | 35996620 | 35996667 | - | retained_intron |
| ENST00000447113.2 | chr19 | 35996636 | 35996667 | 35996620 | 35996667 | - | protein_coding |
| ENST00000489395.1 | chr19 | 35996636 | 35996667 | 35996620 | 35996667 | - | retained_intron |
| ENST00000392206.2 | chr19 | 35996636 | 35996667 | 35996620 | 35996667 | - | protein_coding |
| ENST00000440396.1 | chr19 | 35996636 | 35996667 | 35996620 | 35996667 | - | protein_coding |
| ENST00000458071.1 | chr19 | 35996636 | 35996667 | 35996620 | 35996667 | - | protein_coding |
| ENST00000418261.1 | chr19 | 35996636 | 35996667 | 35996620 | 35996667 | - | protein_coding |
| ENST00000424570.2 | chr19 | 35996636 | 35996667 | 35996620 | 35996667 | - | protein_coding |
| ENST00000451297.2 | chr19 | 35996636 | 35996667 | 35996620 | 35996667 | - | protein_coding |
| ENST00000450261.1 | chr19 | 35996636 | 35996667 | 35996620 | 35996667 | - | protein_coding |
| ENST00000516857.1 | chr19 | 36066585 | 36066628 | 36066627 | 36066721 | + | miRNA |
| ENST00000516857.1 | chr19 | 36066707 | 36066740 | 36066627 | 36066721 | + | miRNA |
| ENST00000420124.1 | chr19 | 36229688 | 36229719 | 36229183 | 36229779 | + | protein_coding |
| ENST00000443387.1 | chr19 | 36727171 | 36727202 | 36726561 | 36729673 | + | protein_coding |
| ENST00000456324.1 | chr19 | 36727171 | 36727202 | 36726561 | 36729673 | + | protein_coding |
| ENST00000443387.1 | chr19 | 36729350 | 36729374 | 36726561 | 36729673 | + | protein_coding |
| ENST00000456324.1 | chr19 | 36729350 | 36729374 | 36726561 | 36729673 | + | protein_coding |
| ENST00000304239.7 | chr19 | 37618227 | 37618258 | 37618030 | 37619304 | + | protein_coding |
| ENST00000337995.2 | chr19 | 37618227 | 37618258 | 37618030 | 37620662 | + | protein_coding |
| ENST00000392138.1 | chr19 | 38229914 | 38229945 | 38229203 | 38231095 | - | protein_coding |
| ENST00000536220.1 | chr19 | 38229914 | 38229945 | 38229203 | 38231095 | - | protein_coding |
| ENST00000357309.3 | chr19 | 38229914 | 38229945 | 38229203 | 38231095 | - | protein_coding |
| ENST00000339503.4 | chr19 | 38229914 | 38229945 | 38229203 | 38231095 | - | protein_coding |
| ENST00000427026.1 | chr19 | 38229914 | 38229945 | 38229393 | 38231111 | - | protein_coding |
| ENST00000301244.5 | chr19 | 38780780 | 38780811 | 38780759 | 38780920 | + | protein_coding |
| ENST00000454580.2 | chr19 | 38780780 | 38780811 | 38780759 | 38780920 | + | protein_coding |
| ENST00000252699.2 | chr19 | 39191240 | 39191271 | 39191240 | 39191354 | + | protein_coding |
| ENST00000445727.2 | chr19 | 39191240 | 39191271 | 39191240 | 39191354 | + | protein_coding |
| ENST00000424234.2 | chr19 | 39191240 | 39191271 | 39191240 | 39191344 | + | protein_coding |
| ENST00000495553.1 | chr19 | 39191240 | 39191271 | 39191240 | 39191354 | + | processed_transcript |
| ENST00000221419.4 | chr19 | 39327187 | 39327218 | 39327029 | 39327420 | - | protein_coding |
| ENST00000388749.2 | chr19 | 39327187 | 39327218 | 39327029 | 39327420 | - | protein_coding |
| ENST00000221419.4 | chr19 | 39327294 | 39327325 | 39327029 | 39327420 | - | protein_coding |
| ENST00000388749.2 | chr19 | 39327294 | 39327325 | 39327029 | 39327420 | - | protein_coding |
| ENST00000388750.5 | chr19 | 39327294 | 39327325 | 39327277 | 39327420 | - | protein_coding |
| ENST00000314471.5 | chr19 | 39847378 | 39847409 | 39847329 | 39847729 | + | protein_coding |
| ENST00000429637.2 | chr19 | 39847378 | 39847409 | 39847329 | 39847729 | + | protein_coding |
| ENST00000315588.4 | chr19 | 39889084 | 39889115 | 39888097 | 39891201 | + | protein_coding |
| ENST00000435462.2 | chr19 | 39889084 | 39889115 | 39888097 | 39889267 | + | protein_coding |
| ENST00000251453.2 | chr19 | 39923867 | 39923902 | 39923852 | 39924058 | - | protein_coding |
| ENST00000339471.3 | chr19 | 39923867 | 39923902 | 39923852 | 39924058 | - | protein_coding |
| ENST00000221801.2 | chr19 | 40328348 | 40328379 | 40328351 | 40328483 | - | protein_coding |
| ENST00000157812.1 | chr19 | 40487278 | 40487309 | 40487099 | 40487351 | + | protein_coding |
| ENST00000455878.1 | chr19 | 40487278 | 40487309 | 40487099 | 40487351 | + | protein_coding |
| ENST00000450241.2 | chr19 | 40578998 | 40579029 | 40575059 | 40582116 | - | protein_coding |
| ENST00000455521.1 | chr19 | 40578998 | 40579029 | 40578900 | 40582116 | - | protein_coding |
| ENST00000384899.1 | chr19 | 40788502 | 40788533 | 40788450 | 40788548 | - | miRNA |
| ENST00000243563.2 | chr19 | 41257292 | 41257323 | 41256759 | 41257386 | + | protein_coding |
| ENST00000545469.1 | chr19 | 41257292 | 41257323 | 41257126 | 41257350 | + | protein_coding |
| ENST00000243563.2 | chr19 | 41257334 | 41257365 | 41256759 | 41257386 | + | protein_coding |
| ENST00000545469.1 | chr19 | 41257334 | 41257365 | 41257126 | 41257350 | + | protein_coding |
| ENST00000352456.2 | chr19 | 41812421 | 41812453 | 41812354 | 41813811 | + | protein_coding |
| ENST00000392006.2 | chr19 | 41812421 | 41812453 | 41812354 | 41813811 | + | protein_coding |
| ENST00000270069.6 | chr19 | 41812421 | 41812453 | 41812354 | 41812943 | + | protein_coding |
| ENST00000378215.3 | chr19 | 41812421 | 41812453 | 41812354 | 41813811 | + | protein_coding |
| ENST00000263367.3 | chr19 | 41812421 | 41812453 | 41812354 | 41812924 | + | protein_coding |
| ENST00000352456.2 | chr19 | 41813464 | 41813495 | 41812354 | 41813811 | + | protein_coding |
| ENST00000392006.2 | chr19 | 41813464 | 41813495 | 41812354 | 41813811 | + | protein_coding |
| ENST00000378215.3 | chr19 | 41813464 | 41813495 | 41812354 | 41813811 | + | protein_coding |
| ENST00000491995.1 | chr19 | 43586275 | 43586310 | 43585939 | 43586423 | - | processed_transcript |
| ENST00000244333.2 | chr19 | 43965024 | 43965055 | 43964946 | 43965999 | - | protein_coding |
| ENST00000422989.1 | chr19 | 44090971 | 44091002 | 44088521 | 44097519 | - | protein_coding |
| ENST00000291187.3 | chr19 | 44930574 | 44930605 | 44930426 | 44934717 | - | protein_coding |
| ENST00000221327.3 | chr19 | 44980648 | 44980679 | 44979861 | 44982363 | - | protein_coding |
| ENST00000391956.3 | chr19 | 44980648 | 44980679 | 44979861 | 44982363 | - | protein_coding |
| ENST00000396737.1 | chr19 | 46005571 | 46005599 | 46005275 | 46005600 | + | protein_coding |
| ENST00000396734.2 | chr19 | 46005571 | 46005599 | 46005275 | 46005762 | + | protein_coding |
| ENST00000415077.1 | chr19 | 46005571 | 46005599 | 46005275 | 46005600 | + | nonsense_mediated_decay |
| ENST00000396735.2 | chr19 | 46005571 | 46005599 | 46005275 | 46005768 | + | protein_coding |
| ENST00000477815.1 | chr19 | 46005571 | 46005599 | 46005275 | 46005599 | + | processed_transcript |
| ENST00000391909.2 | chr19 | 47258745 | 47258776 | 47258669 | 47261832 | + | protein_coding |
| ENST00000318584.4 | chr19 | 47258745 | 47258776 | 47258669 | 47261832 | + | protein_coding |
| ENST00000542575.1 | chr19 | 47278375 | 47278389 | 47278140 | 47279004 | - | protein_coding |
| ENST00000434726.1 | chr19 | 47278375 | 47278389 | 47278140 | 47279004 | - | protein_coding |
| ENST00000412532.1 | chr19 | 47278375 | 47278389 | 47278140 | 47279004 | - | protein_coding |
| ENST00000306894.4 | chr19 | 47278375 | 47278389 | 47278142 | 47279004 | - | protein_coding |
| ENST00000330720.2 | chr19 | 48886132 | 48886151 | 48885827 | 48886584 | - | protein_coding |
| ENST00000330720.2 | chr19 | 48894559 | 48894590 | 48894525 | 48894810 | - | protein_coding |
| ENST00000552347.1 | chr19 | 49120626 | 49120657 | 49119336 | 49120680 | - | retained_intron |
| ENST00000549920.1 | chr19 | 49120626 | 49120657 | 49120573 | 49120680 | - | protein_coding |
| ENST00000547897.1 | chr19 | 49120626 | 49120657 | 49120573 | 49120680 | - | processed_transcript |
| ENST00000084795.5 | chr19 | 49120626 | 49120657 | 49120573 | 49120680 | - | protein_coding |
| ENST00000550645.1 | chr19 | 49120626 | 49120657 | 49120573 | 49120680 | - | protein_coding |
| ENST00000552588.1 | chr19 | 49120626 | 49120657 | 49120573 | 49120680 | - | protein_coding |
| ENST00000551749.1 | chr19 | 49120626 | 49120657 | 49119336 | 49120680 | - | retained_intron |
| ENST00000547892.1 | chr19 | 49120626 | 49120657 | 49118589 | 49122432 | - | retained_intron |
| ENST00000546623.1 | chr19 | 49120626 | 49120657 | 49120573 | 49120680 | - | protein_coding |
| ENST00000549370.1 | chr19 | 49120626 | 49120657 | 49120573 | 49120672 | - | nonsense_mediated_decay |
| ENST00000549273.1 | chr19 | 49120626 | 49120657 | 49120573 | 49120680 | - | protein_coding |
| ENST00000550973.1 | chr19 | 49120626 | 49120657 | 49120573 | 49120680 | - | protein_coding |
| ENST00000552705.1 | chr19 | 49120626 | 49120657 | 49120573 | 49120680 | - | retained_intron |
| ENST00000552851.1 | chr19 | 49120626 | 49120657 | 49120573 | 49120680 | - | retained_intron |
| ENST00000450952.2 | chr19 | 49120626 | 49120657 | 49120573 | 49120680 | - | protein_coding |
| ENST00000550671.1 | chr19 | 49120626 | 49120657 | 49120415 | 49121134 | - | retained_intron |
| ENST00000252825.3 | chr19 | 49657703 | 49657739 | 49656664 | 49658681 | - | protein_coding |
| ENST00000434964.1 | chr19 | 49657703 | 49657739 | 49657043 | 49658284 | - | protein_coding |
| ENST00000488946.1 | chr19 | 49993224 | 49993255 | 49993107 | 49994804 | + | retained_intron |
| ENST00000484279.1 | chr19 | 49993224 | 49993255 | 49993107 | 49993347 | + | processed_transcript |
| ENST00000486930.1 | chr19 | 49993224 | 49993255 | 49993107 | 49993554 | + | retained_intron |
| ENST00000476268.1 | chr19 | 49993224 | 49993255 | 49993104 | 49994523 | + | retained_intron |
| ENST00000364805.1 | chr19 | 49993224 | 49993255 | 49993222 | 49993305 | + | snoRNA |
| ENST00000384345.1 | chr19 | 50069951 | 50069982 | 50069929 | 50070029 | + | misc_RNA |
| ENST00000360565.2 | chr19 | 50161747 | 50161778 | 50161465 | 50161905 | + | protein_coding |
| ENST00000354957.2 | chr19 | 52393184 | 52393214 | 52392489 | 52395150 | - | protein_coding |
| ENST00000243644.3 | chr19 | 52467737 | 52467768 | 52467594 | 52469467 | - | protein_coding |
| ENST00000270649.5 | chr19 | 52519348 | 52519376 | 52516577 | 52520612 | - | protein_coding |
| ENST00000384913.1 | chr19 | 54169983 | 54170014 | 54169927 | 54170024 | + | miRNA |
| ENST00000385134.1 | chr19 | 54177484 | 54177515 | 54177451 | 54177574 | + | miRNA |
| ENST00000384867.1 | chr19 | 54179018 | 54179049 | 54178965 | 54179051 | + | miRNA |
| ENST00000384884.1 | chr19 | 54182270 | 54182301 | 54182257 | 54182339 | + | miRNA |
| ENST00000384883.1 | chr19 | 54188276 | 54188307 | 54188263 | 54188345 | + | miRNA |
| ENST00000385053.1 | chr19 | 54189738 | 54189769 | 54189723 | 54189809 | + | miRNA |
| ENST00000384862.1 | chr19 | 54194187 | 54194218 | 54194135 | 54194219 | + | miRNA |
| ENST00000385090.1 | chr19 | 54198479 | 54198510 | 54198467 | 54198547 | + | miRNA |
| ENST00000385090.1 | chr19 | 54198517 | 54198548 | 54198467 | 54198547 | + | miRNA |
| ENST00000384978.1 | chr19 | 54200801 | 54200832 | 54200787 | 54200871 | + | miRNA |
| ENST00000385281.1 | chr19 | 54201654 | 54201685 | 54201639 | 54201725 | + | miRNA |
| ENST00000385001.1 | chr19 | 54215575 | 54215606 | 54215522 | 54215608 | + | miRNA |
| ENST00000385102.1 | chr19 | 54224372 | 54224403 | 54224330 | 54224396 | + | miRNA |
| ENST00000385190.1 | chr19 | 54228711 | 54228742 | 54228696 | 54228780 | + | miRNA |
| ENST00000385014.1 | chr19 | 54238147 | 54238177 | 54238131 | 54238217 | + | miRNA |
| ENST00000385211.1 | chr19 | 54240114 | 54240145 | 54240099 | 54240188 | + | miRNA |
| ENST00000385071.1 | chr19 | 54254480 | 54254511 | 54254465 | 54254551 | + | miRNA |
| ENST00000385257.1 | chr19 | 54255665 | 54255696 | 54255651 | 54255735 | + | miRNA |
| ENST00000263095.5 | chr19 | 57726696 | 57726727 | 57722722 | 57734212 | + | protein_coding |
| ENST00000414468.1 | chr19 | 57772822 | 57772850 | 57764441 | 57774106 | + | protein_coding |
| ENST00000354309.3 | chr19 | 57772822 | 57772850 | 57764441 | 57774106 | + | protein_coding |
| ENST00000537645.1 | chr19 | 57802533 | 57802564 | 57802067 | 57803902 | + | protein_coding |
| ENST00000360338.2 | chr19 | 57802533 | 57802564 | 57802067 | 57805436 | + | protein_coding |
| ENST00000537645.1 | chr19 | 57802874 | 57802905 | 57802067 | 57803902 | + | protein_coding |
| ENST00000360338.2 | chr19 | 57802874 | 57802905 | 57802067 | 57805436 | + | protein_coding |
| ENST00000537645.1 | chr19 | 57803024 | 57803055 | 57802067 | 57803902 | + | protein_coding |
| ENST00000360338.2 | chr19 | 57803024 | 57803055 | 57802067 | 57805436 | + | protein_coding |
| ENST00000537645.1 | chr19 | 57803622 | 57803643 | 57802067 | 57803902 | + | protein_coding |
| ENST00000360338.2 | chr19 | 57803622 | 57803643 | 57802067 | 57805436 | + | protein_coding |
| ENST00000360338.2 | chr19 | 57804811 | 57804842 | 57802067 | 57805436 | + | protein_coding |
| ENST00000418193.2 | chr19 | 58132845 | 58132876 | 58131528 | 58132950 | + | protein_coding |
| ENST00000541849.1 | chr19 | 58132845 | 58132876 | 58131764 | 58132950 | + | protein_coding |
| ENST00000396161.4 | chr19 | 58132845 | 58132876 | 58131528 | 58134721 | + | protein_coding |
| ENST00000418193.2 | chr19 | 58132935 | 58132966 | 58131528 | 58132950 | + | protein_coding |
| ENST00000541849.1 | chr19 | 58132935 | 58132966 | 58131764 | 58132950 | + | protein_coding |
| ENST00000396161.4 | chr19 | 58132935 | 58132966 | 58131528 | 58134721 | + | protein_coding |
| ENST00000396161.4 | chr19 | 58134273 | 58134305 | 58131528 | 58134721 | + | protein_coding |
| ENST00000317178.5 | chr19 | 58266001 | 58266032 | 58264659 | 58269527 | + | protein_coding |
| ENST00000489376.1 | chr19 | 58266001 | 58266032 | 58265412 | 58267059 | + | processed_transcript |
| ENST00000317178.5 | chr19 | 58267998 | 58268029 | 58264659 | 58269527 | + | protein_coding |
| ENST00000485393.1 | chr1 | 564878 | 564910 | 564879 | 564950 | - | Mt_tRNA_pseudogene |
| ENST00000469752.1 | chr1 | 566130 | 566162 | 566137 | 566205 | - | Mt_tRNA_pseudogene |
| ENST00000469752.1 | chr1 | 566174 | 566205 | 566137 | 566205 | - | Mt_tRNA_pseudogene |
| ENST00000459059.1 | chr1 | 566174 | 566205 | 566187 | 566265 | - | miRNA |
| ENST00000459059.1 | chr1 | 566240 | 566279 | 566187 | 566265 | - | miRNA |
| ENST00000479998.1 | chr1 | 566240 | 566279 | 566207 | 566279 | - | Mt_tRNA_pseudogene |
| ENST00000491051.1 | chr1 | 566417 | 566445 | 566376 | 566441 | - | Mt_tRNA_pseudogene |
| ENST00000414273.1 | chr1 | 567359 | 567386 | 566454 | 567996 | + | unprocessed_pseudogene |
| ENST00000485927.1 | chr1 | 568034 | 568037 | 567996 | 568067 | - | Mt_tRNA_pseudogene |
| ENST00000484771.1 | chr1 | 568882 | 568907 | 568844 | 568913 | + | Mt_tRNA_pseudogene |
| ENST00000327044.6 | chr1 | 886569 | 886600 | 886507 | 886618 | - | protein_coding |
| ENST00000477976.1 | chr1 | 886569 | 886600 | 886507 | 886618 | - | processed_transcript |
| ENST00000344843.7 | chr1 | 1337306 | 1337336 | 1337288 | 1337636 | - | protein_coding |
| ENST00000493287.1 | chr1 | 1337306 | 1337336 | 1337293 | 1337636 | - | processed_transcript |
| ENST00000487659.1 | chr1 | 1337306 | 1337336 | 1337293 | 1337636 | - | processed_transcript |
| ENST00000378785.2 | chr1 | 1404481 | 1404516 | 1403764 | 1405538 | + | protein_coding |
| ENST00000234875.4 | chr1 | 6246224 | 6246235 | 6245080 | 6246876 | - | protein_coding |
| ENST00000343813.5 | chr1 | 6284515 | 6284546 | 6281253 | 6285322 | - | protein_coding |
| ENST00000489498.1 | chr1 | 6284515 | 6284546 | 6283551 | 6285322 | - | processed_transcript |
| ENST00000535068.1 | chr1 | 6284515 | 6284546 | 6283552 | 6285322 | - | protein_coding |
| ENST00000303635.6 | chr1 | 7827135 | 7827166 | 7826519 | 7827903 | + | protein_coding |
| ENST00000495233.1 | chr1 | 7827135 | 7827166 | 7826519 | 7827916 | + | protein_coding |
| ENST00000490905.1 | chr1 | 7827135 | 7827166 | 7826519 | 7827916 | + | protein_coding |
| ENST00000377482.5 | chr1 | 8073212 | 8073243 | 8071779 | 8074456 | - | protein_coding |
| ENST00000377223.1 | chr1 | 9989364 | 9989395 | 9988997 | 9990515 | - | protein_coding |
| ENST00000240185.3 | chr1 | 11083394 | 11083425 | 11082181 | 11084100 | + | protein_coding |
| ENST00000315091.3 | chr1 | 11083394 | 11083425 | 11083250 | 11084240 | + | protein_coding |
| ENST00000475189.1 | chr1 | 11114659 | 11114690 | 11114641 | 11114942 | - | processed_transcript |
| ENST00000376957.2 | chr1 | 11114659 | 11114690 | 11114646 | 11114942 | - | protein_coding |
| ENST00000489073.1 | chr1 | 11114659 | 11114690 | 11114646 | 11114942 | - | processed_transcript |
| ENST00000383925.1 | chr1 | 16840617 | 16840650 | 16840617 | 16840780 | - | snRNA |
| ENST00000383925.1 | chr1 | 16840662 | 16840662 | 16840617 | 16840780 | - | snRNA |
| ENST00000383925.1 | chr1 | 16840674 | 16840678 | 16840617 | 16840780 | - | snRNA |
| ENST00000383925.1 | chr1 | 16840699 | 16840704 | 16840617 | 16840780 | - | snRNA |
| ENST00000384782.1 | chr1 | 16993280 | 16993312 | 16993280 | 16993443 | - | snRNA |
| ENST00000384659.1 | chr1 | 17067079 | 17067082 | 17067011 | 17067174 | + | snRNA |
| ENST00000384659.1 | chr1 | 17067145 | 17067170 | 17067011 | 17067174 | + | snRNA |
| ENST00000384278.1 | chr1 | 17222605 | 17222641 | 17222475 | 17222638 | + | snRNA |
| ENST00000321556.4 | chr1 | 20971012 | 20971043 | 20970983 | 20971165 | + | protein_coding |
| ENST00000492302.1 | chr1 | 20971012 | 20971043 | 20970983 | 20971165 | + | processed_transcript |
| ENST00000335648.3 | chr1 | 23696748 | 23696778 | 23695729 | 23698275 | + | protein_coding |
| ENST00000364535.1 | chr1 | 23696748 | 23696778 | 23696747 | 23696839 | + | misc_RNA |
| ENST00000374550.3 | chr1 | 24021170 | 24021202 | 24021150 | 24021281 | + | protein_coding |
| ENST00000482370.1 | chr1 | 24021170 | 24021202 | 24021150 | 24021281 | + | processed_transcript |
| ENST00000443624.1 | chr1 | 24021170 | 24021202 | 24021150 | 24021885 | + | protein_coding |
| ENST00000458455.1 | chr1 | 24021170 | 24021202 | 24021150 | 24021281 | + | protein_coding |
| ENST00000455785.2 | chr1 | 26227109 | 26227134 | 26226593 | 26227578 | - | protein_coding |
| ENST00000399728.1 | chr1 | 26227109 | 26227134 | 26226609 | 26227578 | - | protein_coding |
| ENST00000465604.1 | chr1 | 26227109 | 26227134 | 26227059 | 26227578 | - | processed_transcript |
| ENST00000357865.2 | chr1 | 26227109 | 26227134 | 26227086 | 26227578 | - | protein_coding |
| ENST00000517138.1 | chr1 | 28160912 | 28160943 | 28160912 | 28161077 | + | snoRNA |
| ENST00000542507.1 | chr1 | 29095771 | 29095789 | 29095441 | 29096286 | + | protein_coding |
| ENST00000373812.3 | chr1 | 29095771 | 29095789 | 29095441 | 29096287 | + | protein_coding |
| ENST00000541996.1 | chr1 | 29095771 | 29095789 | 29095441 | 29096286 | + | protein_coding |
| ENST00000336798.7 | chr1 | 31342386 | 31342417 | 31342314 | 31346224 | - | protein_coding |
| ENST00000424085.2 | chr1 | 31404684 | 31404715 | 31404353 | 31406189 | - | protein_coding |
| ENST00000257075.5 | chr1 | 31404684 | 31404715 | 31404353 | 31406189 | - | protein_coding |
| ENST00000373747.3 | chr1 | 31404684 | 31404715 | 31404353 | 31406189 | - | protein_coding |
| ENST00000373749.4 | chr1 | 31404684 | 31404715 | 31404353 | 31406189 | - | protein_coding |
| ENST00000344035.5 | chr1 | 32384584 | 32384615 | 32384571 | 32385259 | - | protein_coding |
| ENST00000356536.3 | chr1 | 32384584 | 32384615 | 32384571 | 32385259 | - | protein_coding |
| ENST00000457805.2 | chr1 | 32384584 | 32384615 | 32384571 | 32384969 | - | protein_coding |
| ENST00000470404.1 | chr1 | 32384584 | 32384615 | 32384571 | 32384724 | - | protein_coding |
| ENST00000532001.1 | chr1 | 32384584 | 32384615 | 32384571 | 32384960 | - | nonsense_mediated_decay |
| ENST00000528253.1 | chr1 | 32384584 | 32384615 | 32384571 | 32385065 | - | nonsense_mediated_decay |
| ENST00000484483.1 | chr1 | 32384584 | 32384615 | 32384571 | 32384773 | - | retained_intron |
| ENST00000468531.1 | chr1 | 32384584 | 32384615 | 32384571 | 32385259 | - | protein_coding |
| ENST00000534796.1 | chr1 | 32384584 | 32384615 | 32384601 | 32385259 | - | protein_coding |
| ENST00000373625.3 | chr1 | 32637046 | 32637079 | 32636308 | 32642169 | + | protein_coding |
| ENST00000471599.1 | chr1 | 32637046 | 32637079 | 32636308 | 32638774 | + | processed_transcript |
| ENST00000373617.5 | chr1 | 32637046 | 32637079 | 32636308 | 32638774 | + | protein_coding |
| ENST00000373586.1 | chr1 | 32694383 | 32694414 | 32694328 | 32694417 | + | protein_coding |
| ENST00000471486.1 | chr1 | 32694383 | 32694414 | 32694328 | 32694415 | + | processed_transcript |
| ENST00000483517.1 | chr1 | 32694383 | 32694414 | 32694328 | 32694417 | + | processed_transcript |
| ENST00000474371.1 | chr1 | 32694383 | 32694414 | 32694328 | 32694417 | + | processed_transcript |
| ENST00000414241.3 | chr1 | 33134633 | 33134664 | 33134573 | 33134733 | + | protein_coding |
| ENST00000373493.4 | chr1 | 33134633 | 33134664 | 33134573 | 33134733 | + | protein_coding |
| ENST00000527118.1 | chr1 | 33134633 | 33134664 | 33134573 | 33134689 | + | nonsense_mediated_decay |
| ENST00000524393.1 | chr1 | 33134633 | 33134664 | 33134573 | 33134703 | + | processed_transcript |
| ENST00000526193.1 | chr1 | 33134633 | 33134664 | 33134573 | 33134733 | + | nonsense_mediated_decay |
| ENST00000492348.2 | chr1 | 33134633 | 33134664 | 33134573 | 33134733 | + | nonsense_mediated_decay |
| ENST00000544435.1 | chr1 | 33134633 | 33134664 | 33134573 | 33134733 | + | protein_coding |
| ENST00000373485.1 | chr1 | 33134633 | 33134664 | 33134573 | 33134733 | + | protein_coding |
| ENST00000458695.2 | chr1 | 33134633 | 33134664 | 33134573 | 33134733 | + | protein_coding |
| ENST00000475321.1 | chr1 | 33134633 | 33134664 | 33134646 | 33134733 | + | protein_coding |
| ENST00000531983.1 | chr1 | 33134633 | 33134664 | 33134573 | 33134733 | + | retained_intron |
| ENST00000251195.5 | chr1 | 36230147 | 36230178 | 36229867 | 36230315 | - | protein_coding |
| ENST00000318121.3 | chr1 | 36230147 | 36230178 | 36229867 | 36230315 | - | protein_coding |
| ENST00000373220.3 | chr1 | 36230147 | 36230178 | 36229867 | 36230315 | - | protein_coding |
| ENST00000520551.1 | chr1 | 36230147 | 36230178 | 36229867 | 36230315 | - | protein_coding |
| ENST00000544356.1 | chr1 | 36230147 | 36230178 | 36229867 | 36230315 | - | protein_coding |
| ENST00000373206.1 | chr1 | 36354102 | 36354133 | 36354028 | 36354211 | + | protein_coding |
| ENST00000373204.3 | chr1 | 36354102 | 36354133 | 36354028 | 36354211 | + | protein_coding |
| ENST00000373206.1 | chr1 | 36354159 | 36354170 | 36354028 | 36354211 | + | protein_coding |
| ENST00000373204.3 | chr1 | 36354159 | 36354170 | 36354028 | 36354211 | + | protein_coding |
| ENST00000373206.1 | chr1 | 36360787 | 36360818 | 36360723 | 36360870 | + | protein_coding |
| ENST00000373204.3 | chr1 | 36360787 | 36360818 | 36360723 | 36360870 | + | protein_coding |
| ENST00000373206.1 | chr1 | 36372678 | 36372690 | 36372536 | 36372720 | + | protein_coding |
| ENST00000373204.3 | chr1 | 36372678 | 36372690 | 36372536 | 36372720 | + | protein_coding |
| ENST00000373206.1 | chr1 | 36379560 | 36379591 | 36379443 | 36379602 | + | protein_coding |
| ENST00000373204.3 | chr1 | 36379560 | 36379591 | 36379443 | 36379602 | + | protein_coding |
| ENST00000373206.1 | chr1 | 36379813 | 36379841 | 36379785 | 36379875 | + | protein_coding |
| ENST00000373204.3 | chr1 | 36379813 | 36379841 | 36379785 | 36379875 | + | protein_coding |
| ENST00000373206.1 | chr1 | 36381054 | 36381085 | 36380949 | 36381143 | + | protein_coding |
| ENST00000373204.3 | chr1 | 36381054 | 36381085 | 36380949 | 36381143 | + | protein_coding |
| ENST00000373206.1 | chr1 | 36383194 | 36383225 | 36383194 | 36383328 | + | protein_coding |
| ENST00000373204.3 | chr1 | 36383194 | 36383225 | 36383194 | 36383328 | + | protein_coding |
| ENST00000373206.1 | chr1 | 36384774 | 36384776 | 36384656 | 36384855 | + | protein_coding |
| ENST00000373204.3 | chr1 | 36384774 | 36384776 | 36384656 | 36384855 | + | protein_coding |
| ENST00000373191.3 | chr1 | 36474565 | 36474597 | 36474498 | 36474645 | + | protein_coding |
| ENST00000246314.6 | chr1 | 36474565 | 36474597 | 36474498 | 36474645 | + | protein_coding |
| ENST00000373191.3 | chr1 | 36505494 | 36505519 | 36505391 | 36505585 | + | protein_coding |
| ENST00000246314.6 | chr1 | 36505494 | 36505519 | 36505391 | 36505585 | + | protein_coding |
| ENST00000373191.3 | chr1 | 36520619 | 36520639 | 36520547 | 36520746 | + | protein_coding |
| ENST00000246314.6 | chr1 | 36520619 | 36520639 | 36520547 | 36520746 | + | protein_coding |
| ENST00000471099.1 | chr1 | 36520619 | 36520639 | 36520547 | 36520863 | + | processed_transcript |
| ENST00000373129.3 | chr1 | 36851449 | 36851480 | 36851324 | 36851485 | - | protein_coding |
| ENST00000373130.3 | chr1 | 36851449 | 36851480 | 36851324 | 36851489 | - | protein_coding |
| ENST00000373132.3 | chr1 | 36851449 | 36851480 | 36851324 | 36851497 | - | protein_coding |
| ENST00000482458.1 | chr1 | 36851449 | 36851480 | 36851324 | 36851480 | - | processed_transcript |
| ENST00000460017.1 | chr1 | 36851449 | 36851480 | 36851324 | 36851477 | - | processed_transcript |
| ENST00000373036.3 | chr1 | 38300759 | 38300790 | 38300751 | 38300887 | - | protein_coding |
| ENST00000543396.1 | chr1 | 38300759 | 38300790 | 38300751 | 38300887 | - | protein_coding |
| ENST00000432648.3 | chr1 | 39463943 | 39463974 | 39463843 | 39463983 | + | protein_coding |
| ENST00000372975.4 | chr1 | 39463943 | 39463974 | 39463843 | 39463983 | + | protein_coding |
| ENST00000529075.1 | chr1 | 39463943 | 39463974 | 39463843 | 39463983 | + | nonsense_mediated_decay |
| ENST00000446189.2 | chr1 | 39463943 | 39463974 | 39463843 | 39463983 | + | protein_coding |
| ENST00000531822.1 | chr1 | 39463943 | 39463974 | 39463843 | 39463983 | + | protein_coding |
| ENST00000397332.2 | chr1 | 40362970 | 40362971 | 40361098 | 40363642 | - | protein_coding |
| ENST00000372816.2 | chr1 | 40362970 | 40362971 | 40362899 | 40363642 | - | protein_coding |
| ENST00000372718.3 | chr1 | 40888505 | 40888533 | 40887648 | 40888998 | + | protein_coding |
| ENST00000372708.1 | chr1 | 40888505 | 40888533 | 40887648 | 40888998 | + | protein_coding |
| ENST00000362104.1 | chr1 | 41220043 | 41220074 | 41220027 | 41220118 | + | miRNA |
| ENST00000385227.1 | chr1 | 41222972 | 41223003 | 41222956 | 41223044 | + | miRNA |
| ENST00000372621.4 | chr1 | 41449034 | 41449065 | 41448950 | 41449128 | + | protein_coding |
| ENST00000470271.1 | chr1 | 41449034 | 41449065 | 41448950 | 41449128 | + | processed_transcript |
| ENST00000475060.1 | chr1 | 41449034 | 41449065 | 41448950 | 41449117 | + | processed_transcript |
| ENST00000479480.1 | chr1 | 41449034 | 41449065 | 41448950 | 41449128 | + | processed_transcript |
| ENST00000543104.1 | chr1 | 41449034 | 41449065 | 41448950 | 41449128 | + | protein_coding |
| ENST00000372616.1 | chr1 | 41449034 | 41449065 | 41448833 | 41449128 | + | protein_coding |
| ENST00000466944.1 | chr1 | 42779388 | 42779419 | 42779299 | 42779471 | + | rRNA_pseudogene |
| ENST00000372224.4 | chr1 | 45233247 | 45233278 | 45232769 | 45233439 | + | protein_coding |
| ENST00000372218.4 | chr1 | 45233247 | 45233278 | 45232769 | 45233436 | + | protein_coding |
| ENST00000372217.1 | chr1 | 45233247 | 45233278 | 45232769 | 45233439 | + | protein_coding |
| ENST00000484599.1 | chr1 | 45244100 | 45244128 | 45243672 | 45244335 | + | processed_transcript |
| ENST00000384690.1 | chr1 | 45244100 | 45244128 | 45244062 | 45244128 | + | snoRNA |
| ENST00000262746.1 | chr1 | 45981434 | 45981465 | 45981326 | 45981479 | - | protein_coding |
| ENST00000319248.8 | chr1 | 45981434 | 45981465 | 45981326 | 45981479 | - | protein_coding |
| ENST00000447184.1 | chr1 | 45981434 | 45981465 | 45981326 | 45981479 | - | protein_coding |
| ENST00000424390.1 | chr1 | 45981434 | 45981465 | 45981326 | 45981479 | - | protein_coding |
| ENST00000483583.1 | chr1 | 45981434 | 45981465 | 45981326 | 45981479 | - | processed_transcript |
| ENST00000371933.3 | chr1 | 47142088 | 47142119 | 47140831 | 47144308 | - | protein_coding |
| ENST00000371873.5 | chr1 | 47843698 | 47843729 | 47842376 | 47844511 | + | protein_coding |
| ENST00000313334.8 | chr1 | 52553475 | 52553506 | 52552384 | 52556388 | + | protein_coding |
| ENST00000472944.2 | chr1 | 52553475 | 52553506 | 52552384 | 52554090 | + | protein_coding |
| ENST00000357206.2 | chr1 | 52811941 | 52811972 | 52811732 | 52812187 | + | protein_coding |
| ENST00000287727.3 | chr1 | 52811941 | 52811972 | 52811732 | 52812357 | + | protein_coding |
| ENST00000371591.1 | chr1 | 52811941 | 52811972 | 52811732 | 52812358 | + | protein_coding |
| ENST00000516209.1 | chr1 | 53220014 | 53220051 | 53219994 | 53220134 | + | snRNA |
| ENST00000371429.3 | chr1 | 54262430 | 54262462 | 54262362 | 54262477 | - | protein_coding |
| ENST00000360494.4 | chr1 | 54262430 | 54262462 | 54262362 | 54262471 | - | protein_coding |
| ENST00000540001.1 | chr1 | 54262430 | 54262462 | 54262421 | 54262477 | - | protein_coding |
| ENST00000537333.1 | chr1 | 54262430 | 54262462 | 54262362 | 54262477 | - | protein_coding |
| ENST00000234725.8 | chr1 | 54262430 | 54262462 | 54262362 | 54262477 | - | protein_coding |
| ENST00000516036.1 | chr1 | 55950592 | 55950623 | 55950544 | 55950645 | - | misc_RNA |
| ENST00000407417.2 | chr1 | 61924175 | 61924204 | 61920975 | 61928459 | + | protein_coding |
| ENST00000371189.3 | chr1 | 61924175 | 61924204 | 61920975 | 61928459 | + | protein_coding |
| ENST00000403491.3 | chr1 | 61924175 | 61924204 | 61920975 | 61928465 | + | protein_coding |
| ENST00000371017.3 | chr1 | 67447462 | 67447493 | 67447462 | 67447601 | + | protein_coding |
| ENST00000371018.3 | chr1 | 67447462 | 67447493 | 67447462 | 67447601 | + | protein_coding |
| ENST00000355977.6 | chr1 | 67447462 | 67447493 | 67447462 | 67447601 | + | protein_coding |
| ENST00000357692.2 | chr1 | 67447462 | 67447493 | 67447462 | 67447601 | + | protein_coding |
| ENST00000401041.1 | chr1 | 67447462 | 67447493 | 67447462 | 67447601 | + | protein_coding |
| ENST00000371016.1 | chr1 | 67447462 | 67447493 | 67447462 | 67447601 | + | protein_coding |
| ENST00000371014.1 | chr1 | 67447462 | 67447493 | 67447462 | 67447601 | + | protein_coding |
| ENST00000401042.3 | chr1 | 67447462 | 67447493 | 67447462 | 67447601 | + | protein_coding |
| ENST00000355356.3 | chr1 | 67447462 | 67447493 | 67447462 | 67447601 | + | protein_coding |
| ENST00000371017.3 | chr1 | 67453071 | 67453085 | 67450255 | 67454302 | + | protein_coding |
| ENST00000371018.3 | chr1 | 67453071 | 67453085 | 67452086 | 67454302 | + | protein_coding |
| ENST00000355977.6 | chr1 | 67453071 | 67453085 | 67452086 | 67454302 | + | protein_coding |
| ENST00000357692.2 | chr1 | 67453071 | 67453085 | 67450255 | 67453714 | + | protein_coding |
| ENST00000355356.3 | chr1 | 67453071 | 67453085 | 67450255 | 67453713 | + | protein_coding |
| ENST00000370994.4 | chr1 | 67874118 | 67874149 | 67873493 | 67878946 | - | protein_coding |
| ENST00000370994.4 | chr1 | 67876775 | 67876806 | 67873493 | 67878946 | - | protein_coding |
| ENST00000370995.2 | chr1 | 67876775 | 67876806 | 67876721 | 67878946 | - | protein_coding |
| ENST00000370994.4 | chr1 | 67877150 | 67877181 | 67873493 | 67878946 | - | protein_coding |
| ENST00000370995.2 | chr1 | 67877150 | 67877181 | 67876721 | 67878946 | - | protein_coding |
| ENST00000370982.3 | chr1 | 68169575 | 68169606 | 68167149 | 68171259 | - | protein_coding |
| ENST00000357731.4 | chr1 | 72748278 | 72748310 | 72748002 | 72748417 | - | protein_coding |
| ENST00000370721.1 | chr1 | 82456844 | 82456869 | 82456075 | 82457052 | + | protein_coding |
| ENST00000370728.1 | chr1 | 82456844 | 82456869 | 82456075 | 82458106 | + | protein_coding |
| ENST00000370730.1 | chr1 | 82456844 | 82456869 | 82456075 | 82458106 | + | protein_coding |
| ENST00000370727.1 | chr1 | 82456844 | 82456869 | 82456075 | 82457052 | + | protein_coding |
| ENST00000370725.1 | chr1 | 82456844 | 82456869 | 82456075 | 82457052 | + | protein_coding |
| ENST00000370723.1 | chr1 | 82456844 | 82456869 | 82456075 | 82457052 | + | protein_coding |
| ENST00000469377.1 | chr1 | 82456844 | 82456869 | 82456075 | 82458107 | + | processed_transcript |
| ENST00000359929.2 | chr1 | 82456844 | 82456869 | 82456075 | 82458107 | + | protein_coding |
| ENST00000370715.1 | chr1 | 82456844 | 82456869 | 82456075 | 82457880 | + | protein_coding |
| ENST00000370713.1 | chr1 | 82456844 | 82456869 | 82456075 | 82457880 | + | protein_coding |
| ENST00000319517.6 | chr1 | 82456844 | 82456869 | 82456075 | 82457880 | + | protein_coding |
| ENST00000370717.2 | chr1 | 82456844 | 82456869 | 82456075 | 82458106 | + | protein_coding |
| ENST00000394879.1 | chr1 | 82456844 | 82456869 | 82456075 | 82458106 | + | protein_coding |
| ENST00000271029.4 | chr1 | 82456844 | 82456869 | 82456075 | 82458106 | + | protein_coding |
| ENST00000335786.5 | chr1 | 82456844 | 82456869 | 82456075 | 82458106 | + | protein_coding |
| ENST00000449420.1 | chr1 | 82456844 | 82456869 | 82456075 | 82457052 | + | protein_coding |
| ENST00000402328.2 | chr1 | 82456844 | 82456869 | 82456075 | 82458107 | + | protein_coding |
| ENST00000370654.5 | chr1 | 84961656 | 84961687 | 84961565 | 84961746 | + | protein_coding |
| ENST00000487806.1 | chr1 | 84964122 | 84964153 | 84964008 | 84964231 | - | processed_transcript |
| ENST00000370645.4 | chr1 | 84964122 | 84964153 | 84964008 | 84964231 | - | protein_coding |
| ENST00000370641.3 | chr1 | 84964122 | 84964153 | 84964012 | 84964231 | - | protein_coding |
| ENST00000370608.3 | chr1 | 85392131 | 85392162 | 85391268 | 85392404 | - | protein_coding |
| ENST00000463065.1 | chr1 | 85392131 | 85392162 | 85391591 | 85392404 | - | nonsense_mediated_decay |
| ENST00000530971.1 | chr1 | 85392131 | 85392162 | 85392109 | 85392404 | - | retained_intron |
| ENST00000321792.5 | chr1 | 89449309 | 89449340 | 89445139 | 89449749 | - | protein_coding |
| ENST00000399794.2 | chr1 | 89449309 | 89449340 | 89445142 | 89449749 | - | protein_coding |
| ENST00000413769.1 | chr1 | 89449309 | 89449340 | 89449034 | 89449749 | - | processed_transcript |
| ENST00000330947.2 | chr1 | 90061792 | 90061823 | 90058330 | 90063423 | + | protein_coding |
| ENST00000439853.1 | chr1 | 90061792 | 90061823 | 90058330 | 90063418 | + | protein_coding |
| ENST00000370454.4 | chr1 | 90178700 | 90178731 | 90178268 | 90185092 | + | protein_coding |
| ENST00000370440.1 | chr1 | 91381880 | 91381911 | 91380859 | 91382547 | - | protein_coding |
| ENST00000347275.5 | chr1 | 91381880 | 91381911 | 91380861 | 91382547 | - | protein_coding |
| ENST00000337393.5 | chr1 | 91381880 | 91381911 | 91380897 | 91382547 | - | protein_coding |
| ENST00000361321.5 | chr1 | 91381880 | 91381911 | 91381742 | 91382547 | - | protein_coding |
| ENST00000370238.3 | chr1 | 94354044 | 94354070 | 94350761 | 94354715 | - | protein_coding |
| ENST00000385223.1 | chr1 | 98511638 | 98511669 | 98511626 | 98511727 | - | miRNA |
| ENST00000370112.4 | chr1 | 101427429 | 101427445 | 101427384 | 101427474 | + | protein_coding |
| ENST00000357650.4 | chr1 | 101427429 | 101427445 | 101427384 | 101427474 | + | protein_coding |
| ENST00000370111.3 | chr1 | 101427429 | 101427445 | 101427384 | 101427474 | + | protein_coding |
| ENST00000516385.1 | chr1 | 108113446 | 108113477 | 108113461 | 108113552 | - | miRNA |
| ENST00000338366.5 | chr1 | 109606935 | 109606966 | 109606762 | 109607315 | - | protein_coding |
| ENST00000369779.4 | chr1 | 110906248 | 110906279 | 110905470 | 110906515 | - | protein_coding |
| ENST00000472422.2 | chr1 | 110906248 | 110906279 | 110905482 | 110906515 | - | protein_coding |
| ENST00000369781.4 | chr1 | 110906248 | 110906279 | 110905482 | 110906515 | - | protein_coding |
| ENST00000461647.2 | chr1 | 110906248 | 110906279 | 110905482 | 110906515 | - | nonsense_mediated_decay |
| ENST00000528649.1 | chr1 | 110906248 | 110906279 | 110906248 | 110906535 | - | nonsense_mediated_decay |
| ENST00000437429.2 | chr1 | 110906248 | 110906279 | 110906248 | 110906535 | - | protein_coding |
| ENST00000263168.3 | chr1 | 113213601 | 113213632 | 113212614 | 113214241 | + | protein_coding |
| ENST00000355485.2 | chr1 | 116206706 | 116206751 | 116206282 | 116206889 | + | protein_coding |
| ENST00000369510.3 | chr1 | 116206706 | 116206751 | 116206282 | 116206889 | + | protein_coding |
| ENST00000310260.3 | chr1 | 116206706 | 116206751 | 116206282 | 116206889 | + | protein_coding |
| ENST00000369509.1 | chr1 | 116206706 | 116206751 | 116206282 | 116206889 | + | protein_coding |
| ENST00000369486.3 | chr1 | 117117886 | 117117917 | 117117031 | 117120184 | - | protein_coding |
| ENST00000369483.1 | chr1 | 117117886 | 117117917 | 117117031 | 117120184 | - | protein_coding |
| ENST00000390209.2 | chr1 | 117214409 | 117214440 | 117214371 | 117214449 | + | miRNA |
| ENST00000401111.1 | chr1 | 117637279 | 117637310 | 117637265 | 117637350 | + | miRNA |
| ENST00000463780.1 | chr1 | 120544000 | 120544031 | 120543874 | 120544125 | - | rRNA_pseudogene |
| ENST00000483932.1 | chr1 | 145277344 | 145277374 | 145277250 | 145277501 | + | rRNA_pseudogene |
| ENST00000384687.1 | chr1 | 145969344 | 145969365 | 145969270 | 145969433 | - | snRNA |
| ENST00000383858.1 | chr1 | 147511107 | 147511114 | 147511004 | 147511167 | - | snRNA |
| ENST00000461559.1 | chr1 | 147781070 | 147781101 | 147781030 | 147781101 | - | tRNA_pseudogene |
| ENST00000384770.1 | chr1 | 148241539 | 148241569 | 148241465 | 148241628 | + | snRNA |
| ENST00000384010.1 | chr1 | 149224143 | 149224152 | 149224058 | 149224221 | - | snRNA |
| ENST00000486830.1 | chr1 | 149680249 | 149680280 | 149680210 | 149680280 | - | tRNA_pseudogene |
| ENST00000545683.1 | chr1 | 149783887 | 149783918 | 149783502 | 149783928 | - | protein_coding |
| ENST00000427880.2 | chr1 | 149783887 | 149783918 | 149783502 | 149783925 | - | protein_coding |
| ENST00000469483.1 | chr1 | 149783887 | 149783918 | 149783502 | 149783893 | - | processed_transcript |
| ENST00000369167.1 | chr1 | 149783887 | 149783918 | 149783283 | 149783914 | - | protein_coding |
| ENST00000369158.1 | chr1 | 149812268 | 149812312 | 149811110 | 149812765 | - | protein_coding |
| ENST00000369159.2 | chr1 | 149823122 | 149823153 | 149822643 | 149823191 | + | protein_coding |
| ENST00000403683.1 | chr1 | 149824634 | 149824678 | 149824181 | 149825836 | + | protein_coding |
| ENST00000392933.1 | chr1 | 149832593 | 149832594 | 149832268 | 149832704 | - | protein_coding |
| ENST00000369157.1 | chr1 | 149832593 | 149832594 | 149832268 | 149832704 | - | protein_coding |
| ENST00000392932.3 | chr1 | 149832593 | 149832594 | 149831516 | 149832714 | - | protein_coding |
| ENST00000271628.8 | chr1 | 149898569 | 149898600 | 149898268 | 149898810 | - | protein_coding |
| ENST00000457312.1 | chr1 | 149898569 | 149898600 | 149898274 | 149898810 | - | protein_coding |
| ENST00000314136.8 | chr1 | 150191118 | 150191149 | 150190717 | 150193063 | - | protein_coding |
| ENST00000369119.3 | chr1 | 150191118 | 150191149 | 150190855 | 150193063 | - | protein_coding |
| ENST00000369116.4 | chr1 | 150191118 | 150191149 | 150190861 | 150193063 | - | protein_coding |
| ENST00000314136.8 | chr1 | 150192835 | 150192866 | 150190717 | 150193063 | - | protein_coding |
| ENST00000369119.3 | chr1 | 150192835 | 150192866 | 150190855 | 150193063 | - | protein_coding |
| ENST00000369116.4 | chr1 | 150192835 | 150192866 | 150190861 | 150193063 | - | protein_coding |
| ENST00000436748.2 | chr1 | 150192835 | 150192866 | 150192574 | 150193063 | - | protein_coding |
| ENST00000534437.1 | chr1 | 150192835 | 150192866 | 150192680 | 150193063 | - | protein_coding |
| ENST00000369114.5 | chr1 | 150192835 | 150192866 | 150192694 | 150193060 | - | protein_coding |
| ENST00000369115.2 | chr1 | 150192835 | 150192866 | 150192763 | 150193063 | - | protein_coding |
| ENST00000533654.1 | chr1 | 150192835 | 150192866 | 150192840 | 150193063 | - | protein_coding |
| ENST00000410290.1 | chr1 | 150209469 | 150209502 | 150209315 | 150209504 | + | snRNA |
| ENST00000368830.3 | chr1 | 151732347 | 151732378 | 151732119 | 151732657 | - | protein_coding |
| ENST00000467306.1 | chr1 | 151732347 | 151732378 | 151732123 | 151732657 | - | processed_transcript |
| ENST00000468006.1 | chr1 | 151732347 | 151732378 | 151732130 | 151732507 | - | processed_transcript |
| ENST00000368829.3 | chr1 | 151732347 | 151732378 | 151732351 | 151732657 | - | protein_coding |
| ENST00000495867.1 | chr1 | 151732347 | 151732378 | 151732360 | 151732657 | - | processed_transcript |
| ENST00000368830.3 | chr1 | 151734910 | 151734941 | 151734852 | 151734976 | - | protein_coding |
| ENST00000467306.1 | chr1 | 151734910 | 151734941 | 151734852 | 151734976 | - | processed_transcript |
| ENST00000368829.3 | chr1 | 151734910 | 151734941 | 151734852 | 151734976 | - | protein_coding |
| ENST00000478926.1 | chr1 | 151734910 | 151734941 | 151734852 | 151734976 | - | processed_transcript |
| ENST00000486707.1 | chr1 | 151734910 | 151734941 | 151734852 | 151734976 | - | processed_transcript |
| ENST00000492684.1 | chr1 | 151734910 | 151734941 | 151734852 | 151734976 | - | processed_transcript |
| ENST00000481777.1 | chr1 | 151734910 | 151734941 | 151734852 | 151734976 | - | processed_transcript |
| ENST00000461182.1 | chr1 | 151734910 | 151734941 | 151734794 | 151734976 | - | processed_transcript |
| ENST00000303569.6 | chr1 | 153924630 | 153924661 | 153924494 | 153924738 | - | nonsense_mediated_decay |
| ENST00000461638.1 | chr1 | 153924630 | 153924661 | 153924494 | 153924738 | - | nonsense_mediated_decay |
| ENST00000368633.1 | chr1 | 153924630 | 153924661 | 153924494 | 153924738 | - | protein_coding |
| ENST00000487235.1 | chr1 | 153924630 | 153924661 | 153924494 | 153924738 | - | retained_intron |
| ENST00000467860.1 | chr1 | 153924630 | 153924661 | 153924608 | 153924738 | - | retained_intron |
| ENST00000493909.1 | chr1 | 153924630 | 153924661 | 153924632 | 153924738 | - | retained_intron |
| ENST00000350592.3 | chr1 | 154180019 | 154180050 | 154179196 | 154180124 | - | protein_coding |
| ENST00000368521.4 | chr1 | 154180019 | 154180050 | 154179196 | 154180124 | - | protein_coding |
| ENST00000362076.4 | chr1 | 154180019 | 154180050 | 154179196 | 154180124 | - | protein_coding |
| ENST00000368519.1 | chr1 | 154180019 | 154180050 | 154179652 | 154180124 | - | protein_coding |
| ENST00000483282.1 | chr1 | 154180019 | 154180050 | 154179706 | 154180124 | - | processed_transcript |
| ENST00000493814.1 | chr1 | 154180019 | 154180050 | 154180016 | 154180124 | - | processed_transcript |
| ENST00000343815.6 | chr1 | 154223660 | 154223691 | 154223517 | 154223794 | + | protein_coding |
| ENST00000428931.1 | chr1 | 154223660 | 154223691 | 154223517 | 154223794 | + | protein_coding |
| ENST00000456955.2 | chr1 | 154223660 | 154223691 | 154223517 | 154223794 | + | protein_coding |
| ENST00000433006.2 | chr1 | 154223660 | 154223691 | 154223517 | 154223794 | + | protein_coding |
| ENST00000271877.7 | chr1 | 154223660 | 154223691 | 154223517 | 154223794 | + | protein_coding |
| ENST00000361546.2 | chr1 | 154223660 | 154223691 | 154223517 | 154223794 | + | protein_coding |
| ENST00000368439.1 | chr1 | 154947228 | 154947259 | 154947129 | 154947388 | + | protein_coding |
| ENST00000368436.1 | chr1 | 154947228 | 154947259 | 154947159 | 154947280 | + | protein_coding |
| ENST00000308987.5 | chr1 | 154947228 | 154947259 | 154947175 | 154947280 | + | protein_coding |
| ENST00000368346.3 | chr1 | 155447734 | 155447765 | 155447677 | 155452240 | - | protein_coding |
| ENST00000392403.3 | chr1 | 155447734 | 155447765 | 155447677 | 155452240 | - | protein_coding |
| ENST00000368346.3 | chr1 | 155448209 | 155448240 | 155447677 | 155452240 | - | protein_coding |
| ENST00000392403.3 | chr1 | 155448209 | 155448240 | 155447677 | 155452240 | - | protein_coding |
| ENST00000361040.5 | chr1 | 155734217 | 155734245 | 155734218 | 155736500 | - | protein_coding |
| ENST00000466224.1 | chr1 | 155754150 | 155754186 | 155753777 | 155754180 | - | processed_transcript |
| ENST00000467009.1 | chr1 | 155754150 | 155754186 | 155754166 | 155755165 | - | processed_transcript |
| ENST00000368020.1 | chr1 | 161009078 | 161009094 | 161009041 | 161009799 | - | protein_coding |
| ENST00000368021.3 | chr1 | 161009078 | 161009094 | 161009041 | 161009799 | - | protein_coding |
| ENST00000435396.1 | chr1 | 161009078 | 161009094 | 161009042 | 161009799 | - | protein_coding |
| ENST00000472217.1 | chr1 | 161009078 | 161009094 | 161009085 | 161009799 | - | retained_intron |
| ENST00000479948.1 | chr1 | 161176229 | 161176260 | 161176197 | 161176383 | + | processed_transcript |
| ENST00000496133.1 | chr1 | 161176229 | 161176260 | 161176197 | 161176387 | + | processed_transcript |
| ENST00000367993.3 | chr1 | 161176229 | 161176260 | 161176197 | 161176387 | + | protein_coding |
| ENST00000496553.1 | chr1 | 161176229 | 161176260 | 161176197 | 161176387 | + | processed_transcript |
| ENST00000392179.4 | chr1 | 161176229 | 161176260 | 161176197 | 161176387 | + | protein_coding |
| ENST00000467295.1 | chr1 | 161176229 | 161176260 | 161176197 | 161176387 | + | processed_transcript |
| ENST00000476409.2 | chr1 | 161176229 | 161176260 | 161176197 | 161176387 | + | protein_coding |
| ENST00000478866.1 | chr1 | 161176229 | 161176260 | 161176197 | 161176387 | + | processed_transcript |
| ENST00000546154.1 | chr1 | 161176229 | 161176260 | 161176197 | 161176387 | + | protein_coding |
| ENST00000473321.1 | chr1 | 161176229 | 161176260 | 161176199 | 161176387 | + | processed_transcript |
| ENST00000474862.1 | chr1 | 161390661 | 161390692 | 161390661 | 161390733 | + | tRNA_pseudogene |
| ENST00000467525.1 | chr1 | 161493641 | 161493672 | 161493080 | 161493803 | + | processed_transcript |
| ENST00000467525.1 | chr1 | 161493676 | 161493714 | 161493080 | 161493803 | + | processed_transcript |
| ENST00000545294.1 | chr1 | 162469763 | 162469794 | 162469745 | 162470037 | + | protein_coding |
| ENST00000538489.1 | chr1 | 162469763 | 162469794 | 162469745 | 162470037 | + | protein_coding |
| ENST00000489294.1 | chr1 | 162469763 | 162469794 | 162469745 | 162470037 | + | protein_coding |
| ENST00000282169.8 | chr1 | 162469763 | 162469794 | 162469745 | 162470894 | + | processed_transcript |
| ENST00000538489.1 | chr1 | 162498355 | 162498386 | 162492194 | 162499419 | + | protein_coding |
| ENST00000489294.1 | chr1 | 162498355 | 162498386 | 162492194 | 162499400 | + | protein_coding |
| ENST00000367879.4 | chr1 | 165877252 | 165877283 | 165876921 | 165877346 | + | protein_coding |
| ENST00000372212.4 | chr1 | 165877252 | 165877283 | 165876921 | 165877607 | + | protein_coding |
| ENST00000462329.1 | chr1 | 165877252 | 165877283 | 165876921 | 165877337 | + | processed_transcript |
| ENST00000469256.1 | chr1 | 165877252 | 165877283 | 165876921 | 165877311 | + | processed_transcript |
| ENST00000463772.1 | chr1 | 165877252 | 165877283 | 165876921 | 165877346 | + | processed_transcript |
| ENST00000479872.1 | chr1 | 165877252 | 165877283 | 165876921 | 165880855 | + | processed_transcript |
| ENST00000367866.1 | chr1 | 167386453 | 167386484 | 167384737 | 167396582 | + | protein_coding |
| ENST00000367865.1 | chr1 | 167386453 | 167386484 | 167384737 | 167396582 | + | protein_coding |
| ENST00000271411.4 | chr1 | 167386453 | 167386484 | 167384737 | 167396582 | + | nonsense_mediated_decay |
| ENST00000367862.4 | chr1 | 167386453 | 167386484 | 167384737 | 167396582 | + | protein_coding |
| ENST00000385239.1 | chr1 | 168344822 | 168344853 | 168344762 | 168344859 | + | miRNA |
| ENST00000367816.1 | chr1 | 169100842 | 169100873 | 169100530 | 169101960 | + | protein_coding |
| ENST00000367815.3 | chr1 | 169100842 | 169100873 | 169100530 | 169101960 | + | protein_coding |
| ENST00000499679.3 | chr1 | 169100842 | 169100873 | 169100530 | 169101553 | + | protein_coding |
| ENST00000367813.3 | chr1 | 169100842 | 169100873 | 169100530 | 169101254 | + | protein_coding |
| ENST00000367742.3 | chr1 | 171562191 | 171562222 | 171560726 | 171562650 | + | protein_coding |
| ENST00000338920.4 | chr1 | 171562191 | 171562222 | 171560726 | 171562650 | + | protein_coding |
| ENST00000495585.1 | chr1 | 171562191 | 171562222 | 171560726 | 171562650 | + | protein_coding |
| ENST00000236192.7 | chr1 | 171670582 | 171670613 | 171669300 | 171673674 | - | protein_coding |
| ENST00000415773.1 | chr1 | 171670582 | 171670613 | 171669312 | 171671224 | - | protein_coding |
| ENST00000474047.1 | chr1 | 171670582 | 171670613 | 171670232 | 171671224 | - | nonsense_mediated_decay |
| ENST00000344529.4 | chr1 | 172411623 | 172411648 | 172410597 | 172411970 | - | protein_coding |
| ENST00000367728.1 | chr1 | 172411623 | 172411648 | 172410597 | 172413226 | - | protein_coding |
| ENST00000258324.1 | chr1 | 172411623 | 172411648 | 172410598 | 172411993 | - | protein_coding |
| ENST00000478184.1 | chr1 | 172411623 | 172411648 | 172411471 | 172411970 | - | processed_transcript |
| ENST00000363859.1 | chr1 | 173834490 | 173834518 | 173834486 | 173834570 | - | snoRNA |
| ENST00000383880.1 | chr1 | 173835110 | 173835138 | 173835106 | 173835166 | - | snoRNA |
| ENST00000364084.1 | chr1 | 173836817 | 173836844 | 173836809 | 173836888 | - | snoRNA |
| ENST00000502732.1 | chr1 | 179071132 | 179071166 | 179068462 | 179078576 | - | protein_coding |
| ENST00000408940.3 | chr1 | 179071132 | 179071166 | 179068463 | 179078576 | - | protein_coding |
| ENST00000331872.6 | chr1 | 182360894 | 182360925 | 182360814 | 182361341 | - | protein_coding |
| ENST00000311223.5 | chr1 | 182360894 | 182360925 | 182360814 | 182361341 | - | protein_coding |
| ENST00000339526.4 | chr1 | 182360894 | 182360925 | 182359632 | 182360927 | - | protein_coding |
| ENST00000484996.1 | chr1 | 182360894 | 182360925 | 182360310 | 182360908 | - | processed_transcript |
| ENST00000489818.1 | chr1 | 182360894 | 182360925 | 182360814 | 182360925 | - | processed_transcript |
| ENST00000462444.1 | chr1 | 182360894 | 182360925 | 182360814 | 182360909 | - | processed_transcript |
| ENST00000418655.1 | chr1 | 189961337 | 189961344 | 189958664 | 189962227 | - | processed_pseudogene |
| ENST00000418655.1 | chr1 | 189961346 | 189961378 | 189958664 | 189962227 | - | processed_pseudogene |
| ENST00000385240.1 | chr1 | 198828045 | 198828076 | 198828002 | 198828111 | - | miRNA |
| ENST00000385026.1 | chr1 | 198828194 | 198828259 | 198828173 | 198828282 | - | miRNA |
| ENST00000495527.1 | chr1 | 203768175 | 203768206 | 203767974 | 203772144 | + | nonsense_mediated_decay |
| ENST00000550078.1 | chr1 | 203768175 | 203768206 | 203765437 | 203769686 | + | protein_coding |
| ENST00000332127.4 | chr1 | 203822149 | 203822180 | 203821269 | 203823252 | + | protein_coding |
| ENST00000495527.1 | chr1 | 203822149 | 203822180 | 203821269 | 203823162 | + | nonsense_mediated_decay |
| ENST00000545588.1 | chr1 | 203822149 | 203822180 | 203821269 | 203823246 | + | protein_coding |
| ENST00000367210.1 | chr1 | 203822149 | 203822180 | 203821269 | 203823246 | + | protein_coding |
| ENST00000367188.4 | chr1 | 204374891 | 204374922 | 204372515 | 204375441 | - | protein_coding |
| ENST00000367182.3 | chr1 | 204526155 | 204526186 | 204518241 | 204527248 | + | protein_coding |
| ENST00000363538.1 | chr1 | 204531552 | 204531586 | 204531541 | 204531649 | - | rRNA |
| ENST00000539916.1 | chr1 | 205589108 | 205589139 | 205589094 | 205589966 | - | protein_coding |
| ENST00000357992.4 | chr1 | 205589108 | 205589139 | 205589094 | 205589966 | - | protein_coding |
| ENST00000289703.4 | chr1 | 205589108 | 205589139 | 205588396 | 205589966 | - | protein_coding |
| ENST00000367142.4 | chr1 | 205683014 | 205683045 | 205681947 | 205687607 | - | protein_coding |
| ENST00000367142.4 | chr1 | 205685302 | 205685333 | 205681947 | 205687607 | - | protein_coding |
| ENST00000367142.4 | chr1 | 205686773 | 205686804 | 205681947 | 205687607 | - | protein_coding |
| ENST00000483628.1 | chr1 | 206580431 | 206580462 | 206579736 | 206580621 | + | processed_transcript |
| ENST00000467419.1 | chr1 | 206580431 | 206580462 | 206579736 | 206581391 | + | processed_transcript |
| ENST00000451736.1 | chr1 | 206869382 | 206869413 | 206869182 | 206869614 | - | processed_pseudogene |
| ENST00000385231.1 | chr1 | 207975200 | 207975231 | 207975197 | 207975284 | - | miRNA |
| ENST00000384891.1 | chr1 | 209605511 | 209605542 | 209605478 | 209605587 | + | miRNA |
| ENST00000261455.4 | chr1 | 212538456 | 212538456 | 212537273 | 212538718 | - | protein_coding |
| ENST00000535273.1 | chr1 | 212538456 | 212538456 | 212537824 | 212538718 | - | protein_coding |
| ENST00000366971.4 | chr1 | 213070198 | 213070230 | 213068558 | 213072705 | + | protein_coding |
| ENST00000366934.3 | chr1 | 217781865 | 217781896 | 217781526 | 217783534 | - | protein_coding |
| ENST00000366932.3 | chr1 | 218504994 | 218505032 | 218504290 | 218511325 | + | protein_coding |
| ENST00000366923.3 | chr1 | 220184366 | 220184391 | 220184288 | 220184398 | - | protein_coding |
| ENST00000536921.1 | chr1 | 220184366 | 220184391 | 220184288 | 220184398 | - | protein_coding |
| ENST00000435048.1 | chr1 | 220184366 | 220184391 | 220184288 | 220184398 | - | protein_coding |
| ENST00000384858.1 | chr1 | 220291247 | 220291278 | 220291195 | 220291304 | - | miRNA |
| ENST00000384892.1 | chr1 | 220291538 | 220291569 | 220291499 | 220291583 | - | miRNA |
| ENST00000408479.1 | chr1 | 224444741 | 224444772 | 224444706 | 224444843 | - | miRNA |
| ENST00000414423.2 | chr1 | 224577192 | 224577223 | 224572845 | 224577562 | - | protein_coding |
| ENST00000295024.6 | chr1 | 224577192 | 224577223 | 224572851 | 224577562 | - | protein_coding |
| ENST00000486652.1 | chr1 | 224577192 | 224577223 | 224576838 | 224577562 | - | nonsense_mediated_decay |
| ENST00000366852.2 | chr1 | 224577192 | 224577223 | 224576839 | 224577562 | - | protein_coding |
| ENST00000414423.2 | chr1 | 224577458 | 224577490 | 224572845 | 224577562 | - | protein_coding |
| ENST00000295024.6 | chr1 | 224577458 | 224577490 | 224572851 | 224577562 | - | protein_coding |
| ENST00000486652.1 | chr1 | 224577458 | 224577490 | 224576838 | 224577562 | - | nonsense_mediated_decay |
| ENST00000366852.2 | chr1 | 224577458 | 224577490 | 224576839 | 224577562 | - | protein_coding |
| ENST00000272163.4 | chr1 | 225589619 | 225589649 | 225589204 | 225591165 | - | protein_coding |
| ENST00000338179.2 | chr1 | 225589619 | 225589649 | 225589205 | 225591165 | - | protein_coding |
| ENST00000304786.7 | chr1 | 225971012 | 225971044 | 225971002 | 225971070 | + | protein_coding |
| ENST00000366839.4 | chr1 | 225971012 | 225971044 | 225971002 | 225971070 | + | protein_coding |
| ENST00000366838.1 | chr1 | 225971012 | 225971044 | 225971002 | 225971070 | + | protein_coding |
| ENST00000304786.7 | chr1 | 225977626 | 225977655 | 225976942 | 225978168 | + | protein_coding |
| ENST00000366839.4 | chr1 | 225977626 | 225977655 | 225976942 | 225978168 | + | protein_coding |
| ENST00000272091.7 | chr1 | 226172761 | 226172793 | 226170403 | 226173224 | - | protein_coding |
| ENST00000366818.3 | chr1 | 226172761 | 226172793 | 226170406 | 226173224 | - | protein_coding |
| ENST00000366812.5 | chr1 | 226334235 | 226334253 | 226332380 | 226334522 | - | protein_coding |
| ENST00000366794.5 | chr1 | 226549660 | 226549691 | 226549670 | 226549784 | - | protein_coding |
| ENST00000490921.1 | chr1 | 226549660 | 226549691 | 226549670 | 226549784 | - | processed_transcript |
| ENST00000463968.1 | chr1 | 226549660 | 226549691 | 226549670 | 226549784 | - | processed_transcript |
| ENST00000491816.1 | chr1 | 226549660 | 226549691 | 226549670 | 226549953 | - | processed_transcript |
| ENST00000468608.1 | chr1 | 226549660 | 226549691 | 226549409 | 226549784 | - | processed_transcript |
| ENST00000365394.1 | chr1 | 227748973 | 227749004 | 227748882 | 227749001 | + | rRNA |
| ENST00000477821.1 | chr1 | 228284964 | 228284995 | 228284779 | 228284968 | + | processed_transcript |
| ENST00000470670.1 | chr1 | 228284964 | 228284995 | 228284779 | 228285153 | + | processed_transcript |
| ENST00000362482.1 | chr1 | 228746004 | 228746064 | 228746015 | 228746133 | - | rRNA |
| ENST00000362482.1 | chr1 | 228746079 | 228746133 | 228746015 | 228746133 | - | rRNA |
| ENST00000364718.1 | chr1 | 228748245 | 228748304 | 228748256 | 228748374 | - | rRNA |
| ENST00000364718.1 | chr1 | 228748322 | 228748374 | 228748256 | 228748374 | - | rRNA |
| ENST00000362467.1 | chr1 | 228750486 | 228750548 | 228750497 | 228750615 | - | rRNA |
| ENST00000362467.1 | chr1 | 228750565 | 228750615 | 228750497 | 228750615 | - | rRNA |
| ENST00000363511.1 | chr1 | 228752727 | 228752789 | 228752738 | 228752856 | - | rRNA |
| ENST00000363511.1 | chr1 | 228752809 | 228752856 | 228752738 | 228752856 | - | rRNA |
| ENST00000362526.1 | chr1 | 228754967 | 228755028 | 228754979 | 228755097 | - | rRNA |
| ENST00000362526.1 | chr1 | 228755045 | 228755047 | 228754979 | 228755097 | - | rRNA |
| ENST00000362526.1 | chr1 | 228755051 | 228755097 | 228754979 | 228755097 | - | rRNA |
| ENST00000365651.1 | chr1 | 228759402 | 228759454 | 228759414 | 228759532 | - | rRNA |
| ENST00000365651.1 | chr1 | 228759489 | 228759532 | 228759414 | 228759532 | - | rRNA |
| ENST00000363473.1 | chr1 | 228761646 | 228761708 | 228761656 | 228761774 | - | rRNA |
| ENST00000363473.1 | chr1 | 228761722 | 228761733 | 228761656 | 228761774 | - | rRNA |
| ENST00000363473.1 | chr1 | 228761742 | 228761774 | 228761656 | 228761774 | - | rRNA |
| ENST00000364451.1 | chr1 | 228763884 | 228763961 | 228763895 | 228764013 | - | rRNA |
| ENST00000364451.1 | chr1 | 228763963 | 228764013 | 228763895 | 228764013 | - | rRNA |
| ENST00000363040.1 | chr1 | 228766127 | 228766189 | 228766137 | 228766255 | - | rRNA |
| ENST00000363040.1 | chr1 | 228766199 | 228766255 | 228766137 | 228766255 | - | rRNA |
| ENST00000362464.1 | chr1 | 228768367 | 228768418 | 228768378 | 228768496 | - | rRNA |
| ENST00000362464.1 | chr1 | 228768445 | 228768496 | 228768378 | 228768496 | - | rRNA |
| ENST00000362400.1 | chr1 | 228770609 | 228770668 | 228770618 | 228770736 | - | rRNA |
| ENST00000362400.1 | chr1 | 228770683 | 228770736 | 228770618 | 228770736 | - | rRNA |
| ENST00000364485.1 | chr1 | 228775075 | 228775137 | 228775084 | 228775202 | - | rRNA |
| ENST00000364485.1 | chr1 | 228775143 | 228775202 | 228775084 | 228775202 | - | rRNA |
| ENST00000363500.1 | chr1 | 228781778 | 228781836 | 228781787 | 228781905 | - | rRNA |
| ENST00000363500.1 | chr1 | 228781855 | 228781905 | 228781787 | 228781905 | - | rRNA |
| ENST00000366676.1 | chr1 | 229738370 | 229738401 | 229737942 | 229738666 | - | protein_coding |
| ENST00000258281.2 | chr1 | 229738370 | 229738401 | 229737942 | 229738666 | - | protein_coding |
| ENST00000366675.3 | chr1 | 229738370 | 229738401 | 229734941 | 229738666 | - | protein_coding |
| ENST00000497753.1 | chr1 | 231131602 | 231131633 | 231131506 | 231131730 | + | processed_transcript |
| ENST00000310256.2 | chr1 | 231131602 | 231131633 | 231131506 | 231131730 | + | protein_coding |
| ENST00000366658.2 | chr1 | 231131602 | 231131633 | 231131506 | 231131730 | + | protein_coding |
| ENST00000480519.1 | chr1 | 231131602 | 231131633 | 231131506 | 231131730 | + | nonsense_mediated_decay |
| ENST00000450711.1 | chr1 | 231131602 | 231131633 | 231131506 | 231131730 | + | protein_coding |
| ENST00000435927.1 | chr1 | 231131602 | 231131633 | 231131506 | 231131730 | + | protein_coding |
| ENST00000459891.1 | chr1 | 231131602 | 231131633 | 231131506 | 231131730 | + | nonsense_mediated_decay |
| ENST00000366639.3 | chr1 | 231700721 | 231700742 | 231700274 | 231702270 | + | protein_coding |
| ENST00000475168.1 | chr1 | 231700721 | 231700742 | 231700274 | 231702270 | + | processed_transcript |
| ENST00000366607.4 | chr1 | 235274881 | 235274912 | 235272651 | 235275423 | - | protein_coding |
| ENST00000450593.1 | chr1 | 235714136 | 235714166 | 235710987 | 235715537 | - | protein_coding |
| ENST00000391854.2 | chr1 | 235714136 | 235714166 | 235710987 | 235715537 | - | protein_coding |
| ENST00000366598.4 | chr1 | 235714136 | 235714166 | 235714132 | 235715537 | - | protein_coding |
| ENST00000366597.1 | chr1 | 235714136 | 235714166 | 235714132 | 235715537 | - | protein_coding |
| ENST00000354619.5 | chr1 | 236379485 | 236379516 | 236378855 | 236381876 | - | protein_coding |
| ENST00000516017.1 | chr1 | 237766465 | 237766540 | 237766512 | 237766585 | + | miRNA |
| ENST00000516017.1 | chr1 | 237766546 | 237766599 | 237766512 | 237766585 | + | miRNA |
| ENST00000358704.4 | chr1 | 244220605 | 244220619 | 244217090 | 244220778 | + | protein_coding |
| ENST00000302550.11 | chr1 | 244872225 | 244872257 | 244868858 | 244872335 | + | protein_coding |
| ENST00000411948.2 | chr1 | 245006518 | 245006549 | 245006343 | 245008359 | + | protein_coding |
| ENST00000366528.3 | chr1 | 245006518 | 245006549 | 245006343 | 245006554 | + | protein_coding |
| ENST00000391839.2 | chr1 | 245006518 | 245006549 | 245006343 | 245007182 | + | processed_transcript |
| ENST00000498262.1 | chr1 | 245006518 | 245006549 | 245006343 | 245008359 | + | processed_transcript |
| ENST00000464757.1 | chr1 | 245006518 | 245006549 | 245006343 | 245006582 | + | processed_transcript |
| ENST00000444376.2 | chr1 | 245017704 | 245017730 | 245014468 | 245017805 | - | protein_coding |
| ENST00000283179.9 | chr1 | 245017704 | 245017730 | 245016690 | 245017805 | - | protein_coding |
| ENST00000427948.2 | chr1 | 245017704 | 245017730 | 245016691 | 245017805 | - | protein_coding |
| ENST00000366525.3 | chr1 | 245017704 | 245017730 | 245017435 | 245017805 | - | retained_intron |
| ENST00000535105.1 | chr1 | 247263575 | 247263605 | 247263553 | 247264621 | - | protein_coding |
| ENST00000448299.2 | chr1 | 247263575 | 247263605 | 247263264 | 247264621 | - | protein_coding |
| ENST00000358785.4 | chr1 | 247263575 | 247263605 | 247263264 | 247264621 | - | protein_coding |
| ENST00000343381.6 | chr1 | 247263575 | 247263605 | 247263293 | 247264621 | - | protein_coding |
| ENST00000382352.3 | chr20 | 279949 | 279968 | 277737 | 280965 | + | protein_coding |
| ENST00000500893.2 | chr20 | 279949 | 279968 | 278868 | 280961 | + | pseudogene |
| ENST00000382291.3 | chr20 | 334446 | 334477 | 333854 | 335512 | + | protein_coding |
| ENST00000382285.2 | chr20 | 334446 | 334477 | 333854 | 334473 | + | protein_coding |
| ENST00000381342.2 | chr20 | 2442331 | 2442362 | 2442280 | 2442585 | - | protein_coding |
| ENST00000438552.2 | chr20 | 2442331 | 2442362 | 2442280 | 2442439 | - | protein_coding |
| ENST00000303103.6 | chr20 | 2442331 | 2442362 | 2442289 | 2442446 | - | protein_coding |
| ENST00000474384.1 | chr20 | 2442331 | 2442362 | 2442306 | 2442439 | - | processed_transcript |
| ENST00000380266.3 | chr20 | 3128497 | 3128528 | 3127165 | 3129906 | - | protein_coding |
| ENST00000362154.1 | chr20 | 3898188 | 3898219 | 3898141 | 3898218 | + | miRNA |
| ENST00000336095.5 | chr20 | 3913674 | 3913705 | 3912068 | 3914848 | - | protein_coding |
| ENST00000358395.6 | chr20 | 3913674 | 3913705 | 3912068 | 3914848 | - | protein_coding |
| ENST00000545616.1 | chr20 | 3913674 | 3913705 | 3912071 | 3914848 | - | protein_coding |
| ENST00000432261.2 | chr20 | 3913674 | 3913705 | 3912071 | 3914848 | - | protein_coding |
| ENST00000379143.5 | chr20 | 5095959 | 5095990 | 5095599 | 5096008 | - | protein_coding |
| ENST00000379160.3 | chr20 | 5095959 | 5095990 | 5095599 | 5096008 | - | protein_coding |
| ENST00000434910.1 | chr20 | 6194816 | 6194844 | 6194346 | 6196028 | + | processed_pseudogene |
| ENST00000434910.1 | chr20 | 6195064 | 6195095 | 6194346 | 6196028 | + | processed_pseudogene |
| ENST00000254977.3 | chr20 | 11904295 | 11904326 | 11903282 | 11907257 | + | protein_coding |
| ENST00000399006.2 | chr20 | 11904295 | 11904326 | 11903282 | 11907257 | + | protein_coding |
| ENST00000405977.1 | chr20 | 11904295 | 11904326 | 11903282 | 11907257 | + | protein_coding |
| ENST00000378226.2 | chr20 | 11904295 | 11904326 | 11903282 | 11907243 | + | protein_coding |
| ENST00000254977.3 | chr20 | 11906979 | 11907006 | 11903282 | 11907257 | + | protein_coding |
| ENST00000399006.2 | chr20 | 11906979 | 11907006 | 11903282 | 11907257 | + | protein_coding |
| ENST00000405977.1 | chr20 | 11906979 | 11907006 | 11903282 | 11907257 | + | protein_coding |
| ENST00000378226.2 | chr20 | 11906979 | 11907006 | 11903282 | 11907243 | + | protein_coding |
| ENST00000377943.5 | chr20 | 16721825 | 16721856 | 16721491 | 16722421 | + | protein_coding |
| ENST00000246071.6 | chr20 | 16721825 | 16721856 | 16721491 | 16722421 | + | protein_coding |
| ENST00000390930.1 | chr20 | 17943358 | 17943383 | 17943353 | 17943589 | - | snoRNA |
| ENST00000516594.1 | chr20 | 18309718 | 18309747 | 18309650 | 18309746 | + | misc_RNA |
| ENST00000339157.5 | chr20 | 25280875 | 25280906 | 25280834 | 25281520 | - | protein_coding |
| ENST00000526543.2 | chr20 | 25280875 | 25280906 | 25280850 | 25281520 | - | protein_coding |
| ENST00000262460.4 | chr20 | 25427488 | 25427519 | 25426559 | 25429191 | + | protein_coding |
| ENST00000375238.4 | chr20 | 32247355 | 32247390 | 32247303 | 32247356 | - | protein_coding |
| ENST00000246190.6 | chr20 | 32247355 | 32247390 | 32247303 | 32247538 | - | protein_coding |
| ENST00000488489.1 | chr20 | 32247355 | 32247390 | 32247303 | 32247356 | - | processed_transcript |
| ENST00000483813.1 | chr20 | 32247355 | 32247390 | 32247303 | 32247538 | - | processed_transcript |
| ENST00000485399.1 | chr20 | 32247355 | 32247390 | 32247303 | 32247356 | - | processed_transcript |
| ENST00000478237.1 | chr20 | 32247355 | 32247390 | 32247303 | 32247356 | - | processed_transcript |
| ENST00000494174.1 | chr20 | 32247355 | 32247390 | 32247303 | 32247356 | - | processed_transcript |
| ENST00000480994.1 | chr20 | 32247355 | 32247390 | 32247303 | 32247356 | - | processed_transcript |
| ENST00000374980.2 | chr20 | 32677664 | 32677695 | 32676104 | 32677711 | - | protein_coding |
| ENST00000364628.1 | chr20 | 32972015 | 32972046 | 32971971 | 32972071 | - | misc_RNA |
| ENST00000352393.4 | chr20 | 34215246 | 34215277 | 34215202 | 34215335 | - | protein_coding |
| ENST00000473373.1 | chr20 | 34215246 | 34215277 | 34215202 | 34215262 | - | retained_intron |
| ENST00000498814.1 | chr20 | 34215246 | 34215277 | 34215202 | 34215321 | - | processed_transcript |
| ENST00000415920.1 | chr20 | 34215246 | 34215277 | 34215202 | 34215335 | - | protein_coding |
| ENST00000317677.5 | chr20 | 34215246 | 34215277 | 34215202 | 34215335 | - | protein_coding |
| ENST00000317619.3 | chr20 | 34215246 | 34215277 | 34215202 | 34215335 | - | protein_coding |
| ENST00000397446.1 | chr20 | 34215246 | 34215277 | 34215202 | 34215335 | - | protein_coding |
| ENST00000397445.1 | chr20 | 34215246 | 34215277 | 34215202 | 34215335 | - | protein_coding |
| ENST00000397443.1 | chr20 | 34215246 | 34215277 | 34215202 | 34215335 | - | protein_coding |
| ENST00000483359.1 | chr20 | 34215246 | 34215277 | 34215202 | 34215335 | - | retained_intron |
| ENST00000397442.1 | chr20 | 34215246 | 34215277 | 34215202 | 34215335 | - | protein_coding |
| ENST00000483495.1 | chr20 | 34215246 | 34215277 | 34215202 | 34215335 | - | retained_intron |
| ENST00000401607.2 | chr20 | 34215246 | 34215277 | 34215202 | 34215335 | - | nonsense_mediated_decay |
| ENST00000437340.1 | chr20 | 34215246 | 34215277 | 34215202 | 34215335 | - | protein_coding |
| ENST00000498056.1 | chr20 | 34215246 | 34215277 | 34215202 | 34215453 | - | processed_transcript |
| ENST00000430570.1 | chr20 | 34215246 | 34215277 | 34215207 | 34215335 | - | protein_coding |
| ENST00000317677.5 | chr20 | 34241577 | 34241608 | 34241450 | 34241599 | - | protein_coding |
| ENST00000374114.3 | chr20 | 34241577 | 34241608 | 34236847 | 34243266 | - | protein_coding |
| ENST00000359646.1 | chr20 | 34241577 | 34241608 | 34238064 | 34243266 | - | protein_coding |
| ENST00000374104.3 | chr20 | 34241577 | 34241608 | 34238248 | 34243266 | - | protein_coding |
| ENST00000349942.5 | chr20 | 34241577 | 34241608 | 34240331 | 34242653 | - | protein_coding |
| ENST00000374114.3 | chr20 | 34241871 | 34241902 | 34236847 | 34243266 | - | protein_coding |
| ENST00000359646.1 | chr20 | 34241871 | 34241902 | 34238064 | 34243266 | - | protein_coding |
| ENST00000374104.3 | chr20 | 34241871 | 34241902 | 34238248 | 34243266 | - | protein_coding |
| ENST00000349942.5 | chr20 | 34241871 | 34241902 | 34240331 | 34242653 | - | protein_coding |
| ENST00000373433.4 | chr20 | 36720414 | 36720422 | 36718128 | 36720768 | + | protein_coding |
| ENST00000516384.1 | chr20 | 41933284 | 41933315 | 41933195 | 41933319 | + | snoRNA |
| ENST00000244020.3 | chr20 | 42089735 | 42089766 | 42089343 | 42092245 | + | protein_coding |
| ENST00000483871.1 | chr20 | 42089735 | 42089766 | 42089343 | 42092245 | + | nonsense_mediated_decay |
| ENST00000244020.3 | chr20 | 42089976 | 42090014 | 42089343 | 42092245 | + | protein_coding |
| ENST00000483871.1 | chr20 | 42089976 | 42090014 | 42089343 | 42092245 | + | nonsense_mediated_decay |
| ENST00000353703.4 | chr20 | 43535110 | 43535141 | 43535023 | 43537173 | + | protein_coding |
| ENST00000479421.1 | chr20 | 43535110 | 43535141 | 43535023 | 43535161 | + | processed_transcript |
| ENST00000372839.3 | chr20 | 43535110 | 43535141 | 43535023 | 43537173 | + | protein_coding |
| ENST00000479758.1 | chr20 | 43535110 | 43535141 | 43535023 | 43535161 | + | processed_transcript |
| ENST00000353703.4 | chr20 | 43536173 | 43536204 | 43535023 | 43537173 | + | protein_coding |
| ENST00000372839.3 | chr20 | 43536173 | 43536204 | 43535023 | 43537173 | + | protein_coding |
| ENST00000372806.3 | chr20 | 43708333 | 43708364 | 43703659 | 43708600 | + | protein_coding |
| ENST00000372806.3 | chr20 | 43708531 | 43708557 | 43703659 | 43708600 | + | protein_coding |
| ENST00000372523.1 | chr20 | 44511872 | 44511902 | 44511172 | 44513905 | + | protein_coding |
| ENST00000372520.1 | chr20 | 44511872 | 44511902 | 44511172 | 44513037 | + | protein_coding |
| ENST00000371610.2 | chr20 | 49369909 | 49369940 | 49366196 | 49370278 | + | protein_coding |
| ENST00000371602.2 | chr20 | 49509208 | 49509239 | 49505585 | 49511049 | - | protein_coding |
| ENST00000349014.3 | chr20 | 49509208 | 49509239 | 49505585 | 49511049 | - | protein_coding |
| ENST00000396029.3 | chr20 | 49509208 | 49509239 | 49505585 | 49511049 | - | protein_coding |
| ENST00000396032.1 | chr20 | 49509208 | 49509239 | 49506883 | 49511049 | - | protein_coding |
| ENST00000371602.2 | chr20 | 49509732 | 49509763 | 49505585 | 49511049 | - | protein_coding |
| ENST00000349014.3 | chr20 | 49509732 | 49509763 | 49505585 | 49511049 | - | protein_coding |
| ENST00000396029.3 | chr20 | 49509732 | 49509763 | 49505585 | 49511049 | - | protein_coding |
| ENST00000396032.1 | chr20 | 49509732 | 49509763 | 49506883 | 49511049 | - | protein_coding |
| ENST00000395841.2 | chr20 | 55929840 | 55929844 | 55929768 | 55929872 | + | protein_coding |
| ENST00000371242.2 | chr20 | 55929840 | 55929844 | 55929768 | 55929872 | + | protein_coding |
| ENST00000527947.1 | chr20 | 55929840 | 55929844 | 55929768 | 55929872 | + | protein_coding |
| ENST00000492498.1 | chr20 | 55929840 | 55929844 | 55929768 | 55929872 | + | nonsense_mediated_decay |
| ENST00000411894.1 | chr20 | 55929840 | 55929844 | 55929768 | 55929872 | + | protein_coding |
| ENST00000429339.1 | chr20 | 55929840 | 55929844 | 55929768 | 55929872 | + | protein_coding |
| ENST00000395840.2 | chr20 | 55929840 | 55929844 | 55929768 | 55929872 | + | protein_coding |
| ENST00000452119.1 | chr20 | 55929840 | 55929844 | 55929768 | 55929872 | + | protein_coding |
| ENST00000475243.1 | chr20 | 57020538 | 57020569 | 57019133 | 57026157 | + | protein_coding |
| ENST00000395802.3 | chr20 | 57020538 | 57020569 | 57019133 | 57020608 | + | protein_coding |
| ENST00000265619.2 | chr20 | 57020538 | 57020569 | 57019133 | 57021961 | + | processed_transcript |
| ENST00000463370.1 | chr20 | 57020538 | 57020569 | 57019133 | 57020608 | + | processed_transcript |
| ENST00000476395.1 | chr20 | 57020538 | 57020569 | 57019133 | 57020608 | + | retained_intron |
| ENST00000370744.4 | chr20 | 60883848 | 60883879 | 60883711 | 60883916 | + | protein_coding |
| ENST00000253003.2 | chr20 | 60883848 | 60883879 | 60883711 | 60883918 | + | protein_coding |
| ENST00000370339.3 | chr20 | 61827292 | 61827322 | 61826781 | 61828086 | - | protein_coding |
| ENST00000285667.3 | chr21 | 15745064 | 15745096 | 15743436 | 15746605 | - | protein_coding |
| ENST00000285681.2 | chr21 | 17251760 | 17251791 | 17250622 | 17252377 | + | protein_coding |
| ENST00000400183.2 | chr21 | 17251760 | 17251791 | 17250622 | 17252377 | + | protein_coding |
| ENST00000362160.1 | chr21 | 17912158 | 17912189 | 17912143 | 17912236 | + | miRNA |
| ENST00000284878.7 | chr21 | 18941636 | 18941667 | 18937746 | 18942418 | + | protein_coding |
| ENST00000425147.1 | chr21 | 26734369 | 26734400 | 26734135 | 26734745 | + | processed_pseudogene |
| ENST00000385060.1 | chr21 | 26946295 | 26946326 | 26946292 | 26946356 | + | miRNA |
| ENST00000354828.3 | chr21 | 27142897 | 27142928 | 27141315 | 27144771 | + | protein_coding |
| ENST00000400075.3 | chr21 | 27142897 | 27142928 | 27141315 | 27144771 | + | protein_coding |
| ENST00000434667.2 | chr21 | 33043521 | 33043521 | 33043313 | 33044667 | - | protein_coding |
| ENST00000286835.7 | chr21 | 33043521 | 33043521 | 33043346 | 33044667 | - | protein_coding |
| ENST00000399804.1 | chr21 | 33043521 | 33043521 | 33043346 | 33044667 | - | protein_coding |
| ENST00000404220.2 | chr21 | 34636042 | 34636065 | 34634866 | 34637969 | + | protein_coding |
| ENST00000382241.3 | chr21 | 34636042 | 34636065 | 34635098 | 34636818 | + | protein_coding |
| ENST00000342136.4 | chr21 | 34636042 | 34636065 | 34635098 | 34636830 | + | protein_coding |
| ENST00000342101.3 | chr21 | 34636042 | 34636065 | 34635098 | 34636830 | + | protein_coding |
| ENST00000433395.2 | chr21 | 34652183 | 34652202 | 34652057 | 34652223 | + | protein_coding |
| ENST00000451624.1 | chr21 | 34652183 | 34652202 | 34652057 | 34652223 | + | retained_intron |
| ENST00000432231.1 | chr21 | 34652183 | 34652202 | 34652057 | 34652223 | + | nonsense_mediated_decay |
| ENST00000290200.2 | chr21 | 34652183 | 34652202 | 34652057 | 34652223 | + | protein_coding |
| ENST00000422891.1 | chr21 | 34652183 | 34652202 | 34652057 | 34652223 | + | nonsense_mediated_decay |
| ENST00000539894.1 | chr21 | 34652183 | 34652202 | 34652057 | 34652223 | + | protein_coding |
| ENST00000493295.1 | chr21 | 34652183 | 34652202 | 34652057 | 34652223 | + | retained_intron |
| ENST00000498371.1 | chr21 | 34652183 | 34652202 | 34652057 | 34652223 | + | retained_intron |
| ENST00000451065.1 | chr21 | 34652183 | 34652202 | 34652057 | 34652223 | + | nonsense_mediated_decay |
| ENST00000390235.1 | chr21 | 37093029 | 37093060 | 37093013 | 37093106 | + | miRNA |
| ENST00000412073.1 | chr21 | 37503971 | 37504002 | 37502669 | 37504208 | + | retrotransposed |
| ENST00000437106.1 | chr21 | 37541256 | 37541287 | 37541268 | 37542478 | - | processed_pseudogene |
| ENST00000400485.1 | chr21 | 37748158 | 37748189 | 37747441 | 37748944 | + | protein_coding |
| ENST00000487909.1 | chr21 | 37748158 | 37748189 | 37747441 | 37748944 | + | processed_transcript |
| ENST00000400485.1 | chr21 | 37748532 | 37748563 | 37747441 | 37748944 | + | protein_coding |
| ENST00000487909.1 | chr21 | 37748532 | 37748563 | 37747441 | 37748944 | + | processed_transcript |
| ENST00000338785.3 | chr21 | 38884690 | 38884721 | 38884214 | 38887680 | + | protein_coding |
| ENST00000339659.3 | chr21 | 38884690 | 38884721 | 38884214 | 38887483 | + | protein_coding |
| ENST00000398960.2 | chr21 | 38884690 | 38884721 | 38884214 | 38884889 | + | protein_coding |
| ENST00000455387.2 | chr21 | 38884690 | 38884721 | 38884214 | 38884862 | + | protein_coding |
| ENST00000398505.3 | chr21 | 43413682 | 43413713 | 43412602 | 43414217 | - | protein_coding |
| ENST00000310826.5 | chr21 | 43413682 | 43413713 | 43406940 | 43414217 | - | protein_coding |
| ENST00000398499.1 | chr21 | 43413682 | 43413713 | 43408746 | 43414217 | - | protein_coding |
| ENST00000398511.3 | chr21 | 43413682 | 43413713 | 43408746 | 43414217 | - | protein_coding |
| ENST00000496783.1 | chr21 | 43715445 | 43715487 | 43714652 | 43715462 | + | retained_intron |
| ENST00000459639.1 | chr21 | 44520583 | 44520614 | 44520563 | 44520629 | - | protein_coding |
| ENST00000464750.1 | chr21 | 44520583 | 44520614 | 44520563 | 44520629 | - | nonsense_mediated_decay |
| ENST00000475639.1 | chr21 | 44520583 | 44520614 | 44520563 | 44524512 | - | retained_intron |
| ENST00000291552.4 | chr21 | 44520583 | 44520614 | 44520563 | 44520629 | - | protein_coding |
| ENST00000398137.1 | chr21 | 44520583 | 44520614 | 44520563 | 44520629 | - | protein_coding |
| ENST00000463599.1 | chr21 | 44520583 | 44520614 | 44520563 | 44520629 | - | retained_intron |
| ENST00000496462.1 | chr21 | 44520583 | 44520614 | 44517732 | 44521542 | - | retained_intron |
| ENST00000468039.1 | chr21 | 44520583 | 44520614 | 44520563 | 44520629 | - | retained_intron |
| ENST00000464664.1 | chr21 | 44985366 | 44985397 | 44985120 | 44985500 | + | processed_pseudogene |
| ENST00000385288.1 | chr22 | 20020676 | 20020707 | 20020662 | 20020743 | + | miRNA |
| ENST00000390813.1 | chr22 | 22007314 | 22007345 | 22007270 | 22007347 | + | miRNA |
| ENST00000215832.6 | chr22 | 22160154 | 22160177 | 22160139 | 22160328 | - | protein_coding |
| ENST00000415911.2 | chr22 | 22160154 | 22160177 | 22160139 | 22160328 | - | protein_coding |
| ENST00000398822.3 | chr22 | 22160154 | 22160177 | 22160139 | 22160328 | - | protein_coding |
| ENST00000544786.1 | chr22 | 22160154 | 22160177 | 22160139 | 22160328 | - | protein_coding |
| ENST00000385100.1 | chr22 | 23101200 | 23101231 | 23101185 | 23101279 | + | miRNA |
| ENST00000390312.2 | chr22 | 23101200 | 23101231 | 23101189 | 23101275 | + | IG_V_gene |
| ENST00000397906.2 | chr22 | 28377220 | 28377255 | 28374004 | 28379839 | - | protein_coding |
| ENST00000249064.4 | chr22 | 29184669 | 29184683 | 29182077 | 29185283 | + | protein_coding |
| ENST00000382363.3 | chr22 | 30774871 | 30774900 | 30774806 | 30775611 | - | protein_coding |
| ENST00000382151.2 | chr22 | 32015226 | 32015257 | 32014477 | 32015822 | - | protein_coding |
| ENST00000460723.1 | chr22 | 32015226 | 32015257 | 32014477 | 32015822 | - | retained_intron |
| ENST00000437808.1 | chr22 | 32015226 | 32015257 | 32014497 | 32015822 | - | nonsense_mediated_decay |
| ENST00000266095.5 | chr22 | 32015226 | 32015257 | 32014501 | 32015822 | - | protein_coding |
| ENST00000473770.1 | chr22 | 32015226 | 32015257 | 32014545 | 32015822 | - | retained_intron |
| ENST00000397500.1 | chr22 | 32015226 | 32015257 | 32014600 | 32015822 | - | protein_coding |
| ENST00000435900.1 | chr22 | 32015226 | 32015257 | 32014963 | 32015822 | - | protein_coding |
| ENST00000439502.2 | chr22 | 32015226 | 32015257 | 32015089 | 32015822 | - | protein_coding |
| ENST00000336566.4 | chr22 | 32015226 | 32015257 | 32015089 | 32015822 | - | protein_coding |
| ENST00000397493.2 | chr22 | 32112567 | 32112592 | 32108069 | 32113277 | - | protein_coding |
| ENST00000327423.6 | chr22 | 32112567 | 32112592 | 32108069 | 32113277 | - | protein_coding |
| ENST00000434485.1 | chr22 | 32112567 | 32112592 | 32108069 | 32113277 | - | protein_coding |
| ENST00000216190.8 | chr22 | 36907685 | 36907716 | 36907550 | 36907833 | - | protein_coding |
| ENST00000397177.3 | chr22 | 36907685 | 36907716 | 36907550 | 36907833 | - | protein_coding |
| ENST00000478547.1 | chr22 | 36907685 | 36907716 | 36907550 | 36907688 | - | processed_transcript |
| ENST00000541106.1 | chr22 | 36907685 | 36907716 | 36907550 | 36907833 | - | protein_coding |
| ENST00000462641.1 | chr22 | 36907685 | 36907716 | 36907550 | 36907833 | - | processed_transcript |
| ENST00000405442.1 | chr22 | 36907685 | 36907716 | 36907550 | 36907833 | - | protein_coding |
| ENST00000426531.1 | chr22 | 36907685 | 36907716 | 36907624 | 36907833 | - | protein_coding |
| ENST00000340857.2 | chr22 | 38203369 | 38203400 | 38201114 | 38203442 | + | protein_coding |
| ENST00000215941.4 | chr22 | 38240285 | 38240318 | 38239918 | 38240438 | - | protein_coding |
| ENST00000411961.2 | chr22 | 38240285 | 38240318 | 38239918 | 38240328 | - | protein_coding |
| ENST00000434930.1 | chr22 | 38240285 | 38240318 | 38239918 | 38240330 | - | protein_coding |
| ENST00000385210.1 | chr22 | 38240285 | 38240318 | 38240279 | 38240378 | - | miRNA |
| ENST00000412331.2 | chr22 | 38247358 | 38247389 | 38247287 | 38247497 | + | protein_coding |
| ENST00000425539.2 | chr22 | 38247358 | 38247389 | 38247287 | 38247497 | + | protein_coding |
| ENST00000439997.1 | chr22 | 38247358 | 38247389 | 38247287 | 38247568 | + | nonsense_mediated_decay |
| ENST00000476955.1 | chr22 | 38247358 | 38247389 | 38247287 | 38247497 | + | processed_transcript |
| ENST00000414316.1 | chr22 | 38247358 | 38247389 | 38247287 | 38247497 | + | protein_coding |
| ENST00000436452.2 | chr22 | 38247358 | 38247389 | 38247287 | 38247497 | + | nonsense_mediated_decay |
| ENST00000477256.1 | chr22 | 38247358 | 38247389 | 38247287 | 38247497 | + | retained_intron |
| ENST00000381683.4 | chr22 | 38247358 | 38247389 | 38247287 | 38247497 | + | protein_coding |
| ENST00000262832.7 | chr22 | 38247358 | 38247389 | 38247287 | 38247497 | + | protein_coding |
| ENST00000406934.1 | chr22 | 38247358 | 38247389 | 38247287 | 38247497 | + | protein_coding |
| ENST00000451427.1 | chr22 | 38247358 | 38247389 | 38247287 | 38247497 | + | protein_coding |
| ENST00000396821.3 | chr22 | 38894508 | 38894539 | 38894445 | 38894578 | - | protein_coding |
| ENST00000381633.3 | chr22 | 38894508 | 38894539 | 38894445 | 38894578 | - | protein_coding |
| ENST00000216019.7 | chr22 | 38894508 | 38894539 | 38894445 | 38894578 | - | retained_intron |
| ENST00000403230.1 | chr22 | 38894508 | 38894539 | 38894445 | 38894578 | - | protein_coding |
| ENST00000404499.2 | chr22 | 38894508 | 38894539 | 38894445 | 38894578 | - | protein_coding |
| ENST00000432525.1 | chr22 | 38894508 | 38894539 | 38894445 | 38894578 | - | processed_transcript |
| ENST00000396811.2 | chr22 | 39069814 | 39069838 | 39069164 | 39069851 | + | protein_coding |
| ENST00000216029.3 | chr22 | 39069814 | 39069838 | 39069164 | 39069859 | + | protein_coding |
| ENST00000467105.1 | chr22 | 39709830 | 39709856 | 39709196 | 39710213 | - | retained_intron |
| ENST00000386745.1 | chr22 | 39709830 | 39709856 | 39709824 | 39709916 | - | snoRNA |
| ENST00000433117.2 | chr22 | 39911811 | 39911842 | 39909522 | 39912385 | + | nonsense_mediated_decay |
| ENST00000325301.2 | chr22 | 39911811 | 39911842 | 39909522 | 39914137 | + | protein_coding |
| ENST00000404569.1 | chr22 | 39911811 | 39911842 | 39909522 | 39912183 | + | protein_coding |
| ENST00000404241.2 | chr22 | 39918019 | 39918046 | 39917778 | 39918688 | + | protein_coding |
| ENST00000337304.2 | chr22 | 39918019 | 39918046 | 39917778 | 39918688 | + | protein_coding |
| ENST00000396680.1 | chr22 | 39918019 | 39918046 | 39917778 | 39918691 | + | protein_coding |
| ENST00000301923.9 | chr22 | 40722900 | 40722931 | 40718858 | 40731811 | + | protein_coding |
| ENST00000454349.2 | chr22 | 40722900 | 40722931 | 40718858 | 40731811 | + | protein_coding |
| ENST00000400140.3 | chr22 | 40722900 | 40722931 | 40718858 | 40731811 | + | protein_coding |
| ENST00000335727.8 | chr22 | 40722900 | 40722931 | 40718858 | 40731809 | + | protein_coding |
| ENST00000360079.3 | chr22 | 42059984 | 42060012 | 42059626 | 42060043 | + | protein_coding |
| ENST00000402580.3 | chr22 | 42059984 | 42060012 | 42059626 | 42060044 | + | protein_coding |
| ENST00000428575.2 | chr22 | 42059984 | 42060012 | 42059626 | 42060042 | + | protein_coding |
| ENST00000359308.4 | chr22 | 42059984 | 42060012 | 42059626 | 42060043 | + | protein_coding |
| ENST00000405878.1 | chr22 | 42059984 | 42060012 | 42059626 | 42060043 | + | protein_coding |
| ENST00000402409.1 | chr22 | 42059984 | 42060012 | 42059645 | 42060010 | + | protein_coding |
| ENST00000405506.1 | chr22 | 42059984 | 42060012 | 42059626 | 42060038 | + | protein_coding |
| ENST00000362512.1 | chr22 | 43011374 | 43011396 | 43011250 | 43011399 | + | snRNA |
| ENST00000347635.3 | chr22 | 45583219 | 45583250 | 45580334 | 45583892 | + | protein_coding |
| ENST00000336156.4 | chr22 | 45589934 | 45589972 | 45588114 | 45593097 | - | protein_coding |
| ENST00000360737.3 | chr22 | 46508632 | 46508663 | 46505644 | 46509808 | + | protein_coding |
| ENST00000362116.1 | chr22 | 46508632 | 46508663 | 46508629 | 46508702 | + | miRNA |
| ENST00000360737.3 | chr22 | 46508669 | 46508700 | 46505644 | 46509808 | + | protein_coding |
| ENST00000362116.1 | chr22 | 46508669 | 46508700 | 46508629 | 46508702 | + | miRNA |
| ENST00000360737.3 | chr22 | 46509571 | 46509602 | 46505644 | 46509808 | + | protein_coding |
| ENST00000385140.1 | chr22 | 46509571 | 46509602 | 46509566 | 46509648 | + | miRNA |
| ENST00000360737.3 | chr22 | 46509621 | 46509652 | 46505644 | 46509808 | + | protein_coding |
| ENST00000385140.1 | chr22 | 46509621 | 46509652 | 46509566 | 46509648 | + | miRNA |
| ENST00000322002.3 | chr2 | 5834799 | 5834830 | 5832799 | 5841516 | + | protein_coding |
| ENST00000497473.1 | chr2 | 9614713 | 9614744 | 9614659 | 9614776 | + | protein_coding |
| ENST00000470914.1 | chr2 | 9614713 | 9614744 | 9614666 | 9614776 | + | protein_coding |
| ENST00000492223.1 | chr2 | 9614713 | 9614744 | 9614668 | 9614776 | + | nonsense_mediated_decay |
| ENST00000495050.1 | chr2 | 9614713 | 9614744 | 9614693 | 9614776 | + | retained_intron |
| ENST00000481367.1 | chr2 | 9614713 | 9614744 | 9614704 | 9614776 | + | protein_coding |
| ENST00000545602.1 | chr2 | 9614713 | 9614744 | 9614704 | 9614776 | + | protein_coding |
| ENST00000351760.7 | chr2 | 9614713 | 9614744 | 9614707 | 9614776 | + | nonsense_mediated_decay |
| ENST00000305883.1 | chr2 | 10194771 | 10194798 | 10192354 | 10194963 | + | protein_coding |
| ENST00000540845.1 | chr2 | 10194771 | 10194798 | 10192354 | 10194963 | + | protein_coding |
| ENST00000535335.1 | chr2 | 10194771 | 10194798 | 10192354 | 10194963 | + | protein_coding |
| ENST00000360566.2 | chr2 | 10269630 | 10269661 | 10269361 | 10271545 | + | protein_coding |
| ENST00000304567.4 | chr2 | 10269630 | 10269661 | 10269361 | 10271545 | + | protein_coding |
| ENST00000360566.2 | chr2 | 10269881 | 10269902 | 10269361 | 10271545 | + | protein_coding |
| ENST00000304567.4 | chr2 | 10269881 | 10269902 | 10269361 | 10271545 | + | protein_coding |
| ENST00000434820.1 | chr2 | 17566514 | 17566545 | 17566358 | 17567684 | + | processed_pseudogene |
| ENST00000175091.4 | chr2 | 20232635 | 20232668 | 20232411 | 20233040 | - | protein_coding |
| ENST00000338086.5 | chr2 | 20497367 | 20497398 | 20497314 | 20497439 | - | protein_coding |
| ENST00000361078.2 | chr2 | 20497367 | 20497398 | 20497314 | 20497439 | - | protein_coding |
| ENST00000319801.5 | chr2 | 20497367 | 20497398 | 20497314 | 20497439 | - | protein_coding |
| ENST00000440577.1 | chr2 | 20497367 | 20497398 | 20497314 | 20497439 | - | protein_coding |
| ENST00000403432.1 | chr2 | 20497367 | 20497398 | 20497314 | 20497439 | - | protein_coding |
| ENST00000536417.1 | chr2 | 20497367 | 20497398 | 20497314 | 20497439 | - | protein_coding |
| ENST00000233468.4 | chr2 | 24290597 | 24290628 | 24290454 | 24290721 | - | protein_coding |
| ENST00000495146.1 | chr2 | 26357929 | 26357961 | 26357807 | 26358216 | + | retained_intron |
| ENST00000264710.4 | chr2 | 26357929 | 26357961 | 26357807 | 26360323 | + | protein_coding |
| ENST00000473035.1 | chr2 | 26357929 | 26357961 | 26357807 | 26357937 | + | processed_transcript |
| ENST00000429494.1 | chr2 | 26997935 | 26997965 | 26997912 | 26998083 | + | nonsense_mediated_decay |
| ENST00000344420.5 | chr2 | 26997935 | 26997965 | 26997912 | 26998083 | + | protein_coding |
| ENST00000264705.3 | chr2 | 27466577 | 27466608 | 27466284 | 27466660 | + | protein_coding |
| ENST00000403525.1 | chr2 | 27466577 | 27466608 | 27466284 | 27466653 | + | protein_coding |
| ENST00000446765.1 | chr2 | 32093102 | 32093133 | 32092878 | 32093561 | - | nonsense_mediated_decay |
| ENST00000295065.4 | chr2 | 32093102 | 32093133 | 32092878 | 32093561 | - | protein_coding |
| ENST00000379383.3 | chr2 | 32093102 | 32093133 | 32092892 | 32093561 | - | protein_coding |
| ENST00000404530.1 | chr2 | 32093102 | 32093133 | 32092892 | 32093561 | - | protein_coding |
| ENST00000490459.1 | chr2 | 32093102 | 32093133 | 32092896 | 32093561 | - | processed_transcript |
| ENST00000426310.2 | chr2 | 32093102 | 32093133 | 32093047 | 32093561 | - | protein_coding |
| ENST00000402297.1 | chr2 | 37429789 | 37429834 | 37429006 | 37431884 | + | protein_coding |
| ENST00000397226.2 | chr2 | 37429789 | 37429834 | 37429006 | 37429835 | + | protein_coding |
| ENST00000477642.1 | chr2 | 38525205 | 38525236 | 38525073 | 38525717 | - | retained_intron |
| ENST00000378954.4 | chr2 | 38525492 | 38525523 | 38525286 | 38525717 | - | protein_coding |
| ENST00000406122.1 | chr2 | 38525492 | 38525523 | 38525286 | 38525717 | - | protein_coding |
| ENST00000402054.1 | chr2 | 38525492 | 38525523 | 38525286 | 38525717 | - | protein_coding |
| ENST00000405384.2 | chr2 | 38525492 | 38525523 | 38525286 | 38525717 | - | nonsense_mediated_decay |
| ENST00000539122.1 | chr2 | 38525492 | 38525523 | 38525286 | 38525717 | - | protein_coding |
| ENST00000332337.4 | chr2 | 38525492 | 38525523 | 38525286 | 38525717 | - | protein_coding |
| ENST00000419554.2 | chr2 | 38525492 | 38525523 | 38525286 | 38525717 | - | protein_coding |
| ENST00000452935.2 | chr2 | 38525492 | 38525523 | 38525286 | 38525717 | - | protein_coding |
| ENST00000546051.1 | chr2 | 38525492 | 38525523 | 38525286 | 38525717 | - | protein_coding |
| ENST00000477642.1 | chr2 | 38525492 | 38525523 | 38525073 | 38525717 | - | retained_intron |
| ENST00000489896.1 | chr2 | 38525492 | 38525523 | 38525360 | 38525717 | - | retained_intron |
| ENST00000306156.3 | chr2 | 46413554 | 46413585 | 46411874 | 46415129 | + | protein_coding |
| ENST00000272298.7 | chr2 | 47387332 | 47387363 | 47387221 | 47387943 | - | protein_coding |
| ENST00000460218.1 | chr2 | 47387332 | 47387363 | 47387303 | 47387943 | - | retained_intron |
| ENST00000432899.1 | chr2 | 47387332 | 47387363 | 47387311 | 47387943 | - | nonsense_mediated_decay |
| ENST00000272298.7 | chr2 | 47387457 | 47387489 | 47387221 | 47387943 | - | protein_coding |
| ENST00000460218.1 | chr2 | 47387457 | 47387489 | 47387303 | 47387943 | - | retained_intron |
| ENST00000432899.1 | chr2 | 47387457 | 47387489 | 47387311 | 47387943 | - | nonsense_mediated_decay |
| ENST00000272298.7 | chr2 | 47387819 | 47387850 | 47387221 | 47387943 | - | protein_coding |
| ENST00000460218.1 | chr2 | 47387819 | 47387850 | 47387303 | 47387943 | - | retained_intron |
| ENST00000432899.1 | chr2 | 47387819 | 47387850 | 47387311 | 47387943 | - | nonsense_mediated_decay |
| ENST00000482532.1 | chr2 | 47387819 | 47387850 | 47387741 | 47387943 | - | retained_intron |
| ENST00000456319.1 | chr2 | 47387819 | 47387850 | 47387776 | 47387943 | - | protein_coding |
| ENST00000272298.7 | chr2 | 47387863 | 47387895 | 47387221 | 47387943 | - | protein_coding |
| ENST00000460218.1 | chr2 | 47387863 | 47387895 | 47387303 | 47387943 | - | retained_intron |
| ENST00000432899.1 | chr2 | 47387863 | 47387895 | 47387311 | 47387943 | - | nonsense_mediated_decay |
| ENST00000482532.1 | chr2 | 47387863 | 47387895 | 47387741 | 47387943 | - | retained_intron |
| ENST00000456319.1 | chr2 | 47387863 | 47387895 | 47387776 | 47387943 | - | protein_coding |
| ENST00000409563.1 | chr2 | 47387863 | 47387895 | 47387878 | 47387943 | - | protein_coding |
| ENST00000233146.2 | chr2 | 47656916 | 47656947 | 47656881 | 47657080 | + | protein_coding |
| ENST00000543555.1 | chr2 | 47656916 | 47656947 | 47656881 | 47657080 | + | protein_coding |
| ENST00000406134.1 | chr2 | 47656916 | 47656947 | 47656881 | 47657080 | + | protein_coding |
| ENST00000419559.1 | chr2 | 47656916 | 47656947 | 47656881 | 47657080 | + | protein_coding |
| ENST00000432737.1 | chr2 | 47656916 | 47656947 | 47656881 | 47657080 | + | protein_coding |
| ENST00000453755.1 | chr2 | 47656916 | 47656947 | 47656881 | 47657080 | + | protein_coding |
| ENST00000448533.1 | chr2 | 47656916 | 47656947 | 47656881 | 47656941 | + | protein_coding |
| ENST00000448533.1 | chr2 | 47656916 | 47656947 | 47656943 | 47657080 | + | protein_coding |
| ENST00000394792.2 | chr2 | 47656916 | 47656947 | 47656881 | 47657080 | + | protein_coding |
| ENST00000422810.1 | chr2 | 47656916 | 47656947 | 47656881 | 47657080 | + | protein_coding |
| ENST00000413880.2 | chr2 | 47656916 | 47656947 | 47656881 | 47657080 | + | protein_coding |
| ENST00000420788.2 | chr2 | 47959034 | 47959066 | 47958541 | 47959174 | - | processed_pseudogene |
| ENST00000384817.1 | chr2 | 56210146 | 56210177 | 56210102 | 56210211 | - | miRNA |
| ENST00000390186.2 | chr2 | 56227889 | 56227920 | 56227849 | 56227930 | - | miRNA |
| ENST00000516964.1 | chr2 | 59549348 | 59549353 | 59549315 | 59549400 | + | miRNA |
| ENST00000516964.1 | chr2 | 59549356 | 59549362 | 59549315 | 59549400 | + | miRNA |
| ENST00000335712.6 | chr2 | 60687211 | 60687231 | 60684333 | 60689559 | - | protein_coding |
| ENST00000477659.1 | chr2 | 60687211 | 60687231 | 60686955 | 60689559 | - | processed_transcript |
| ENST00000401558.2 | chr2 | 61710187 | 61710218 | 61710092 | 61710226 | - | protein_coding |
| ENST00000428210.1 | chr2 | 61710187 | 61710218 | 61710092 | 61710226 | - | nonsense_mediated_decay |
| ENST00000481073.1 | chr2 | 61710187 | 61710218 | 61710092 | 61710226 | - | retained_intron |
| ENST00000404992.2 | chr2 | 61710187 | 61710218 | 61710092 | 61710226 | - | protein_coding |
| ENST00000406957.1 | chr2 | 61710187 | 61710218 | 61710092 | 61710226 | - | protein_coding |
| ENST00000492182.1 | chr2 | 61710187 | 61710218 | 61709760 | 61710226 | - | retained_intron |
| ENST00000461407.1 | chr2 | 61710187 | 61710218 | 61709760 | 61710226 | - | retained_intron |
| ENST00000301998.4 | chr2 | 62450106 | 62450137 | 62449347 | 62451866 | + | protein_coding |
| ENST00000405767.1 | chr2 | 62450106 | 62450137 | 62449347 | 62451684 | + | protein_coding |
| ENST00000366671.3 | chr2 | 63284351 | 63284382 | 63282636 | 63284971 | + | protein_coding |
| ENST00000282549.2 | chr2 | 63284351 | 63284382 | 63282636 | 63284971 | + | protein_coding |
| ENST00000238875.4 | chr2 | 64685950 | 64685981 | 64685419 | 64688515 | + | protein_coding |
| ENST00000409537.2 | chr2 | 64685950 | 64685981 | 64685419 | 64688515 | + | protein_coding |
| ENST00000238875.4 | chr2 | 64686645 | 64686676 | 64685419 | 64688515 | + | protein_coding |
| ENST00000409537.2 | chr2 | 64686645 | 64686676 | 64685419 | 64688515 | + | protein_coding |
| ENST00000313349.3 | chr2 | 64863449 | 64863480 | 64858755 | 64864009 | - | protein_coding |
| ENST00000406334.3 | chr2 | 68413649 | 68413680 | 68413600 | 68413784 | - | nonsense_mediated_decay |
| ENST00000234310.3 | chr2 | 68413649 | 68413680 | 68413600 | 68413784 | - | protein_coding |
| ENST00000409752.1 | chr2 | 68413649 | 68413680 | 68413600 | 68413784 | - | protein_coding |
| ENST00000409377.1 | chr2 | 68413649 | 68413680 | 68413600 | 68413784 | - | protein_coding |
| ENST00000303577.5 | chr2 | 70315723 | 70315754 | 70314585 | 70316332 | + | protein_coding |
| ENST00000244230.2 | chr2 | 71371637 | 71371666 | 71371558 | 71371668 | + | protein_coding |
| ENST00000425650.1 | chr2 | 71371637 | 71371666 | 71371558 | 71371668 | + | protein_coding |
| ENST00000476969.1 | chr2 | 71371637 | 71371666 | 71371370 | 71371668 | + | retained_intron |
| ENST00000540468.1 | chr2 | 73479790 | 73479821 | 73479768 | 73480149 | + | protein_coding |
| ENST00000539919.1 | chr2 | 73479790 | 73479821 | 73479768 | 73480149 | + | protein_coding |
| ENST00000258091.5 | chr2 | 73479790 | 73479821 | 73479768 | 73480132 | + | protein_coding |
| ENST00000398422.2 | chr2 | 73479790 | 73479821 | 73479768 | 73480132 | + | protein_coding |
| ENST00000537131.1 | chr2 | 73479790 | 73479821 | 73479768 | 73480132 | + | protein_coding |
| ENST00000538797.1 | chr2 | 73479790 | 73479821 | 73479768 | 73480134 | + | protein_coding |
| ENST00000409081.1 | chr2 | 73479790 | 73479821 | 73479768 | 73480103 | + | protein_coding |
| ENST00000305799.7 | chr2 | 74275026 | 74275057 | 74273405 | 74275538 | + | protein_coding |
| ENST00000409262.2 | chr2 | 74275026 | 74275057 | 74273450 | 74275538 | + | protein_coding |
| ENST00000233310.5 | chr2 | 74275026 | 74275057 | 74273450 | 74275538 | + | protein_coding |
| ENST00000475405.2 | chr2 | 74275026 | 74275057 | 74274287 | 74275538 | + | retained_intron |
| ENST00000396049.3 | chr2 | 74386230 | 74386261 | 74386218 | 74386381 | - | protein_coding |
| ENST00000497054.1 | chr2 | 74386230 | 74386261 | 74386196 | 74386381 | - | processed_transcript |
| ENST00000495286.1 | chr2 | 74386230 | 74386261 | 74386236 | 74386381 | - | processed_transcript |
| ENST00000425118.1 | chr2 | 74755920 | 74755951 | 74755878 | 74756062 | - | nonsense_mediated_decay |
| ENST00000377526.3 | chr2 | 74755920 | 74755951 | 74755878 | 74756062 | - | protein_coding |
| ENST00000258081.4 | chr2 | 74755920 | 74755951 | 74755878 | 74756110 | - | protein_coding |
| ENST00000463900.1 | chr2 | 74755920 | 74755951 | 74755878 | 74756409 | - | retained_intron |
| ENST00000466894.1 | chr2 | 74755920 | 74755951 | 74755878 | 74756409 | - | retained_intron |
| ENST00000412627.2 | chr2 | 74755920 | 74755951 | 74755878 | 74756062 | - | protein_coding |
| ENST00000464887.1 | chr2 | 74755920 | 74755951 | 74755878 | 74756063 | - | retained_intron |
| ENST00000472800.1 | chr2 | 74755920 | 74755951 | 74755949 | 74756062 | - | retained_intron |
| ENST00000393909.2 | chr2 | 75883749 | 75883780 | 75882190 | 75884986 | + | protein_coding |
| ENST00000377386.3 | chr2 | 85548803 | 85548834 | 85545147 | 85549864 | - | protein_coding |
| ENST00000282120.2 | chr2 | 85548803 | 85548834 | 85545147 | 85549864 | - | protein_coding |
| ENST00000398263.2 | chr2 | 85548803 | 85548834 | 85545150 | 85549864 | - | protein_coding |
| ENST00000306434.3 | chr2 | 85770044 | 85770074 | 85770024 | 85770157 | + | protein_coding |
| ENST00000424323.1 | chr2 | 85770044 | 85770074 | 85770024 | 85770157 | + | protein_coding |
| ENST00000481412.1 | chr2 | 85770044 | 85770074 | 85770024 | 85770341 | + | retained_intron |
| ENST00000409017.1 | chr2 | 85770044 | 85770074 | 85770024 | 85770731 | + | protein_coding |
| ENST00000323853.5 | chr2 | 96949696 | 96949727 | 96949551 | 96949742 | - | protein_coding |
| ENST00000497539.1 | chr2 | 96949696 | 96949727 | 96949551 | 96949742 | - | retained_intron |
| ENST00000540328.1 | chr2 | 96949696 | 96949727 | 96949551 | 96949742 | - | protein_coding |
| ENST00000543553.1 | chr2 | 96949696 | 96949727 | 96949551 | 96949742 | - | protein_coding |
| ENST00000429650.1 | chr2 | 96949696 | 96949727 | 96949551 | 96949742 | - | nonsense_mediated_decay |
| ENST00000480242.1 | chr2 | 96949696 | 96949727 | 96949551 | 96950612 | - | retained_intron |
| ENST00000264963.4 | chr2 | 97373013 | 97373044 | 97371666 | 97373135 | - | protein_coding |
| ENST00000434524.1 | chr2 | 97373013 | 97373044 | 97371712 | 97373135 | - | nonsense_mediated_decay |
| ENST00000377079.4 | chr2 | 97373013 | 97373044 | 97372077 | 97373135 | - | protein_coding |
| ENST00000446780.1 | chr2 | 97373013 | 97373044 | 97372614 | 97373135 | - | nonsense_mediated_decay |
| ENST00000426463.2 | chr2 | 97373013 | 97373044 | 97372614 | 97373135 | - | protein_coding |
| ENST00000434865.1 | chr2 | 97373013 | 97373044 | 97372781 | 97373135 | - | nonsense_mediated_decay |
| ENST00000537039.1 | chr2 | 97373013 | 97373044 | 97372781 | 97373135 | - | protein_coding |
| ENST00000440610.1 | chr2 | 97373013 | 97373044 | 97372854 | 97373135 | - | nonsense_mediated_decay |
| ENST00000534882.1 | chr2 | 97373013 | 97373044 | 97372854 | 97373135 | - | protein_coding |
| ENST00000449221.1 | chr2 | 97373013 | 97373044 | 97372917 | 97373135 | - | nonsense_mediated_decay |
| ENST00000258436.5 | chr2 | 103332506 | 103332535 | 103332299 | 103335666 | - | protein_coding |
| ENST00000258436.5 | chr2 | 103347495 | 103347526 | 103347484 | 103347527 | - | protein_coding |
| ENST00000438943.1 | chr2 | 103347495 | 103347526 | 103347484 | 103347527 | - | nonsense_mediated_decay |
| ENST00000437075.2 | chr2 | 103347495 | 103347526 | 103347484 | 103347527 | - | nonsense_mediated_decay |
| ENST00000411991.1 | chr2 | 103347495 | 103347526 | 103347484 | 103347527 | - | nonsense_mediated_decay |
| ENST00000428085.1 | chr2 | 103347495 | 103347526 | 103347484 | 103347527 | - | nonsense_mediated_decay |
| ENST00000421966.1 | chr2 | 103347495 | 103347526 | 103347484 | 103347527 | - | nonsense_mediated_decay |
| ENST00000393310.1 | chr2 | 109301351 | 109301382 | 109300341 | 109303702 | + | protein_coding |
| ENST00000356454.3 | chr2 | 110374833 | 110374864 | 110371911 | 110376563 | + | protein_coding |
| ENST00000426778.1 | chr2 | 128036898 | 128036929 | 128036749 | 128036951 | - | nonsense_mediated_decay |
| ENST00000285398.2 | chr2 | 128036898 | 128036929 | 128036749 | 128036951 | - | protein_coding |
| ENST00000493187.2 | chr2 | 128036898 | 128036929 | 128036749 | 128036951 | - | protein_coding |
| ENST00000445889.1 | chr2 | 128036898 | 128036929 | 128036749 | 128036951 | - | nonsense_mediated_decay |
| ENST00000322313.4 | chr2 | 128460313 | 128460343 | 128458596 | 128464126 | - | protein_coding |
| ENST00000310981.4 | chr2 | 128460313 | 128460343 | 128458597 | 128461385 | + | protein_coding |
| ENST00000364600.1 | chr2 | 128556540 | 128556572 | 128556477 | 128556572 | - | misc_RNA |
| ENST00000272645.3 | chr2 | 128605389 | 128605420 | 128603896 | 128605759 | - | protein_coding |
| ENST00000409955.1 | chr2 | 128605389 | 128605420 | 128604273 | 128605759 | - | protein_coding |
| ENST00000487079.1 | chr2 | 128605389 | 128605420 | 128604815 | 128605759 | - | processed_transcript |
| ENST00000409698.1 | chr2 | 128605389 | 128605420 | 128605415 | 128605759 | - | protein_coding |
| ENST00000459503.1 | chr2 | 129202644 | 129202677 | 129202655 | 129202772 | - | rRNA |
| ENST00000409158.1 | chr2 | 131883428 | 131883459 | 131883326 | 131883478 | + | protein_coding |
| ENST00000403716.1 | chr2 | 131883428 | 131883459 | 131883326 | 131883478 | + | protein_coding |
| ENST00000234115.6 | chr2 | 131883428 | 131883459 | 131883326 | 131883478 | + | protein_coding |
| ENST00000439822.2 | chr2 | 131883428 | 131883459 | 131883326 | 131883478 | + | protein_coding |
| ENST00000482225.1 | chr2 | 131883428 | 131883459 | 131883326 | 131883478 | + | retained_intron |
| ENST00000438882.2 | chr2 | 131883428 | 131883459 | 131883326 | 131883478 | + | protein_coding |
| ENST00000538982.1 | chr2 | 131883428 | 131883459 | 131883326 | 131883478 | + | protein_coding |
| ENST00000404460.1 | chr2 | 131883428 | 131883459 | 131883326 | 131883478 | + | protein_coding |
| ENST00000409612.1 | chr2 | 131883428 | 131883459 | 131883326 | 131883478 | + | protein_coding |
| ENST00000409279.1 | chr2 | 131883428 | 131883459 | 131883326 | 131883478 | + | protein_coding |
| ENST00000303908.3 | chr2 | 131883428 | 131883459 | 131883326 | 131883478 | + | protein_coding |
| ENST00000516509.1 | chr2 | 133038628 | 133038723 | 133038647 | 133038738 | - | miRNA |
| ENST00000539609.1 | chr2 | 145157003 | 145157034 | 145155868 | 145157837 | - | protein_coding |
| ENST00000303660.2 | chr2 | 145157003 | 145157034 | 145155868 | 145157837 | - | protein_coding |
| ENST00000409487.1 | chr2 | 145157003 | 145157034 | 145155868 | 145157837 | - | protein_coding |
| ENST00000427902.1 | chr2 | 145157003 | 145157034 | 145156647 | 145157837 | - | protein_coding |
| ENST00000410194.1 | chr2 | 149583258 | 149583285 | 149583260 | 149583404 | - | snRNA |
| ENST00000539935.1 | chr2 | 152693336 | 152693367 | 152689290 | 152695893 | - | protein_coding |
| ENST00000360283.6 | chr2 | 152693336 | 152693367 | 152689300 | 152695893 | - | protein_coding |
| ENST00000543269.1 | chr2 | 152693336 | 152693367 | 152689300 | 152695893 | - | protein_coding |
| ENST00000295101.2 | chr2 | 155714400 | 155714438 | 155711239 | 155714863 | + | protein_coding |
| ENST00000454300.1 | chr2 | 160087655 | 160087686 | 160086089 | 160089166 | + | protein_coding |
| ENST00000263635.6 | chr2 | 160087655 | 160087686 | 160086089 | 160089170 | + | protein_coding |
| ENST00000470074.1 | chr2 | 160087655 | 160087686 | 160086089 | 160089166 | + | retained_intron |
| ENST00000496406.1 | chr2 | 160087655 | 160087686 | 160086089 | 160089170 | + | retained_intron |
| ENST00000483316.1 | chr2 | 160318050 | 160318081 | 160314835 | 160321273 | - | processed_transcript |
| ENST00000409175.1 | chr2 | 160604391 | 160604421 | 160604316 | 160605414 | + | protein_coding |
| ENST00000539065.1 | chr2 | 160604391 | 160604421 | 160604316 | 160605414 | + | protein_coding |
| ENST00000259050.3 | chr2 | 160604391 | 160604421 | 160604316 | 160605414 | + | protein_coding |
| ENST00000473749.1 | chr2 | 160604391 | 160604421 | 160604316 | 160604637 | + | processed_transcript |
| ENST00000409591.1 | chr2 | 160604391 | 160604421 | 160604316 | 160605414 | + | protein_coding |
| ENST00000539065.1 | chr2 | 160623908 | 160623925 | 160623795 | 160624193 | + | protein_coding |
| ENST00000259050.3 | chr2 | 160623908 | 160623925 | 160623795 | 160625359 | + | protein_coding |
| ENST00000409591.1 | chr2 | 160623908 | 160623925 | 160623795 | 160623920 | + | protein_coding |
| ENST00000420397.1 | chr2 | 160623908 | 160623925 | 160623795 | 160624471 | + | protein_coding |
| ENST00000478396.1 | chr2 | 160623908 | 160623925 | 160623795 | 160624361 | + | processed_transcript |
| ENST00000409682.3 | chr2 | 162242037 | 162242068 | 162241975 | 162242082 | + | protein_coding |
| ENST00000477232.1 | chr2 | 162242037 | 162242068 | 162237871 | 162242082 | + | retained_intron |
| ENST00000409682.3 | chr2 | 162247638 | 162247669 | 162247615 | 162247689 | + | protein_coding |
| ENST00000477232.1 | chr2 | 162247638 | 162247669 | 162247615 | 162247689 | + | retained_intron |
| ENST00000314499.7 | chr2 | 166537569 | 166537592 | 166535211 | 166545917 | + | protein_coding |
| ENST00000342316.4 | chr2 | 166537569 | 166537592 | 166535211 | 166545917 | + | protein_coding |
| ENST00000260970.3 | chr2 | 170494572 | 170494603 | 170492923 | 170497916 | + | protein_coding |
| ENST00000260956.4 | chr2 | 170668531 | 170668561 | 170668178 | 170668574 | + | protein_coding |
| ENST00000409333.1 | chr2 | 170668531 | 170668561 | 170668178 | 170668574 | + | protein_coding |
| ENST00000410097.1 | chr2 | 170668531 | 170668561 | 170668324 | 170669016 | - | protein_coding |
| ENST00000543391.1 | chr2 | 171609139 | 171609170 | 171608221 | 171609562 | + | processed_pseudogene |
| ENST00000461070.1 | chr2 | 171609139 | 171609170 | 171608315 | 171609298 | + | processed_pseudogene |
| ENST00000264110.2 | chr2 | 175938252 | 175938283 | 175936978 | 175939563 | - | protein_coding |
| ENST00000345739.5 | chr2 | 175938252 | 175938283 | 175936978 | 175939563 | - | protein_coding |
| ENST00000409437.1 | chr2 | 175938252 | 175938283 | 175936993 | 175939563 | - | protein_coding |
| ENST00000385011.1 | chr2 | 177015057 | 177015089 | 177015031 | 177015140 | + | miRNA |
| ENST00000397063.4 | chr2 | 178098032 | 178098063 | 178097978 | 178098067 | - | protein_coding |
| ENST00000397062.3 | chr2 | 178098032 | 178098063 | 178097978 | 178098067 | - | protein_coding |
| ENST00000446151.2 | chr2 | 178098032 | 178098063 | 178097978 | 178098067 | - | protein_coding |
| ENST00000464747.1 | chr2 | 178098032 | 178098063 | 178097978 | 178098067 | - | processed_transcript |
| ENST00000448782.1 | chr2 | 178098032 | 178098063 | 178097978 | 178098067 | - | protein_coding |
| ENST00000421929.1 | chr2 | 178098032 | 178098063 | 178097978 | 178098067 | - | protein_coding |
| ENST00000423513.1 | chr2 | 178098032 | 178098063 | 178097005 | 178098067 | - | protein_coding |
| ENST00000295108.3 | chr2 | 182541840 | 182541873 | 182541194 | 182543598 | - | protein_coding |
| ENST00000409150.3 | chr2 | 191372383 | 191372414 | 191369068 | 191374038 | - | protein_coding |
| ENST00000492032.1 | chr2 | 191402715 | 191402746 | 191402645 | 191402786 | - | rRNA_pseudogene |
| ENST00000307849.3 | chr2 | 192543130 | 192543161 | 192542862 | 192543431 | + | nonsense_mediated_decay |
| ENST00000451500.1 | chr2 | 192543130 | 192543161 | 192542873 | 192543431 | + | nonsense_mediated_decay |
| ENST00000307834.5 | chr2 | 192550534 | 192550565 | 192550325 | 192551665 | + | nonsense_mediated_decay |
| ENST00000307849.3 | chr2 | 192550534 | 192550565 | 192550325 | 192552298 | + | nonsense_mediated_decay |
| ENST00000410026.1 | chr2 | 192550534 | 192550565 | 192550325 | 192553251 | + | protein_coding |
| ENST00000451500.1 | chr2 | 192550534 | 192550565 | 192550325 | 192550883 | + | nonsense_mediated_decay |
| ENST00000409510.1 | chr2 | 192550534 | 192550565 | 192550325 | 192550620 | + | protein_coding |
| ENST00000425611.2 | chr2 | 192550534 | 192550565 | 192550325 | 192551674 | + | protein_coding |
| ENST00000435931.1 | chr2 | 192550534 | 192550565 | 192550325 | 192550903 | + | protein_coding |
| ENST00000461337.1 | chr2 | 200648059 | 200648097 | 200648064 | 200648134 | - | scRNA_pseudogene |
| ENST00000409600.1 | chr2 | 201688097 | 201688131 | 201686870 | 201688569 | + | protein_coding |
| ENST00000426253.1 | chr2 | 201760055 | 201760086 | 201759987 | 201760096 | + | protein_coding |
| ENST00000416651.1 | chr2 | 201760055 | 201760086 | 201759987 | 201760113 | + | protein_coding |
| ENST00000409020.1 | chr2 | 201760055 | 201760086 | 201759987 | 201760113 | + | protein_coding |
| ENST00000359683.4 | chr2 | 201760055 | 201760086 | 201759987 | 201760113 | + | protein_coding |
| ENST00000409357.1 | chr2 | 201760055 | 201760086 | 201759987 | 201760113 | + | protein_coding |
| ENST00000374679.4 | chr2 | 201760055 | 201760086 | 201759987 | 201760086 | + | protein_coding |
| ENST00000409588.1 | chr2 | 201760055 | 201760086 | 201759987 | 201760113 | + | protein_coding |
| ENST00000286181.3 | chr2 | 201845709 | 201845740 | 201843216 | 201846594 | - | nonsense_mediated_decay |
| ENST00000418596.2 | chr2 | 201845709 | 201845740 | 201843772 | 201846594 | - | protein_coding |
| ENST00000473498.1 | chr2 | 203211051 | 203211054 | 203210989 | 203211097 | - | rRNA_pseudogene |
| ENST00000374580.4 | chr2 | 203425132 | 203425161 | 203424419 | 203432474 | + | protein_coding |
| ENST00000374574.2 | chr2 | 203425132 | 203425161 | 203424419 | 203432474 | + | protein_coding |
| ENST00000491122.1 | chr2 | 206218954 | 206218985 | 206218955 | 206219025 | + | tRNA_pseudogene |
| ENST00000384320.1 | chr2 | 207026606 | 207026637 | 207026605 | 207026678 | + | snoRNA |
| ENST00000392133.3 | chr2 | 217057418 | 217057449 | 217057362 | 217057458 | + | protein_coding |
| ENST00000392132.2 | chr2 | 217057418 | 217057449 | 217057362 | 217057458 | + | protein_coding |
| ENST00000460284.1 | chr2 | 217057418 | 217057449 | 217057362 | 217057458 | + | retained_intron |
| ENST00000446716.1 | chr2 | 220084406 | 220084436 | 220082847 | 220085174 | - | nonsense_mediated_decay |
| ENST00000409033.3 | chr2 | 220084406 | 220084436 | 220084102 | 220085174 | - | nonsense_mediated_decay |
| ENST00000396761.2 | chr2 | 220084406 | 220084436 | 220084102 | 220085174 | - | protein_coding |
| ENST00000409618.1 | chr2 | 220084406 | 220084436 | 220084104 | 220085174 | - | protein_coding |
| ENST00000415298.1 | chr2 | 224622916 | 224622947 | 224621706 | 224623467 | - | nonsense_mediated_decay |
| ENST00000334271.3 | chr2 | 224622916 | 224622947 | 224621707 | 224623467 | - | nonsense_mediated_decay |
| ENST00000461045.1 | chr2 | 225798822 | 225798853 | 225798823 | 225798884 | + | tRNA_pseudogene |
| ENST00000490522.1 | chr2 | 230045490 | 230045514 | 230045488 | 230045666 | - | rRNA_pseudogene |
| ENST00000490522.1 | chr2 | 230045516 | 230045519 | 230045488 | 230045666 | - | rRNA_pseudogene |
| ENST00000490522.1 | chr2 | 230045578 | 230045609 | 230045488 | 230045666 | - | rRNA_pseudogene |
| ENST00000490522.1 | chr2 | 230045649 | 230045680 | 230045488 | 230045666 | - | rRNA_pseudogene |
| ENST00000494901.1 | chr2 | 231375189 | 231375246 | 231372861 | 231375881 | + | retained_intron |
| ENST00000493540.1 | chr2 | 231678695 | 231678727 | 231678698 | 231678829 | + | retained_intron |
| ENST00000365530.1 | chr2 | 232325083 | 232325112 | 232325082 | 232325151 | - | snoRNA |
| ENST00000448874.1 | chr2 | 232578017 | 232578048 | 232577514 | 232578175 | + | nonsense_mediated_decay |
| ENST00000409115.3 | chr2 | 232578017 | 232578048 | 232577514 | 232578251 | + | protein_coding |
| ENST00000341369.7 | chr2 | 232578017 | 232578048 | 232577514 | 232578246 | + | protein_coding |
| ENST00000481928.1 | chr2 | 232578017 | 232578048 | 232577514 | 232578216 | + | retained_intron |
| ENST00000409401.3 | chr2 | 232995905 | 232995936 | 232995329 | 232996600 | + | protein_coding |
| ENST00000491769.1 | chr2 | 238252158 | 238252219 | 238250708 | 238253486 | - | retained_intron |
| ENST00000391973.2 | chr2 | 242291872 | 242291903 | 242291382 | 242293439 | + | protein_coding |
| ENST00000360051.3 | chr2 | 242291872 | 242291903 | 242291382 | 242293439 | + | protein_coding |
| ENST00000391971.2 | chr2 | 242291872 | 242291903 | 242291382 | 242293442 | + | protein_coding |
| ENST00000407971.1 | chr2 | 242291872 | 242291903 | 242291382 | 242293442 | + | protein_coding |
| ENST00000402092.2 | chr2 | 242291872 | 242291903 | 242291382 | 242293442 | + | protein_coding |
| ENST00000391972.3 | chr2 | 242291872 | 242291903 | 242291381 | 242293439 | + | protein_coding |
| ENST00000451310.1 | chr2 | 242291872 | 242291903 | 242291358 | 242292041 | + | protein_coding |
| ENST00000384553.1 | chr3 | 5127911 | 5127942 | 5127863 | 5127964 | - | misc_RNA |
| ENST00000256496.3 | chr3 | 5222272 | 5222307 | 5220349 | 5222596 | + | protein_coding |
| ENST00000438743.1 | chr3 | 5222272 | 5222307 | 5220349 | 5222431 | + | protein_coding |
| ENST00000452837.2 | chr3 | 9416271 | 9416302 | 9416200 | 9416330 | + | protein_coding |
| ENST00000345094.3 | chr3 | 9416271 | 9416302 | 9416200 | 9416330 | + | protein_coding |
| ENST00000515662.1 | chr3 | 9416271 | 9416302 | 9416200 | 9416330 | + | protein_coding |
| ENST00000441127.1 | chr3 | 9416271 | 9416302 | 9416200 | 9416330 | + | protein_coding |
| ENST00000416603.1 | chr3 | 9416271 | 9416302 | 9416200 | 9416330 | + | protein_coding |
| ENST00000461636.1 | chr3 | 9416271 | 9416302 | 9416200 | 9416330 | + | retained_intron |
| ENST00000402198.1 | chr3 | 9518026 | 9518057 | 9517167 | 9519838 | + | protein_coding |
| ENST00000402466.1 | chr3 | 9518026 | 9518057 | 9517167 | 9519838 | + | protein_coding |
| ENST00000406341.1 | chr3 | 9518026 | 9518057 | 9517167 | 9519838 | + | protein_coding |
| ENST00000407969.1 | chr3 | 9518026 | 9518057 | 9517167 | 9519838 | + | protein_coding |
| ENST00000302463.6 | chr3 | 9518026 | 9518057 | 9517167 | 9518162 | + | protein_coding |
| ENST00000493918.1 | chr3 | 9518026 | 9518057 | 9517167 | 9518143 | + | retained_intron |
| ENST00000466242.1 | chr3 | 9518026 | 9518057 | 9517167 | 9518143 | + | retained_intron |
| ENST00000401316.1 | chr3 | 10436172 | 10436236 | 10436173 | 10436246 | - | miRNA |
| ENST00000306077.4 | chr3 | 14183917 | 14183948 | 14183093 | 14185179 | + | protein_coding |
| ENST00000253699.3 | chr3 | 15115119 | 15115138 | 15111580 | 15116437 | - | protein_coding |
| ENST00000476527.2 | chr3 | 15115119 | 15115138 | 15115112 | 15116437 | - | protein_coding |
| ENST00000442720.1 | chr3 | 20211918 | 20211949 | 20210978 | 20212307 | - | protein_coding |
| ENST00000412997.1 | chr3 | 20211918 | 20211949 | 20211545 | 20212307 | - | protein_coding |
| ENST00000460637.1 | chr3 | 20211918 | 20211949 | 20211553 | 20212307 | - | processed_transcript |
| ENST00000205636.3 | chr3 | 32524910 | 32524944 | 32522804 | 32525589 | - | protein_coding |
| ENST00000384893.1 | chr3 | 35786019 | 35786050 | 35785968 | 35786051 | + | miRNA |
| ENST00000362205.1 | chr3 | 38010904 | 38010935 | 38010895 | 38010971 | + | miRNA |
| ENST00000405570.1 | chr3 | 41268699 | 41268730 | 41268699 | 41268843 | + | protein_coding |
| ENST00000396183.3 | chr3 | 41268699 | 41268730 | 41268699 | 41268843 | + | protein_coding |
| ENST00000349496.5 | chr3 | 41268699 | 41268730 | 41268699 | 41268843 | + | protein_coding |
| ENST00000453024.1 | chr3 | 41268699 | 41268730 | 41268699 | 41268843 | + | protein_coding |
| ENST00000396185.3 | chr3 | 41268699 | 41268730 | 41268699 | 41268843 | + | protein_coding |
| ENST00000396183.3 | chr3 | 41281313 | 41281344 | 41281310 | 41281934 | + | protein_coding |
| ENST00000349496.5 | chr3 | 41281313 | 41281344 | 41280625 | 41281936 | + | protein_coding |
| ENST00000396185.3 | chr3 | 41281313 | 41281344 | 41281151 | 41281934 | + | protein_coding |
| ENST00000542870.1 | chr3 | 42707449 | 42707480 | 42705729 | 42707643 | + | protein_coding |
| ENST00000457842.3 | chr3 | 42707449 | 42707480 | 42705729 | 42709072 | + | protein_coding |
| ENST00000429705.2 | chr3 | 43391852 | 43391885 | 43388831 | 43392635 | + | protein_coding |
| ENST00000296088.7 | chr3 | 43391852 | 43391885 | 43388831 | 43392630 | + | protein_coding |
| ENST00000385219.1 | chr3 | 44155726 | 44155757 | 44155704 | 44155802 | + | miRNA |
| ENST00000431726.1 | chr3 | 47451149 | 47451180 | 47451145 | 47451324 | + | nonsense_mediated_decay |
| ENST00000456408.2 | chr3 | 47451149 | 47451180 | 47451145 | 47451324 | + | protein_coding |
| ENST00000265562.4 | chr3 | 47451149 | 47451180 | 47451145 | 47451324 | + | protein_coding |
| ENST00000495653.1 | chr3 | 47451149 | 47451180 | 47450788 | 47451324 | + | retained_intron |
| ENST00000408658.1 | chr3 | 47891045 | 47891076 | 47891045 | 47891119 | + | miRNA |
| ENST00000395474.3 | chr3 | 49053266 | 49053297 | 49052258 | 49053386 | + | protein_coding |
| ENST00000354294.5 | chr3 | 49053266 | 49053297 | 49053186 | 49053293 | + | protein_coding |
| ENST00000481001.1 | chr3 | 49053266 | 49053297 | 49053252 | 49053716 | - | retained_intron |
| ENST00000438585.1 | chr3 | 49053266 | 49053297 | 49053252 | 49053305 | - | protein_coding |
| ENST00000441576.2 | chr3 | 49053266 | 49053297 | 49053237 | 49053305 | - | protein_coding |
| ENST00000460505.1 | chr3 | 49053266 | 49053297 | 49053237 | 49053305 | - | retained_intron |
| ENST00000341949.4 | chr3 | 49053266 | 49053297 | 49053237 | 49053305 | - | protein_coding |
| ENST00000498498.1 | chr3 | 49053266 | 49053297 | 49053237 | 49053773 | - | retained_intron |
| ENST00000498794.1 | chr3 | 49053266 | 49053297 | 49052924 | 49054946 | - | retained_intron |
| ENST00000395462.4 | chr3 | 49053266 | 49053297 | 49053237 | 49053305 | - | protein_coding |
| ENST00000440857.1 | chr3 | 49053266 | 49053297 | 49053252 | 49053519 | - | protein_coding |
| ENST00000313778.5 | chr3 | 49053266 | 49053297 | 49053237 | 49053305 | - | protein_coding |
| ENST00000472331.1 | chr3 | 49053266 | 49053297 | 49053237 | 49053773 | - | retained_intron |
| ENST00000467457.1 | chr3 | 49053266 | 49053297 | 49052958 | 49053773 | - | retained_intron |
| ENST00000484831.1 | chr3 | 49053266 | 49053297 | 49053237 | 49053305 | - | retained_intron |
| ENST00000384873.1 | chr3 | 49058095 | 49058127 | 49058051 | 49058142 | - | miRNA |
| ENST00000418115.1 | chr3 | 49396738 | 49396745 | 49396578 | 49397815 | - | protein_coding |
| ENST00000423656.1 | chr3 | 51467520 | 51467551 | 51467485 | 51467664 | - | protein_coding |
| ENST00000364725.1 | chr3 | 51728469 | 51728506 | 51728481 | 51728598 | - | rRNA |
| ENST00000354773.4 | chr3 | 52328277 | 52328308 | 52326276 | 52329272 | + | protein_coding |
| ENST00000411757.1 | chr3 | 52328277 | 52328308 | 52326276 | 52329272 | + | protein_coding |
| ENST00000385191.1 | chr3 | 52328277 | 52328308 | 52328235 | 52328324 | - | miRNA |
| ENST00000490804.1 | chr3 | 52439508 | 52439536 | 52439126 | 52439669 | - | processed_transcript |
| ENST00000462138.1 | chr3 | 53276186 | 53276217 | 53276141 | 53276258 | - | protein_coding |
| ENST00000423525.2 | chr3 | 53276186 | 53276217 | 53276141 | 53276258 | - | protein_coding |
| ENST00000450814.2 | chr3 | 53276186 | 53276217 | 53276141 | 53276258 | - | nonsense_mediated_decay |
| ENST00000423516.1 | chr3 | 53276186 | 53276217 | 53276141 | 53276258 | - | protein_coding |
| ENST00000296289.6 | chr3 | 53276186 | 53276217 | 53276141 | 53276258 | - | protein_coding |
| ENST00000414014.1 | chr3 | 53276186 | 53276217 | 53276141 | 53276258 | - | protein_coding |
| ENST00000469678.1 | chr3 | 53276186 | 53276217 | 53276141 | 53276258 | - | nonsense_mediated_decay |
| ENST00000472528.1 | chr3 | 53276186 | 53276217 | 53276141 | 53276258 | - | nonsense_mediated_decay |
| ENST00000487660.1 | chr3 | 53276186 | 53276217 | 53276141 | 53276215 | - | retained_intron |
| ENST00000483706.1 | chr3 | 53276186 | 53276217 | 53275911 | 53276258 | - | retained_intron |
| ENST00000498251.1 | chr3 | 53713299 | 53713333 | 53713021 | 53713790 | + | processed_transcript |
| ENST00000288266.3 | chr3 | 57304182 | 57304213 | 57303569 | 57307496 | + | protein_coding |
| ENST00000463280.1 | chr3 | 58411587 | 58411618 | 58410479 | 58411748 | + | protein_coding |
| ENST00000479241.1 | chr3 | 58411587 | 58411618 | 58410479 | 58411746 | + | protein_coding |
| ENST00000536750.1 | chr3 | 58411587 | 58411618 | 58410479 | 58411743 | + | protein_coding |
| ENST00000479134.1 | chr3 | 58411587 | 58411618 | 58410479 | 58411601 | + | protein_coding |
| ENST00000302746.6 | chr3 | 58414296 | 58414327 | 58414200 | 58414341 | - | protein_coding |
| ENST00000383714.4 | chr3 | 58414296 | 58414327 | 58414200 | 58414341 | - | protein_coding |
| ENST00000469364.1 | chr3 | 58414296 | 58414327 | 58414200 | 58414341 | - | nonsense_mediated_decay |
| ENST00000485460.1 | chr3 | 58414296 | 58414327 | 58414200 | 58414341 | - | protein_coding |
| ENST00000479945.1 | chr3 | 58414296 | 58414327 | 58413408 | 58414341 | - | retained_intron |
| ENST00000474765.1 | chr3 | 58414296 | 58414327 | 58414079 | 58414341 | - | protein_coding |
| ENST00000461692.1 | chr3 | 58414296 | 58414327 | 58414200 | 58414341 | - | retained_intron |
| ENST00000360997.2 | chr3 | 58551156 | 58551179 | 58549844 | 58552422 | - | protein_coding |
| ENST00000394481.1 | chr3 | 58551156 | 58551179 | 58549844 | 58552422 | - | protein_coding |
| ENST00000295901.4 | chr3 | 63999160 | 63999191 | 63999115 | 63999283 | - | protein_coding |
| ENST00000467853.1 | chr3 | 63999160 | 63999191 | 63999115 | 64003408 | - | retained_intron |
| ENST00000476464.1 | chr3 | 63999160 | 63999191 | 63999115 | 63999283 | - | retained_intron |
| ENST00000492933.1 | chr3 | 63999160 | 63999191 | 63999115 | 63999283 | - | protein_coding |
| ENST00000394431.2 | chr3 | 63999160 | 63999191 | 63999115 | 63999283 | - | protein_coding |
| ENST00000482510.1 | chr3 | 63999160 | 63999191 | 63999115 | 63999283 | - | protein_coding |
| ENST00000420581.2 | chr3 | 69156604 | 69156645 | 69156023 | 69158272 | - | protein_coding |
| ENST00000318789.4 | chr3 | 71005879 | 71005924 | 71003844 | 71008542 | - | protein_coding |
| ENST00000358280.5 | chr3 | 71005879 | 71005924 | 71004736 | 71008542 | - | protein_coding |
| ENST00000318796.5 | chr3 | 71005879 | 71005924 | 71004736 | 71008542 | - | protein_coding |
| ENST00000475937.1 | chr3 | 71005879 | 71005924 | 71005896 | 71008542 | - | protein_coding |
| ENST00000410457.1 | chr3 | 81558650 | 81558678 | 81558627 | 81558815 | + | snRNA |
| ENST00000263780.4 | chr3 | 87304544 | 87304575 | 87302862 | 87304698 | + | protein_coding |
| ENST00000341181.6 | chr3 | 98234362 | 98234393 | 98234317 | 98235723 | - | protein_coding |
| ENST00000437922.1 | chr3 | 98234362 | 98234393 | 98234318 | 98235723 | - | protein_coding |
| ENST00000394180.2 | chr3 | 98234362 | 98234393 | 98234324 | 98235723 | - | protein_coding |
| ENST00000506885.1 | chr3 | 98234362 | 98234393 | 98234327 | 98234676 | - | protein_coding |
| ENST00000503004.1 | chr3 | 98234362 | 98234393 | 98234327 | 98235723 | - | protein_coding |
| ENST00000394185.2 | chr3 | 98234362 | 98234393 | 98234327 | 98235723 | - | protein_coding |
| ENST00000394181.2 | chr3 | 98234362 | 98234393 | 98234327 | 98235723 | - | protein_coding |
| ENST00000513873.1 | chr3 | 98234362 | 98234393 | 98234328 | 98234676 | - | protein_coding |
| ENST00000510545.1 | chr3 | 98234362 | 98234393 | 98234331 | 98235723 | - | protein_coding |
| ENST00000284320.5 | chr3 | 100091523 | 100091553 | 100091450 | 100091566 | - | protein_coding |
| ENST00000544924.1 | chr3 | 100091523 | 100091553 | 100091450 | 100091566 | - | protein_coding |
| ENST00000483945.1 | chr3 | 100091523 | 100091553 | 100090850 | 100091566 | - | retained_intron |
| ENST00000314261.7 | chr3 | 101232058 | 101232089 | 101231935 | 101232085 | - | protein_coding |
| ENST00000348610.3 | chr3 | 101232058 | 101232089 | 101231935 | 101232067 | - | protein_coding |
| ENST00000471694.1 | chr3 | 107763263 | 107763307 | 107762145 | 107766139 | - | processed_transcript |
| ENST00000355354.7 | chr3 | 107763263 | 107763307 | 107762146 | 107766139 | - | protein_coding |
| ENST00000491772.1 | chr3 | 108304013 | 108304044 | 108303912 | 108304059 | - | protein_coding |
| ENST00000295746.7 | chr3 | 108304013 | 108304044 | 108303912 | 108304059 | - | protein_coding |
| ENST00000481530.1 | chr3 | 108304013 | 108304044 | 108303912 | 108304059 | - | nonsense_mediated_decay |
| ENST00000487834.1 | chr3 | 108304013 | 108304044 | 108303912 | 108304059 | - | processed_transcript |
| ENST00000468953.1 | chr3 | 108304013 | 108304044 | 108303912 | 108304059 | - | nonsense_mediated_decay |
| ENST00000461666.1 | chr3 | 108304013 | 108304044 | 108303707 | 108304059 | - | retained_intron |
| ENST00000493013.1 | chr3 | 113351627 | 113351650 | 113351551 | 113351746 | + | scRNA_pseudogene |
| ENST00000485050.1 | chr3 | 113806079 | 113806110 | 113804520 | 113807196 | + | protein_coding |
| ENST00000281273.4 | chr3 | 113806079 | 113806110 | 113804520 | 113807269 | + | protein_coding |
| ENST00000465850.1 | chr3 | 113822199 | 113822230 | 113822096 | 113822475 | - | processed_pseudogene |
| ENST00000409521.2 | chr3 | 116365006 | 116365029 | 116365001 | 116365252 | - | processed_pseudogene |
| ENST00000473537.1 | chr3 | 116365006 | 116365029 | 116364749 | 116365574 | - | processed_pseudogene |
| ENST00000360772.3 | chr3 | 123330290 | 123330326 | 123328896 | 123333196 | - | protein_coding |
| ENST00000243253.3 | chr3 | 127790161 | 127790184 | 127788319 | 127790525 | + | protein_coding |
| ENST00000424880.2 | chr3 | 127790161 | 127790184 | 127788319 | 127790520 | + | protein_coding |
| ENST00000483956.1 | chr3 | 127790161 | 127790184 | 127788319 | 127790526 | + | processed_transcript |
| ENST00000432054.2 | chr3 | 129370062 | 129370093 | 129366635 | 129370638 | - | protein_coding |
| ENST00000393238.3 | chr3 | 129370062 | 129370093 | 129366635 | 129370638 | - | protein_coding |
| ENST00000426664.2 | chr3 | 129370062 | 129370093 | 129366637 | 129370638 | - | protein_coding |
| ENST00000329333.5 | chr3 | 129370062 | 129370093 | 129370018 | 129370638 | - | protein_coding |
| ENST00000312481.7 | chr3 | 130064687 | 130064718 | 130064359 | 130064824 | + | nonsense_mediated_decay |
| ENST00000432398.2 | chr3 | 130064687 | 130064718 | 130064359 | 130064824 | + | protein_coding |
| ENST00000265379.6 | chr3 | 130064687 | 130064718 | 130064359 | 130064824 | + | protein_coding |
| ENST00000264995.2 | chr3 | 131206552 | 131206583 | 131206524 | 131206584 | - | protein_coding |
| ENST00000511168.1 | chr3 | 131206552 | 131206583 | 131206524 | 131206584 | - | protein_coding |
| ENST00000425847.2 | chr3 | 131206552 | 131206583 | 131206524 | 131206584 | - | protein_coding |
| ENST00000507669.1 | chr3 | 131206552 | 131206583 | 131206524 | 131206584 | - | protein_coding |
| ENST00000506487.1 | chr3 | 131206552 | 131206583 | 131205994 | 131206584 | - | retained_intron |
| ENST00000512877.1 | chr3 | 131206552 | 131206583 | 131206525 | 131206584 | - | protein_coding |
| ENST00000264993.3 | chr3 | 133307697 | 133307728 | 133306740 | 133309105 | + | protein_coding |
| ENST00000420115.2 | chr3 | 133307697 | 133307728 | 133306006 | 133309104 | + | protein_coding |
| ENST00000515421.1 | chr3 | 133307697 | 133307728 | 133307197 | 133308006 | + | protein_coding |
| ENST00000364543.1 | chr3 | 134502334 | 134502397 | 134502278 | 134502397 | - | rRNA |
| ENST00000309993.2 | chr3 | 135870440 | 135870471 | 135867764 | 135871580 | - | protein_coding |
| ENST00000434835.2 | chr3 | 135870440 | 135870471 | 135869469 | 135871580 | - | protein_coding |
| ENST00000312960.3 | chr3 | 150459884 | 150459916 | 150458914 | 150460485 | - | protein_coding |
| ENST00000363124.1 | chr3 | 150905886 | 150905973 | 150905886 | 150906010 | + | rRNA |
| ENST00000464316.1 | chr3 | 156871337 | 156871368 | 156869966 | 156872455 | - | retained_intron |
| ENST00000474539.1 | chr3 | 156871337 | 156871368 | 156868071 | 156871617 | - | retained_intron |
| ENST00000479052.1 | chr3 | 156871337 | 156871368 | 156871288 | 156871417 | - | processed_transcript |
| ENST00000364908.1 | chr3 | 156871337 | 156871368 | 156871337 | 156871429 | + | misc_RNA |
| ENST00000385045.1 | chr3 | 160122395 | 160122450 | 160122376 | 160122473 | + | miRNA |
| ENST00000362117.1 | chr3 | 160122542 | 160122608 | 160122533 | 160122613 | + | miRNA |
| ENST00000351193.2 | chr3 | 160956560 | 160956591 | 160956518 | 160956614 | + | protein_coding |
| ENST00000472947.1 | chr3 | 160956560 | 160956591 | 160956518 | 160956614 | + | protein_coding |
| ENST00000460469.1 | chr3 | 160956560 | 160956591 | 160956518 | 160956614 | + | protein_coding |
| ENST00000540137.1 | chr3 | 160956560 | 160956591 | 160956518 | 160956614 | + | protein_coding |
| ENST00000415807.2 | chr3 | 172115954 | 172115985 | 172114954 | 172116561 | + | protein_coding |
| ENST00000336824.4 | chr3 | 172115954 | 172115985 | 172114954 | 172119455 | + | protein_coding |
| ENST00000311417.2 | chr3 | 178740318 | 178740349 | 178735011 | 178743016 | - | protein_coding |
| ENST00000429709.2 | chr3 | 179305855 | 179305886 | 179305718 | 179306196 | + | protein_coding |
| ENST00000450518.2 | chr3 | 179305855 | 179305886 | 179305718 | 179306195 | + | protein_coding |
| ENST00000392662.1 | chr3 | 179305855 | 179305886 | 179305718 | 179306186 | + | protein_coding |
| ENST00000461125.1 | chr3 | 179305855 | 179305886 | 179305718 | 179306029 | + | retained_intron |
| ENST00000364952.1 | chr3 | 179879719 | 179879765 | 179879674 | 179879765 | - | rRNA |
| ENST00000362969.1 | chr3 | 181540656 | 181540695 | 181540660 | 181540778 | - | rRNA |
| ENST00000362969.1 | chr3 | 181540718 | 181540778 | 181540660 | 181540778 | - | rRNA |
| ENST00000453386.2 | chr3 | 185635363 | 185635373 | 185633694 | 185635513 | - | protein_coding |
| ENST00000492417.1 | chr3 | 185635363 | 185635373 | 185634532 | 185635513 | - | retained_intron |
| ENST00000463328.1 | chr3 | 185635363 | 185635373 | 185634537 | 185635513 | - | retained_intron |
| ENST00000259043.7 | chr3 | 185635363 | 185635373 | 185634943 | 185635513 | - | protein_coding |
| ENST00000414862.1 | chr3 | 185635363 | 185635373 | 185635116 | 185635513 | - | protein_coding |
| ENST00000487615.1 | chr3 | 185635363 | 185635373 | 185635117 | 185635513 | - | retained_intron |
| ENST00000466832.1 | chr3 | 185635363 | 185635373 | 185635117 | 185635513 | - | retained_intron |
| ENST00000456380.1 | chr3 | 185635363 | 185635373 | 185635118 | 185635513 | - | nonsense_mediated_decay |
| ENST00000465222.1 | chr3 | 186502620 | 186502652 | 186502221 | 186502831 | + | retained_intron |
| ENST00000491473.1 | chr3 | 186502620 | 186502652 | 186502218 | 186502733 | + | retained_intron |
| ENST00000486805.1 | chr3 | 186502620 | 186502652 | 186502353 | 186502830 | + | retained_intron |
| ENST00000485101.1 | chr3 | 186502620 | 186502652 | 186502353 | 186503840 | + | retained_intron |
| ENST00000459163.1 | chr3 | 186502620 | 186502652 | 186502585 | 186502653 | + | snoRNA |
| ENST00000318037.3 | chr3 | 196198068 | 196198100 | 196195654 | 196199643 | - | protein_coding |
| ENST00000447325.1 | chr3 | 196662924 | 196662954 | 196662277 | 196663953 | - | protein_coding |
| ENST00000321256.5 | chr3 | 196662924 | 196662954 | 196662278 | 196663953 | - | protein_coding |
| ENST00000463783.1 | chr3 | 196662924 | 196662954 | 196662278 | 196666303 | - | retained_intron |
| ENST00000428425.1 | chr3 | 196662924 | 196662954 | 196662280 | 196663953 | - | nonsense_mediated_decay |
| ENST00000427641.2 | chr3 | 196662924 | 196662954 | 196662305 | 196663953 | - | protein_coding |
| ENST00000464167.1 | chr3 | 197680882 | 197680903 | 197680874 | 197681018 | + | protein_coding |
| ENST00000485439.1 | chr3 | 197680882 | 197680903 | 197680874 | 197681322 | + | retained_intron |
| ENST00000329092.8 | chr3 | 197680882 | 197680903 | 197680874 | 197681018 | + | processed_transcript |
| ENST00000448864.1 | chr3 | 197680882 | 197680903 | 197680874 | 197681018 | + | protein_coding |
| ENST00000429437.1 | chr3 | 197680882 | 197680903 | 197680874 | 197681018 | + | nonsense_mediated_decay |
| ENST00000442341.1 | chr3 | 197680882 | 197680903 | 197680874 | 197680991 | + | protein_coding |
| ENST00000439255.1 | chr3 | 197680882 | 197680903 | 197680874 | 197681018 | + | nonsense_mediated_decay |
| ENST00000474640.1 | chr3 | 197680882 | 197680903 | 197680434 | 197681018 | + | retained_intron |
| ENST00000429429.2 | chr4 | 1695154 | 1695185 | 1694527 | 1695440 | - | protein_coding |
| ENST00000489418.1 | chr4 | 1695154 | 1695185 | 1694527 | 1695440 | - | protein_coding |
| ENST00000460392.1 | chr4 | 1695154 | 1695185 | 1694560 | 1695440 | - | protein_coding |
| ENST00000318386.4 | chr4 | 1695154 | 1695185 | 1694585 | 1695440 | - | protein_coding |
| ENST00000483348.1 | chr4 | 1695154 | 1695185 | 1694971 | 1695440 | - | protein_coding |
| ENST00000488267.1 | chr4 | 1695154 | 1695185 | 1695174 | 1695440 | - | protein_coding |
| ENST00000508803.1 | chr4 | 1976665 | 1976696 | 1976590 | 1976731 | + | protein_coding |
| ENST00000353275.5 | chr4 | 1976665 | 1976696 | 1976590 | 1976731 | + | nonsense_mediated_decay |
| ENST00000382891.5 | chr4 | 1976665 | 1976696 | 1976590 | 1976731 | + | protein_coding |
| ENST00000382892.2 | chr4 | 1976665 | 1976696 | 1976590 | 1976731 | + | protein_coding |
| ENST00000312087.6 | chr4 | 1976665 | 1976696 | 1976590 | 1976731 | + | nonsense_mediated_decay |
| ENST00000382895.3 | chr4 | 1976665 | 1976696 | 1976590 | 1976731 | + | protein_coding |
| ENST00000482415.2 | chr4 | 1976665 | 1976696 | 1976590 | 1976731 | + | processed_transcript |
| ENST00000382888.3 | chr4 | 1976665 | 1976696 | 1976590 | 1976731 | + | protein_coding |
| ENST00000515695.1 | chr4 | 1976665 | 1976696 | 1976514 | 1976731 | + | retained_intron |
| ENST00000382723.4 | chr4 | 4865281 | 4865315 | 4864428 | 4865663 | + | protein_coding |
| ENST00000471237.1 | chr4 | 7584238 | 7584272 | 7584187 | 7584364 | + | rRNA_pseudogene |
| ENST00000471237.1 | chr4 | 7584330 | 7584364 | 7584187 | 7584364 | + | rRNA_pseudogene |
| ENST00000385072.1 | chr4 | 8007029 | 8007060 | 8007028 | 8007108 | - | miRNA |
| ENST00000265018.3 | chr4 | 17633668 | 17633710 | 17630929 | 17634248 | - | protein_coding |
| ENST00000384999.1 | chr4 | 20529922 | 20529953 | 20529898 | 20530007 | + | miRNA |
| ENST00000342295.1 | chr4 | 26432812 | 26432838 | 26432314 | 26433278 | + | protein_coding |
| ENST00000355476.3 | chr4 | 26432812 | 26432838 | 26432314 | 26433278 | + | protein_coding |
| ENST00000342320.4 | chr4 | 26432812 | 26432838 | 26432314 | 26436541 | + | protein_coding |
| ENST00000264313.5 | chr4 | 48426001 | 48426032 | 48424028 | 48428076 | + | protein_coding |
| ENST00000512093.1 | chr4 | 48426001 | 48426032 | 48424028 | 48427032 | + | protein_coding |
| ENST00000461294.2 | chr4 | 55144394 | 55144427 | 55144063 | 55144413 | + | retained_intron |
| ENST00000273854.3 | chr4 | 66186035 | 66186069 | 66185281 | 66189937 | - | protein_coding |
| ENST00000474870.1 | chr4 | 70296652 | 70296744 | 70296579 | 70296753 | - | rRNA_pseudogene |
| ENST00000324439.5 | chr4 | 76406977 | 76407008 | 76404247 | 76407876 | - | protein_coding |
| ENST00000451788.1 | chr4 | 76406977 | 76407008 | 76404357 | 76407876 | - | protein_coding |
| ENST00000380840.2 | chr4 | 76406977 | 76407008 | 76406679 | 76407876 | - | protein_coding |
| ENST00000395719.3 | chr4 | 76568619 | 76568642 | 76567966 | 76570886 | - | protein_coding |
| ENST00000359707.4 | chr4 | 76568619 | 76568642 | 76567966 | 76570886 | - | protein_coding |
| ENST00000395719.3 | chr4 | 76570240 | 76570255 | 76567966 | 76570886 | - | protein_coding |
| ENST00000359707.4 | chr4 | 76570240 | 76570255 | 76567966 | 76570886 | - | protein_coding |
| ENST00000357854.3 | chr4 | 76570240 | 76570255 | 76569573 | 76570886 | - | protein_coding |
| ENST00000464639.1 | chr4 | 76807187 | 76807223 | 76807176 | 76807697 | + | rRNA_pseudogene |
| ENST00000264893.6 | chr4 | 77955686 | 77955717 | 77955650 | 77959767 | + | protein_coding |
| ENST00000512575.1 | chr4 | 77955686 | 77955717 | 77955650 | 77955820 | + | processed_transcript |
| ENST00000510515.1 | chr4 | 77955686 | 77955717 | 77955650 | 77955974 | + | protein_coding |
| ENST00000541121.1 | chr4 | 77955686 | 77955717 | 77955650 | 77959767 | + | protein_coding |
| ENST00000316355.5 | chr4 | 78082071 | 78082101 | 78081874 | 78082124 | + | protein_coding |
| ENST00000354403.5 | chr4 | 78082071 | 78082101 | 78081874 | 78082124 | + | protein_coding |
| ENST00000502280.1 | chr4 | 78082071 | 78082101 | 78081874 | 78082124 | + | protein_coding |
| ENST00000497512.1 | chr4 | 78082071 | 78082101 | 78081874 | 78082124 | + | processed_transcript |
| ENST00000395640.1 | chr4 | 78082071 | 78082101 | 78081874 | 78082124 | + | protein_coding |
| ENST00000512918.1 | chr4 | 78082071 | 78082101 | 78081874 | 78082079 | + | protein_coding |
| ENST00000509972.1 | chr4 | 78082071 | 78082101 | 78081874 | 78082124 | + | protein_coding |
| ENST00000504123.1 | chr4 | 78635593 | 78635629 | 78634541 | 78641797 | - | protein_coding |
| ENST00000264903.4 | chr4 | 78635593 | 78635629 | 78634541 | 78641797 | - | protein_coding |
| ENST00000295470.5 | chr4 | 83347700 | 83347705 | 83347616 | 83347786 | - | protein_coding |
| ENST00000502762.1 | chr4 | 83347700 | 83347705 | 83347616 | 83347786 | - | protein_coding |
| ENST00000507721.1 | chr4 | 83347700 | 83347705 | 83347616 | 83347786 | - | retained_intron |
| ENST00000514511.1 | chr4 | 83347700 | 83347705 | 83347616 | 83347786 | - | processed_transcript |
| ENST00000333209.3 | chr4 | 90169555 | 90169586 | 90165429 | 90171384 | - | protein_coding |
| ENST00000515059.1 | chr4 | 96076627 | 96076658 | 96075699 | 96079599 | + | protein_coding |
| ENST00000380158.4 | chr4 | 99363330 | 99363361 | 99363141 | 99363524 | + | protein_coding |
| ENST00000408927.3 | chr4 | 99363330 | 99363361 | 99363141 | 99365012 | + | protein_coding |
| ENST00000453712.2 | chr4 | 99363330 | 99363361 | 99363141 | 99363817 | + | protein_coding |
| ENST00000408900.3 | chr4 | 99363330 | 99363361 | 99363141 | 99363485 | + | protein_coding |
| ENST00000339360.5 | chr4 | 99363330 | 99363361 | 99363141 | 99363485 | + | protein_coding |
| ENST00000394735.1 | chr4 | 106604314 | 106604346 | 106603784 | 106604474 | - | protein_coding |
| ENST00000340139.5 | chr4 | 106604314 | 106604346 | 106603784 | 106604474 | - | protein_coding |
| ENST00000451321.2 | chr4 | 106604314 | 106604346 | 106603784 | 106604474 | - | protein_coding |
| ENST00000493425.1 | chr4 | 106604314 | 106604346 | 106604219 | 106604474 | - | retained_intron |
| ENST00000394684.4 | chr4 | 108816882 | 108816890 | 108816466 | 108817164 | + | protein_coding |
| ENST00000359079.4 | chr4 | 108816882 | 108816890 | 108816466 | 108817164 | + | protein_coding |
| ENST00000394686.3 | chr4 | 108816882 | 108816890 | 108816466 | 108817164 | + | protein_coding |
| ENST00000394684.4 | chr4 | 108820767 | 108820798 | 108820731 | 108820848 | + | protein_coding |
| ENST00000503862.1 | chr4 | 108820767 | 108820798 | 108820731 | 108820848 | + | protein_coding |
| ENST00000359079.4 | chr4 | 108820767 | 108820798 | 108820731 | 108820848 | + | protein_coding |
| ENST00000394686.3 | chr4 | 108820767 | 108820798 | 108820731 | 108820848 | + | protein_coding |
| ENST00000394684.4 | chr4 | 108832270 | 108832301 | 108831506 | 108836203 | + | protein_coding |
| ENST00000385192.1 | chr4 | 113569333 | 113569364 | 113569339 | 113569407 | - | miRNA |
| ENST00000394524.3 | chr4 | 114436273 | 114436304 | 114436225 | 114436347 | - | protein_coding |
| ENST00000454265.2 | chr4 | 114436273 | 114436304 | 114436225 | 114436347 | - | protein_coding |
| ENST00000429180.1 | chr4 | 114436273 | 114436304 | 114436225 | 114436347 | - | protein_coding |
| ENST00000418639.2 | chr4 | 114436273 | 114436304 | 114436225 | 114436347 | - | protein_coding |
| ENST00000394526.2 | chr4 | 114436273 | 114436304 | 114436225 | 114436347 | - | protein_coding |
| ENST00000296402.5 | chr4 | 114436273 | 114436304 | 114436225 | 114436347 | - | protein_coding |
| ENST00000511664.1 | chr4 | 114436273 | 114436304 | 114436225 | 114436347 | - | protein_coding |
| ENST00000342666.5 | chr4 | 114436273 | 114436304 | 114436225 | 114436347 | - | protein_coding |
| ENST00000515496.1 | chr4 | 114436273 | 114436304 | 114436225 | 114436347 | - | protein_coding |
| ENST00000514328.1 | chr4 | 114436273 | 114436304 | 114436225 | 114436347 | - | protein_coding |
| ENST00000394522.3 | chr4 | 114436273 | 114436304 | 114436225 | 114436347 | - | protein_coding |
| ENST00000505990.1 | chr4 | 114436273 | 114436304 | 114436225 | 114436347 | - | protein_coding |
| ENST00000379773.2 | chr4 | 114436273 | 114436304 | 114436225 | 114436347 | - | protein_coding |
| ENST00000508738.1 | chr4 | 114436273 | 114436304 | 114436225 | 114436347 | - | protein_coding |
| ENST00000509594.1 | chr4 | 114436273 | 114436304 | 114436225 | 114436347 | - | protein_coding |
| ENST00000264501.4 | chr4 | 123283637 | 123283669 | 123283186 | 123283907 | + | protein_coding |
| ENST00000388738.3 | chr4 | 123283637 | 123283669 | 123283186 | 123283913 | + | protein_coding |
| ENST00000438707.1 | chr4 | 123283637 | 123283669 | 123283186 | 123283905 | + | protein_coding |
| ENST00000306802.4 | chr4 | 123283637 | 123283669 | 123283186 | 123283907 | + | protein_coding |
| ENST00000431755.2 | chr4 | 123283637 | 123283669 | 123283186 | 123283905 | + | protein_coding |
| ENST00000264498.3 | chr4 | 123814652 | 123814680 | 123813366 | 123819391 | + | protein_coding |
| ENST00000497779.1 | chr4 | 134082979 | 134083010 | 134082915 | 134083021 | - | rRNA_pseudogene |
| ENST00000323570.3 | chr4 | 141471542 | 141471572 | 141471384 | 141474924 | + | protein_coding |
| ENST00000502290.1 | chr4 | 141471542 | 141471572 | 141471384 | 141471589 | + | retained_intron |
| ENST00000510377.1 | chr4 | 144135698 | 144135729 | 144134734 | 144136730 | + | protein_coding |
| ENST00000307017.4 | chr4 | 144135698 | 144135729 | 144134734 | 144136096 | + | protein_coding |
| ENST00000511739.1 | chr4 | 144135698 | 144135729 | 144134734 | 144136096 | + | nonsense_mediated_decay |
| ENST00000296582.3 | chr4 | 148539166 | 148539197 | 148538534 | 148539230 | + | protein_coding |
| ENST00000508208.1 | chr4 | 148539166 | 148539197 | 148538537 | 148539230 | + | protein_coding |
| ENST00000499023.2 | chr4 | 155457681 | 155457712 | 155456158 | 155457896 | - | protein_coding |
| ENST00000512773.1 | chr4 | 155457681 | 155457712 | 155457584 | 155457896 | - | retained_intron |
| ENST00000393905.2 | chr4 | 155457681 | 155457712 | 155457663 | 155457896 | - | protein_coding |
| ENST00000506627.1 | chr4 | 155457681 | 155457712 | 155457673 | 155457896 | - | nonsense_mediated_decay |
| ENST00000462868.1 | chr4 | 167063754 | 167063780 | 167063725 | 167063863 | + | rRNA_pseudogene |
| ENST00000265000.4 | chr4 | 174244751 | 174244782 | 174242731 | 174245118 | + | protein_coding |
| ENST00000296503.5 | chr4 | 174252985 | 174253000 | 174252846 | 174253389 | - | protein_coding |
| ENST00000446922.2 | chr4 | 174252985 | 174253000 | 174252852 | 174253389 | - | protein_coding |
| ENST00000438704.2 | chr4 | 174252985 | 174253000 | 174252852 | 174253389 | - | protein_coding |
| ENST00000511316.1 | chr4 | 174252985 | 174253000 | 174252856 | 174254365 | - | retained_intron |
| ENST00000296503.5 | chr4 | 174253051 | 174253082 | 174252846 | 174253389 | - | protein_coding |
| ENST00000446922.2 | chr4 | 174253051 | 174253082 | 174252852 | 174253389 | - | protein_coding |
| ENST00000438704.2 | chr4 | 174253051 | 174253082 | 174252852 | 174253389 | - | protein_coding |
| ENST00000511316.1 | chr4 | 174253051 | 174253082 | 174252856 | 174254365 | - | retained_intron |
| ENST00000447135.2 | chr4 | 174537456 | 174537487 | 174537087 | 174538057 | - | processed_pseudogene |
| ENST00000434007.1 | chr4 | 174537456 | 174537487 | 174537087 | 174537794 | - | retrotransposed |
| ENST00000393658.2 | chr4 | 176733449 | 176733491 | 176733342 | 176733902 | - | protein_coding |
| ENST00000464416.1 | chr4 | 181405930 | 181405964 | 181405960 | 181406042 | - | snoRNA_pseudogene |
| ENST00000504169.1 | chr4 | 184367887 | 184367918 | 184367241 | 184369351 | + | protein_coding |
| ENST00000302350.4 | chr4 | 184367887 | 184367918 | 184367241 | 184369351 | + | protein_coding |
| ENST00000506835.1 | chr4 | 184367887 | 184367918 | 184367241 | 184367912 | + | processed_transcript |
| ENST00000308394.4 | chr4 | 185550046 | 185550077 | 185548850 | 185550655 | - | protein_coding |
| ENST00000393585.2 | chr4 | 185550046 | 185550077 | 185548850 | 185550655 | - | protein_coding |
| ENST00000523916.1 | chr4 | 185550046 | 185550077 | 185548850 | 185550655 | - | protein_coding |
| ENST00000438467.2 | chr4 | 185550046 | 185550077 | 185548851 | 185550655 | - | protein_coding |
| ENST00000536428.1 | chr4 | 190988793 | 190988827 | 190988796 | 190990406 | + | unprocessed_pseudogene |
| ENST00000399816.3 | chr5 | 6491520 | 6491551 | 6491331 | 6495022 | + | protein_coding |
| ENST00000264670.6 | chr5 | 6609946 | 6609977 | 6609939 | 6610035 | - | protein_coding |
| ENST00000504374.1 | chr5 | 6609946 | 6609977 | 6609939 | 6610035 | - | nonsense_mediated_decay |
| ENST00000539938.1 | chr5 | 6609946 | 6609977 | 6609939 | 6610035 | - | protein_coding |
| ENST00000505892.1 | chr5 | 6609946 | 6609977 | 6609939 | 6610035 | - | retained_intron |
| ENST00000506139.1 | chr5 | 6609946 | 6609977 | 6609939 | 6610035 | - | protein_coding |
| ENST00000264668.2 | chr5 | 7901063 | 7901095 | 7900027 | 7901237 | + | protein_coding |
| ENST00000511461.1 | chr5 | 7901063 | 7901095 | 7900027 | 7901237 | + | nonsense_mediated_decay |
| ENST00000440940.2 | chr5 | 7901063 | 7901095 | 7900027 | 7901237 | + | protein_coding |
| ENST00000510525.1 | chr5 | 7901063 | 7901095 | 7900027 | 7901229 | + | nonsense_mediated_decay |
| ENST00000280326.4 | chr5 | 10250010 | 10250046 | 10250033 | 10250557 | + | protein_coding |
| ENST00000503026.1 | chr5 | 10250010 | 10250046 | 10250041 | 10250157 | + | protein_coding |
| ENST00000487740.1 | chr5 | 10374760 | 10374788 | 10374042 | 10375088 | + | rRNA_pseudogene |
| ENST00000265071.2 | chr5 | 31326785 | 31326815 | 31322925 | 31329253 | + | protein_coding |
| ENST00000265070.6 | chr5 | 32125075 | 32125106 | 32124810 | 32126742 | - | protein_coding |
| ENST00000542582.1 | chr5 | 32125075 | 32125106 | 32124818 | 32126742 | - | protein_coding |
| ENST00000499354.2 | chr5 | 32125075 | 32125106 | 32124827 | 32126742 | - | processed_transcript |
| ENST00000265073.4 | chr5 | 32601807 | 32601838 | 32601111 | 32604185 | + | protein_coding |
| ENST00000511615.1 | chr5 | 32601807 | 32601838 | 32601111 | 32602094 | + | nonsense_mediated_decay |
| ENST00000265073.4 | chr5 | 32602551 | 32602582 | 32601111 | 32604185 | + | protein_coding |
| ENST00000542111.1 | chr5 | 32602551 | 32602582 | 32602321 | 32603326 | + | protein_coding |
| ENST00000336767.5 | chr5 | 34925659 | 34925690 | 34925331 | 34926101 | + | protein_coding |
| ENST00000515798.1 | chr5 | 34925659 | 34925690 | 34924952 | 34925785 | + | retained_intron |
| ENST00000364498.1 | chr5 | 35429166 | 35429197 | 35429166 | 35429378 | + | snoRNA |
| ENST00000231498.3 | chr5 | 37351352 | 37351383 | 37351292 | 37351458 | - | protein_coding |
| ENST00000381843.2 | chr5 | 37351352 | 37351383 | 37351292 | 37351458 | - | protein_coding |
| ENST00000434056.2 | chr5 | 37351352 | 37351383 | 37351292 | 37351458 | - | protein_coding |
| ENST00000513532.1 | chr5 | 37351352 | 37351383 | 37351292 | 37351458 | - | protein_coding |
| ENST00000337702.4 | chr5 | 40716090 | 40716121 | 40714577 | 40716600 | - | protein_coding |
| ENST00000503936.2 | chr5 | 40716090 | 40716121 | 40715238 | 40716600 | - | processed_transcript |
| ENST00000504251.2 | chr5 | 40716090 | 40716121 | 40716046 | 40716600 | - | processed_transcript |
| ENST00000397128.2 | chr5 | 40762973 | 40763003 | 40759481 | 40763124 | - | protein_coding |
| ENST00000354209.3 | chr5 | 40762973 | 40763003 | 40762872 | 40763124 | - | protein_coding |
| ENST00000381647.2 | chr5 | 41917363 | 41917392 | 41917118 | 41921738 | + | protein_coding |
| ENST00000507110.1 | chr5 | 44815333 | 44815364 | 44815015 | 44815624 | + | protein_coding |
| ENST00000230914.4 | chr5 | 44815333 | 44815364 | 44815321 | 44815665 | + | processed_transcript |
| ENST00000381226.3 | chr5 | 56216085 | 56216115 | 56215429 | 56219412 | - | protein_coding |
| ENST00000381213.3 | chr5 | 56216085 | 56216115 | 56215430 | 56219412 | - | protein_coding |
| ENST00000381199.3 | chr5 | 56216085 | 56216115 | 56215431 | 56219412 | - | protein_coding |
| ENST00000452157.1 | chr5 | 56216085 | 56216115 | 56215431 | 56219412 | - | nonsense_mediated_decay |
| ENST00000381226.3 | chr5 | 56216982 | 56217013 | 56215429 | 56219412 | - | protein_coding |
| ENST00000381213.3 | chr5 | 56216982 | 56217013 | 56215430 | 56219412 | - | protein_coding |
| ENST00000381199.3 | chr5 | 56216982 | 56217013 | 56215431 | 56219412 | - | protein_coding |
| ENST00000452157.1 | chr5 | 56216982 | 56217013 | 56215431 | 56219412 | - | nonsense_mediated_decay |
| ENST00000381226.3 | chr5 | 56218511 | 56218542 | 56215429 | 56219412 | - | protein_coding |
| ENST00000381213.3 | chr5 | 56218511 | 56218542 | 56215430 | 56219412 | - | protein_coding |
| ENST00000381199.3 | chr5 | 56218511 | 56218542 | 56215431 | 56219412 | - | protein_coding |
| ENST00000452157.1 | chr5 | 56218511 | 56218542 | 56215431 | 56219412 | - | nonsense_mediated_decay |
| ENST00000409421.1 | chr5 | 56218511 | 56218542 | 56218411 | 56219412 | - | protein_coding |
| ENST00000454432.2 | chr5 | 56559609 | 56559638 | 56558421 | 56560505 | + | protein_coding |
| ENST00000538707.1 | chr5 | 56559609 | 56559638 | 56558421 | 56560505 | + | protein_coding |
| ENST00000381103.2 | chr5 | 61681777 | 61681808 | 61681311 | 61682210 | + | protein_coding |
| ENST00000399438.3 | chr5 | 64961872 | 64961903 | 64960331 | 64962060 | + | protein_coding |
| ENST00000512009.1 | chr5 | 64961872 | 64961903 | 64960331 | 64961901 | + | retained_intron |
| ENST00000231526.4 | chr5 | 64961872 | 64961903 | 64960331 | 64961896 | + | protein_coding |
| ENST00000545191.1 | chr5 | 64961872 | 64961903 | 64960331 | 64961892 | + | protein_coding |
| ENST00000381007.4 | chr5 | 64961872 | 64961903 | 64961755 | 64966184 | - | protein_coding |
| ENST00000256442.5 | chr5 | 68473069 | 68473097 | 68473066 | 68473176 | + | protein_coding |
| ENST00000506572.1 | chr5 | 68473069 | 68473097 | 68473066 | 68473185 | + | protein_coding |
| ENST00000513102.1 | chr5 | 68473069 | 68473097 | 68472888 | 68473176 | + | retained_intron |
| ENST00000489202.1 | chr5 | 71146778 | 71146778 | 71146740 | 71146942 | + | rRNA_pseudogene |
| ENST00000489202.1 | chr5 | 71146783 | 71146783 | 71146740 | 71146942 | + | rRNA_pseudogene |
| ENST00000489202.1 | chr5 | 71146785 | 71146806 | 71146740 | 71146942 | + | rRNA_pseudogene |
| ENST00000489202.1 | chr5 | 71146897 | 71146925 | 71146740 | 71146942 | + | rRNA_pseudogene |
| ENST00000296755.7 | chr5 | 71482519 | 71482550 | 71482441 | 71482581 | + | protein_coding |
| ENST00000513526.2 | chr5 | 71482519 | 71482550 | 71482441 | 71482581 | + | nonsense_mediated_decay |
| ENST00000511641.2 | chr5 | 71482519 | 71482550 | 71482441 | 71482581 | + | protein_coding |
| ENST00000504492.1 | chr5 | 71482519 | 71482550 | 71482441 | 71482581 | + | protein_coding |
| ENST00000296755.7 | chr5 | 71489805 | 71489843 | 71489693 | 71496194 | + | protein_coding |
| ENST00000513526.2 | chr5 | 71489805 | 71489843 | 71489693 | 71491249 | + | nonsense_mediated_decay |
| ENST00000511641.2 | chr5 | 71489805 | 71489843 | 71489693 | 71491141 | + | protein_coding |
| ENST00000504492.1 | chr5 | 71489805 | 71489843 | 71489693 | 71491267 | + | protein_coding |
| ENST00000296755.7 | chr5 | 71493146 | 71493209 | 71489693 | 71496194 | + | protein_coding |
| ENST00000296755.7 | chr5 | 71502612 | 71502643 | 71500911 | 71505395 | + | protein_coding |
| ENST00000296755.7 | chr5 | 71503271 | 71503302 | 71500911 | 71505395 | + | protein_coding |
| ENST00000296755.7 | chr5 | 71505298 | 71505329 | 71500911 | 71505395 | + | protein_coding |
| ENST00000296755.7 | chr5 | 71505349 | 71505363 | 71500911 | 71505395 | + | protein_coding |
| ENST00000337273.5 | chr5 | 72211948 | 72211978 | 72204536 | 72212560 | + | protein_coding |
| ENST00000296805.3 | chr5 | 74017522 | 74017553 | 74017029 | 74017608 | - | protein_coding |
| ENST00000345239.2 | chr5 | 74017522 | 74017553 | 74017031 | 74017608 | - | protein_coding |
| ENST00000546082.1 | chr5 | 74017522 | 74017553 | 74017034 | 74017608 | - | protein_coding |
| ENST00000509430.1 | chr5 | 74017522 | 74017553 | 74017083 | 74017608 | - | protein_coding |
| ENST00000515125.1 | chr5 | 74017522 | 74017553 | 74017098 | 74017608 | - | processed_transcript |
| ENST00000255198.2 | chr5 | 76370169 | 76370181 | 76367897 | 76373720 | - | protein_coding |
| ENST00000255198.2 | chr5 | 76372593 | 76372624 | 76367897 | 76373720 | - | protein_coding |
| ENST00000282259.5 | chr5 | 78617686 | 78617717 | 78617429 | 78623036 | + | protein_coding |
| ENST00000396137.4 | chr5 | 78617686 | 78617717 | 78617429 | 78623038 | + | protein_coding |
| ENST00000515800.2 | chr5 | 86687243 | 86687274 | 86686617 | 86687736 | + | nonsense_mediated_decay |
| ENST00000456692.2 | chr5 | 86687243 | 86687274 | 86686617 | 86687748 | + | protein_coding |
| ENST00000384838.1 | chr5 | 87962710 | 87962742 | 87962671 | 87962757 | - | miRNA |
| ENST00000520612.1 | chr5 | 93905170 | 93905224 | 93904634 | 93905175 | - | processed_pseudogene |
| ENST00000463081.1 | chr5 | 93905170 | 93905224 | 93905172 | 93905240 | - | Mt_tRNA_pseudogene |
| ENST00000308234.7 | chr5 | 98130956 | 98130987 | 98128789 | 98132198 | + | protein_coding |
| ENST00000491579.1 | chr5 | 99384656 | 99384687 | 99384620 | 99384688 | - | Mt_tRNA_pseudogene |
| ENST00000333274.6 | chr5 | 106716076 | 106716112 | 106712590 | 106717077 | - | protein_coding |
| ENST00000503970.2 | chr5 | 109203708 | 109203739 | 109202547 | 109205326 | + | retained_intron |
| ENST00000515408.1 | chr5 | 112350508 | 112350539 | 112349018 | 112355820 | + | protein_coding |
| ENST00000389063.2 | chr5 | 112350508 | 112350539 | 112349018 | 112356667 | + | protein_coding |
| ENST00000509365.1 | chr5 | 121007144 | 121007172 | 121007068 | 121007567 | - | processed_pseudogene |
| ENST00000304043.5 | chr5 | 130498322 | 130498352 | 130498265 | 130498369 | - | protein_coding |
| ENST00000511475.1 | chr5 | 130498322 | 130498352 | 130498265 | 130498369 | - | nonsense_mediated_decay |
| ENST00000508495.1 | chr5 | 130498322 | 130498352 | 130498265 | 130498369 | - | nonsense_mediated_decay |
| ENST00000513345.1 | chr5 | 130498322 | 130498352 | 130498265 | 130498369 | - | nonsense_mediated_decay |
| ENST00000506207.1 | chr5 | 130498322 | 130498352 | 130498265 | 130498369 | - | processed_transcript |
| ENST00000504202.1 | chr5 | 130498322 | 130498352 | 130498265 | 130498369 | - | nonsense_mediated_decay |
| ENST00000508488.1 | chr5 | 130498322 | 130498352 | 130498265 | 130498369 | - | protein_coding |
| ENST00000506908.1 | chr5 | 130498322 | 130498352 | 130497936 | 130498369 | - | protein_coding |
| ENST00000513012.1 | chr5 | 130498322 | 130498352 | 130497984 | 130498369 | - | protein_coding |
| ENST00000304858.2 | chr5 | 132425265 | 132425296 | 132425254 | 132425387 | + | protein_coding |
| ENST00000537974.1 | chr5 | 132425265 | 132425296 | 132425254 | 132425356 | + | protein_coding |
| ENST00000304858.2 | chr5 | 132441785 | 132441831 | 132439925 | 132442141 | + | protein_coding |
| ENST00000481195.1 | chr5 | 133530804 | 133530835 | 133530025 | 133533535 | - | protein_coding |
| ENST00000481195.1 | chr5 | 133532918 | 133532942 | 133530025 | 133533535 | - | protein_coding |
| ENST00000398844.2 | chr5 | 134061920 | 134061951 | 134060670 | 134063513 | + | protein_coding |
| ENST00000537371.1 | chr5 | 134132861 | 134132892 | 134132851 | 134132882 | + | protein_coding |
| ENST00000473313.1 | chr5 | 134262316 | 134262347 | 134262280 | 134262348 | - | Mt_tRNA_pseudogene |
| ENST00000498999.2 | chr5 | 134262846 | 134262895 | 134262350 | 134263726 | - | processed_pseudogene |
| ENST00000498999.2 | chr5 | 134263423 | 134263452 | 134262350 | 134263726 | - | processed_pseudogene |
| ENST00000454092.1 | chr5 | 134263853 | 134263873 | 134263720 | 134264016 | - | processed_pseudogene |
| ENST00000491516.1 | chr5 | 134264017 | 134264048 | 134264017 | 134264081 | - | Mt_tRNA_pseudogene |
| ENST00000337225.5 | chr5 | 134907433 | 134907474 | 134906373 | 134907558 | - | protein_coding |
| ENST00000512158.1 | chr5 | 134907433 | 134907474 | 134906376 | 134907558 | - | protein_coding |
| ENST00000365160.1 | chr5 | 135416256 | 135416287 | 135416186 | 135416286 | - | misc_RNA |
| ENST00000514641.1 | chr5 | 135510147 | 135510178 | 135510065 | 135510321 | + | polymorphic_pseudogene |
| ENST00000545279.1 | chr5 | 135510147 | 135510178 | 135510065 | 135510321 | + | protein_coding |
| ENST00000545620.1 | chr5 | 135510147 | 135510178 | 135510065 | 135510321 | + | protein_coding |
| ENST00000514777.1 | chr5 | 135510147 | 135510178 | 135510065 | 135510321 | + | processed_transcript |
| ENST00000513418.1 | chr5 | 135510147 | 135510178 | 135510158 | 135510321 | + | nonsense_mediated_decay |
| ENST00000514641.1 | chr5 | 135513999 | 135514030 | 135513026 | 135517935 | + | polymorphic_pseudogene |
| ENST00000545279.1 | chr5 | 135513999 | 135514030 | 135513088 | 135518422 | + | protein_coding |
| ENST00000545620.1 | chr5 | 135513999 | 135514030 | 135513088 | 135518422 | + | protein_coding |
| ENST00000314940.4 | chr5 | 137087837 | 137087862 | 137087075 | 137090039 | - | protein_coding |
| ENST00000505625.1 | chr5 | 138647440 | 138647474 | 138645732 | 138650425 | + | retained_intron |
| ENST00000363120.1 | chr5 | 140090859 | 140090892 | 140090860 | 140090958 | + | misc_RNA |
| ENST00000363120.1 | chr5 | 140090927 | 140090962 | 140090860 | 140090958 | + | misc_RNA |
| ENST00000365241.1 | chr5 | 140098510 | 140098564 | 140098510 | 140098598 | + | misc_RNA |
| ENST00000365241.1 | chr5 | 140098566 | 140098599 | 140098510 | 140098598 | + | misc_RNA |
| ENST00000313368.5 | chr5 | 140698080 | 140698094 | 140698057 | 140700330 | - | protein_coding |
| ENST00000305264.3 | chr5 | 141008748 | 141008779 | 141008740 | 141008873 | - | protein_coding |
| ENST00000469550.2 | chr5 | 141008748 | 141008779 | 141008740 | 141008873 | - | retained_intron |
| ENST00000492407.1 | chr5 | 141008748 | 141008779 | 141008740 | 141008825 | - | retained_intron |
| ENST00000523088.1 | chr5 | 141008748 | 141008779 | 141008740 | 141008873 | - | protein_coding |
| ENST00000519474.1 | chr5 | 141008748 | 141008779 | 141008740 | 141008873 | - | nonsense_mediated_decay |
| ENST00000476739.1 | chr5 | 141008748 | 141008779 | 141008740 | 141008791 | - | retained_intron |
| ENST00000495485.1 | chr5 | 141008748 | 141008779 | 141008740 | 141008873 | - | retained_intron |
| ENST00000490808.1 | chr5 | 141008748 | 141008779 | 141008740 | 141009309 | - | retained_intron |
| ENST00000253814.3 | chr5 | 141532313 | 141532344 | 141531296 | 141534008 | + | protein_coding |
| ENST00000503388.1 | chr5 | 141532313 | 141532344 | 141531296 | 141533760 | + | retained_intron |
| ENST00000394434.2 | chr5 | 145537147 | 145537178 | 145536966 | 145537191 | - | protein_coding |
| ENST00000545646.1 | chr5 | 145537147 | 145537178 | 145536966 | 145537191 | - | protein_coding |
| ENST00000510191.1 | chr5 | 145537147 | 145537178 | 145536966 | 145537191 | - | protein_coding |
| ENST00000274562.9 | chr5 | 145537147 | 145537178 | 145536966 | 145537191 | - | protein_coding |
| ENST00000511505.1 | chr5 | 145537147 | 145537178 | 145536960 | 145537191 | - | processed_transcript |
| ENST00000261798.5 | chr5 | 148874686 | 148874717 | 148871760 | 148876423 | - | protein_coding |
| ENST00000401695.3 | chr5 | 149826442 | 149826473 | 149826365 | 149826526 | - | protein_coding |
| ENST00000407193.1 | chr5 | 149826442 | 149826473 | 149826365 | 149826526 | - | protein_coding |
| ENST00000521466.1 | chr5 | 149826442 | 149826473 | 149826365 | 149826526 | - | protein_coding |
| ENST00000518139.1 | chr5 | 149826442 | 149826473 | 149826365 | 149826526 | - | nonsense_mediated_decay |
| ENST00000312037.5 | chr5 | 149826442 | 149826473 | 149826365 | 149826526 | - | protein_coding |
| ENST00000519690.1 | chr5 | 149826442 | 149826473 | 149826365 | 149829290 | - | retained_intron |
| ENST00000519855.1 | chr5 | 149826442 | 149826473 | 149826365 | 149826500 | - | protein_coding |
| ENST00000447998.2 | chr5 | 150089595 | 150089626 | 150088002 | 150090924 | - | protein_coding |
| ENST00000424236.1 | chr5 | 150089595 | 150089626 | 150088310 | 150090924 | - | protein_coding |
| ENST00000394123.3 | chr5 | 151184739 | 151184770 | 151183446 | 151192346 | + | protein_coding |
| ENST00000356245.3 | chr5 | 151184739 | 151184770 | 151183446 | 151184911 | + | protein_coding |
| ENST00000520177.1 | chr5 | 151184739 | 151184770 | 151183446 | 151184910 | + | nonsense_mediated_decay |
| ENST00000274596.6 | chr5 | 151184739 | 151184770 | 151183446 | 151184903 | + | protein_coding |
| ENST00000394123.3 | chr5 | 151184908 | 151184938 | 151183446 | 151192346 | + | protein_coding |
| ENST00000356245.3 | chr5 | 151184908 | 151184938 | 151183446 | 151184911 | + | protein_coding |
| ENST00000520177.1 | chr5 | 151184908 | 151184938 | 151183446 | 151184910 | + | nonsense_mediated_decay |
| ENST00000394123.3 | chr5 | 151186279 | 151186310 | 151183446 | 151192346 | + | protein_coding |
| ENST00000363405.1 | chr5 | 155272447 | 155272478 | 155272447 | 155272538 | + | rRNA |
| ENST00000274542.2 | chr5 | 158603778 | 158603809 | 158603640 | 158603875 | - | protein_coding |
| ENST00000519865.1 | chr5 | 158603778 | 158603809 | 158603640 | 158603875 | - | protein_coding |
| ENST00000424310.2 | chr5 | 158603778 | 158603809 | 158603640 | 158603875 | - | protein_coding |
| ENST00000521606.1 | chr5 | 158603778 | 158603809 | 158603640 | 158603875 | - | protein_coding |
| ENST00000413445.2 | chr5 | 158603778 | 158603809 | 158603640 | 158603875 | - | protein_coding |
| ENST00000518802.1 | chr5 | 158603778 | 158603809 | 158603640 | 158603875 | - | protein_coding |
| ENST00000535312.1 | chr5 | 158603778 | 158603809 | 158603640 | 158603875 | - | protein_coding |
| ENST00000520638.1 | chr5 | 158603778 | 158603809 | 158603640 | 158603875 | - | protein_coding |
| ENST00000393975.3 | chr5 | 159774751 | 159774778 | 159774758 | 159776788 | - | protein_coding |
| ENST00000362165.1 | chr5 | 167987900 | 167987931 | 167987897 | 167987982 | - | miRNA |
| ENST00000517671.1 | chr5 | 170837591 | 170837622 | 170837531 | 170837887 | + | protein_coding |
| ENST00000296930.5 | chr5 | 170837591 | 170837622 | 170837531 | 170838141 | + | protein_coding |
| ENST00000351986.6 | chr5 | 170837591 | 170837622 | 170837531 | 170837888 | + | protein_coding |
| ENST00000524204.1 | chr5 | 170837591 | 170837622 | 170837531 | 170837888 | + | retained_intron |
| ENST00000329198.4 | chr5 | 172662354 | 172662394 | 172661753 | 172662360 | - | protein_coding |
| ENST00000298569.4 | chr5 | 175773629 | 175773660 | 175773064 | 175774811 | - | protein_coding |
| ENST00000393728.2 | chr5 | 175773629 | 175773660 | 175773065 | 175774811 | - | processed_transcript |
| ENST00000393725.2 | chr5 | 175773629 | 175773660 | 175773065 | 175774811 | - | protein_coding |
| ENST00000261942.6 | chr5 | 175923578 | 175923609 | 175923487 | 175923664 | + | protein_coding |
| ENST00000504983.1 | chr5 | 175923578 | 175923609 | 175923487 | 175923664 | + | retained_intron |
| ENST00000513627.1 | chr5 | 175923578 | 175923609 | 175923487 | 175923987 | + | retained_intron |
| ENST00000261942.6 | chr5 | 175935985 | 175936016 | 175933769 | 175937075 | + | protein_coding |
| ENST00000274811.4 | chr5 | 175953998 | 175954029 | 175953698 | 175956091 | - | protein_coding |
| ENST00000423571.2 | chr5 | 176734225 | 176734251 | 176733818 | 176734965 | - | protein_coding |
| ENST00000439742.2 | chr5 | 176734225 | 176734251 | 176734206 | 176734704 | - | protein_coding |
| ENST00000503782.1 | chr5 | 176734225 | 176734251 | 176734209 | 176734704 | - | retained_intron |
| ENST00000303165.5 | chr5 | 176734225 | 176734251 | 176734215 | 176734704 | - | protein_coding |
| ENST00000261951.4 | chr5 | 180004885 | 180004916 | 180000988 | 180005353 | + | protein_coding |
| ENST00000393356.1 | chr5 | 180004885 | 180004916 | 180000988 | 180005405 | + | protein_coding |
| ENST00000502844.1 | chr5 | 180670316 | 180670344 | 180669554 | 180670370 | - | protein_coding |
| ENST00000513027.1 | chr5 | 180670316 | 180670344 | 180670286 | 180670405 | - | protein_coding |
| ENST00000503170.1 | chr5 | 180670316 | 180670344 | 180670254 | 180670405 | - | processed_transcript |
| ENST00000514318.1 | chr5 | 180670316 | 180670344 | 180670042 | 180670405 | - | processed_transcript |
| ENST00000511473.1 | chr5 | 180670709 | 180670740 | 180670680 | 180670916 | - | nonsense_mediated_decay |
| ENST00000504325.1 | chr5 | 180670709 | 180670740 | 180670692 | 180670889 | - | nonsense_mediated_decay |
| ENST00000511566.1 | chr5 | 180670709 | 180670740 | 180670692 | 180670880 | - | protein_coding |
| ENST00000508682.1 | chr5 | 180670709 | 180670740 | 180670692 | 180670902 | - | nonsense_mediated_decay |
| ENST00000376817.4 | chr5 | 180670709 | 180670740 | 180670692 | 180670903 | - | protein_coding |
| ENST00000512805.1 | chr5 | 180670709 | 180670740 | 180670692 | 180671209 | - | protein_coding |
| ENST00000504726.1 | chr5 | 180670709 | 180670740 | 180670692 | 180670891 | - | protein_coding |
| ENST00000511900.1 | chr5 | 180670709 | 180670740 | 180670692 | 180670880 | - | protein_coding |
| ENST00000513060.1 | chr5 | 180670709 | 180670740 | 180670692 | 180670947 | - | retained_intron |
| ENST00000512968.1 | chr5 | 180670709 | 180670740 | 180670692 | 180670893 | - | protein_coding |
| ENST00000514183.1 | chr5 | 180670709 | 180670740 | 180670692 | 180670742 | - | retained_intron |
| ENST00000502548.1 | chr5 | 180670709 | 180670740 | 180670680 | 180670904 | - | retained_intron |
| ENST00000503494.1 | chr5 | 180670709 | 180670740 | 180670692 | 180670881 | - | nonsense_mediated_decay |
| ENST00000506312.1 | chr5 | 180670709 | 180670740 | 180670692 | 180670902 | - | nonsense_mediated_decay |
| ENST00000503081.1 | chr5 | 180670709 | 180670740 | 180670692 | 180670906 | - | protein_coding |
| ENST00000515417.1 | chr5 | 180670709 | 180670740 | 180670692 | 180670872 | - | retained_intron |
| ENST00000513027.1 | chr5 | 180670709 | 180670740 | 180670692 | 180670904 | - | protein_coding |
| ENST00000456394.2 | chr5 | 180670709 | 180670740 | 180670692 | 180670914 | - | protein_coding |
| ENST00000505461.1 | chr5 | 180670709 | 180670740 | 180670692 | 180670914 | - | processed_transcript |
| ENST00000507261.1 | chr5 | 180670709 | 180670740 | 180670692 | 180670902 | - | retained_intron |
| ENST00000508963.1 | chr5 | 180670709 | 180670740 | 180670692 | 180670903 | - | retained_intron |
| ENST00000503170.1 | chr5 | 180670709 | 180670740 | 180670692 | 180670904 | - | processed_transcript |
| ENST00000514318.1 | chr5 | 180670709 | 180670740 | 180670692 | 180670908 | - | processed_transcript |
| ENST00000380874.2 | chr6 | 1613830 | 1613860 | 1610681 | 1614127 | + | protein_coding |
| ENST00000380773.4 | chr6 | 2770403 | 2770434 | 2770354 | 2770595 | + | protein_coding |
| ENST00000380771.4 | chr6 | 2770403 | 2770434 | 2770429 | 2770595 | + | protein_coding |
| ENST00000380769.3 | chr6 | 2770403 | 2770434 | 2770354 | 2770595 | + | protein_coding |
| ENST00000380764.1 | chr6 | 2770403 | 2770434 | 2770354 | 2770595 | + | protein_coding |
| ENST00000364315.1 | chr6 | 4428197 | 4428228 | 4428197 | 4428315 | + | rRNA |
| ENST00000364315.1 | chr6 | 4428242 | 4428257 | 4428197 | 4428315 | + | rRNA |
| ENST00000364315.1 | chr6 | 4428264 | 4428324 | 4428197 | 4428315 | + | rRNA |
| ENST00000397457.2 | chr6 | 8015521 | 8015552 | 8013800 | 8016061 | - | protein_coding |
| ENST00000244777.2 | chr6 | 8015521 | 8015552 | 8014211 | 8016061 | - | nonsense_mediated_decay |
| ENST00000543936.1 | chr6 | 8015521 | 8015552 | 8015296 | 8016061 | - | protein_coding |
| ENST00000475998.1 | chr6 | 8015521 | 8015552 | 8015318 | 8016061 | - | processed_transcript |
| ENST00000379433.5 | chr6 | 11199872 | 11199910 | 11198541 | 11201279 | - | protein_coding |
| ENST00000379388.2 | chr6 | 12164278 | 12164309 | 12163516 | 12165232 | + | protein_coding |
| ENST00000399469.2 | chr6 | 12164278 | 12164309 | 12163516 | 12164808 | + | nonsense_mediated_decay |
| ENST00000541134.1 | chr6 | 12164278 | 12164309 | 12163516 | 12164808 | + | protein_coding |
| ENST00000542327.1 | chr6 | 12164278 | 12164309 | 12163516 | 12165231 | + | protein_coding |
| ENST00000391491.2 | chr6 | 12514879 | 12514910 | 12514342 | 12514956 | + | processed_pseudogene |
| ENST00000450992.2 | chr6 | 12514879 | 12514910 | 12514804 | 12515006 | + | pseudogene |
| ENST00000383975.1 | chr6 | 13214405 | 13214434 | 13214288 | 13214451 | + | snRNA |
| ENST00000341776.2 | chr6 | 15521742 | 15521773 | 15520300 | 15522252 | + | protein_coding |
| ENST00000349606.4 | chr6 | 16148094 | 16148111 | 16146893 | 16148479 | + | protein_coding |
| ENST00000349606.4 | chr6 | 16148309 | 16148340 | 16146893 | 16148479 | + | protein_coding |
| ENST00000259963.3 | chr6 | 17609179 | 17609210 | 17608426 | 17611950 | + | protein_coding |
| ENST00000397239.3 | chr6 | 18225025 | 18225056 | 18224099 | 18225961 | - | protein_coding |
| ENST00000397239.3 | chr6 | 18225566 | 18225597 | 18224099 | 18225961 | - | protein_coding |
| ENST00000467883.1 | chr6 | 20422014 | 20422045 | 20421880 | 20422118 | - | scRNA_pseudogene |
| ENST00000378646.3 | chr6 | 20493638 | 20493670 | 20490587 | 20493945 | + | protein_coding |
| ENST00000346618.3 | chr6 | 20493638 | 20493670 | 20490399 | 20493941 | + | protein_coding |
| ENST00000535432.1 | chr6 | 20493638 | 20493670 | 20490399 | 20493789 | + | protein_coding |
| ENST00000244745.1 | chr6 | 21597263 | 21597265 | 21593972 | 21598847 | + | protein_coding |
| ENST00000244745.1 | chr6 | 21597485 | 21597516 | 21593972 | 21598847 | + | protein_coding |
| ENST00000244745.1 | chr6 | 21597644 | 21597675 | 21593972 | 21598847 | + | protein_coding |
| ENST00000244745.1 | chr6 | 21597784 | 21597815 | 21593972 | 21598847 | + | protein_coding |
| ENST00000244745.1 | chr6 | 21598039 | 21598070 | 21593972 | 21598847 | + | protein_coding |
| ENST00000244745.1 | chr6 | 21598272 | 21598303 | 21593972 | 21598847 | + | protein_coding |
| ENST00000244745.1 | chr6 | 21598635 | 21598666 | 21593972 | 21598847 | + | protein_coding |
| ENST00000244661.2 | chr6 | 26031832 | 26031869 | 26031817 | 26032288 | - | protein_coding |
| ENST00000343677.2 | chr6 | 26056403 | 26056434 | 26055968 | 26056699 | - | protein_coding |
| ENST00000314332.4 | chr6 | 26123855 | 26123886 | 26123743 | 26124154 | - | protein_coding |
| ENST00000396984.1 | chr6 | 26123855 | 26123886 | 26123584 | 26124154 | - | protein_coding |
| ENST00000304218.3 | chr6 | 26156573 | 26156600 | 26156559 | 26157343 | + | protein_coding |
| ENST00000377777.4 | chr6 | 26158644 | 26158675 | 26158349 | 26158871 | + | protein_coding |
| ENST00000289316.2 | chr6 | 26158644 | 26158675 | 26158374 | 26158787 | + | protein_coding |
| ENST00000377777.4 | chr6 | 26158793 | 26158812 | 26158349 | 26158871 | + | protein_coding |
| ENST00000356530.3 | chr6 | 26184270 | 26184301 | 26183958 | 26184454 | + | protein_coding |
| ENST00000360441.4 | chr6 | 26205149 | 26205180 | 26204858 | 26206266 | + | protein_coding |
| ENST00000377600.2 | chr6 | 26458862 | 26458893 | 26458835 | 26458946 | + | nonsense_mediated_decay |
| ENST00000312541.5 | chr6 | 26458862 | 26458893 | 26458835 | 26458946 | + | protein_coding |
| ENST00000429381.1 | chr6 | 26458862 | 26458893 | 26458835 | 26458946 | + | protein_coding |
| ENST00000265424.4 | chr6 | 26458862 | 26458893 | 26458835 | 26458946 | + | protein_coding |
| ENST00000469185.1 | chr6 | 26458862 | 26458893 | 26458835 | 26458946 | + | protein_coding |
| ENST00000541790.1 | chr6 | 27100252 | 27100283 | 27100146 | 27100541 | - | protein_coding |
| ENST00000339812.2 | chr6 | 27100252 | 27100283 | 27100149 | 27100529 | - | protein_coding |
| ENST00000396891.4 | chr6 | 27114300 | 27114331 | 27114188 | 27114619 | - | protein_coding |
| ENST00000356950.1 | chr6 | 27114300 | 27114331 | 27114197 | 27114577 | - | protein_coding |
| ENST00000396891.4 | chr6 | 27114565 | 27114585 | 27114188 | 27114619 | - | protein_coding |
| ENST00000356950.1 | chr6 | 27114565 | 27114585 | 27114197 | 27114577 | - | protein_coding |
| ENST00000396891.4 | chr6 | 27114590 | 27114603 | 27114188 | 27114619 | - | protein_coding |
| ENST00000377459.1 | chr6 | 27115142 | 27115167 | 27114861 | 27115317 | + | protein_coding |
| ENST00000458896.1 | chr6 | 27599187 | 27599254 | 27599188 | 27599311 | + | miRNA |
| ENST00000458896.1 | chr6 | 27599274 | 27599275 | 27599188 | 27599311 | + | miRNA |
| ENST00000458896.1 | chr6 | 27599280 | 27599312 | 27599188 | 27599311 | + | miRNA |
| ENST00000377401.2 | chr6 | 27775407 | 27775438 | 27775257 | 27775709 | - | protein_coding |
| ENST00000449538.2 | chr6 | 27806686 | 27806717 | 27806323 | 27806843 | + | protein_coding |
| ENST00000396980.3 | chr6 | 27806686 | 27806717 | 27806440 | 27806888 | + | protein_coding |
| ENST00000303806.4 | chr6 | 27861487 | 27861518 | 27861203 | 27861669 | + | protein_coding |
| ENST00000524745.1 | chr6 | 28557156 | 28557187 | 28557137 | 28557480 | - | processed_transcript |
| ENST00000481474.1 | chr6 | 28871990 | 28872021 | 28870779 | 28876617 | - | retained_intron |
| ENST00000414543.1 | chr6 | 28871990 | 28872021 | 28870779 | 28872442 | - | protein_coding |
| ENST00000377199.3 | chr6 | 28871990 | 28872021 | 28870779 | 28872442 | - | protein_coding |
| ENST00000377194.3 | chr6 | 28871990 | 28872021 | 28870779 | 28872050 | - | protein_coding |
| ENST00000376185.1 | chr6 | 31504310 | 31504341 | 31504277 | 31504460 | - | nonsense_mediated_decay |
| ENST00000376177.2 | chr6 | 31504310 | 31504341 | 31504277 | 31504460 | - | protein_coding |
| ENST00000462256.1 | chr6 | 31504310 | 31504341 | 31504277 | 31504460 | - | retained_intron |
| ENST00000458640.1 | chr6 | 31504310 | 31504341 | 31504277 | 31504460 | - | protein_coding |
| ENST00000396172.1 | chr6 | 31504310 | 31504341 | 31504277 | 31504460 | - | protein_coding |
| ENST00000417556.2 | chr6 | 31504310 | 31504341 | 31504277 | 31504460 | - | protein_coding |
| ENST00000415382.2 | chr6 | 31504310 | 31504341 | 31504277 | 31504460 | - | protein_coding |
| ENST00000482195.1 | chr6 | 31504310 | 31504341 | 31504277 | 31504460 | - | retained_intron |
| ENST00000431908.1 | chr6 | 31504310 | 31504341 | 31504277 | 31504460 | - | protein_coding |
| ENST00000481456.1 | chr6 | 31504310 | 31504341 | 31504277 | 31504460 | - | retained_intron |
| ENST00000427214.1 | chr6 | 31504310 | 31504341 | 31504277 | 31504460 | - | protein_coding |
| ENST00000453105.2 | chr6 | 31504310 | 31504341 | 31504277 | 31504460 | - | protein_coding |
| ENST00000428098.1 | chr6 | 31504310 | 31504341 | 31504277 | 31504460 | - | protein_coding |
| ENST00000419338.1 | chr6 | 31504310 | 31504341 | 31504277 | 31504460 | - | protein_coding |
| ENST00000456662.1 | chr6 | 31504310 | 31504341 | 31504277 | 31504460 | - | protein_coding |
| ENST00000449074.2 | chr6 | 31504310 | 31504341 | 31504332 | 31504460 | - | protein_coding |
| ENST00000374685.4 | chr6 | 33165621 | 33165652 | 33165539 | 33165718 | - | protein_coding |
| ENST00000374680.3 | chr6 | 33165621 | 33165652 | 33165539 | 33165718 | - | protein_coding |
| ENST00000483281.1 | chr6 | 33165621 | 33165652 | 33165539 | 33165718 | - | nonsense_mediated_decay |
| ENST00000544186.1 | chr6 | 33165621 | 33165652 | 33165539 | 33165718 | - | protein_coding |
| ENST00000481441.1 | chr6 | 33165621 | 33165652 | 33165539 | 33165718 | - | retained_intron |
| ENST00000413614.2 | chr6 | 33165621 | 33165652 | 33165539 | 33165718 | - | protein_coding |
| ENST00000365571.1 | chr6 | 33167378 | 33167409 | 33167378 | 33167472 | + | misc_RNA |
| ENST00000244520.5 | chr6 | 34735708 | 34735739 | 34735685 | 34735774 | + | protein_coding |
| ENST00000374018.1 | chr6 | 34735708 | 34735739 | 34735685 | 34735774 | + | protein_coding |
| ENST00000374017.3 | chr6 | 34735708 | 34735739 | 34735685 | 34735774 | + | protein_coding |
| ENST00000474635.1 | chr6 | 34735708 | 34735739 | 34735685 | 34735774 | + | processed_transcript |
| ENST00000229812.7 | chr6 | 36462172 | 36462203 | 36461669 | 36463691 | - | protein_coding |
| ENST00000229812.7 | chr6 | 36462888 | 36462919 | 36461669 | 36463691 | - | protein_coding |
| ENST00000373699.5 | chr6 | 36823332 | 36823363 | 36822603 | 36823809 | - | protein_coding |
| ENST00000483552.1 | chr6 | 36823332 | 36823363 | 36822603 | 36823809 | - | processed_transcript |
| ENST00000423336.1 | chr6 | 37012902 | 37012933 | 37012607 | 37013177 | + | protein_coding |
| ENST00000314100.6 | chr6 | 38139624 | 38139660 | 38136227 | 38142958 | - | protein_coding |
| ENST00000481247.1 | chr6 | 38139624 | 38139660 | 38136227 | 38142958 | - | protein_coding |
| ENST00000470973.1 | chr6 | 38644719 | 38644750 | 38643719 | 38645159 | - | processed_transcript |
| ENST00000373365.3 | chr6 | 38644719 | 38644750 | 38643719 | 38645159 | - | protein_coding |
| ENST00000483998.1 | chr6 | 42852445 | 42852476 | 42852351 | 42852488 | + | processed_transcript |
| ENST00000493763.1 | chr6 | 42852445 | 42852476 | 42852351 | 42852488 | + | protein_coding |
| ENST00000397415.3 | chr6 | 42852445 | 42852476 | 42852351 | 42852488 | + | processed_transcript |
| ENST00000304734.5 | chr6 | 42852445 | 42852476 | 42852351 | 42852488 | + | protein_coding |
| ENST00000424341.2 | chr6 | 42852445 | 42852476 | 42852351 | 42852488 | + | protein_coding |
| ENST00000462348.1 | chr6 | 42852445 | 42852476 | 42852351 | 42852488 | + | processed_transcript |
| ENST00000326974.4 | chr6 | 42985320 | 42985351 | 42985257 | 42985433 | + | protein_coding |
| ENST00000432243.1 | chr6 | 42985320 | 42985351 | 42985257 | 42985433 | + | protein_coding |
| ENST00000244670.8 | chr6 | 42985320 | 42985351 | 42985257 | 42985433 | + | protein_coding |
| ENST00000394096.4 | chr6 | 42985320 | 42985351 | 42985257 | 42985433 | + | protein_coding |
| ENST00000426116.2 | chr6 | 42985320 | 42985351 | 42985257 | 42985433 | + | protein_coding |
| ENST00000372067.3 | chr6 | 43752555 | 43752591 | 43752278 | 43754224 | + | protein_coding |
| ENST00000425836.2 | chr6 | 43752555 | 43752591 | 43752278 | 43752686 | + | protein_coding |
| ENST00000372064.4 | chr6 | 43752555 | 43752591 | 43752278 | 43754224 | + | protein_coding |
| ENST00000372077.4 | chr6 | 43752555 | 43752591 | 43752278 | 43754224 | + | protein_coding |
| ENST00000520948.1 | chr6 | 43752555 | 43752591 | 43752278 | 43752686 | + | protein_coding |
| ENST00000523950.1 | chr6 | 43752555 | 43752591 | 43752278 | 43752639 | + | protein_coding |
| ENST00000230480.6 | chr6 | 43752555 | 43752591 | 43752278 | 43752672 | + | protein_coding |
| ENST00000480614.1 | chr6 | 43752555 | 43752591 | 43746626 | 43754221 | + | retained_intron |
| ENST00000497139.1 | chr6 | 43752555 | 43752591 | 43752278 | 43754176 | + | retained_intron |
| ENST00000441736.1 | chr6 | 44216442 | 44216473 | 44216367 | 44216513 | + | protein_coding |
| ENST00000371646.5 | chr6 | 44216442 | 44216473 | 44216367 | 44216513 | + | protein_coding |
| ENST00000353801.3 | chr6 | 44216442 | 44216473 | 44216354 | 44216513 | + | protein_coding |
| ENST00000371565.5 | chr6 | 44216442 | 44216473 | 44216367 | 44216513 | + | protein_coding |
| ENST00000371562.5 | chr6 | 44216442 | 44216473 | 44216367 | 44216513 | + | protein_coding |
| ENST00000371556.5 | chr6 | 44216442 | 44216473 | 44216367 | 44216513 | + | protein_coding |
| ENST00000371554.1 | chr6 | 44216442 | 44216473 | 44216367 | 44216513 | + | protein_coding |
| ENST00000371646.5 | chr6 | 44218810 | 44218841 | 44218785 | 44218950 | + | protein_coding |
| ENST00000353801.3 | chr6 | 44218810 | 44218841 | 44218785 | 44218950 | + | protein_coding |
| ENST00000371565.5 | chr6 | 44218810 | 44218841 | 44218785 | 44218950 | + | protein_coding |
| ENST00000371562.5 | chr6 | 44218810 | 44218841 | 44218785 | 44218950 | + | protein_coding |
| ENST00000371556.5 | chr6 | 44218810 | 44218841 | 44218785 | 44218950 | + | protein_coding |
| ENST00000371554.1 | chr6 | 44218810 | 44218841 | 44218785 | 44218950 | + | protein_coding |
| ENST00000371477.3 | chr6 | 44416729 | 44416760 | 44414344 | 44418163 | + | protein_coding |
| ENST00000371383.1 | chr6 | 46128722 | 46128753 | 46128152 | 46129490 | - | protein_coding |
| ENST00000230565.3 | chr6 | 46128722 | 46128753 | 46128152 | 46129490 | - | protein_coding |
| ENST00000442253.2 | chr6 | 52272479 | 52272508 | 52267960 | 52272575 | + | protein_coding |
| ENST00000360726.3 | chr6 | 52272479 | 52272508 | 52267960 | 52272489 | + | protein_coding |
| ENST00000182527.3 | chr6 | 52363048 | 52363078 | 52362200 | 52368068 | - | protein_coding |
| ENST00000370963.4 | chr6 | 52849320 | 52849351 | 52849262 | 52849403 | - | protein_coding |
| ENST00000477599.1 | chr6 | 52849320 | 52849351 | 52849262 | 52849403 | - | processed_transcript |
| ENST00000486559.1 | chr6 | 52849320 | 52849351 | 52849262 | 52849403 | - | processed_transcript |
| ENST00000541324.1 | chr6 | 52849320 | 52849351 | 52849262 | 52849403 | - | protein_coding |
| ENST00000370960.1 | chr6 | 52849320 | 52849351 | 52849262 | 52849403 | - | protein_coding |
| ENST00000370959.1 | chr6 | 52849320 | 52849351 | 52849262 | 52849403 | - | protein_coding |
| ENST00000457564.1 | chr6 | 52849320 | 52849351 | 52849262 | 52849403 | - | protein_coding |
| ENST00000370577.3 | chr6 | 70410687 | 70410719 | 70410657 | 70410761 | - | protein_coding |
| ENST00000370570.1 | chr6 | 70410687 | 70410719 | 70410657 | 70410761 | - | protein_coding |
| ENST00000472827.1 | chr6 | 70410687 | 70410719 | 70410657 | 70410694 | - | nonsense_mediated_decay |
| ENST00000362224.1 | chr6 | 72086698 | 72086728 | 72086663 | 72086734 | - | miRNA |
| ENST00000385092.1 | chr6 | 72113288 | 72113319 | 72113254 | 72113324 | - | miRNA |
| ENST00000316292.9 | chr6 | 74229063 | 74229094 | 74229060 | 74229239 | - | protein_coding |
| ENST00000358190.6 | chr6 | 74229063 | 74229094 | 74229060 | 74229239 | - | protein_coding |
| ENST00000491404.1 | chr6 | 74229063 | 74229094 | 74229060 | 74229140 | - | processed_transcript |
| ENST00000309268.6 | chr6 | 74229063 | 74229094 | 74229060 | 74229239 | - | protein_coding |
| ENST00000490569.1 | chr6 | 74229063 | 74229094 | 74229060 | 74229239 | - | retained_intron |
| ENST00000331523.2 | chr6 | 74229063 | 74229094 | 74229060 | 74229239 | - | protein_coding |
| ENST00000391977.3 | chr6 | 74229063 | 74229094 | 74229060 | 74229239 | - | protein_coding |
| ENST00000495333.1 | chr6 | 74229063 | 74229094 | 74228655 | 74229239 | - | retained_intron |
| ENST00000488500.1 | chr6 | 74229063 | 74229094 | 74229060 | 74229239 | - | retained_intron |
| ENST00000356303.2 | chr6 | 74229063 | 74229094 | 74229060 | 74229239 | - | protein_coding |
| ENST00000455918.1 | chr6 | 74229063 | 74229094 | 74229064 | 74229239 | - | protein_coding |
| ENST00000355238.6 | chr6 | 86320203 | 86320234 | 86318053 | 86322641 | - | protein_coding |
| ENST00000355238.6 | chr6 | 86322436 | 86322467 | 86318053 | 86322641 | - | protein_coding |
| ENST00000355238.6 | chr6 | 86333744 | 86333775 | 86333695 | 86333830 | - | protein_coding |
| ENST00000369622.3 | chr6 | 86333744 | 86333775 | 86333695 | 86333830 | - | protein_coding |
| ENST00000411069.1 | chr6 | 89773292 | 89773317 | 89773219 | 89773409 | - | snRNA |
| ENST00000369239.5 | chr6 | 99846675 | 99846706 | 99845927 | 99849506 | - | protein_coding |
| ENST00000369239.5 | chr6 | 99846842 | 99846873 | 99845927 | 99849506 | - | protein_coding |
| ENST00000482541.2 | chr6 | 100005234 | 100005265 | 100005116 | 100006424 | - | protein_coding |
| ENST00000363444.1 | chr6 | 106902771 | 106902796 | 106902703 | 106902804 | + | misc_RNA |
| ENST00000436639.2 | chr6 | 109308248 | 109308279 | 109307640 | 109308833 | - | protein_coding |
| ENST00000302071.2 | chr6 | 109308248 | 109308279 | 109307645 | 109308833 | - | protein_coding |
| ENST00000356644.7 | chr6 | 109308248 | 109308279 | 109307645 | 109308833 | - | protein_coding |
| ENST00000415861.2 | chr6 | 109688426 | 109688441 | 109687717 | 109690220 | - | retained_intron |
| ENST00000413644.2 | chr6 | 109688426 | 109688441 | 109687717 | 109689528 | - | protein_coding |
| ENST00000368961.5 | chr6 | 109688426 | 109688441 | 109687719 | 109690220 | - | protein_coding |
| ENST00000324953.5 | chr6 | 109688426 | 109688441 | 109687721 | 109690220 | - | protein_coding |
| ENST00000310786.4 | chr6 | 109688426 | 109688441 | 109687721 | 109690220 | - | protein_coding |
| ENST00000275080.7 | chr6 | 109688426 | 109688441 | 109687721 | 109690220 | - | protein_coding |
| ENST00000415861.2 | chr6 | 109689700 | 109689731 | 109687717 | 109690220 | - | retained_intron |
| ENST00000368961.5 | chr6 | 109689700 | 109689731 | 109687719 | 109690220 | - | protein_coding |
| ENST00000324953.5 | chr6 | 109689700 | 109689731 | 109687721 | 109690220 | - | protein_coding |
| ENST00000310786.4 | chr6 | 109689700 | 109689731 | 109687721 | 109690220 | - | protein_coding |
| ENST00000275080.7 | chr6 | 109689700 | 109689731 | 109687721 | 109690220 | - | protein_coding |
| ENST00000499860.2 | chr6 | 109689700 | 109689731 | 109688482 | 109690220 | - | retained_intron |
| ENST00000504373.1 | chr6 | 109689700 | 109689731 | 109689268 | 109690220 | - | protein_coding |
| ENST00000230122.3 | chr6 | 109787438 | 109787469 | 109783797 | 109787777 | - | protein_coding |
| ENST00000516899.1 | chr6 | 110074386 | 110074417 | 110074372 | 110074462 | - | miRNA |
| ENST00000516899.1 | chr6 | 110074431 | 110074435 | 110074372 | 110074462 | - | miRNA |
| ENST00000368885.3 | chr6 | 111196114 | 111196145 | 111195973 | 111196418 | + | protein_coding |
| ENST00000368882.3 | chr6 | 111196114 | 111196145 | 111195973 | 111196418 | + | protein_coding |
| ENST00000451850.2 | chr6 | 111196114 | 111196145 | 111195987 | 111196418 | + | protein_coding |
| ENST00000368877.5 | chr6 | 111196114 | 111196145 | 111195988 | 111196418 | + | protein_coding |
| ENST00000422377.1 | chr6 | 111696717 | 111696748 | 111693799 | 111697960 | - | nonsense_mediated_decay |
| ENST00000434009.1 | chr6 | 111696717 | 111696748 | 111693799 | 111697960 | - | nonsense_mediated_decay |
| ENST00000368802.3 | chr6 | 111696717 | 111696748 | 111693799 | 111697960 | - | protein_coding |
| ENST00000368805.1 | chr6 | 111696717 | 111696748 | 111693799 | 111697960 | - | protein_coding |
| ENST00000358835.3 | chr6 | 111696717 | 111696748 | 111693799 | 111697960 | - | protein_coding |
| ENST00000435970.1 | chr6 | 111696717 | 111696748 | 111693799 | 111697960 | - | protein_coding |
| ENST00000543871.1 | chr6 | 111696717 | 111696748 | 111693799 | 111697960 | - | protein_coding |
| ENST00000368635.4 | chr6 | 114181514 | 114181545 | 114180859 | 114184648 | + | protein_coding |
| ENST00000368608.3 | chr6 | 116598240 | 116598271 | 116597741 | 116601066 | - | protein_coding |
| ENST00000368608.3 | chr6 | 116598351 | 116598382 | 116597741 | 116601066 | - | protein_coding |
| ENST00000368608.3 | chr6 | 116599149 | 116599162 | 116597741 | 116601066 | - | protein_coding |
| ENST00000368564.1 | chr6 | 117053883 | 117053914 | 117053299 | 117056807 | + | protein_coding |
| ENST00000360388.4 | chr6 | 118636589 | 118636617 | 118635191 | 118638839 | + | protein_coding |
| ENST00000229595.5 | chr6 | 119228678 | 119228709 | 119228567 | 119230332 | + | protein_coding |
| ENST00000229595.5 | chr6 | 119230008 | 119230031 | 119228567 | 119230332 | + | protein_coding |
| ENST00000492060.1 | chr6 | 120583445 | 120583501 | 120583432 | 120583547 | + | rRNA_pseudogene |
| ENST00000363440.1 | chr6 | 121007810 | 121007841 | 121007767 | 121007888 | + | rRNA |
| ENST00000229633.5 | chr6 | 126301036 | 126301080 | 126298790 | 126301387 | + | protein_coding |
| ENST00000516111.1 | chr6 | 126301036 | 126301080 | 126300958 | 126301080 | - | rRNA |
| ENST00000368314.1 | chr6 | 127607833 | 127607864 | 127607761 | 127609703 | + | protein_coding |
| ENST00000476956.1 | chr6 | 127607833 | 127607864 | 127607761 | 127608036 | + | processed_transcript |
| ENST00000356799.2 | chr6 | 127607833 | 127607864 | 127607761 | 127609712 | + | protein_coding |
| ENST00000477776.1 | chr6 | 127607833 | 127607864 | 127607761 | 127607994 | + | processed_transcript |
| ENST00000309649.3 | chr6 | 127607833 | 127607864 | 127607761 | 127609502 | + | protein_coding |
| ENST00000489534.1 | chr6 | 127607833 | 127607864 | 127607761 | 127608098 | + | processed_transcript |
| ENST00000480444.1 | chr6 | 127607833 | 127607864 | 127607761 | 127608048 | + | processed_transcript |
| ENST00000556132.1 | chr6 | 127764377 | 127764408 | 127759551 | 127765427 | - | protein_coding |
| ENST00000368281.1 | chr6 | 127764377 | 127764408 | 127761488 | 127765427 | - | retained_intron |
| ENST00000465254.2 | chr6 | 127764377 | 127764408 | 127762193 | 127765427 | - | protein_coding |
| ENST00000363664.1 | chr6 | 133138426 | 133138486 | 133138358 | 133138487 | + | snoRNA |
| ENST00000341911.5 | chr6 | 135539994 | 135540025 | 135539002 | 135540305 | + | protein_coding |
| ENST00000367812.2 | chr6 | 135539994 | 135540025 | 135539002 | 135540305 | + | nonsense_mediated_decay |
| ENST00000463282.2 | chr6 | 135539994 | 135540025 | 135539002 | 135540305 | + | nonsense_mediated_decay |
| ENST00000525477.1 | chr6 | 135539994 | 135540025 | 135539002 | 135540305 | + | nonsense_mediated_decay |
| ENST00000533837.1 | chr6 | 135539994 | 135540025 | 135539002 | 135540305 | + | nonsense_mediated_decay |
| ENST00000339290.5 | chr6 | 135539994 | 135540025 | 135539002 | 135540305 | + | nonsense_mediated_decay |
| ENST00000442647.2 | chr6 | 135539994 | 135540025 | 135539002 | 135540309 | + | protein_coding |
| ENST00000316528.8 | chr6 | 135539994 | 135540025 | 135539002 | 135540309 | + | protein_coding |
| ENST00000237302.4 | chr6 | 135539994 | 135540025 | 135538999 | 135540305 | + | protein_coding |
| ENST00000367814.4 | chr6 | 135539994 | 135540025 | 135539002 | 135540311 | + | protein_coding |
| ENST00000525369.1 | chr6 | 135539994 | 135540025 | 135539002 | 135540310 | + | protein_coding |
| ENST00000438901.2 | chr6 | 135539994 | 135540025 | 135539002 | 135540305 | + | nonsense_mediated_decay |
| ENST00000525514.1 | chr6 | 135539994 | 135540025 | 135539002 | 135540305 | + | nonsense_mediated_decay |
| ENST00000531519.1 | chr6 | 135539994 | 135540025 | 135539002 | 135540305 | + | nonsense_mediated_decay |
| ENST00000526320.1 | chr6 | 135539994 | 135540025 | 135539002 | 135540305 | + | nonsense_mediated_decay |
| ENST00000526889.1 | chr6 | 135539994 | 135540025 | 135539002 | 135540305 | + | nonsense_mediated_decay |
| ENST00000533808.1 | chr6 | 135539994 | 135540025 | 135539002 | 135540305 | + | nonsense_mediated_decay |
| ENST00000529586.1 | chr6 | 135539994 | 135540025 | 135539002 | 135540305 | + | nonsense_mediated_decay |
| ENST00000531845.1 | chr6 | 135539994 | 135540025 | 135539002 | 135540309 | + | processed_transcript |
| ENST00000531224.1 | chr6 | 136599227 | 136599258 | 136599003 | 136599914 | - | protein_coding |
| ENST00000353331.4 | chr6 | 136599227 | 136599258 | 136599003 | 136599908 | - | protein_coding |
| ENST00000527536.1 | chr6 | 136599227 | 136599258 | 136599003 | 136599914 | - | protein_coding |
| ENST00000527613.1 | chr6 | 136599227 | 136599258 | 136599003 | 136599914 | - | nonsense_mediated_decay |
| ENST00000530767.1 | chr6 | 136599227 | 136599258 | 136599003 | 136599914 | - | protein_coding |
| ENST00000527759.1 | chr6 | 136599227 | 136599258 | 136599003 | 136599908 | - | protein_coding |
| ENST00000534269.1 | chr6 | 136599227 | 136599258 | 136599003 | 136599908 | - | nonsense_mediated_decay |
| ENST00000532384.1 | chr6 | 136599227 | 136599258 | 136599003 | 136599914 | - | nonsense_mediated_decay |
| ENST00000530429.1 | chr6 | 136599227 | 136599258 | 136599003 | 136599908 | - | nonsense_mediated_decay |
| ENST00000392348.2 | chr6 | 136599227 | 136599258 | 136599003 | 136599908 | - | protein_coding |
| ENST00000529826.1 | chr6 | 136599227 | 136599258 | 136599003 | 136599914 | - | protein_coding |
| ENST00000533621.1 | chr6 | 136599227 | 136599258 | 136599003 | 136599241 | - | nonsense_mediated_decay |
| ENST00000367739.4 | chr6 | 137518816 | 137518847 | 137518621 | 137519776 | - | protein_coding |
| ENST00000418947.2 | chr6 | 137518816 | 137518847 | 137518622 | 137519776 | - | protein_coding |
| ENST00000421351.2 | chr6 | 138412613 | 138412644 | 138411923 | 138413405 | - | protein_coding |
| ENST00000265603.8 | chr6 | 138412613 | 138412644 | 138411926 | 138413405 | - | protein_coding |
| ENST00000367467.3 | chr6 | 148872879 | 148872910 | 148869431 | 148873186 | + | protein_coding |
| ENST00000537769.1 | chr6 | 148872879 | 148872910 | 148872561 | 148873184 | + | protein_coding |
| ENST00000543571.1 | chr6 | 150023114 | 150023145 | 150022915 | 150023402 | - | protein_coding |
| ENST00000253339.5 | chr6 | 150023114 | 150023145 | 150022915 | 150023490 | - | protein_coding |
| ENST00000441107.1 | chr6 | 150023114 | 150023145 | 150022915 | 150023402 | - | nonsense_mediated_decay |
| ENST00000392273.3 | chr6 | 150023114 | 150023145 | 150022915 | 150023402 | - | protein_coding |
| ENST00000542720.1 | chr6 | 150023114 | 150023145 | 150022915 | 150023402 | - | nonsense_mediated_decay |
| ENST00000464882.1 | chr6 | 150665171 | 150665197 | 150665159 | 150665282 | + | rRNA_pseudogene |
| ENST00000516678.1 | chr6 | 151620037 | 151620068 | 151619976 | 151620068 | - | misc_RNA |
| ENST00000354675.6 | chr6 | 151646985 | 151647016 | 151646823 | 151647020 | + | protein_coding |
| ENST00000367097.3 | chr6 | 158932131 | 158932163 | 158927610 | 158932860 | + | protein_coding |
| ENST00000453779.2 | chr6 | 163990924 | 163990955 | 163986978 | 163991177 | + | protein_coding |
| ENST00000275262.7 | chr6 | 163990924 | 163990955 | 163985699 | 163991177 | + | protein_coding |
| ENST00000392127.2 | chr6 | 163990924 | 163990955 | 163984452 | 163991177 | + | protein_coding |
| ENST00000362653.1 | chr6 | 165823084 | 165823115 | 165823051 | 165823165 | + | rRNA |
| ENST00000332290.2 | chr6 | 170103816 | 170103847 | 170102257 | 170106401 | + | protein_coding |
| ENST00000262193.6 | chr6 | 170852704 | 170852738 | 170852689 | 170852818 | - | protein_coding |
| ENST00000462957.1 | chr6 | 170852704 | 170852738 | 170852689 | 170852818 | - | processed_transcript |
| ENST00000392093.3 | chr6 | 170852704 | 170852738 | 170852689 | 170852818 | - | protein_coding |
| ENST00000402802.3 | chr7 | 537578 | 537610 | 536895 | 538211 | - | protein_coding |
| ENST00000456758.2 | chr7 | 881604 | 881635 | 881583 | 881767 | + | protein_coding |
| ENST00000389574.3 | chr7 | 881604 | 881635 | 881583 | 881767 | + | protein_coding |
| ENST00000427969.1 | chr7 | 881604 | 881635 | 881583 | 881767 | + | processed_transcript |
| ENST00000450538.1 | chr7 | 881604 | 881635 | 881583 | 881767 | + | processed_transcript |
| ENST00000457378.2 | chr7 | 881604 | 881635 | 881583 | 881767 | + | protein_coding |
| ENST00000452783.2 | chr7 | 881604 | 881635 | 881583 | 881767 | + | protein_coding |
| ENST00000435699.1 | chr7 | 881604 | 881635 | 881583 | 881767 | + | protein_coding |
| ENST00000439679.1 | chr7 | 881604 | 881635 | 881583 | 881723 | + | protein_coding |
| ENST00000457598.1 | chr7 | 881604 | 881635 | 881583 | 881672 | + | protein_coding |
| ENST00000421580.1 | chr7 | 881604 | 881635 | 881583 | 881656 | + | protein_coding |
| ENST00000405266.1 | chr7 | 881604 | 881635 | 881583 | 881767 | + | protein_coding |
| ENST00000340926.3 | chr7 | 881604 | 881635 | 881583 | 881767 | + | retained_intron |
| ENST00000413171.2 | chr7 | 881604 | 881635 | 881583 | 881767 | + | retained_intron |
| ENST00000401592.1 | chr7 | 881604 | 881635 | 881583 | 881767 | + | protein_coding |
| ENST00000403868.1 | chr7 | 881604 | 881635 | 881583 | 881767 | + | protein_coding |
| ENST00000467483.1 | chr7 | 881604 | 881635 | 881583 | 881767 | + | retained_intron |
| ENST00000297445.5 | chr7 | 881604 | 881635 | 881583 | 881767 | + | protein_coding |
| ENST00000425407.2 | chr7 | 881604 | 881635 | 881583 | 881767 | + | protein_coding |
| ENST00000469755.1 | chr7 | 881604 | 881635 | 881583 | 881710 | + | processed_transcript |
| ENST00000450881.1 | chr7 | 881604 | 881635 | 881583 | 881767 | + | protein_coding |
| ENST00000493681.1 | chr7 | 881604 | 881635 | 881453 | 881767 | + | retained_intron |
| ENST00000477950.1 | chr7 | 881604 | 881635 | 881549 | 881767 | + | retained_intron |
| ENST00000362153.1 | chr7 | 1062582 | 1062613 | 1062569 | 1062662 | - | miRNA |
| ENST00000258711.6 | chr7 | 2474216 | 2474250 | 2472198 | 2474242 | + | protein_coding |
| ENST00000275364.3 | chr7 | 2768205 | 2768236 | 2767746 | 2771384 | - | protein_coding |
| ENST00000491117.1 | chr7 | 2768205 | 2768236 | 2768195 | 2771384 | - | processed_transcript |
| ENST00000407904.3 | chr7 | 2768205 | 2768236 | 2768195 | 2771384 | - | protein_coding |
| ENST00000407653.1 | chr7 | 2768205 | 2768236 | 2768200 | 2771384 | - | protein_coding |
| ENST00000396960.3 | chr7 | 2768205 | 2768236 | 2768200 | 2771384 | - | protein_coding |
| ENST00000382368.3 | chr7 | 5520180 | 5520211 | 5520189 | 5521562 | - | protein_coding |
| ENST00000385238.1 | chr7 | 5535457 | 5535488 | 5535450 | 5535548 | - | miRNA |
| ENST00000331789.5 | chr7 | 5568882 | 5568913 | 5568792 | 5569031 | - | protein_coding |
| ENST00000425660.1 | chr7 | 5568882 | 5568913 | 5568792 | 5569031 | - | nonsense_mediated_decay |
| ENST00000462494.1 | chr7 | 5568882 | 5568913 | 5567912 | 5569031 | - | retained_intron |
| ENST00000445914.2 | chr7 | 5568882 | 5568913 | 5568792 | 5569031 | - | protein_coding |
| ENST00000400179.3 | chr7 | 5568882 | 5568913 | 5568792 | 5568935 | - | protein_coding |
| ENST00000320713.6 | chr7 | 5568882 | 5568913 | 5568792 | 5568933 | - | protein_coding |
| ENST00000493945.1 | chr7 | 5568882 | 5568913 | 5568792 | 5569031 | - | retained_intron |
| ENST00000484841.1 | chr7 | 5568882 | 5568913 | 5568792 | 5569031 | - | retained_intron |
| ENST00000477812.1 | chr7 | 5568882 | 5568913 | 5568792 | 5569294 | - | retained_intron |
| ENST00000432588.1 | chr7 | 5568882 | 5568913 | 5568792 | 5569031 | - | protein_coding |
| ENST00000480301.1 | chr7 | 5568882 | 5568913 | 5568698 | 5569294 | - | retained_intron |
| ENST00000443528.1 | chr7 | 5568882 | 5568913 | 5568847 | 5569031 | - | protein_coding |
| ENST00000417101.1 | chr7 | 5568882 | 5568913 | 5568866 | 5569031 | - | protein_coding |
| ENST00000199389.6 | chr7 | 6077065 | 6077096 | 6077055 | 6077155 | - | protein_coding |
| ENST00000536084.1 | chr7 | 6077065 | 6077096 | 6077055 | 6077155 | - | protein_coding |
| ENST00000426957.1 | chr7 | 6077065 | 6077096 | 6077055 | 6077155 | - | protein_coding |
| ENST00000348035.4 | chr7 | 6442969 | 6443000 | 6441947 | 6443608 | + | protein_coding |
| ENST00000348035.4 | chr7 | 6443432 | 6443463 | 6441947 | 6443608 | + | protein_coding |
| ENST00000258739.4 | chr7 | 6505713 | 6505744 | 6505702 | 6505954 | - | protein_coding |
| ENST00000454368.2 | chr7 | 6505713 | 6505744 | 6505702 | 6505954 | - | retained_intron |
| ENST00000405731.3 | chr7 | 6621807 | 6621838 | 6621704 | 6621882 | + | protein_coding |
| ENST00000396713.2 | chr7 | 6621807 | 6621838 | 6621704 | 6621882 | + | protein_coding |
| ENST00000396707.2 | chr7 | 6621807 | 6621838 | 6621704 | 6621882 | + | protein_coding |
| ENST00000335965.6 | chr7 | 6621807 | 6621838 | 6621704 | 6621882 | + | protein_coding |
| ENST00000493944.1 | chr7 | 6621807 | 6621838 | 6621704 | 6621952 | + | retained_intron |
| ENST00000474097.1 | chr7 | 6621807 | 6621838 | 6621704 | 6621882 | + | retained_intron |
| ENST00000396709.1 | chr7 | 6621807 | 6621838 | 6621704 | 6621882 | + | protein_coding |
| ENST00000483589.1 | chr7 | 6621807 | 6621838 | 6621704 | 6621882 | + | protein_coding |
| ENST00000396706.2 | chr7 | 6621807 | 6621838 | 6621704 | 6621882 | + | protein_coding |
| ENST00000429911.1 | chr7 | 7273936 | 7273965 | 7273934 | 7274170 | + | protein_coding |
| ENST00000419721.1 | chr7 | 7273936 | 7273965 | 7273934 | 7274170 | + | protein_coding |
| ENST00000436587.2 | chr7 | 7273936 | 7273965 | 7273934 | 7274170 | + | protein_coding |
| ENST00000476068.1 | chr7 | 7273936 | 7273965 | 7273934 | 7274928 | + | retained_intron |
| ENST00000223122.2 | chr7 | 7273936 | 7273965 | 7273889 | 7274170 | + | protein_coding |
| ENST00000402468.3 | chr7 | 7273936 | 7273965 | 7273951 | 7274170 | + | protein_coding |
| ENST00000262067.4 | chr7 | 16823453 | 16823484 | 16823042 | 16824161 | + | protein_coding |
| ENST00000222567.5 | chr7 | 19739699 | 19739728 | 19739695 | 19739903 | - | protein_coding |
| ENST00000462263.1 | chr7 | 19739699 | 19739728 | 19739695 | 19739958 | - | retained_intron |
| ENST00000313367.2 | chr7 | 24836883 | 24836914 | 24836158 | 24839898 | - | protein_coding |
| ENST00000352860.1 | chr7 | 24836883 | 24836914 | 24836166 | 24839898 | - | protein_coding |
| ENST00000353930.1 | chr7 | 24836883 | 24836914 | 24836166 | 24839898 | - | protein_coding |
| ENST00000431825.2 | chr7 | 24836883 | 24836914 | 24836166 | 24839898 | - | protein_coding |
| ENST00000305786.2 | chr7 | 25162937 | 25162965 | 25159710 | 25163468 | - | protein_coding |
| ENST00000409409.1 | chr7 | 25162937 | 25162965 | 25162926 | 25163468 | - | protein_coding |
| ENST00000354667.4 | chr7 | 26231863 | 26231895 | 26229547 | 26231958 | - | protein_coding |
| ENST00000463181.1 | chr7 | 26231863 | 26231895 | 26229924 | 26232197 | - | retained_intron |
| ENST00000356674.7 | chr7 | 26231863 | 26231895 | 26231464 | 26231958 | - | protein_coding |
| ENST00000409814.2 | chr7 | 26231863 | 26231895 | 26231467 | 26231958 | - | protein_coding |
| ENST00000490912.1 | chr7 | 26231863 | 26231895 | 26231469 | 26231958 | - | retained_intron |
| ENST00000337620.4 | chr7 | 26251905 | 26251936 | 26251702 | 26252976 | + | protein_coding |
| ENST00000396386.2 | chr7 | 26251905 | 26251936 | 26251702 | 26252971 | + | protein_coding |
| ENST00000409747.1 | chr7 | 26251905 | 26251936 | 26251702 | 26252976 | + | protein_coding |
| ENST00000481057.1 | chr7 | 26251905 | 26251936 | 26251702 | 26252043 | + | retained_intron |
| ENST00000337620.4 | chr7 | 26252603 | 26252634 | 26251702 | 26252976 | + | protein_coding |
| ENST00000396386.2 | chr7 | 26252603 | 26252634 | 26251702 | 26252971 | + | protein_coding |
| ENST00000409747.1 | chr7 | 26252603 | 26252634 | 26251702 | 26252976 | + | protein_coding |
| ENST00000222726.3 | chr7 | 27180996 | 27181027 | 27180671 | 27181704 | - | protein_coding |
| ENST00000354032.4 | chr7 | 27209112 | 27209168 | 27209116 | 27209199 | - | protein_coding |
| ENST00000354032.4 | chr7 | 27209112 | 27209168 | 27208913 | 27209114 | - | protein_coding |
| ENST00000384852.1 | chr7 | 27209112 | 27209168 | 27209099 | 27209183 | - | miRNA |
| ENST00000283921.4 | chr7 | 27213861 | 27213892 | 27212968 | 27213925 | - | protein_coding |
| ENST00000381834.4 | chr7 | 27213861 | 27213892 | 27213548 | 27213914 | - | protein_coding |
| ENST00000222753.4 | chr7 | 27236487 | 27236518 | 27234637 | 27238061 | - | protein_coding |
| ENST00000396299.2 | chr7 | 28859070 | 28859101 | 28858733 | 28859617 | + | protein_coding |
| ENST00000357727.2 | chr7 | 28859070 | 28859101 | 28858733 | 28865511 | + | protein_coding |
| ENST00000396300.2 | chr7 | 28859070 | 28859101 | 28858733 | 28859617 | + | protein_coding |
| ENST00000409603.1 | chr7 | 28859070 | 28859101 | 28858733 | 28859254 | + | protein_coding |
| ENST00000396298.2 | chr7 | 28859070 | 28859101 | 28858733 | 28859617 | + | protein_coding |
| ENST00000498316.2 | chr7 | 28859070 | 28859101 | 28858733 | 28859176 | + | protein_coding |
| ENST00000406581.2 | chr7 | 44161533 | 44161563 | 44161433 | 44161794 | - | protein_coding |
| ENST00000223361.3 | chr7 | 44161533 | 44161563 | 44161433 | 44161795 | - | protein_coding |
| ENST00000452185.1 | chr7 | 44161533 | 44161563 | 44161433 | 44161881 | - | protein_coding |
| ENST00000436844.1 | chr7 | 44161533 | 44161563 | 44161433 | 44161708 | - | protein_coding |
| ENST00000433715.1 | chr7 | 44161533 | 44161563 | 44161433 | 44161708 | - | protein_coding |
| ENST00000463464.1 | chr7 | 44161533 | 44161563 | 44161433 | 44161708 | - | retained_intron |
| ENST00000470867.1 | chr7 | 44161533 | 44161563 | 44161433 | 44161708 | - | retained_intron |
| ENST00000456038.1 | chr7 | 44161533 | 44161563 | 44161433 | 44161708 | - | protein_coding |
| ENST00000418438.1 | chr7 | 44161533 | 44161563 | 44161433 | 44161708 | - | protein_coding |
| ENST00000496539.1 | chr7 | 44161533 | 44161563 | 44161433 | 44161708 | - | retained_intron |
| ENST00000457475.1 | chr7 | 44258942 | 44258969 | 44256749 | 44259121 | - | protein_coding |
| ENST00000395749.2 | chr7 | 44258942 | 44258969 | 44256755 | 44259121 | - | protein_coding |
| ENST00000425809.1 | chr7 | 44258942 | 44258969 | 44258812 | 44259121 | - | protein_coding |
| ENST00000440254.2 | chr7 | 44258942 | 44258969 | 44258891 | 44259121 | - | protein_coding |
| ENST00000358707.3 | chr7 | 44258942 | 44258969 | 44258945 | 44259121 | - | protein_coding |
| ENST00000466584.1 | chr7 | 44258942 | 44258969 | 44258948 | 44259121 | - | retained_intron |
| ENST00000353625.4 | chr7 | 44258942 | 44258969 | 44258950 | 44259121 | - | protein_coding |
| ENST00000497584.1 | chr7 | 44258942 | 44258969 | 44258950 | 44259201 | - | processed_transcript |
| ENST00000489429.1 | chr7 | 44258942 | 44258969 | 44258964 | 44259121 | - | processed_transcript |
| ENST00000457408.2 | chr7 | 44617612 | 44617643 | 44617493 | 44619227 | - | protein_coding |
| ENST00000289577.5 | chr7 | 44617612 | 44617643 | 44617493 | 44619227 | - | protein_coding |
| ENST00000349299.3 | chr7 | 44874010 | 44874041 | 44873827 | 44874161 | - | protein_coding |
| ENST00000308153.4 | chr7 | 44874010 | 44874041 | 44873827 | 44874161 | - | protein_coding |
| ENST00000350771.3 | chr7 | 44874010 | 44874041 | 44873827 | 44874161 | - | protein_coding |
| ENST00000381124.5 | chr7 | 44874010 | 44874041 | 44873838 | 44874161 | - | protein_coding |
| ENST00000521529.1 | chr7 | 44874010 | 44874041 | 44873909 | 44874161 | - | protein_coding |
| ENST00000395699.2 | chr7 | 44920455 | 44920486 | 44915896 | 44924960 | - | protein_coding |
| ENST00000356889.4 | chr7 | 50514703 | 50514735 | 50511831 | 50514995 | - | protein_coding |
| ENST00000395556.2 | chr7 | 50514703 | 50514735 | 50511831 | 50514995 | - | protein_coding |
| ENST00000433017.1 | chr7 | 50514703 | 50514735 | 50512175 | 50514995 | - | protein_coding |
| ENST00000419119.1 | chr7 | 50514703 | 50514735 | 50512840 | 50514995 | - | protein_coding |
| ENST00000435566.1 | chr7 | 50514703 | 50514735 | 50514485 | 50514714 | - | protein_coding |
| ENST00000436590.1 | chr7 | 50514703 | 50514735 | 50514610 | 50514995 | - | protein_coding |
| ENST00000422854.1 | chr7 | 50514703 | 50514735 | 50514679 | 50514995 | - | protein_coding |
| ENST00000440350.1 | chr7 | 50514703 | 50514735 | 50514693 | 50514995 | - | protein_coding |
| ENST00000489672.1 | chr7 | 64384163 | 64384194 | 64383764 | 64384809 | + | processed_transcript |
| ENST00000545510.1 | chr7 | 64384163 | 64384194 | 64383764 | 64384595 | + | protein_coding |
| ENST00000328747.7 | chr7 | 64864024 | 64864055 | 64863254 | 64866038 | + | protein_coding |
| ENST00000431504.1 | chr7 | 64864024 | 64864055 | 64863254 | 64866038 | + | protein_coding |
| ENST00000357512.2 | chr7 | 64864024 | 64864055 | 64863254 | 64865997 | + | protein_coding |
| ENST00000450302.2 | chr7 | 64864024 | 64864055 | 64863254 | 64865994 | + | protein_coding |
| ENST00000503687.1 | chr7 | 66274477 | 66274508 | 66273873 | 66276174 | + | nonsense_mediated_decay |
| ENST00000380827.3 | chr7 | 66274477 | 66274508 | 66273873 | 66276446 | + | protein_coding |
| ENST00000380828.2 | chr7 | 66274477 | 66274508 | 66273873 | 66276446 | + | protein_coding |
| ENST00000510829.2 | chr7 | 66274477 | 66274508 | 66273873 | 66276446 | + | protein_coding |
| ENST00000284957.4 | chr7 | 66274477 | 66274508 | 66273873 | 66276423 | + | protein_coding |
| ENST00000439720.1 | chr7 | 66274477 | 66274508 | 66273873 | 66274563 | + | protein_coding |
| ENST00000437078.2 | chr7 | 66274477 | 66274508 | 66273873 | 66274563 | + | protein_coding |
| ENST00000484547.1 | chr7 | 66274477 | 66274508 | 66273873 | 66276174 | + | processed_transcript |
| ENST00000516439.1 | chr7 | 68527415 | 68527445 | 68527371 | 68527457 | + | miRNA |
| ENST00000517158.1 | chr7 | 68527603 | 68527655 | 68527589 | 68527662 | + | miRNA |
| ENST00000485000.1 | chr7 | 74071971 | 74072002 | 74071994 | 74072394 | + | processed_transcript |
| ENST00000413643.1 | chr7 | 87536794 | 87536823 | 87536503 | 87537691 | + | nonsense_mediated_decay |
| ENST00000265728.1 | chr7 | 87536794 | 87536823 | 87536503 | 87538856 | + | protein_coding |
| ENST00000431138.1 | chr7 | 87536794 | 87536823 | 87536503 | 87537688 | + | nonsense_mediated_decay |
| ENST00000287916.4 | chr7 | 90045208 | 90045239 | 90041958 | 90045268 | + | protein_coding |
| ENST00000535571.1 | chr7 | 90045208 | 90045239 | 90041951 | 90045268 | + | protein_coding |
| ENST00000394605.1 | chr7 | 90045208 | 90045239 | 90041958 | 90045268 | + | protein_coding |
| ENST00000482108.1 | chr7 | 94294772 | 94294803 | 94292646 | 94299007 | + | protein_coding |
| ENST00000488574.1 | chr7 | 94294772 | 94294803 | 94292646 | 94295010 | + | protein_coding |
| ENST00000482108.1 | chr7 | 94296725 | 94296743 | 94292646 | 94299007 | + | protein_coding |
| ENST00000482108.1 | chr7 | 94296926 | 94296956 | 94292646 | 94299007 | + | protein_coding |
| ENST00000175506.4 | chr7 | 97481721 | 97481747 | 97481430 | 97481780 | - | protein_coding |
| ENST00000394309.3 | chr7 | 97481721 | 97481747 | 97481440 | 97481780 | - | protein_coding |
| ENST00000454046.1 | chr7 | 97481721 | 97481747 | 97481443 | 97481780 | - | nonsense_mediated_decay |
| ENST00000437628.1 | chr7 | 97481721 | 97481747 | 97481443 | 97481780 | - | protein_coding |
| ENST00000394308.3 | chr7 | 97481721 | 97481747 | 97481443 | 97481780 | - | protein_coding |
| ENST00000422745.1 | chr7 | 97481721 | 97481747 | 97481458 | 97481780 | - | protein_coding |
| ENST00000455086.1 | chr7 | 97481721 | 97481747 | 97481464 | 97481780 | - | protein_coding |
| ENST00000444334.1 | chr7 | 97481721 | 97481747 | 97481516 | 97481780 | - | protein_coding |
| ENST00000487714.1 | chr7 | 97481721 | 97481747 | 97481626 | 97481780 | - | retained_intron |
| ENST00000297293.5 | chr7 | 97821087 | 97821118 | 97820926 | 97823884 | + | protein_coding |
| ENST00000359863.4 | chr7 | 98507686 | 98507718 | 98507679 | 98508042 | + | protein_coding |
| ENST00000355540.3 | chr7 | 98507686 | 98507718 | 98507679 | 98508042 | + | protein_coding |
| ENST00000446306.2 | chr7 | 98507686 | 98507718 | 98507679 | 98508039 | + | protein_coding |
| ENST00000456197.1 | chr7 | 98507686 | 98507718 | 98507679 | 98508042 | + | protein_coding |
| ENST00000324306.5 | chr7 | 99632550 | 99632582 | 99630928 | 99636631 | + | protein_coding |
| ENST00000535170.1 | chr7 | 99632550 | 99632582 | 99630928 | 99634338 | + | protein_coding |
| ENST00000324306.5 | chr7 | 99633646 | 99633677 | 99630928 | 99636631 | + | protein_coding |
| ENST00000535170.1 | chr7 | 99633646 | 99633677 | 99630928 | 99634338 | + | protein_coding |
| ENST00000489841.1 | chr7 | 99690432 | 99690463 | 99690351 | 99690756 | - | retained_intron |
| ENST00000343023.6 | chr7 | 99690432 | 99690463 | 99690403 | 99690739 | - | protein_coding |
| ENST00000485286.1 | chr7 | 99690432 | 99690463 | 99690404 | 99690756 | - | retained_intron |
| ENST00000303887.5 | chr7 | 99690432 | 99690463 | 99690404 | 99690756 | - | protein_coding |
| ENST00000542483.1 | chr7 | 99690432 | 99690463 | 99690404 | 99690756 | - | protein_coding |
| ENST00000362082.3 | chr7 | 99690432 | 99690463 | 99690404 | 99690756 | - | protein_coding |
| ENST00000354230.3 | chr7 | 99690432 | 99690463 | 99690404 | 99690756 | - | protein_coding |
| ENST00000385024.1 | chr7 | 99691401 | 99691460 | 99691391 | 99691470 | - | miRNA |
| ENST00000385301.1 | chr7 | 99691655 | 99691686 | 99691616 | 99691697 | - | miRNA |
| ENST00000223076.2 | chr7 | 100168468 | 100168505 | 100167754 | 100168759 | + | processed_pseudogene |
| ENST00000539430.1 | chr7 | 100279164 | 100279214 | 100277130 | 100279394 | - | protein_coding |
| ENST00000275732.4 | chr7 | 100279164 | 100279214 | 100278172 | 100279394 | - | protein_coding |
| ENST00000303151.4 | chr7 | 100305079 | 100305110 | 100304444 | 100305118 | + | protein_coding |
| ENST00000360264.3 | chr7 | 101894466 | 101894497 | 101891692 | 101901513 | + | protein_coding |
| ENST00000292535.7 | chr7 | 101894466 | 101894497 | 101891692 | 101901513 | + | protein_coding |
| ENST00000360264.3 | chr7 | 101901021 | 101901052 | 101891692 | 101901513 | + | protein_coding |
| ENST00000292535.7 | chr7 | 101901021 | 101901052 | 101891692 | 101901513 | + | protein_coding |
| ENST00000455385.2 | chr7 | 105731112 | 105731143 | 105730949 | 105732385 | - | protein_coding |
| ENST00000011473.2 | chr7 | 105731112 | 105731143 | 105730951 | 105732385 | - | protein_coding |
| ENST00000455385.2 | chr7 | 105731500 | 105731531 | 105730949 | 105732385 | - | protein_coding |
| ENST00000011473.2 | chr7 | 105731500 | 105731531 | 105730951 | 105732385 | - | protein_coding |
| ENST00000455385.2 | chr7 | 105732102 | 105732133 | 105730949 | 105732385 | - | protein_coding |
| ENST00000011473.2 | chr7 | 105732102 | 105732133 | 105730951 | 105732385 | - | protein_coding |
| ENST00000464029.1 | chr7 | 105732102 | 105732133 | 105731853 | 105732385 | - | protein_coding |
| ENST00000470347.1 | chr7 | 105732102 | 105732133 | 105731903 | 105732385 | - | protein_coding |
| ENST00000523505.1 | chr7 | 106300309 | 106300332 | 106297211 | 106301442 | - | protein_coding |
| ENST00000468410.1 | chr7 | 106842722 | 106842753 | 106841859 | 106842953 | + | protein_coding |
| ENST00000222574.4 | chr7 | 106842722 | 106842753 | 106841859 | 106842974 | + | protein_coding |
| ENST00000347053.2 | chr7 | 106842722 | 106842753 | 106842189 | 106844075 | - | protein_coding |
| ENST00000205402.3 | chr7 | 107543932 | 107543963 | 107543923 | 107543992 | + | protein_coding |
| ENST00000415325.1 | chr7 | 107543932 | 107543963 | 107543923 | 107543992 | + | nonsense_mediated_decay |
| ENST00000417551.1 | chr7 | 107543932 | 107543963 | 107543923 | 107543992 | + | protein_coding |
| ENST00000537148.1 | chr7 | 107543932 | 107543963 | 107543923 | 107543992 | + | protein_coding |
| ENST00000440410.1 | chr7 | 107543932 | 107543963 | 107543923 | 107543992 | + | protein_coding |
| ENST00000437604.2 | chr7 | 107543932 | 107543963 | 107543923 | 107543992 | + | protein_coding |
| ENST00000451081.1 | chr7 | 107543932 | 107543963 | 107543923 | 107543992 | + | nonsense_mediated_decay |
| ENST00000453354.1 | chr7 | 107543932 | 107543963 | 107543923 | 107543992 | + | nonsense_mediated_decay |
| ENST00000450038.1 | chr7 | 107543932 | 107543963 | 107543923 | 107543992 | + | nonsense_mediated_decay |
| ENST00000460577.1 | chr7 | 107543932 | 107543963 | 107543923 | 107544221 | + | retained_intron |
| ENST00000494441.1 | chr7 | 107543932 | 107543963 | 107543923 | 107543992 | + | processed_transcript |
| ENST00000539590.1 | chr7 | 107543932 | 107543963 | 107543923 | 107543992 | + | protein_coding |
| ENST00000478414.1 | chr7 | 107543932 | 107543963 | 107543948 | 107543992 | + | retained_intron |
| ENST00000249356.3 | chr7 | 108214658 | 108214689 | 108213343 | 108215294 | + | protein_coding |
| ENST00000343743.4 | chr7 | 112161006 | 112161047 | 112160543 | 112161206 | + | processed_pseudogene |
| ENST00000160373.3 | chr7 | 117424383 | 117424409 | 117424305 | 117424508 | - | protein_coding |
| ENST00000446636.1 | chr7 | 117424383 | 117424409 | 117424305 | 117424508 | - | protein_coding |
| ENST00000441556.1 | chr7 | 117424383 | 117424409 | 117424305 | 117424508 | - | nonsense_mediated_decay |
| ENST00000223023.4 | chr7 | 123324412 | 123324444 | 123321989 | 123324634 | - | protein_coding |
| ENST00000536685.1 | chr7 | 123324412 | 123324444 | 123322001 | 123324634 | - | protein_coding |
| ENST00000457231.1 | chr7 | 124675061 | 124675092 | 124674908 | 124675355 | + | processed_pseudogene |
| ENST00000384856.1 | chr7 | 127721928 | 127721959 | 127721913 | 127722012 | + | miRNA |
| ENST00000489835.2 | chr7 | 127999856 | 127999887 | 127999394 | 128000117 | - | protein_coding |
| ENST00000446477.2 | chr7 | 127999856 | 127999887 | 127999394 | 128000117 | - | protein_coding |
| ENST00000535159.1 | chr7 | 127999856 | 127999887 | 127999394 | 128000117 | - | protein_coding |
| ENST00000435512.1 | chr7 | 127999856 | 127999887 | 127999394 | 128000117 | - | protein_coding |
| ENST00000489517.1 | chr7 | 127999856 | 127999887 | 127999394 | 128000117 | - | protein_coding |
| ENST00000464607.1 | chr7 | 127999856 | 127999887 | 127999421 | 128000117 | - | protein_coding |
| ENST00000495931.1 | chr7 | 127999856 | 127999887 | 127999429 | 128000117 | - | protein_coding |
| ENST00000364171.1 | chr7 | 128337591 | 128337614 | 128337493 | 128337610 | + | rRNA |
| ENST00000517057.1 | chr7 | 128337763 | 128337790 | 128337764 | 128337798 | + | rRNA |
| ENST00000542996.1 | chr7 | 128410443 | 128410462 | 128409117 | 128411530 | + | protein_coding |
| ENST00000535623.1 | chr7 | 128410443 | 128410462 | 128409117 | 128411530 | + | protein_coding |
| ENST00000538394.1 | chr7 | 128410443 | 128410462 | 128409117 | 128411530 | + | protein_coding |
| ENST00000537667.1 | chr7 | 128410443 | 128410462 | 128409135 | 128411530 | + | protein_coding |
| ENST00000535011.1 | chr7 | 128410443 | 128410462 | 128409117 | 128411530 | + | protein_coding |
| ENST00000537014.1 | chr7 | 128410443 | 128410462 | 128409117 | 128411530 | + | protein_coding |
| ENST00000538546.1 | chr7 | 128410443 | 128410462 | 128409117 | 128411530 | + | protein_coding |
| ENST00000249364.4 | chr7 | 128410443 | 128410462 | 128409117 | 128411528 | + | protein_coding |
| ENST00000449187.2 | chr7 | 128410443 | 128410462 | 128409117 | 128410630 | + | protein_coding |
| ENST00000342367.6 | chr7 | 128410443 | 128410462 | 128409117 | 128410630 | + | protein_coding |
| ENST00000249344.2 | chr7 | 129126838 | 129126871 | 129125420 | 129128240 | + | protein_coding |
| ENST00000385255.1 | chr7 | 129410279 | 129410310 | 129410223 | 129410332 | - | miRNA |
| ENST00000362288.1 | chr7 | 129414570 | 129414601 | 129414532 | 129414609 | - | miRNA |
| ENST00000362173.1 | chr7 | 130135967 | 130135998 | 130135952 | 130136045 | + | miRNA |
| ENST00000362111.1 | chr7 | 130561497 | 130561528 | 130561506 | 130561570 | - | miRNA |
| ENST00000428680.2 | chr7 | 135072941 | 135072974 | 135070903 | 135073646 | - | protein_coding |
| ENST00000315544.5 | chr7 | 135072941 | 135072974 | 135070923 | 135073646 | - | protein_coding |
| ENST00000343526.4 | chr7 | 138235862 | 138235893 | 138235808 | 138235925 | + | protein_coding |
| ENST00000536822.1 | chr7 | 138235862 | 138235893 | 138235808 | 138235925 | + | protein_coding |
| ENST00000415680.2 | chr7 | 138235862 | 138235893 | 138235808 | 138235925 | + | protein_coding |
| ENST00000476785.1 | chr7 | 138235862 | 138235893 | 138235808 | 138235925 | + | processed_transcript |
| ENST00000497516.1 | chr7 | 138235862 | 138235893 | 138235808 | 138235925 | + | processed_transcript |
| ENST00000378381.2 | chr7 | 138235862 | 138235893 | 138235808 | 138235925 | + | protein_coding |
| ENST00000242351.5 | chr7 | 138730641 | 138730673 | 138728266 | 138732599 | - | protein_coding |
| ENST00000464606.1 | chr7 | 138730641 | 138730673 | 138730662 | 138732599 | - | protein_coding |
| ENST00000473989.2 | chr7 | 138968350 | 138968372 | 138967770 | 138969320 | + | protein_coding |
| ENST00000288561.8 | chr7 | 138968350 | 138968372 | 138967770 | 138969320 | + | protein_coding |
| ENST00000468383.1 | chr7 | 139030283 | 139030319 | 139030247 | 139030541 | + | processed_transcript |
| ENST00000481123.1 | chr7 | 139030283 | 139030319 | 139030247 | 139030536 | + | processed_transcript |
| ENST00000482181.1 | chr7 | 139030283 | 139030319 | 139030247 | 139030523 | + | processed_transcript |
| ENST00000488886.1 | chr7 | 139030283 | 139030319 | 139030247 | 139030541 | + | processed_transcript |
| ENST00000297534.6 | chr7 | 139030283 | 139030319 | 139030247 | 139031065 | + | protein_coding |
| ENST00000541515.1 | chr7 | 139030283 | 139030319 | 139030247 | 139030367 | + | protein_coding |
| ENST00000365602.1 | chr7 | 140086571 | 140086607 | 140086581 | 140086707 | - | rRNA |
| ENST00000322764.5 | chr7 | 143088132 | 143088163 | 143087671 | 143088204 | + | protein_coding |
| ENST00000354434.4 | chr7 | 143088132 | 143088163 | 143087671 | 143088202 | + | protein_coding |
| ENST00000449423.2 | chr7 | 143088132 | 143088163 | 143087671 | 143088202 | + | protein_coding |
| ENST00000392910.2 | chr7 | 143088132 | 143088163 | 143087671 | 143088202 | + | protein_coding |
| ENST00000492114.1 | chr7 | 145694455 | 145694488 | 145694421 | 145694483 | + | Mt_tRNA_pseudogene |
| ENST00000516501.1 | chr7 | 148638580 | 148638613 | 148638580 | 148638658 | + | misc_RNA |
| ENST00000516501.1 | chr7 | 148638626 | 148638661 | 148638580 | 148638658 | + | misc_RNA |
| ENST00000516507.1 | chr7 | 148660407 | 148660438 | 148660407 | 148660502 | + | misc_RNA |
| ENST00000516507.1 | chr7 | 148660466 | 148660504 | 148660407 | 148660502 | + | misc_RNA |
| ENST00000365484.1 | chr7 | 148680846 | 148680878 | 148680847 | 148680948 | + | misc_RNA |
| ENST00000365484.1 | chr7 | 148680918 | 148680950 | 148680847 | 148680948 | + | misc_RNA |
| ENST00000364228.1 | chr7 | 148684307 | 148684340 | 148684228 | 148684340 | - | misc_RNA |
| ENST00000495645.1 | chr7 | 150935535 | 150935566 | 150934460 | 150935905 | + | protein_coding |
| ENST00000035307.2 | chr7 | 150935535 | 150935566 | 150934460 | 150935908 | + | protein_coding |
| ENST00000390183.1 | chr7 | 150935535 | 150935566 | 150935507 | 150935624 | + | miRNA |
| ENST00000495645.1 | chr7 | 150935574 | 150935605 | 150934460 | 150935905 | + | protein_coding |
| ENST00000035307.2 | chr7 | 150935574 | 150935605 | 150934460 | 150935908 | + | protein_coding |
| ENST00000390183.1 | chr7 | 150935574 | 150935605 | 150935507 | 150935624 | + | miRNA |
| ENST00000348165.5 | chr7 | 156974317 | 156974350 | 156974212 | 156974365 | + | protein_coding |
| ENST00000389103.4 | chr7 | 156974317 | 156974350 | 156974212 | 156974365 | + | protein_coding |
| ENST00000429029.2 | chr7 | 157178644 | 157178672 | 157178235 | 157179018 | + | protein_coding |
| ENST00000487480.1 | chr7 | 157178644 | 157178672 | 157178235 | 157179012 | + | retained_intron |
| ENST00000285518.6 | chr8 | 6616978 | 6617009 | 6614684 | 6617184 | + | protein_coding |
| ENST00000384868.1 | chr8 | 10892721 | 10892752 | 10892716 | 10892812 | - | miRNA |
| ENST00000516370.1 | chr8 | 13393538 | 13393569 | 13393538 | 13393644 | + | rRNA |
| ENST00000361272.4 | chr8 | 17088187 | 17088218 | 17086737 | 17088357 | - | protein_coding |
| ENST00000518541.1 | chr8 | 17088187 | 17088218 | 17088077 | 17088357 | - | protein_coding |
| ENST00000308511.4 | chr8 | 22477864 | 22477895 | 22477151 | 22479027 | + | protein_coding |
| ENST00000389279.3 | chr8 | 22477864 | 22477895 | 22477151 | 22477984 | + | protein_coding |
| ENST00000520861.1 | chr8 | 22477864 | 22477895 | 22476663 | 22477981 | + | protein_coding |
| ENST00000520738.1 | chr8 | 22477864 | 22477895 | 22477151 | 22477981 | + | protein_coding |
| ENST00000521436.1 | chr8 | 22477864 | 22477895 | 22477151 | 22477981 | + | retained_intron |
| ENST00000520536.1 | chr8 | 22477864 | 22477895 | 22477151 | 22478477 | + | retained_intron |
| ENST00000523752.1 | chr8 | 22879993 | 22880024 | 22877646 | 22880497 | - | retained_intron |
| ENST00000276431.4 | chr8 | 22879993 | 22880024 | 22877646 | 22880497 | - | protein_coding |
| ENST00000347739.3 | chr8 | 22879993 | 22880024 | 22879824 | 22880497 | - | protein_coding |
| ENST00000523504.1 | chr8 | 22879993 | 22880024 | 22879987 | 22880497 | - | nonsense_mediated_decay |
| ENST00000542226.1 | chr8 | 22879993 | 22880024 | 22879987 | 22880497 | - | protein_coding |
| ENST00000519973.1 | chr8 | 23429718 | 23429762 | 23428848 | 23432976 | + | protein_coding |
| ENST00000518881.1 | chr8 | 23429718 | 23429762 | 23428848 | 23429734 | + | retained_intron |
| ENST00000524174.1 | chr8 | 23492778 | 23492810 | 23492389 | 23493020 | - | processed_pseudogene |
| ENST00000523949.1 | chr8 | 26363146 | 26363177 | 26362823 | 26363152 | + | protein_coding |
| ENST00000522362.1 | chr8 | 26363146 | 26363177 | 26362217 | 26366759 | - | protein_coding |
| ENST00000517322.1 | chr8 | 26922393 | 26922414 | 26921513 | 26922795 | - | processed_pseudogene |
| ENST00000365541.1 | chr8 | 28387989 | 28388000 | 28387928 | 28388036 | + | rRNA |
| ENST00000240093.3 | chr8 | 28424030 | 28424061 | 28420315 | 28431775 | + | protein_coding |
| ENST00000221138.4 | chr8 | 30643261 | 30643292 | 30643126 | 30643823 | - | protein_coding |
| ENST00000518564.1 | chr8 | 30643261 | 30643292 | 30643132 | 30643505 | - | protein_coding |
| ENST00000406655.1 | chr8 | 30643261 | 30643292 | 30643137 | 30643823 | - | protein_coding |
| ENST00000221138.4 | chr8 | 30657113 | 30657144 | 30657062 | 30657271 | - | protein_coding |
| ENST00000406655.1 | chr8 | 30657113 | 30657144 | 30657062 | 30657271 | - | protein_coding |
| ENST00000518243.1 | chr8 | 30657113 | 30657144 | 30657062 | 30657164 | - | protein_coding |
| ENST00000520056.1 | chr8 | 30657113 | 30657144 | 30657062 | 30657271 | - | protein_coding |
| ENST00000520500.1 | chr8 | 30657113 | 30657144 | 30657097 | 30657271 | - | processed_transcript |
| ENST00000256246.2 | chr8 | 30694843 | 30694874 | 30694319 | 30695577 | - | protein_coding |
| ENST00000331569.4 | chr8 | 37555743 | 37555774 | 37554663 | 37557537 | + | protein_coding |
| ENST00000331569.4 | chr8 | 37556106 | 37556137 | 37554663 | 37557537 | + | protein_coding |
| ENST00000397235.1 | chr8 | 37556106 | 37556137 | 37555999 | 37556192 | + | protein_coding |
| ENST00000331569.4 | chr8 | 37556357 | 37556401 | 37554663 | 37557537 | + | protein_coding |
| ENST00000487273.2 | chr8 | 38961732 | 38961763 | 38961126 | 38962520 | + | protein_coding |
| ENST00000481873.3 | chr8 | 38961732 | 38961763 | 38961126 | 38962656 | + | nonsense_mediated_decay |
| ENST00000468065.1 | chr8 | 38961732 | 38961763 | 38961126 | 38962663 | + | nonsense_mediated_decay |
| ENST00000379917.3 | chr8 | 38961732 | 38961763 | 38961126 | 38962663 | + | nonsense_mediated_decay |
| ENST00000254250.3 | chr8 | 42693054 | 42693085 | 42691817 | 42693479 | - | protein_coding |
| ENST00000345117.2 | chr8 | 42693054 | 42693085 | 42691819 | 42693479 | - | protein_coding |
| ENST00000517759.1 | chr8 | 42981517 | 42981548 | 42981138 | 42981540 | - | processed_pseudogene |
| ENST00000498323.1 | chr8 | 46951839 | 46951866 | 46951571 | 46952175 | - | rRNA_pseudogene |
| ENST00000360540.5 | chr8 | 52732392 | 52732423 | 52730140 | 52733278 | - | protein_coding |
| ENST00000544451.1 | chr8 | 52732392 | 52732423 | 52730140 | 52733278 | - | protein_coding |
| ENST00000260102.4 | chr8 | 55060410 | 55060442 | 55059942 | 55060461 | + | protein_coding |
| ENST00000522521.1 | chr8 | 55060410 | 55060442 | 55059942 | 55060418 | + | nonsense_mediated_decay |
| ENST00000498581.1 | chr8 | 56755034 | 56755067 | 56755033 | 56755341 | - | rRNA_pseudogene |
| ENST00000316981.3 | chr8 | 57074313 | 57074344 | 57073463 | 57080062 | - | protein_coding |
| ENST00000260130.4 | chr8 | 59494467 | 59494470 | 59494245 | 59495419 | + | protein_coding |
| ENST00000422546.2 | chr8 | 59494467 | 59494470 | 59494245 | 59495418 | + | protein_coding |
| ENST00000447182.2 | chr8 | 59494467 | 59494470 | 59494245 | 59495405 | + | protein_coding |
| ENST00000413219.2 | chr8 | 59494467 | 59494470 | 59494245 | 59495389 | + | protein_coding |
| ENST00000424270.2 | chr8 | 59494467 | 59494470 | 59494245 | 59494615 | + | protein_coding |
| ENST00000523483.1 | chr8 | 59494467 | 59494470 | 59494245 | 59495417 | + | protein_coding |
| ENST00000520168.1 | chr8 | 59494467 | 59494470 | 59494245 | 59494612 | + | protein_coding |
| ENST00000517961.1 | chr8 | 67025669 | 67025692 | 67025522 | 67025831 | + | protein_coding |
| ENST00000310421.4 | chr8 | 67546740 | 67546768 | 67540722 | 67547607 | - | protein_coding |
| ENST00000483850.1 | chr8 | 68497701 | 68497734 | 68497633 | 68497703 | - | Mt_tRNA_pseudogene |
| ENST00000466170.1 | chr8 | 68497701 | 68497734 | 68497704 | 68497762 | - | Mt_tRNA_pseudogene |
| ENST00000491379.1 | chr8 | 69629526 | 69629557 | 69629526 | 69629594 | + | tRNA_pseudogene |
| ENST00000452400.2 | chr8 | 71068902 | 71068933 | 71068206 | 71069475 | - | protein_coding |
| ENST00000518287.1 | chr8 | 71068902 | 71068933 | 71068206 | 71069475 | - | nonsense_mediated_decay |
| ENST00000517608.1 | chr8 | 74701544 | 74701577 | 74698455 | 74706350 | - | protein_coding |
| ENST00000312184.5 | chr8 | 74893849 | 74893880 | 74893390 | 74895018 | + | protein_coding |
| ENST00000517662.1 | chr8 | 75515550 | 75515581 | 75515479 | 75516479 | - | processed_pseudogene |
| ENST00000521891.1 | chr8 | 77778143 | 77778175 | 77775330 | 77779516 | + | protein_coding |
| ENST00000399468.3 | chr8 | 77778143 | 77778175 | 77775330 | 77779520 | + | protein_coding |
| ENST00000455469.2 | chr8 | 77778143 | 77778175 | 77775330 | 77779520 | + | protein_coding |
| ENST00000518282.1 | chr8 | 77778143 | 77778175 | 77775330 | 77778159 | + | protein_coding |
| ENST00000379091.4 | chr8 | 81431841 | 81431872 | 81431458 | 81432039 | + | protein_coding |
| ENST00000430430.1 | chr8 | 81431841 | 81431872 | 81431458 | 81438500 | + | protein_coding |
| ENST00000455036.2 | chr8 | 81431841 | 81431872 | 81431458 | 81434007 | + | protein_coding |
| ENST00000426744.1 | chr8 | 81431841 | 81431872 | 81431458 | 81434610 | + | protein_coding |
| ENST00000519370.1 | chr8 | 81431841 | 81431872 | 81431458 | 81432370 | + | protein_coding |
| ENST00000430430.1 | chr8 | 81433008 | 81433039 | 81431458 | 81438500 | + | protein_coding |
| ENST00000455036.2 | chr8 | 81433008 | 81433039 | 81431458 | 81434007 | + | protein_coding |
| ENST00000426744.1 | chr8 | 81433008 | 81433039 | 81431458 | 81434610 | + | protein_coding |
| ENST00000220597.3 | chr8 | 81888674 | 81888705 | 81886402 | 81889141 | - | protein_coding |
| ENST00000418930.2 | chr8 | 86126501 | 86126530 | 86125988 | 86126753 | + | protein_coding |
| ENST00000256117.5 | chr8 | 86126501 | 86126530 | 86125988 | 86126751 | + | protein_coding |
| ENST00000416274.2 | chr8 | 86126501 | 86126530 | 86125988 | 86126750 | + | protein_coding |
| ENST00000519128.1 | chr8 | 86126501 | 86126530 | 86124392 | 86128536 | + | processed_transcript |
| ENST00000521429.1 | chr8 | 86126501 | 86126530 | 86125988 | 86126725 | + | protein_coding |
| ENST00000520225.1 | chr8 | 86126501 | 86126530 | 86126395 | 86126753 | + | protein_coding |
| ENST00000417663.2 | chr8 | 86126501 | 86126530 | 86126311 | 86126860 | - | protein_coding |
| ENST00000431163.2 | chr8 | 86126501 | 86126530 | 86126312 | 86126860 | - | protein_coding |
| ENST00000520099.1 | chr8 | 92130581 | 92130622 | 92130452 | 92130780 | + | nonsense_mediated_decay |
| ENST00000516719.1 | chr8 | 93179723 | 93179766 | 93179692 | 93179785 | - | miRNA |
| ENST00000399300.2 | chr8 | 94744448 | 94744479 | 94741584 | 94748666 | - | protein_coding |
| ENST00000517700.1 | chr8 | 94744448 | 94744479 | 94742492 | 94746416 | - | protein_coding |
| ENST00000399300.2 | chr8 | 94745275 | 94745306 | 94741584 | 94748666 | - | protein_coding |
| ENST00000517700.1 | chr8 | 94745275 | 94745306 | 94742492 | 94746416 | - | protein_coding |
| ENST00000315367.3 | chr8 | 96168370 | 96168401 | 96166259 | 96168912 | + | protein_coding |
| ENST00000401250.1 | chr8 | 100549014 | 100549045 | 100549014 | 100549089 | - | miRNA |
| ENST00000522754.1 | chr8 | 102705546 | 102705579 | 102705019 | 102706617 | - | retained_intron |
| ENST00000285407.6 | chr8 | 103661051 | 103661082 | 103661007 | 103662619 | - | protein_coding |
| ENST00000395884.3 | chr8 | 103661051 | 103661082 | 103661014 | 103662619 | - | protein_coding |
| ENST00000239690.4 | chr8 | 110254407 | 110254438 | 110253148 | 110255530 | - | protein_coding |
| ENST00000466509.1 | chr8 | 111117520 | 111117558 | 111117305 | 111118193 | - | processed_pseudogene |
| ENST00000395715.3 | chr8 | 116617112 | 116617142 | 116616100 | 116617229 | - | protein_coding |
| ENST00000220888.5 | chr8 | 116617112 | 116617142 | 116616100 | 116617229 | - | protein_coding |
| ENST00000517323.1 | chr8 | 116617112 | 116617142 | 116616878 | 116617229 | - | nonsense_mediated_decay |
| ENST00000519076.1 | chr8 | 116617112 | 116617142 | 116616878 | 116617229 | - | protein_coding |
| ENST00000520276.1 | chr8 | 116617112 | 116617142 | 116616100 | 116617229 | - | protein_coding |
| ENST00000519674.1 | chr8 | 116617112 | 116617142 | 116616100 | 116617229 | - | protein_coding |
| ENST00000471086.1 | chr8 | 120874970 | 120874978 | 120874903 | 120875102 | + | scRNA_pseudogene |
| ENST00000314393.4 | chr8 | 123986394 | 123986424 | 123985482 | 123986750 | + | protein_coding |
| ENST00000362808.1 | chr8 | 124057016 | 124057050 | 124056957 | 124057052 | + | misc_RNA |
| ENST00000522746.1 | chr8 | 130854380 | 130854411 | 130853716 | 130854451 | - | protein_coding |
| ENST00000523509.1 | chr8 | 130854380 | 130854411 | 130853716 | 130854451 | - | protein_coding |
| ENST00000401979.2 | chr8 | 130854380 | 130854411 | 130853716 | 130854451 | - | protein_coding |
| ENST00000519110.1 | chr8 | 130854380 | 130854411 | 130853716 | 130854451 | - | protein_coding |
| ENST00000523288.1 | chr8 | 130854380 | 130854411 | 130853716 | 130854451 | - | retained_intron |
| ENST00000522250.1 | chr8 | 130854380 | 130854411 | 130853730 | 130854451 | - | protein_coding |
| ENST00000519824.1 | chr8 | 130854380 | 130854411 | 130853730 | 130854451 | - | protein_coding |
| ENST00000517654.1 | chr8 | 130854380 | 130854411 | 130853730 | 130854451 | - | protein_coding |
| ENST00000519540.1 | chr8 | 130854380 | 130854411 | 130853743 | 130854451 | - | protein_coding |
| ENST00000522941.1 | chr8 | 130854380 | 130854411 | 130853749 | 130854451 | - | protein_coding |
| ENST00000520887.1 | chr8 | 130854380 | 130854411 | 130854326 | 130854451 | - | retained_intron |
| ENST00000362283.1 | chr8 | 135817152 | 135817183 | 135817119 | 135817188 | - | miRNA |
| ENST00000220592.5 | chr8 | 141569537 | 141569550 | 141569494 | 141569628 | - | protein_coding |
| ENST00000519980.1 | chr8 | 141569537 | 141569550 | 141569494 | 141569628 | - | protein_coding |
| ENST00000523609.1 | chr8 | 141569537 | 141569550 | 141569494 | 141569628 | - | nonsense_mediated_decay |
| ENST00000220592.5 | chr8 | 141582978 | 141583009 | 141582911 | 141583031 | - | protein_coding |
| ENST00000519980.1 | chr8 | 141582978 | 141583009 | 141582911 | 141583031 | - | protein_coding |
| ENST00000523609.1 | chr8 | 141582978 | 141583009 | 141582911 | 141583031 | - | nonsense_mediated_decay |
| ENST00000517293.1 | chr8 | 141582978 | 141583009 | 141582911 | 141583031 | - | processed_transcript |
| ENST00000524328.1 | chr8 | 141582978 | 141583009 | 141583001 | 141583031 | - | protein_coding |
| ENST00000349721.2 | chr9 | 2193493 | 2193525 | 2192704 | 2193624 | + | protein_coding |
| ENST00000357248.2 | chr9 | 2193493 | 2193525 | 2192704 | 2193624 | + | protein_coding |
| ENST00000382203.1 | chr9 | 2193493 | 2193525 | 2192704 | 2193624 | + | protein_coding |
| ENST00000382194.1 | chr9 | 2193493 | 2193525 | 2192704 | 2193577 | + | protein_coding |
| ENST00000302401.3 | chr9 | 2193493 | 2193525 | 2192704 | 2193621 | + | protein_coding |
| ENST00000382185.1 | chr9 | 2193493 | 2193525 | 2192704 | 2193571 | + | protein_coding |
| ENST00000382182.1 | chr9 | 2193493 | 2193525 | 2192704 | 2193577 | + | protein_coding |
| ENST00000259569.5 | chr9 | 6012239 | 6012270 | 6011043 | 6015618 | - | protein_coding |
| ENST00000380338.4 | chr9 | 20343184 | 20343216 | 20341663 | 20346572 | - | protein_coding |
| ENST00000401147.1 | chr9 | 28863624 | 28863655 | 28863624 | 28863704 | - | miRNA |
| ENST00000330899.4 | chr9 | 33039457 | 33039488 | 33038683 | 33039905 | + | protein_coding |
| ENST00000379731.4 | chr9 | 33111252 | 33111281 | 33110635 | 33113584 | - | protein_coding |
| ENST00000361264.3 | chr9 | 34088435 | 34088466 | 34086522 | 34088506 | - | protein_coding |
| ENST00000363271.1 | chr9 | 34194814 | 34194846 | 34194810 | 34194905 | - | misc_RNA |
| ENST00000363046.1 | chr9 | 35657743 | 35657778 | 35657751 | 35658014 | - | misc_RNA |
| ENST00000363046.1 | chr9 | 35657985 | 35657990 | 35657751 | 35658014 | - | misc_RNA |
| ENST00000410533.1 | chr9 | 72803488 | 72803520 | 72803487 | 72803666 | - | snRNA |
| ENST00000459091.1 | chr9 | 72926528 | 72926552 | 72926523 | 72926617 | - | misc_RNA |
| ENST00000516350.1 | chr9 | 77462481 | 77462520 | 77462422 | 77462517 | + | misc_RNA |
| ENST00000463508.1 | chr9 | 79186662 | 79186717 | 79186649 | 79186950 | + | rRNA_pseudogene |
| ENST00000463508.1 | chr9 | 79186804 | 79186885 | 79186649 | 79186950 | + | rRNA_pseudogene |
| ENST00000463508.1 | chr9 | 79186914 | 79186953 | 79186649 | 79186950 | + | rRNA_pseudogene |
| ENST00000286548.4 | chr9 | 80335357 | 80335388 | 80331003 | 80336429 | - | protein_coding |
| ENST00000421149.2 | chr9 | 80944234 | 80944265 | 80943897 | 80944340 | + | protein_coding |
| ENST00000347159.2 | chr9 | 80944234 | 80944265 | 80943897 | 80944370 | + | protein_coding |
| ENST00000376588.3 | chr9 | 80944234 | 80944265 | 80943897 | 80945009 | + | protein_coding |
| ENST00000376344.3 | chr9 | 86553773 | 86553804 | 86553226 | 86554658 | - | protein_coding |
| ENST00000314700.1 | chr9 | 86553773 | 86553804 | 86553229 | 86554658 | - | protein_coding |
| ENST00000481820.1 | chr9 | 86584689 | 86584749 | 86584411 | 86585246 | - | processed_transcript |
| ENST00000384871.1 | chr9 | 86584689 | 86584749 | 86584663 | 86584772 | - | miRNA |
| ENST00000298743.7 | chr9 | 89559867 | 89559899 | 89559279 | 89562104 | - | protein_coding |
| ENST00000356884.6 | chr9 | 95481020 | 95481051 | 95480821 | 95481864 | - | protein_coding |
| ENST00000375512.3 | chr9 | 95481020 | 95481051 | 95480821 | 95481864 | - | protein_coding |
| ENST00000277165.5 | chr9 | 96326729 | 96326760 | 96326511 | 96328389 | + | protein_coding |
| ENST00000333936.5 | chr9 | 96326729 | 96326760 | 96326511 | 96328397 | + | protein_coding |
| ENST00000340893.4 | chr9 | 96326729 | 96326760 | 96326511 | 96328397 | + | protein_coding |
| ENST00000427765.1 | chr9 | 96326729 | 96326760 | 96326511 | 96327047 | + | protein_coding |
| ENST00000277165.5 | chr9 | 96327300 | 96327331 | 96326511 | 96328389 | + | protein_coding |
| ENST00000333936.5 | chr9 | 96327300 | 96327331 | 96326511 | 96328397 | + | protein_coding |
| ENST00000340893.4 | chr9 | 96327300 | 96327331 | 96326511 | 96328397 | + | protein_coding |
| ENST00000362295.1 | chr9 | 96938244 | 96938278 | 96938239 | 96938318 | + | miRNA |
| ENST00000362295.1 | chr9 | 96938291 | 96938297 | 96938239 | 96938318 | + | miRNA |
| ENST00000362202.1 | chr9 | 96938635 | 96938666 | 96938629 | 96938715 | + | miRNA |
| ENST00000362202.1 | chr9 | 96938687 | 96938710 | 96938629 | 96938715 | + | miRNA |
| ENST00000362263.1 | chr9 | 96941123 | 96941154 | 96941116 | 96941202 | + | miRNA |
| ENST00000297979.5 | chr9 | 97767410 | 97767441 | 97767440 | 97767502 | + | protein_coding |
| ENST00000375315.2 | chr9 | 97767410 | 97767441 | 97767440 | 97767502 | + | protein_coding |
| ENST00000424143.1 | chr9 | 97767410 | 97767441 | 97767440 | 97767502 | + | protein_coding |
| ENST00000428313.1 | chr9 | 97767410 | 97767441 | 97767440 | 97767502 | + | protein_coding |
| ENST00000375316.2 | chr9 | 97767410 | 97767441 | 97767440 | 97767502 | + | protein_coding |
| ENST00000462125.1 | chr9 | 97767410 | 97767441 | 97767440 | 97767502 | + | processed_transcript |
| ENST00000479161.1 | chr9 | 97767410 | 97767441 | 97767440 | 97767502 | + | processed_transcript |
| ENST00000451893.1 | chr9 | 97767410 | 97767441 | 97767440 | 97767502 | + | protein_coding |
| ENST00000460573.1 | chr9 | 97767410 | 97767441 | 97767440 | 97767502 | + | processed_transcript |
| ENST00000478603.1 | chr9 | 97767410 | 97767441 | 97767440 | 97767502 | + | processed_transcript |
| ENST00000425634.2 | chr9 | 97767410 | 97767441 | 97767440 | 97767502 | + | protein_coding |
| ENST00000471978.1 | chr9 | 97767410 | 97767441 | 97767440 | 97767502 | + | processed_transcript |
| ENST00000445181.1 | chr9 | 97767410 | 97767441 | 97767440 | 97767502 | + | protein_coding |
| ENST00000375314.1 | chr9 | 97767410 | 97767441 | 97767440 | 97767502 | + | protein_coding |
| ENST00000478473.1 | chr9 | 97767410 | 97767441 | 97767440 | 97767502 | + | nonsense_mediated_decay |
| ENST00000496567.1 | chr9 | 97767410 | 97767441 | 97767440 | 97767502 | + | processed_transcript |
| ENST00000463372.1 | chr9 | 97767410 | 97767441 | 97767379 | 97767502 | + | processed_transcript |
| ENST00000384832.1 | chr9 | 97847547 | 97847578 | 97847490 | 97847586 | + | miRNA |
| ENST00000385129.1 | chr9 | 97847787 | 97847818 | 97847727 | 97847823 | + | miRNA |
| ENST00000478473.1 | chr9 | 97848346 | 97848377 | 97848212 | 97848401 | + | nonsense_mediated_decay |
| ENST00000384885.1 | chr9 | 97848346 | 97848377 | 97848303 | 97848370 | + | miRNA |
| ENST00000269395.5 | chr9 | 107090858 | 107090896 | 107090607 | 107091470 | - | processed_pseudogene |
| ENST00000374723.1 | chr9 | 108061547 | 108061578 | 108061501 | 108061590 | + | protein_coding |
| ENST00000470972.1 | chr9 | 108061547 | 108061578 | 108061501 | 108061590 | + | nonsense_mediated_decay |
| ENST00000374720.3 | chr9 | 108061547 | 108061578 | 108061501 | 108061590 | + | protein_coding |
| ENST00000374724.1 | chr9 | 108061547 | 108061578 | 108061501 | 108061590 | + | protein_coding |
| ENST00000415694.1 | chr9 | 109933149 | 109933173 | 109931850 | 109933429 | - | processed_pseudogene |
| ENST00000374279.3 | chr9 | 114696363 | 114696394 | 114695107 | 114697649 | + | protein_coding |
| ENST00000385079.1 | chr9 | 123007251 | 123007282 | 123007257 | 123007328 | - | miRNA |
| ENST00000384863.1 | chr9 | 127454759 | 127454790 | 127454721 | 127454830 | + | miRNA |
| ENST00000384863.1 | chr9 | 127454792 | 127454823 | 127454721 | 127454830 | + | miRNA |
| ENST00000385004.1 | chr9 | 127456004 | 127456035 | 127455989 | 127456077 | + | miRNA |
| ENST00000324460.6 | chr9 | 127998576 | 127998607 | 127997132 | 127999433 | - | protein_coding |
| ENST00000384849.1 | chr9 | 131007053 | 131007084 | 131007000 | 131007109 | - | miRNA |
| ENST00000372692.4 | chr9 | 131457070 | 131457101 | 131456920 | 131458679 | + | protein_coding |
| ENST00000322030.8 | chr9 | 131457070 | 131457101 | 131456920 | 131458667 | + | protein_coding |
| ENST00000494141.1 | chr9 | 131457070 | 131457101 | 131456920 | 131457792 | + | processed_transcript |
| ENST00000372688.4 | chr9 | 131457070 | 131457101 | 131456920 | 131457421 | + | protein_coding |
| ENST00000372686.5 | chr9 | 131457070 | 131457101 | 131456920 | 131457173 | + | protein_coding |
| ENST00000372447.3 | chr9 | 132589890 | 132589922 | 132589569 | 132590531 | - | protein_coding |
| ENST00000435276.2 | chr9 | 132589890 | 132589922 | 132589576 | 132590531 | - | protein_coding |
| ENST00000372146.4 | chr9 | 135554266 | 135554275 | 135553364 | 135555190 | + | protein_coding |
| ENST00000483873.1 | chr9 | 135554266 | 135554275 | 135554265 | 135554697 | + | nonsense_mediated_decay |
| ENST00000372146.4 | chr9 | 135564900 | 135564931 | 135564268 | 135570342 | + | protein_coding |
| ENST00000483873.1 | chr9 | 135564900 | 135564931 | 135564268 | 135570342 | + | nonsense_mediated_decay |
| ENST00000371785.1 | chr9 | 138395778 | 138395809 | 138395388 | 138396519 | + | protein_coding |
| ENST00000241600.5 | chr9 | 138395778 | 138395809 | 138395388 | 138396516 | + | protein_coding |
| ENST00000488610.1 | chr9 | 138395778 | 138395809 | 138395388 | 138395806 | + | processed_transcript |
| ENST00000485333.1 | chr9 | 138395778 | 138395809 | 138395388 | 138395822 | + | processed_transcript |
| ENST00000472852.1 | chr9 | 138395778 | 138395809 | 138395388 | 138395801 | + | processed_transcript |
| ENST00000453385.1 | chr9 | 138395778 | 138395809 | 138395388 | 138395856 | + | protein_coding |
| ENST00000472946.1 | chr9 | 138395778 | 138395809 | 138395388 | 138395832 | + | processed_transcript |
| ENST00000389532.4 | chr9 | 138701868 | 138701899 | 138700333 | 138703457 | - | protein_coding |
| ENST00000312405.6 | chr9 | 138701868 | 138701899 | 138701427 | 138703457 | - | protein_coding |
| ENST00000362291.1 | chr9 | 139565105 | 139565136 | 139565054 | 139565138 | + | miRNA |
| ENST00000465160.1 | chr9 | 140287075 | 140287106 | 140286271 | 140287687 | - | protein_coding |
| ENST00000384960.1 | chr9 | 140732886 | 140732917 | 140732871 | 140732968 | + | miRNA |
| ENST00000387314.1 | chrM | 579 | 610 | 577 | 647 | + | Mt_tRNA |
| ENST00000387314.1 | chrM | 614 | 654 | 577 | 647 | + | Mt_tRNA |
| ENST00000389680.2 | chrM | 614 | 654 | 648 | 1601 | + | Mt_rRNA |
| ENST00000389680.2 | chrM | 712 | 760 | 648 | 1601 | + | Mt_rRNA |
| ENST00000389680.2 | chrM | 960 | 991 | 648 | 1601 | + | Mt_rRNA |
| ENST00000389680.2 | chrM | 1005 | 1036 | 648 | 1601 | + | Mt_rRNA |
| ENST00000389680.2 | chrM | 1273 | 1288 | 648 | 1601 | + | Mt_rRNA |
| ENST00000389680.2 | chrM | 1319 | 1343 | 648 | 1601 | + | Mt_rRNA |
| ENST00000389680.2 | chrM | 1367 | 1389 | 648 | 1601 | + | Mt_rRNA |
| ENST00000389680.2 | chrM | 1391 | 1391 | 648 | 1601 | + | Mt_rRNA |
| ENST00000389680.2 | chrM | 1394 | 1398 | 648 | 1601 | + | Mt_rRNA |
| ENST00000389680.2 | chrM | 1403 | 1458 | 648 | 1601 | + | Mt_rRNA |
| ENST00000389680.2 | chrM | 1460 | 1469 | 648 | 1601 | + | Mt_rRNA |
| ENST00000389680.2 | chrM | 1520 | 1551 | 648 | 1601 | + | Mt_rRNA |
| ENST00000387342.1 | chrM | 1617 | 1635 | 1602 | 1670 | + | Mt_tRNA |
| ENST00000387342.1 | chrM | 1637 | 1705 | 1602 | 1670 | + | Mt_tRNA |
| ENST00000387347.2 | chrM | 1637 | 1705 | 1671 | 3229 | + | Mt_rRNA |
| ENST00000387347.2 | chrM | 1758 | 1759 | 1671 | 3229 | + | Mt_rRNA |
| ENST00000387347.2 | chrM | 1824 | 1855 | 1671 | 3229 | + | Mt_rRNA |
| ENST00000387347.2 | chrM | 1871 | 1909 | 1671 | 3229 | + | Mt_rRNA |
| ENST00000387347.2 | chrM | 1963 | 1991 | 1671 | 3229 | + | Mt_rRNA |
| ENST00000387347.2 | chrM | 2141 | 2144 | 1671 | 3229 | + | Mt_rRNA |
| ENST00000387347.2 | chrM | 2150 | 2165 | 1671 | 3229 | + | Mt_rRNA |
| ENST00000387347.2 | chrM | 2247 | 2298 | 1671 | 3229 | + | Mt_rRNA |
| ENST00000387347.2 | chrM | 2340 | 2367 | 1671 | 3229 | + | Mt_rRNA |
| ENST00000387347.2 | chrM | 2400 | 2427 | 1671 | 3229 | + | Mt_rRNA |
| ENST00000387347.2 | chrM | 2761 | 2769 | 1671 | 3229 | + | Mt_rRNA |
| ENST00000387347.2 | chrM | 2788 | 2819 | 1671 | 3229 | + | Mt_rRNA |
| ENST00000387347.2 | chrM | 2864 | 2899 | 1671 | 3229 | + | Mt_rRNA |
| ENST00000387347.2 | chrM | 2987 | 2999 | 1671 | 3229 | + | Mt_rRNA |
| ENST00000387347.2 | chrM | 3070 | 3084 | 1671 | 3229 | + | Mt_rRNA |
| ENST00000387347.2 | chrM | 3088 | 3099 | 1671 | 3229 | + | Mt_rRNA |
| ENST00000387347.2 | chrM | 3121 | 3124 | 1671 | 3229 | + | Mt_rRNA |
| ENST00000387347.2 | chrM | 3128 | 3152 | 1671 | 3229 | + | Mt_rRNA |
| ENST00000387347.2 | chrM | 3195 | 3226 | 1671 | 3229 | + | Mt_rRNA |
| ENST00000386347.1 | chrM | 3248 | 3317 | 3230 | 3304 | + | Mt_tRNA |
| ENST00000361390.2 | chrM | 3248 | 3317 | 3307 | 4262 | + | protein_coding |
| ENST00000361390.2 | chrM | 3498 | 3529 | 3307 | 4262 | + | protein_coding |
| ENST00000387365.1 | chrM | 4302 | 4358 | 4263 | 4331 | + | Mt_tRNA |
| ENST00000387372.1 | chrM | 4302 | 4358 | 4329 | 4400 | - | Mt_tRNA |
| ENST00000387372.1 | chrM | 4370 | 4401 | 4329 | 4400 | - | Mt_tRNA |
| ENST00000387377.1 | chrM | 4439 | 4471 | 4402 | 4469 | + | Mt_tRNA |
| ENST00000361453.3 | chrM | 4439 | 4471 | 4470 | 5511 | + | protein_coding |
| ENST00000387392.1 | chrM | 5581 | 5613 | 5587 | 5655 | - | Mt_tRNA |
| ENST00000387400.1 | chrM | 5699 | 5730 | 5657 | 5729 | - | Mt_tRNA |
| ENST00000361624.2 | chrM | 6213 | 6244 | 5904 | 7445 | + | protein_coding |
| ENST00000361624.2 | chrM | 6252 | 6262 | 5904 | 7445 | + | protein_coding |
| ENST00000361624.2 | chrM | 6425 | 6459 | 5904 | 7445 | + | protein_coding |
| ENST00000361624.2 | chrM | 6475 | 6489 | 5904 | 7445 | + | protein_coding |
| ENST00000361624.2 | chrM | 6805 | 6840 | 5904 | 7445 | + | protein_coding |
| ENST00000387416.2 | chrM | 7484 | 7515 | 7446 | 7514 | - | Mt_tRNA |
| ENST00000387419.1 | chrM | 7519 | 7550 | 7518 | 7585 | + | Mt_tRNA |
| ENST00000361739.1 | chrM | 8002 | 8034 | 7586 | 8269 | + | protein_coding |
| ENST00000361227.2 | chrM | 10073 | 10092 | 10059 | 10404 | + | protein_coding |
| ENST00000361227.2 | chrM | 10182 | 10213 | 10059 | 10404 | + | protein_coding |
| ENST00000387439.1 | chrM | 10406 | 10437 | 10405 | 10469 | + | Mt_tRNA |
| ENST00000387439.1 | chrM | 10439 | 10470 | 10405 | 10469 | + | Mt_tRNA |
| ENST00000361335.1 | chrM | 10495 | 10502 | 10470 | 10766 | + | protein_coding |
| ENST00000361335.1 | chrM | 10753 | 10771 | 10470 | 10766 | + | protein_coding |
| ENST00000361381.2 | chrM | 10753 | 10771 | 10760 | 12137 | + | protein_coding |
| ENST00000361381.2 | chrM | 10954 | 10968 | 10760 | 12137 | + | protein_coding |
| ENST00000361381.2 | chrM | 10980 | 10985 | 10760 | 12137 | + | protein_coding |
| ENST00000361381.2 | chrM | 11316 | 11331 | 10760 | 12137 | + | protein_coding |
| ENST00000361381.2 | chrM | 11593 | 11628 | 10760 | 12137 | + | protein_coding |
| ENST00000361381.2 | chrM | 11801 | 11832 | 10760 | 12137 | + | protein_coding |
| ENST00000387441.1 | chrM | 12140 | 12225 | 12138 | 12206 | + | Mt_tRNA |
| ENST00000387449.1 | chrM | 12140 | 12225 | 12207 | 12265 | + | Mt_tRNA |
| ENST00000387449.1 | chrM | 12228 | 12267 | 12207 | 12265 | + | Mt_tRNA |
| ENST00000387456.1 | chrM | 12228 | 12267 | 12266 | 12336 | + | Mt_tRNA |
| ENST00000361567.2 | chrM | 12338 | 12369 | 12337 | 14148 | + | protein_coding |
| ENST00000361567.2 | chrM | 12473 | 12480 | 12337 | 14148 | + | protein_coding |
| ENST00000361567.2 | chrM | 12726 | 12757 | 12337 | 14148 | + | protein_coding |
| ENST00000361567.2 | chrM | 13211 | 13242 | 12337 | 14148 | + | protein_coding |
| ENST00000361567.2 | chrM | 13245 | 13258 | 12337 | 14148 | + | protein_coding |
| ENST00000361567.2 | chrM | 13440 | 13452 | 12337 | 14148 | + | protein_coding |
| ENST00000361567.2 | chrM | 13560 | 13591 | 12337 | 14148 | + | protein_coding |
| ENST00000361567.2 | chrM | 13972 | 14003 | 12337 | 14148 | + | protein_coding |
| ENST00000361681.2 | chrM | 14667 | 14714 | 14149 | 14673 | - | protein_coding |
| ENST00000387459.1 | chrM | 14667 | 14714 | 14674 | 14742 | - | Mt_tRNA |
| ENST00000387460.2 | chrM | 15889 | 16026 | 15888 | 15953 | + | Mt_tRNA |
| ENST00000381578.1 | chrX | 606124 | 606158 | 605126 | 607558 | + | protein_coding |
| ENST00000554971.1 | chrX | 606124 | 606158 | 605126 | 606303 | + | protein_coding |
| ENST00000384896.1 | chrX | 8095021 | 8095052 | 8095006 | 8095102 | + | miRNA |
| ENST00000464506.1 | chrX | 13726859 | 13726890 | 13726840 | 13728625 | + | protein_coding |
| ENST00000243325.5 | chrX | 13726859 | 13726890 | 13726840 | 13727473 | + | processed_transcript |
| ENST00000380122.5 | chrX | 16861756 | 16861791 | 16859551 | 16862642 | + | protein_coding |
| ENST00000398155.4 | chrX | 16861756 | 16861791 | 16859551 | 16862640 | + | protein_coding |
| ENST00000485153.1 | chrX | 16861756 | 16861791 | 16859551 | 16862642 | + | processed_transcript |
| ENST00000380045.3 | chrX | 17772604 | 17772634 | 17771378 | 17773105 | + | protein_coding |
| ENST00000380043.3 | chrX | 17772604 | 17772634 | 17771378 | 17773105 | + | protein_coding |
| ENST00000398080.1 | chrX | 17772604 | 17772634 | 17771378 | 17773104 | + | protein_coding |
| ENST00000487842.1 | chrX | 17772604 | 17772634 | 17771378 | 17773105 | + | retained_intron |
| ENST00000328046.8 | chrX | 24004506 | 24004537 | 24001837 | 24007147 | - | protein_coding |
| ENST00000328046.8 | chrX | 24004805 | 24004838 | 24001837 | 24007147 | - | protein_coding |
| ENST00000328046.8 | chrX | 24005161 | 24005192 | 24001837 | 24007147 | - | protein_coding |
| ENST00000328046.8 | chrX | 24005861 | 24005893 | 24001837 | 24007147 | - | protein_coding |
| ENST00000328046.8 | chrX | 24024316 | 24024347 | 24024106 | 24024817 | - | protein_coding |
| ENST00000253039.4 | chrX | 24095142 | 24095172 | 24094839 | 24096088 | + | protein_coding |
| ENST00000364486.1 | chrX | 29000958 | 29000994 | 29000969 | 29001087 | - | rRNA |
| ENST00000424932.2 | chrX | 30635577 | 30635608 | 30635571 | 30635810 | + | processed_pseudogene |
| ENST00000383950.1 | chrX | 40352115 | 40352146 | 40352092 | 40352192 | + | misc_RNA |
| ENST00000362694.1 | chrX | 41152099 | 41152123 | 41152051 | 41152169 | + | rRNA |
| ENST00000302548.4 | chrX | 41589003 | 41589035 | 41586246 | 41589388 | + | protein_coding |
| ENST00000334516.4 | chrX | 44922718 | 44922749 | 44922667 | 44923062 | + | protein_coding |
| ENST00000377967.4 | chrX | 44922718 | 44922749 | 44922667 | 44923062 | + | protein_coding |
| ENST00000536777.1 | chrX | 44922718 | 44922749 | 44922667 | 44923062 | + | protein_coding |
| ENST00000382899.4 | chrX | 44922718 | 44922749 | 44922667 | 44923062 | + | protein_coding |
| ENST00000543216.1 | chrX | 44922718 | 44922749 | 44922667 | 44923062 | + | protein_coding |
| ENST00000542299.1 | chrX | 44922718 | 44922749 | 44922667 | 44923062 | + | protein_coding |
| ENST00000451692.1 | chrX | 44922718 | 44922749 | 44922667 | 44922921 | + | protein_coding |
| ENST00000414389.1 | chrX | 44922718 | 44922749 | 44922667 | 44923062 | + | protein_coding |
| ENST00000535688.1 | chrX | 44922718 | 44922749 | 44922667 | 44922855 | + | protein_coding |
| ENST00000433797.1 | chrX | 44922718 | 44922749 | 44922667 | 44923062 | + | protein_coding |
| ENST00000385135.1 | chrX | 45605599 | 45605630 | 45605585 | 45605694 | - | miRNA |
| ENST00000384992.1 | chrX | 45606431 | 45606462 | 45606421 | 45606530 | - | miRNA |
| ENST00000385235.1 | chrX | 49777864 | 49777895 | 49777849 | 49777945 | + | miRNA |
| ENST00000553557.1 | chrX | 53111748 | 53111771 | 53111549 | 53112487 | + | retained_intron |
| ENST00000375442.4 | chrX | 53111748 | 53111771 | 53111549 | 53112487 | + | protein_coding |
| ENST00000362552.1 | chrX | 53489149 | 53489182 | 53489150 | 53489251 | + | misc_RNA |
| ENST00000342160.3 | chrX | 53560151 | 53560182 | 53559057 | 53560372 | - | protein_coding |
| ENST00000427052.1 | chrX | 53560151 | 53560182 | 53559057 | 53560372 | - | protein_coding |
| ENST00000426907.1 | chrX | 53560151 | 53560182 | 53559057 | 53560372 | - | protein_coding |
| ENST00000262854.6 | chrX | 53560151 | 53560182 | 53559105 | 53560372 | - | protein_coding |
| ENST00000385277.1 | chrX | 53584197 | 53584228 | 53584153 | 53584235 | - | miRNA |
| ENST00000357988.5 | chrX | 53966797 | 53966828 | 53966613 | 53966949 | - | protein_coding |
| ENST00000338154.6 | chrX | 53966797 | 53966828 | 53966613 | 53966949 | - | protein_coding |
| ENST00000338946.6 | chrX | 53966797 | 53966828 | 53966613 | 53966949 | - | protein_coding |
| ENST00000396277.3 | chrX | 53966797 | 53966828 | 53966613 | 53966949 | - | protein_coding |
| ENST00000396282.2 | chrX | 53966797 | 53966828 | 53966613 | 53966949 | - | protein_coding |
| ENST00000470103.1 | chrX | 53966797 | 53966828 | 53966802 | 53966949 | - | retained_intron |
| ENST00000338222.5 | chrX | 56592248 | 56592261 | 56590026 | 56593443 | + | protein_coding |
| ENST00000535171.1 | chrX | 56592248 | 56592261 | 56591974 | 56592545 | + | protein_coding |
| ENST00000374888.1 | chrX | 57620960 | 57620991 | 57618269 | 57623906 | + | protein_coding |
| ENST00000358697.4 | chrX | 57934110 | 57934142 | 57931864 | 57937067 | - | protein_coding |
| ENST00000358697.4 | chrX | 57934357 | 57934388 | 57931864 | 57937067 | - | protein_coding |
| ENST00000457579.1 | chrX | 62103968 | 62103999 | 62102082 | 62104312 | - | processed_pseudogene |
| ENST00000364816.1 | chrX | 68892413 | 68892451 | 68892323 | 68892441 | + | rRNA |
| ENST00000535149.1 | chrX | 70519839 | 70519843 | 70519792 | 70521016 | + | protein_coding |
| ENST00000276079.8 | chrX | 70519839 | 70519843 | 70519792 | 70521018 | + | protein_coding |
| ENST00000373856.3 | chrX | 70519839 | 70519843 | 70519792 | 70520119 | + | protein_coding |
| ENST00000373841.1 | chrX | 70519839 | 70519843 | 70519792 | 70521018 | + | protein_coding |
| ENST00000474431.1 | chrX | 70519839 | 70519843 | 70519792 | 70520119 | + | processed_transcript |
| ENST00000490044.1 | chrX | 70519839 | 70519843 | 70519792 | 70521018 | + | processed_transcript |
| ENST00000473525.1 | chrX | 70519839 | 70519843 | 70519792 | 70520166 | + | processed_transcript |
| ENST00000362881.1 | chrX | 70710964 | 70710995 | 70710916 | 70711017 | - | misc_RNA |
| ENST00000316084.6 | chrX | 71492452 | 71492484 | 71491892 | 71492622 | - | protein_coding |
| ENST00000419932.1 | chrX | 108297376 | 108297444 | 108297361 | 108297792 | - | processed_pseudogene |
| ENST00000419932.1 | chrX | 108297449 | 108297495 | 108297361 | 108297792 | - | processed_pseudogene |
| ENST00000419932.1 | chrX | 108297503 | 108297572 | 108297361 | 108297792 | - | processed_pseudogene |
| ENST00000516235.1 | chrX | 108297503 | 108297572 | 108297553 | 108297626 | + | miRNA |
| ENST00000419932.1 | chrX | 108297582 | 108297589 | 108297361 | 108297792 | - | processed_pseudogene |
| ENST00000516235.1 | chrX | 108297582 | 108297589 | 108297553 | 108297626 | + | miRNA |
| ENST00000419932.1 | chrX | 108297604 | 108297677 | 108297361 | 108297792 | - | processed_pseudogene |
| ENST00000516235.1 | chrX | 108297604 | 108297677 | 108297553 | 108297626 | + | miRNA |
| ENST00000419932.1 | chrX | 108297724 | 108297776 | 108297361 | 108297792 | - | processed_pseudogene |
| ENST00000434365.1 | chrX | 114937678 | 114937709 | 114937131 | 114938440 | + | processed_pseudogene |
| ENST00000363421.1 | chrX | 117048277 | 117048308 | 117048207 | 117048308 | - | misc_RNA |
| ENST00000371558.2 | chrX | 118718289 | 118718320 | 118717090 | 118718381 | + | protein_coding |
| ENST00000371569.5 | chrX | 118718289 | 118718320 | 118717090 | 118718381 | + | protein_coding |
| ENST00000371527.1 | chrX | 118724304 | 118724335 | 118722300 | 118725272 | - | protein_coding |
| ENST00000304449.5 | chrX | 118724304 | 118724335 | 118722300 | 118725272 | - | protein_coding |
| ENST00000542113.1 | chrX | 118724304 | 118724335 | 118722304 | 118725272 | - | protein_coding |
| ENST00000276201.2 | chrX | 118979169 | 118979199 | 118979161 | 118979259 | - | protein_coding |
| ENST00000345865.2 | chrX | 118979169 | 118979199 | 118979161 | 118979259 | - | protein_coding |
| ENST00000439808.2 | chrX | 118979169 | 118979199 | 118979161 | 118979259 | - | protein_coding |
| ENST00000478840.1 | chrX | 118979169 | 118979199 | 118979161 | 118979217 | - | processed_transcript |
| ENST00000326624.2 | chrX | 119387877 | 119387902 | 119387269 | 119392253 | + | protein_coding |
| ENST00000540105.1 | chrX | 119387877 | 119387902 | 119387269 | 119392251 | + | protein_coding |
| ENST00000557385.1 | chrX | 119387877 | 119387902 | 119387269 | 119389289 | + | protein_coding |
| ENST00000326624.2 | chrX | 119389637 | 119389668 | 119387269 | 119392253 | + | protein_coding |
| ENST00000540105.1 | chrX | 119389637 | 119389668 | 119387269 | 119392251 | + | protein_coding |
| ENST00000371335.4 | chrX | 119571380 | 119571411 | 119570349 | 119573148 | - | protein_coding |
| ENST00000218089.9 | chrX | 123094453 | 123094497 | 123094369 | 123094716 | + | protein_coding |
| ENST00000455404.1 | chrX | 123094453 | 123094497 | 123094369 | 123094716 | + | protein_coding |
| ENST00000483575.1 | chrX | 123094453 | 123094497 | 123094410 | 123094716 | + | processed_transcript |
| ENST00000218089.9 | chrX | 123235123 | 123235154 | 123234424 | 123235348 | + | protein_coding |
| ENST00000371160.1 | chrX | 123235123 | 123235154 | 123234424 | 123236506 | + | protein_coding |
| ENST00000371157.3 | chrX | 123235123 | 123235154 | 123234424 | 123236506 | + | protein_coding |
| ENST00000385077.2 | chrX | 133303704 | 133303735 | 133303701 | 133303797 | - | miRNA |
| ENST00000384977.1 | chrX | 133303871 | 133303902 | 133303839 | 133303907 | - | miRNA |
| ENST00000518153.1 | chrX | 134185220 | 134185251 | 134184962 | 134185549 | - | processed_transcript |
| ENST00000370775.2 | chrX | 134185220 | 134185251 | 134184962 | 134186205 | - | protein_coding |
| ENST00000520964.1 | chrX | 134185220 | 134185251 | 134184970 | 134185506 | - | processed_transcript |
| ENST00000522309.1 | chrX | 134185220 | 134185251 | 134185088 | 134185398 | - | processed_transcript |
| ENST00000431446.2 | chrX | 135961191 | 135961222 | 135961176 | 135961282 | - | protein_coding |
| ENST00000464781.1 | chrX | 135961191 | 135961222 | 135961176 | 135961282 | - | processed_transcript |
| ENST00000320676.7 | chrX | 135961191 | 135961222 | 135961176 | 135961282 | - | protein_coding |
| ENST00000449161.2 | chrX | 135961191 | 135961222 | 135961176 | 135961219 | - | protein_coding |
| ENST00000315930.6 | chrX | 137714487 | 137714519 | 137713735 | 137715147 | - | protein_coding |
| ENST00000305414.4 | chrX | 137714487 | 137714519 | 137713735 | 137715147 | - | protein_coding |
| ENST00000441825.2 | chrX | 137714487 | 137714519 | 137713740 | 137715147 | - | protein_coding |
| ENST00000385065.1 | chrX | 137749886 | 137749941 | 137749872 | 137749954 | - | miRNA |
| ENST00000474745.1 | chrX | 139858492 | 139858510 | 139858393 | 139858607 | + | rRNA_pseudogene |
| ENST00000381779.4 | chrX | 142715897 | 142715943 | 142710596 | 142718974 | - | protein_coding |
| ENST00000401237.1 | chrX | 145109349 | 145109380 | 145109312 | 145109390 | - | miRNA |
| ENST00000384909.1 | chrX | 146353886 | 146353917 | 146353853 | 146353926 | - | miRNA |
| ENST00000330374.6 | chrX | 150574522 | 150574554 | 150573388 | 150577836 | + | protein_coding |
| ENST00000370353.3 | chrX | 150872329 | 150872384 | 150868978 | 150874396 | + | protein_coding |
| ENST00000486255.1 | chrX | 151127092 | 151127123 | 151124180 | 151128195 | - | retained_intron |
| ENST00000476016.1 | chrX | 151127092 | 151127123 | 151127118 | 151127621 | - | retained_intron |
| ENST00000384889.1 | chrX | 151127092 | 151127123 | 151127050 | 151127130 | - | miRNA |
| ENST00000474786.1 | chrX | 153628205 | 153628236 | 153628144 | 153628507 | + | retained_intron |
| ENST00000369817.2 | chrX | 153628205 | 153628236 | 153628144 | 153628282 | + | protein_coding |
| ENST00000424325.2 | chrX | 153628205 | 153628236 | 153628144 | 153628282 | + | protein_coding |
| ENST00000436473.1 | chrX | 153628205 | 153628236 | 153628144 | 153628282 | + | protein_coding |
| ENST00000344746.4 | chrX | 153628205 | 153628236 | 153628144 | 153628282 | + | protein_coding |
| ENST00000491035.1 | chrX | 153628205 | 153628236 | 153628144 | 153628967 | + | retained_intron |
| ENST00000489200.1 | chrX | 153628205 | 153628236 | 153628144 | 153629239 | + | retained_intron |
| ENST00000458500.1 | chrX | 153628205 | 153628236 | 153628144 | 153628282 | + | protein_coding |
| ENST00000485196.1 | chrX | 153628205 | 153628236 | 153628144 | 153628282 | + | retained_intron |
| ENST00000482732.1 | chrX | 153628205 | 153628236 | 153627679 | 153629254 | + | retained_intron |
| ENST00000492572.1 | chrX | 153628205 | 153628236 | 153628144 | 153629238 | + | retained_intron |
| ENST00000467168.1 | chrX | 153628205 | 153628236 | 153628144 | 153628282 | + | retained_intron |
| ENST00000406022.2 | chrX | 153628205 | 153628236 | 153628144 | 153628282 | + | protein_coding |
| ENST00000451365.1 | chrX | 153628205 | 153628236 | 153628144 | 153628282 | + | protein_coding |
| ENST00000427682.1 | chrX | 153628205 | 153628236 | 153628224 | 153628282 | + | protein_coding |
| ENST00000449494.1 | chrX | 153628205 | 153628236 | 153628224 | 153628282 | + | protein_coding |
| ENST00000428169.1 | chrX | 153628205 | 153628236 | 153628224 | 153628282 | + | protein_coding |
| ENST00000491154.1 | chrX | 153670832 | 153670863 | 153670458 | 153671812 | + | retained_intron |
| ENST00000460984.1 | chrX | 153670832 | 153670863 | 153670721 | 153671065 | + | retained_intron |
| ENST00000447750.2 | chrX | 153671070 | 153671101 | 153670867 | 153671814 | + | protein_coding |
| ENST00000369741.5 | chrX | 153671070 | 153671101 | 153670867 | 153671159 | + | protein_coding |
| ENST00000491154.1 | chrX | 153671070 | 153671101 | 153670458 | 153671812 | + | retained_intron |
| ENST00000468483.1 | chrX | 153671070 | 153671101 | 153670867 | 153671812 | + | retained_intron |
| ENST00000465640.1 | chrX | 153671070 | 153671101 | 153670867 | 153671075 | + | processed_transcript |
| ENSTR0000381578.1 | chrY | 556126 | 556157 | 555126 | 557558 | + | protein_coding |
| ENSTR0000554971.1 | chrY | 556126 | 556157 | 555126 | 556303 | + | protein_coding |
| ENST00000445125.1 | chrY | 10036137 | 10036188 | 10036113 | 10036711 | - | processed_pseudogene |
| ENST00000445125.1 | chrY | 10036193 | 10036228 | 10036113 | 10036711 | - | processed_pseudogene |
| ENST00000445125.1 | chrY | 10036235 | 10036268 | 10036113 | 10036711 | - | processed_pseudogene |
| ENST00000515896.1 | chrY | 10037831 | 10037880 | 10037764 | 10037915 | + | rRNA |